\documentclass[journal,onecolumn]{IEEEtran}
\usepackage{cite}

\ifCLASSINFOpdf
  \usepackage[pdftex]{graphicx}
  \graphicspath{{Figure/}}
  \DeclareGraphicsExtensions{.pdf,.jpg,.png}
\else
  \usepackage[dvips]{graphicx}
  \graphicspath{{Figure}}
  \DeclareGraphicsExtensions{.eps}
\fi
\usepackage{amsmath}
\usepackage{array}
\usepackage{dblfloatfix}
\usepackage{url}

\newcommand{\pushright}[1]{\ifmeasuring@#1\else\omit\hfill$\displaystyle#1$\fi\ignorespaces}
\makeatother
\newcommand {\ie} {{\em i.e., }}
\newcommand {\eg} {{\em e.g., }}
\newcommand {\beq} {\begin{equation}}
\newcommand {\eeq} {\end{equation}}
\newcommand {\bequn} {\begin{equation*}}
\newcommand {\eequn} {\end{equation*}}
\newcommand {\bear} {\begin{eqnarray}}
\newcommand {\eear} {\end{eqnarray}}
\newcommand {\bearun} {\begin{eqnarray*}}
\newcommand {\eearun} {\end{eqnarray*}}
\newcommand {\fig}[1]{Fig.~\ref{#1}}
\newcommand{\figs}[2]{Figs.~\ref{#1} and~\ref{#2}}
\newcommand {\Eqref}[1]{Eq.~(\ref{#1})}
\newcommand{\Eqrefs}[2]{Eqs.~(\ref{#1}) and~(\ref{#2})}

\usepackage[utf8]{inputenc}
\usepackage{amsfonts}
\usepackage{amssymb}
\usepackage{amsthm}
\usepackage{bm}
\usepackage{comment}
\usepackage{paralist}
\usepackage[shortlabels]{enumitem}
\usepackage{color}

\usepackage{breakurl}
\usepackage[normalem]{ulem}
\usepackage{longtable}
\usepackage{subcaption}
\usepackage{algorithm}
\usepackage{algorithmic}
\usepackage{multirow}

\usepackage{lipsum}
\usepackage{mathtools}
\usepackage{cuted}

\usepackage{placeins} 
\usepackage{balance} 
\usepackage{caption} 

\newtheorem{theorem}{Theorem}
\newtheorem{lemma}[theorem]{Lemma}

\newtheorem{proposition}[theorem]{Proposition}

\makeatletter
\newcommand{\vast}{\bBigg@{3}}
\newcommand{\Vast}{\bBigg@{4}}
\makeatother

\usepackage{graphicx,environ}
\usepackage{hyperref}

\NewEnviron{sourcefigure}[1][htbp]{%
  {\let\caption\relax\let\ref\relax
   \renewcommand{\label}[1]{%
    \gdef\sfname{sf:##1}}%
    \setbox1=\hbox{\BODY}}
    \global\expandafter\let\csname\sfname\endcsname\BODY
  \begin{figure}[#1]
    \BODY
  \end{figure}
}
\newcommand{\reusefigure}[2][htbp]{%
  {\addtocounter{figure}{-1}%
   \renewcommand{\theHfigure}{dupe-fig}
   \renewcommand{\thefigure}{\ref{#2}}
   \renewcommand{\addcontentsline}[3]{}
   \renewcommand{\label}[1]{}
   \begin{figure}[#1] \csname sf:#2\endcsname \end{figure}}
}

 \newcommand{\useColor}{clean}  
\usepackage{xstring}
\usepackage{soul,xcolor} 
\setstcolor{red}

\newcommand{\highlight}[2]{%
    \IfEqCase{#1}{%
        {color}{\textcolor{blue}{#2}}%
        {clean}{#2}%
    }[\PackageError{highlight}{Undefined option to highlight: #1}{}]%
}%

\newcommand{\strikenOut}[2]{%
    \IfEqCase{#1}{%
        {color}{\st{#2}}%
        {clean}{}%
    }[\PackageError{strikenOut}{Undefined option to strikenOut: #1}{}]%
}%

\newcommand{\useOnline}{clean}  

\newcommand{\onlineVersion}[2]{%
    \IfEqCase{#1}{%
        {notonline}{\textcolor{black}{#2}}%
        {clean}{}%
    }[\PackageError{onlineVersion}{Undefined option to onlineVersion: #1}{}]%
}%

\newcommand{\add}[1]{{\color{black}{#1}}}
\newcommand{\addnew}[1]{{\color{black}{#1}}}

\usepackage{supertabular}
\usepackage{amsbsy}

\begin{document}
%
\title{On the Benefits of Traffic ``Reprofiling'' \\
The Multiple Hops Case -- Part~I}
%
%
%

\author{Jiaming~Qiu,~\IEEEmembership{Member,~IEEE,}
        Jiayi~Song,~\IEEEmembership{}
        Roch~Gu\'{e}rin,~\IEEEmembership{Fellow,~ACM,~IEEE,} and~Henry~Sariowan,~\IEEEmembership{Member,~IEEE}
\thanks{J. Qiu and R. Gu\'{e}rin are with the Computer Science and Engineering department at Washington University in St.~Louis, Saint Louis, MO 63130, USA, e-mail: {\tt \{qiujiaming,guerin\}@wustl.edu}.}
\thanks{J. Song was with the Computer Science and Engineering department at Washington University in St.~Louis and is now with ByteDance Inc., San Jose, CA 95110, USA, e-mail: {\tt jiayisong@wustl.edu}.}
\thanks{H. Sariowan is with Google, Mountain View, CA 94043, USA, e-mail {\tt hsariowan@google.com}.}
\thanks{This work was supported by NSF grant CNS 2006530 and a gift from Google.}
}

\maketitle

\begin{abstract}
This paper considers networks where user traffic is regulated through deterministic \emph{traffic profiles}, \eg token buckets, and \highlight{\useColor}{requires}\strikenOut{clean}{guaranteed} hard delay bounds. The network's goal is to minimize the resources it needs to meet those \strikenOut{clean}{requirements}\highlight{\useColor}{bounds}.  The paper explores how \emph{reprofiling}, \ie proactively modifying how user traffic enters the network, can be of benefit.  Reprofiling produces ``smoother'' flows but introduces an up-front access delay that forces tighter network delays. The paper explores this trade-off and demonstrates \highlight{\useColor}{that, unlike what holds in the single-hop case, reprofiling can be of benefit} even when \highlight{\useColor}{``optimal''}\strikenOut{clean}{sophisticated} schedulers are available \highlight{\useColor}{at each hop}.

\bigskip

\hspace{-1.3cm}\fbox{%
	\parbox{\textwidth}%
	 {\textcopyright 2024 IEEE.  Personal use of this material is permitted.  Permission from IEEE must be obtained for all other uses, in any current or future media, including reprinting/republishing this material for advertising or promotional purposes, creating new collective works, for resale or redistribution to servers or lists, or reuse of any copyrighted component of this work in other works.
	}%
}
\end{abstract}


%
\IEEEpeerreviewmaketitle

\section{Introduction}
\label{sec:intro}

Networks in domains as diverse as automotive, avionics, manufacturing, smart grids, and datacenters~\cite{automotive21,afdx,factory20,smartgrid23,aws22,google-netw21,msft21} serve users and applications with known traffic profiles and with expectations for delay guarantees, often in the form of hard end-to-end delay bounds.  This has been reflected in standardization efforts such as the Time Sensitive Networking (TSN) and Deterministic Networking standards~\cite{tsn18,parsons22,seol21,detnet} that target hard delay bounds for deterministic traffic profiles under various schedulers.  This is the environment this paper assumes.

Specifically, we consider networks where user flows are characterized using a \emph{token bucket}, a common traffic regulator~\cite{leboudec18} that limits long-term transmission rates and flows' burstiness, and where performance guarantees are in the form of deterministic end-to-end delay bounds. The problem we target is that of meeting those bounds while \emph{using the least amount of overall network bandwidth}\footnote{The formulation minimizes \emph{total} bandwidth, but other objectives such as maximum or weighted sum of link bandwidth are possible. It can also be applied to the dual problem of maximizing the admitted traffic.}.  For ease of presentation, the results are derived under a \emph{fluid} model.  They hold with minor adjustments in a packet setting.  Network topology and routing are assumed fixed. 

The paper's focuses on whether \emph{reprofiling} flows, \ie modifying on ingress the traffic profiles specified by users, can help reduce the bandwidth needed to meet delay bounds, inclusive of any reprofiling delay this introduces. 
The answer obviously depends on the type of scheduler used in the network. The base one-hop case was investigated in~\cite{onehop21} that established the benefits of reprofiling with \highlight{\useColor}{First-In-First-Out (FIFO)} and static priority schedulers, and confirmed the known result~\cite{Georgiadis97,liebeherr96} that reprofiling is unnecessary when an (optimal) Earliest Deadline-First (EDF) scheduler is used.  This paper explores the extent to which this latter result remains true in a multi-hop network setting, \highlight{\useColor}{with an extension considering the simpler FIFO and static priority schedulers left to a sequel (Part~II).}

We show that, in a \highlight{\useColor}{multi-hop} setting, reprofiling can help even with EDF-based schedulers.  The intuition is that, while reprofiling has a cost (delay), it is incurred \emph{once} (on ingress), but its benefits (from smoother traffic) accrue at \emph{every hop} a flow traverses. Of note is the fact that the optimal solution often lies between the two extremes of ``No Reprofiling'' (NR) and ``Full Reprofiling'' (FR). NR keeps the traffic profile unchanged and allocates the entire delay budget to the network for maximum scheduling flexibility.  Conversely, FR spends as much as possible of the delay budget on making flows smoother on ingress, at the cost of much tighter network delay budgets and, consequently, more limited scheduling flexibility.   

The paper develops approaches for determining the best reprofiling solutions under a general service curve scheduler, SCED~\cite{sced}, and makes the following contributions:
\begin{enumerate}[nosep,label=(\roman*)]
\item Demonstrates the benefit of ingress reprofiling in meeting network delay bounds with less bandwidth;
\item Develops and evaluates exact and approximate solutions for several scenarios.
\end{enumerate}
\highlight{\useColor}{Those contributions notwithstanding,
the aspect of most interest is \emph{establishing} that reprofiling can help even in the presence of powerful schedulers such as SCED.
}

The paper is structured as follows.  Section~\ref{sec:background} reviews network calculus and scheduling results we rely on. Section~\ref{sec:formulation} presents our optimization framework.
An exact but high complexity algorithm is introduced in Section~\ref{sec:solution}, followed by a simpler yet effective approximate algorithm in Section~\ref{sec:greedy}. The performance and run time of the two algorithms are compared in Section~\ref{sec:algorithm_comparison}.  Section~\ref{sec:evaluation} illustrates the bandwidth benefits of reprofiling for several scenarios.  Section~\ref{sec:related} reviews related work. Finally, Section~\ref{sec:conclusion} summarizes the paper's contributions and discusses extensions.  For clarity of presentation, proofs and ancillary results are relegated to appendices\onlineVersion{\useOnline}{~of~\cite{multihop22}}.  \highlight{\useColor}{For ease of reproducibility, both the reprofiling algorithms derived in the paper and the scenarios used to evaluate them are available at~\url{https://github.com/qiujiaming315/traffic-reprofiling}.}

\section{Background}
\label{sec:background}

This section reviews relevant concepts from Network Calculus~\cite{nc} (NC) and several results we rely on.
Although the NC framework applies to packet and fluid models, for ease of exposition, we use a fluid model in the paper.  The results can be extended to a packet model through standard approaches\highlight{\useColor}{~\cite[Section~1.7]{nc}}, but do not yield further insight\footnote{\highlight{\useColor}{See Appendix~\ref{app:packet_model}\onlineVersion{\useOnline}{~of~\cite{multihop22}} for a detailed discussion.}}.

\subsection{Network Calculus}
\label{sec:NC}

\subsubsection{Arrival Curves}
\label{sec:ac}
They constrain the amount of traffic a flow is allowed to transmit over time.  A flow with a cumulative arrival function $A(t)$ (\ie the traffic sent between $0$ and $t$) conforms to the arrival curve $\alpha$ if:
\begin{equation*}
A(t+\tau) - A(t) \leq \alpha(\tau), \quad \forall \tau, t \geq 0.
\end{equation*}
In other words, $\alpha$ upper bounds the data sent in any time interval.
The \emph{token bucket} is such an arrival curve that bounds a flow's long-term rate $r$ and burst size $b$:
\begin{equation*}
\alpha(t) = 
\begin{cases}
0 & t = 0, \\
b + rt & t > 0.
\end{cases}
\end{equation*}
We assume that, the traffic of every flow~$i$ is \highlight{\useColor}{initially} regulated by a token bucket $\alpha_i=(r_i,b_i)$ as it enters the network. \highlight{\useColor}{We explore how modifying this original token bucket profile $\alpha_i$ can allow a network to offer given end-to-end latency guarantees to flows, and do so with less bandwidth.}

\subsubsection{Service Curves}
\label{sec:sc}
They specify minimum service guarantees to flows, namely, if a flow with arrival function $A$ is guaranteed a service curve $\beta$, then under relatively mild conditions\footnote{$\beta$ is continuous, $\beta(0)=0$, and $A$ is left-continuous.} that we assume hold, the amount of service $S(t)$ received by the flow by time~$t$ is such that there always exists a time $t-\tau, \tau\geq 0$ such that~\cite[Section 1.3]{nc}:
\begin{equation*}
S(t)\geq A(t-\tau)+\beta(\tau).
\end{equation*}

\subsubsection{Delay and Buffer Bounds}
\label{sec:bounds}
Given a flow with arrival and service curves $\alpha$ and $\beta$, its delay can be shown~\cite[Section 1.4]{nc} to be upper-bounded by the maximum horizontal distance $\Delta(\alpha, \beta)$ between $\alpha$ and $\beta$\highlight{\useColor}{, where:}
\highlight{\useColor}{
\begin{equation}
\label{eq:delay}
\Delta(\alpha, \beta) = \sup_{t \geq 0 }\{\inf\{\tau \geq 0: \alpha(t) \leq \beta(t + \tau)\}\}.
\end{equation}
}
\highlight{\useColor}{
Conversely, the maximum vertical distance $\Theta(\alpha, \beta)$ yields buffer bounds:
\highlight{\useColor}{
\begin{equation}
\label{eq:buffer}
\Theta(\alpha, \beta) = \sup_{t \geq 0 }\{\alpha(t) - \beta(t)\}.
\end{equation}
}
}
Another important result, known as \textit{Pay Burst Only Once} (PBOO), that derives from those concepts is that the \emph{end-to-end delay} of a flow traversing $n$ hops and provided with service curve $\beta_j$ at hop $j$ can be upper-bounded using a \emph{concatenated service curve} of the form~\cite[Section 3.1]{nc}:
\begin{equation*}
\bigotimes_{j=1}^{n}\beta_{j} = \beta_{1} \otimes \beta_{2} \otimes \cdots \otimes \beta_{n},
\end{equation*}
where $\otimes$ is the min-plus convolution operator defined as:
\begin{equation}
\label{eq:concat}
(\beta_{1} \otimes \beta_{2})(t) = \inf_{0 \leq \tau \leq t}\big\{ \beta_1(\tau) + \beta_2(t-\tau) \big\},
\end{equation}
with $\Delta(\alpha, \bigotimes_{j=1}^{n}\beta_{j})$ the end-to-end delay upper-bound.

Note that this upper-bound needs not be tight\footnote{See \cite[Section 1.2.3]{bouillard16} for a pertinent discussion of this issue.}. A service curve is a \emph{lower bound} for the service a flow is guaranteed to receive, and its realization by the scheduler can deviate from it. This affects the flow's \emph{departure curve} from the scheduler.  In a multi-hop setting, this departure curve is the flow's arrival curve at the next hop and affects its delay there. Accurately characterizing departure curves is challenging for most schedulers, \highlight{\useColor}{and hence so is tightening the upper bound.}

\subsubsection{Traffic Shaping and Greedy Shaper}
\label{sec:shaper}

A traffic shaper $\sigma$ \emph{enforces} conformance of incoming traffic with the arrival curve specified by~$\sigma$.  Shapers are typically realized as \emph{greedy shapers}~\cite[Section 1.5.3]{nc} that release traffic at the earliest possible time that guarantees conformance with $\sigma$. \highlight{\useColor}{Because the term shaper implies a non-work-conserving behavior, we opt to use the more general term ``profiler'' to account for the fact that, at least with sophisticated schedulers such as SCED, reprofiling can also be realized in a work-conserving manner\footnote{\highlight{\useColor}{Under SCED, the  (reprofiling) ``delay'' of a work-conserving reprofiler is simply integrated in the deadline SCED uses for the flow's transmissions.}}. We use the notation $\sigma$ to indicate a flow's new (re)profile, with $\alpha$ denoting its original token bucket profile.}

\highlight{\useColor}{The goal of reprofiling is to make flows ``smoother'', and a flow with initial token bucket profile $\alpha$ is reprofiled to $\sigma \leq \alpha$. Recall that a profiler/shaper enforces an arrival curve, so that a smaller profile imposes a corresponding reprofiling delay\footnote{\highlight{\useColor}{Under work-conserving operation, this delay is not necessarily incurred.}} Given $\alpha$ and $\sigma$, this delay is given by $D=\Delta(\alpha,\sigma)$, where $\Delta(\cdot,\cdot)$ is as per \Eqref{eq:delay}.  Ensuring that the flow's end-to-end deadline $d$ is still met, inclusive of this reprofiling delay, therefore, calls for reducing the flow's network delay by a commensurate amount, \ie from $d$ down to $d-D$.}

\highlight{\useColor}{We note that, as articulated in~\cite[Corollary~$1.5.1$]{nc} and in the use of \Eqref{eq:delay} to compute $D$, there is a close relation between greedy (re)profilers and service curves. The former  upper bound the traffic a flow can transmit, while the latter give a lower bound on the service (transmission opportunities) the flow can receive.  Hence, when, at a given hop, a flow is assigned a service curve equal to its profile, the flow experiences a delay of~$0$ (under a fluid model) at this hop.} 

\highlight{\useColor}{In this paper, we explore} whether making flows smoother \highlight{\useColor}{outweighs} the tighter network delays this forces. We also note that once a flow has been reprofiled according to $\sigma$, the same reprofiling can be applied at every hop without causing additional delays~\cite[Section 1.5.3]{nc}.  \highlight{\useColor}{This ensures that the benefits of reprofiling remain present at every hop.}

\subsubsection{SCED Scheduling Policy}
\label{sec:sced}
\highlight{\useColor}{Service Curve Earliest Deadline first (SCED)~\cite{sced} is a scheduling policy that extends EDF and that, when schedulable (see Lemma~\ref{lemma:min_bandwidth}), provides service curve guarantees to flows. SCED operates as follows: Given $m$~flows with flow~$i=1,\ldots,m,$ having arrival curve $A_i(t)$ and service curve $\beta_i(t)$, a bit arriving at time $t$ is assigned deadline $t'$, where $t'$ is the latest possible time that meets the flow's service curve guarantee and is computed as follows:
\begin{equation*}
    A_i(t) = (A_i \otimes \beta_{i})(t').
\end{equation*}
Bits are scheduled for transmission in order of their deadlines.}

\highlight{\useColor}{Given a set of service curves, the link bandwidth $C^*$ that ensures SCED's schedulability is characterized in Lemma~\ref{lemma:min_bandwidth}:
\begin{lemma}
\label{lemma:min_bandwidth}
Given a set of service curves $\beta_i(t), i=1,\ldots,m$, any scheduling mechanism requires a link bandwidth of at least:
\begin{equation*}
    C^* = \sup_{t \geq 0} \frac{\sum_i^m \beta_{i}(t)}{t},
\end{equation*}
to guarantee those service curves. SCED realizes those service curves with a link bandwidth of exactly $C^*$.
\end{lemma}
The proof is in Appendix~\ref{app:min_bandwidth}\onlineVersion{\useOnline}{~of~\cite{multihop22}}.  We note that $\sum_i^m \beta_{i}(t)$ represents an \emph{aggregate service curve} that captures the minimum amount of service SCED needs to provide to guarantee service curves $\beta_{i}(t)$ to all flows $i=1,\ldots,m$.
}

\highlight{\useColor}{The flexibility of} EDF schedulers \highlight{\useColor}{(and therefore SCED) notwithstanding, they} are complex even if efficient implementations are increasingly available~\cite{sharma20,sivaraman16}.  However, they are optimal when it comes to minimizing bandwidth on \emph{individual} links~\cite{Georgiadis97,liebeherr96,onehop21}, and therefore represent a baseline when evaluating the potential benefits of reprofiling.  To that end, we assume the availability of SCED schedulers capable of allocating dedicated service curves to flows at every hop.

\subsection{Two-Slope Rate-Latency Service Curve}
\label{sec:2slope}

Although SCED can accommodate arbitrary service curves, for tractability we limit our investigation to a subset of piece-wise-linear service curves we term \emph{two-slope rate-latency service curves} (2SRLSC).
\begin{figure}[!h]
\centering
\begin{subfigure}{0.27\linewidth}
  \centering
  \includegraphics[width=\linewidth]{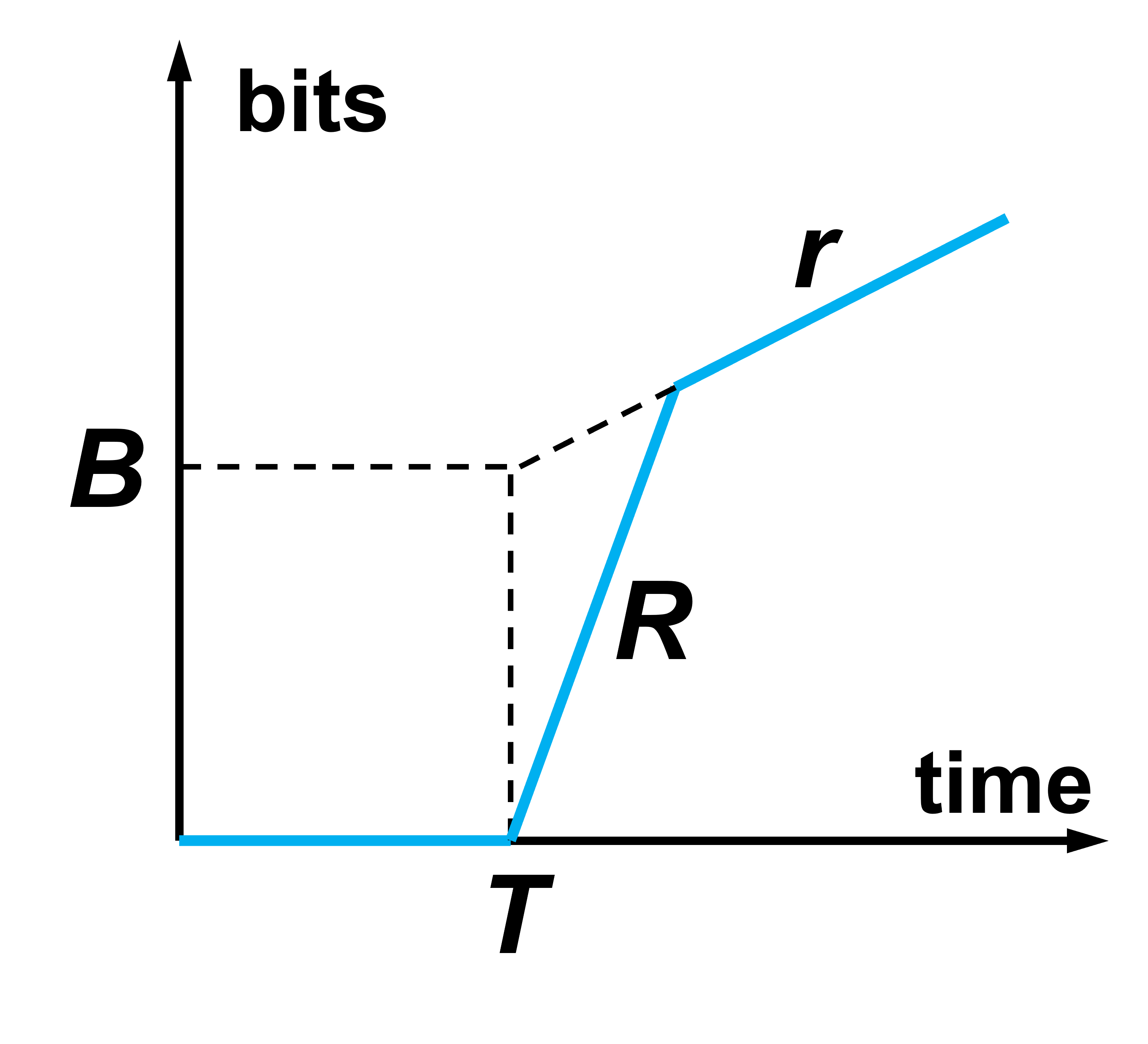}
  \caption{2SRLSC $\beta$}
  \label{fig:2srlsc}
\end{subfigure}
\begin{subfigure}{0.05\linewidth}
$\mbox{\Large $=$}$
\vspace{1.2cm}
\end{subfigure}
\begin{subfigure}{0.27\linewidth}
  \centering
  \includegraphics[width=\linewidth]{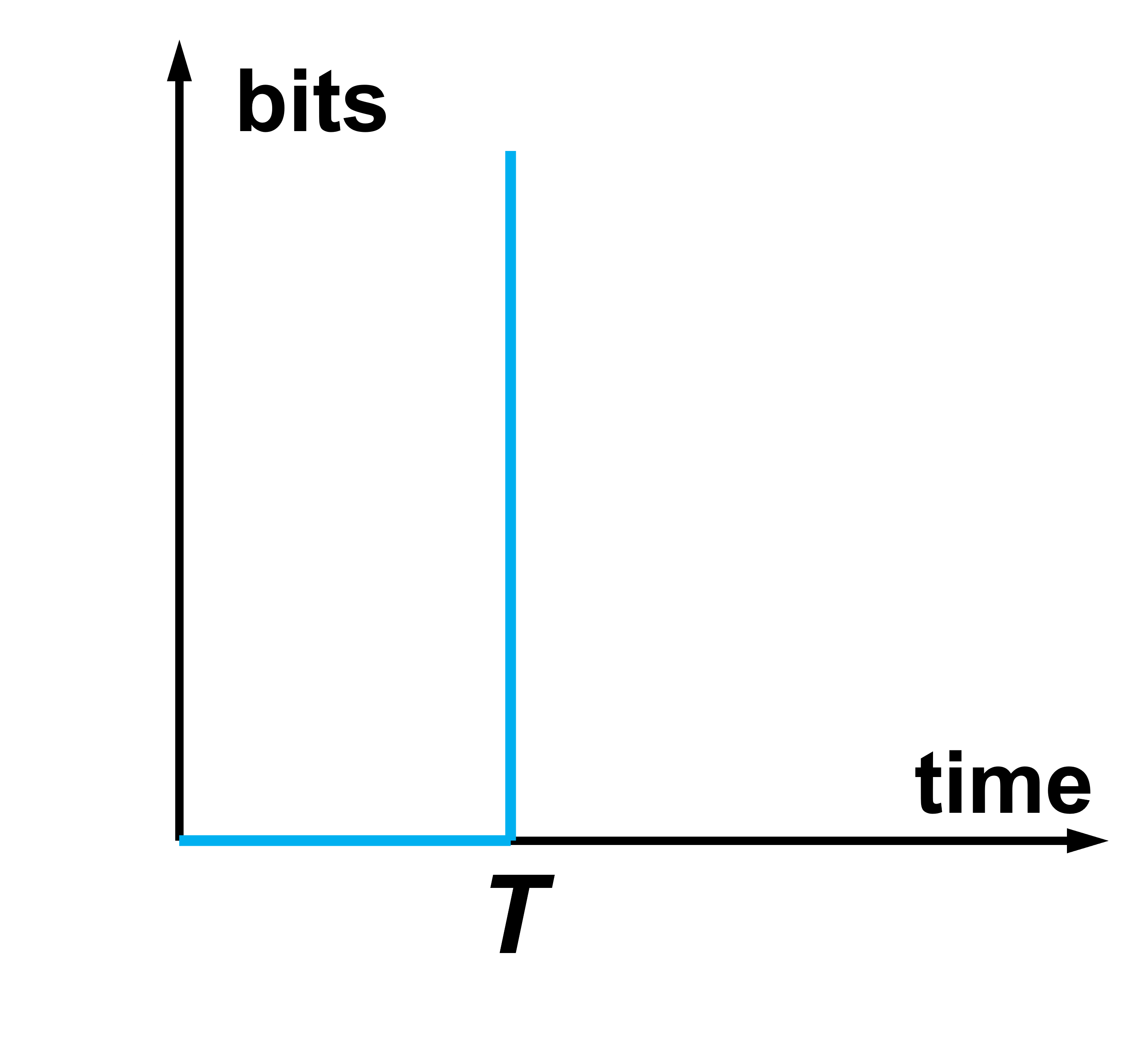}
  \caption{DESC $\delta$}
  \label{fig:desc}
\end{subfigure}
\begin{subfigure}{0.05\linewidth}
$\mbox{\Large $\otimes$}$
\vspace{1.2cm}
\end{subfigure}
\begin{subfigure}{0.27\linewidth}
  \centering
  \includegraphics[width=\linewidth]{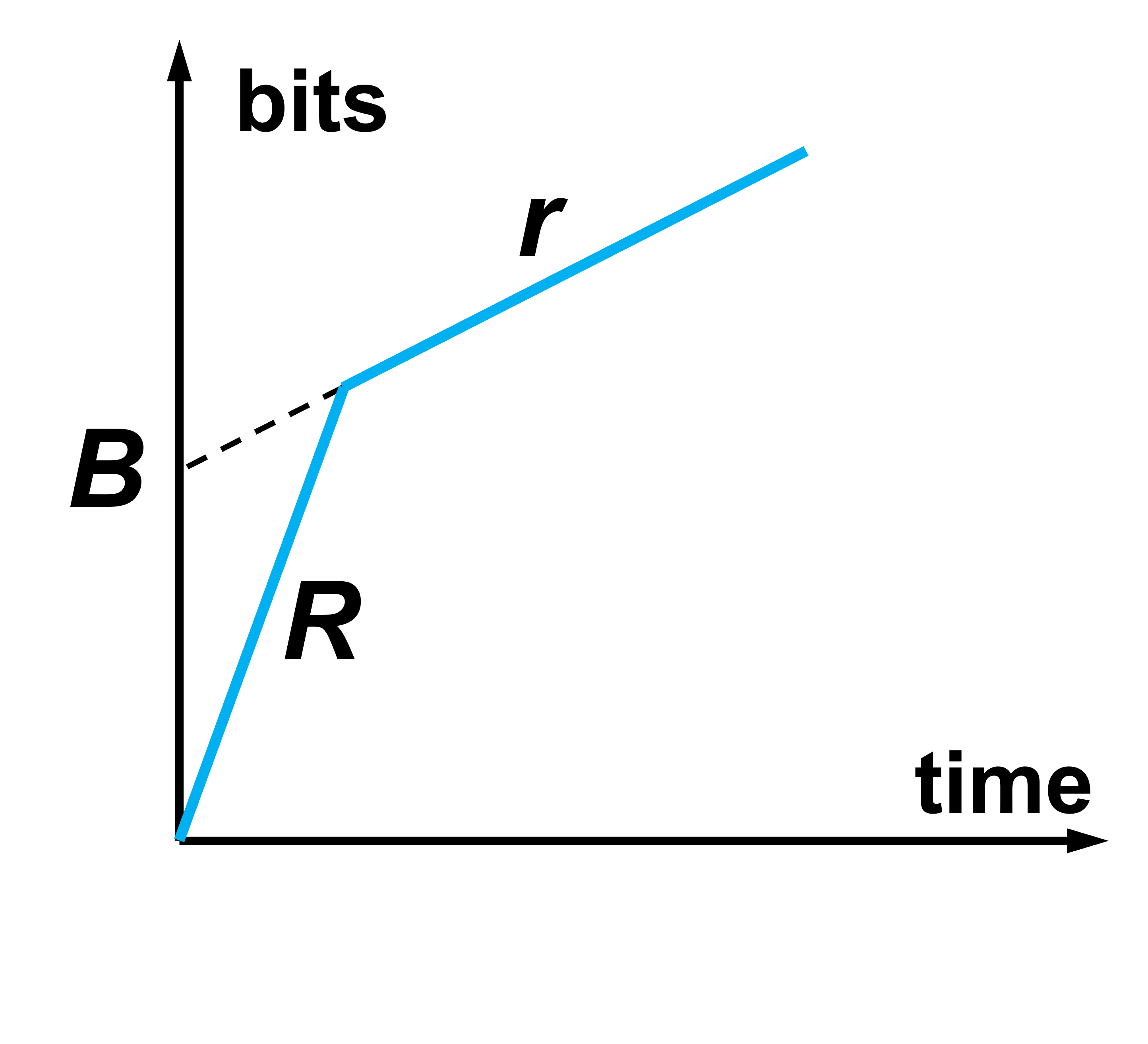}
  \caption{2SRC $\sigma$}
  \label{fig:2src}
\end{subfigure}
\caption{2SRLSC and its two components.}
\label{fig:service_curve}
\end{figure}
\fig{fig:2srlsc} illustrates the generic structure of the 2SRLSC $\beta=(T,R,B,r)$ allocated to a flow at a given hop. It consists of three segments and 
can be decomposed into the concatenation of a Delay Element Service Curve (DESC) $\delta$, and a Two-Slope Reprofiling Curve (2SRC) $\sigma$, \ie $\beta = \delta \otimes \sigma$.
\highlight{\useColor}{There are two primary motivations for our choice.
\begin{enumerate}[(i),nosep]
\item The use of separate latency and rate components (DESC \& 2SRC) acknowledges the efficacy of decoupling delay and rate guarantees, as embodied in the rate-latency service curves behind most practical schedulers~\cite[p.~21]{nc};
\item The use of a 2SRC stems from our reliance on reprofiling.  As each flow is reprofiled, assigning it a 2SRC equal to its profile ensures that its delay at each hop is just the delay~$T$ of the delay element. Further, as shown in Section~\ref{sec:2src}, restricting reprofiling to only two slopes is not restrictive and can minimize the required bandwidth.
\end{enumerate}
}

\subsubsection{Delay Element Service Curve}
\label{sec:desc}
A DESC $\delta$ is shown in~\fig{fig:desc}. It ensures that arriving traffic is transmitted before a \emph{deadline} $T$. Operationally, it maps to an EDF scheduler (assuming schedulability), and enforces \emph{local deadlines} for flows at every hop.

\subsubsection{Two-Slope Reprofiling Curve}
\label{sec:2src}
A 2SRC $\sigma=(R,B,r)$, is defined by three parameters: a short-term rate $R\geq r$, a parameter $B$ that determines the duration $t=B/(R-r)$ of transmission at $R$, and a long-term rate $r$. \highlight{\useColor}{We note that $\sigma$ can be realized through the concatenation of two token buckets $\alpha_1^*=(R,0)$ and $\alpha_2^*(r,B)$, \ie $\sigma=\alpha_1^*\otimes\alpha_2^*$. 
The next lemma formalizes the extent to which this is not a limitation.}
\highlight{\useColor}{
\begin{lemma}
\label{lemma:optimal_reprofiler}
Consider a flow with token bucket profile $\alpha=(r, b)$ that is reprofiled using $\sigma=\bigotimes_{j=1}^n\widetilde{\alpha}_j$, where $\widetilde{\alpha}_j=(r_j,b_j), j=1,\ldots,n,$ are two-parameters token buckets, and $\sigma$ is such that $\Delta(\alpha,\sigma)=D$.  Let $\sigma^{*}=\alpha_1^*\otimes\alpha_2^*$, where $\alpha_1^*$ and $\alpha_2^*$ are token buckets with profiles $\alpha_1^*=(R^{*}= b/D,0)$ and $\alpha_2^*=(r,B^{*}=b-rD)$, then $\Delta(\alpha,\sigma^*)=D$ and $\sigma(t) \geq \sigma^{*}(t), \forall t \geq 0$.
\end{lemma}
The proof is in Appendix~\ref{app:reprofiler}\onlineVersion{\useOnline}{~of~\cite{multihop22}}, and the lemma establishes that, among all concatenations of token buckets that realize the (reprofiling) delay target~$D$, $\sigma^*$ does so while being the most frugal when it comes to bandwidth, \ie $\sigma^*(t) \leq \sigma(t)$.}

\highlight{\useColor}{
Combining Lemma~\ref{lemma:optimal_reprofiler} with  Lemma~\ref{lemma:min_bandwidth} establishes that, when using SCED to realize a delay target of $D$ for an arrival curve $\alpha$, the 2SRC $\sigma^*$ requires the least amount of bandwidth among all concave, piece-wise-linear service curves, \ie service curves realized by a concatenation of token buckets and an EDF scheduler.  Hence, $\sigma^{*}$ is \emph{optimal} among this family of service curves.  \fig{fig:optimal_reprofiler}, illustrates how $\sigma^{*}$ realizes the (reprofiling) delay $D$ for an arrival curve $\alpha$.
\begin{figure}[!h]
\centering
\includegraphics[width=0.5\linewidth]{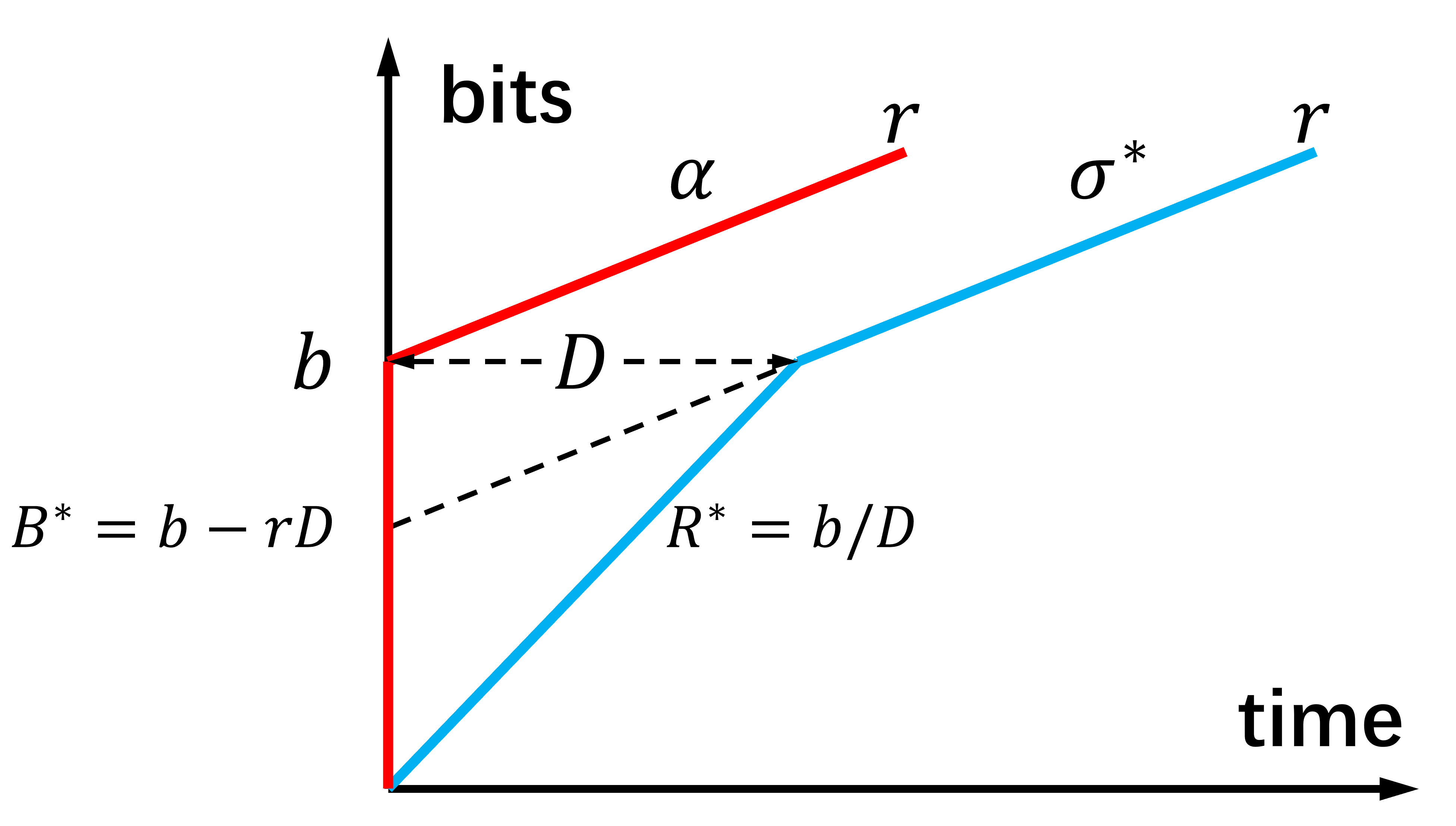}
\caption{Optimal reprofiler given reprofiling delay.}
\label{fig:optimal_reprofiler}
\end{figure}
}

\section{Problem Formulation}
\label{sec:formulation}

\subsection{Problem Setting}
\label{sec:setting}

\begin{figure}[!h]
\centering
\includegraphics[width=0.7\linewidth]{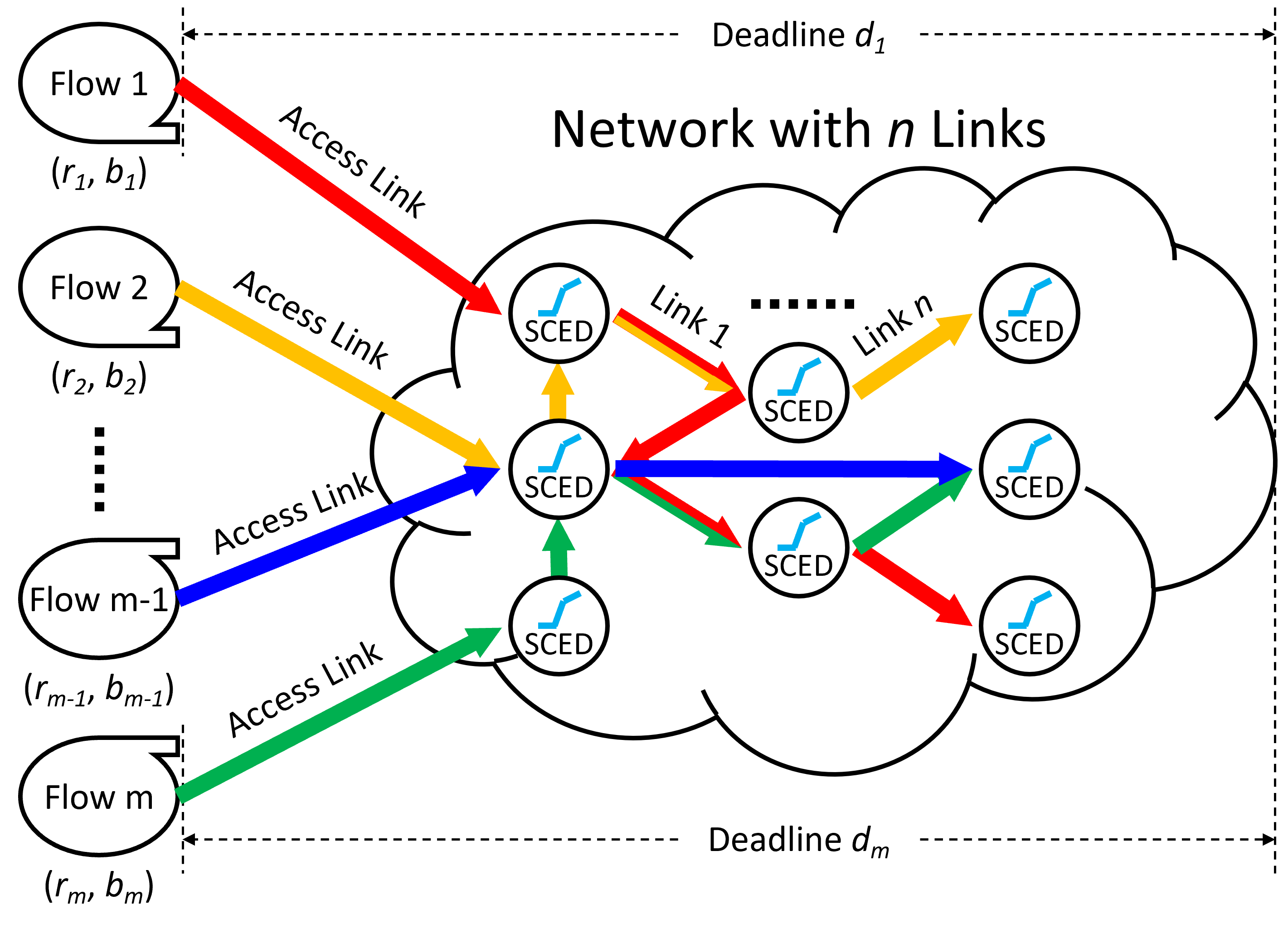}
\caption{Network with $m$ flows and $n$ links.}
\label{fig:overview}
\end{figure}

Consider a network with $n$ links identified by their index $j, 1\leq j\leq n$, and carrying $m$ flows, as illustrated in \fig{fig:overview}.  Flow~$i, 1\leq i\leq m$, is characterized as follows:
\begin{itemize}[wide=0pt]
\item The path of flow $i$
from source to destination is assumed acyclic and specified by an ordered set of distinct link indices $\mathcal{P}_i=\{j_{i_1},j_{i_2},\ldots , j_{i_k}\}$ for a $k$~hops path.  The directed graph resulting from the union of the flow paths is allowed a general structure, \ie either acyclic or cyclic.
\item Traffic from flow~$i$ conforms to token bucket $\alpha_{i}=(r_i,b_i)$, and has an end-to-end packet-level latency target\footnote{Exclusive of propagation and processing delays.}
$d_i$.  Together, $\alpha_i$ and $d_i$ define the flow's profile $(r_i, b_i, d_i)$.

\end{itemize}

Flows connect to the network through dedicated, high-speed access links and at each hop are assigned a 2SRLSC as defined in Section~\ref{sec:2slope}.
Further, network buffers are assumed large enough to ensure lossless operation. \highlight{\useColor}{Buffer bounds can be readily derived from basic network calculus, and are presented in Section~\ref{sec:buffer_bound} for a non-work-conserving setting.}

\subsection{Optimization Framework}

\subsubsection{Inputs}
They consist of the vector triplet $(\mathbf{r}, \mathbf{b}, \mathbf{d})$ and the path matrix $\pmb{\mathcal{P}} = (\mathcal{P}_1, \mathcal{P}_2, \ldots, \mathcal{P}_m)$, where $\mathbf{r} = (r_1, r_2, \ldots, r_m)$, $\mathbf{b} = (b_1, b_2, \ldots, b_m)$, and $\mathbf{d} = (d_1, d_2, \ldots, d_m)$ specify the flows' token rates, burst sizes, and deadlines, respectively.

\subsubsection{Variables}

Given our reliance on SCED, link bandwidths depend on the 2SRLSCs assigned to the flows sharing a link, with their parameters the variables that capture our bandwidth minimization goal.  Given a link~$j$ and denoting as $\mathcal{F}_j$ the set of flows whose path traverses link~$j$, the 2SRLSC $\beta_{ij}$ of a flow in $\mathcal{F}_j$ is specified through four parameters $(T_{ij}, R_{ij}, B_{ij}, r_i)$, where we have taken advantage of the fact that flows are regulated with a token rate that can, therefore, be assigned as the long-term rate of their 2SRLSCs.  Hence, in addition to satisfying the stability condition $C_j\geq \sum_{i\in\mathcal{F}_j}r_i$, the bandwidth $C_j$ \addnew{needed on} link~$j$ \addnew{to accommodate the} 2SRLSCs of the flows in $\mathcal{F}_j$ depends only on the variables $(T_{ij}, R_{ij}, B_{ij})$.  

In general, the 2SRLSCs of flow~$i$ are captured in three vectors $\mathbf{T}_i$, $\mathbf{R}_i$, and $\mathbf{B}_i$ that specify the variables $(T_{ij}, R_{ij}, B_{ij})$ for the links of path $\mathcal{P}_i$.  The vectors for all $m$~flows are then the variables\footnote{Section~\ref{sec:reduction} shows how the number of variables can be reduced.} that determine the vector of link bandwidths $\mathbf{C} = (C_1, C_2, \ldots, C_{n})$ the flows' 2SRLSCs require. 

\subsubsection{Constraints}

The end-to-end latency bounds are the constraints of our optimization, which, recalling the results of Section~\ref{sec:sc}, can be expressed as:
\beq
\label{eq:const}
\Delta(\alpha_i, \bigotimes_{j \in \mathcal{P}_i}\beta_{ij}) \leq d_i, \quad \forall\, 1 \leq i \leq m,
\eeq
In other words, for all flows, the sum over all hops of their reprofiling and scheduling delays (ingress reprofiling is assumed done at the first hop) is no larger than their end-to-end latency target.

\subsubsection{Objective Function}
\label{sec:obj}
Our goal is to meet latency targets while minimizing \textit{total bandwidth} (again, other wide-sense increasing functions such as maximum link bandwidth or weighted sum of link bandwidth can be used).  Formally, the optimization objective function  (\textbf{OPT}) is:
\beq\label{eq:opt}
\text{\bf{OPT}}: \quad 
\min_{\substack{\mathbf{R}_i, \mathbf{T}_i, \mathbf{B}_i \\ 1\leq i \leq m }}
\sum_{j=1}^{n} C_j,
\eeq
subject to the constraints of \Eqref{eq:const}.

\subsection{Discussion}
\label{sec:discuss}

We note that \textbf{OPT} relies on the upper-bounds of Section~\ref{sec:sc} through \Eqref{eq:const}.  This has implications when it comes to attributing causes to the bandwidth improvements its solutions afford. Specifically, improvements come from three possible sources: (i) reprofiling; (ii) the assignments of distinct per-hop deadlines; {\bf and} (iii) the tighter delay bounds that are themselves a consequence of reprofiling.  

Our focus is on reprofiling, but local deadline assignments can also reduce bandwidth by leveraging heterogeneity across hops.  Further, making flows smoother, as reprofiling does, reduces the possible ``gap'' between arrival and service curves. This allows tighter end-to-end bounds (see Section~\ref{sec:bounds}) as it restricts deviations between departure curves and service curves\footnote{Note that under the \emph{linear} service curve of full reprofiling, departure and service curves are identical and the bound is tight.}.  As a result, accurately quantifying how much of the improvements we observe come from reprofiling alone is challenging.  Nevertheless, the results we present in Section~\ref{sec:evaluation} provide evidence of the benefits of reprofiling, even if precisely assessing their relative contribution remains elusive.

\section{Solving Problem \textbf{OPT}}
\label{sec:solution}

\highlight{\useColor}{The main challenge in solving \textbf{OPT} comes from the large number of variables associated with having individual 2SRLSCs for each flow at each hop.} Next, We introduce steps to \highlight{\useColor}{decrease this complexity by reducing} the number of variables involved \highlight{\useColor}{while preserving the solution's optimality}.

\subsection{OPT Reduction}
\label{sec:reduction}

We first establish that, for each flow, it is enough to consider the same 2SRC at each hop.  We term this 2SRC
the flow's \emph{minimum reprofiler}, and 
show that for any optimal solution of \textbf{OPT}, it can be specified through a single parameter, the flow's \emph{reprofiling delay}.

\subsubsection{Minimum Reprofiler}
\label{sec:min_reprofiler}
\highlight{\useColor}{This first lemma is a direct consequence of applying NC's concatenation result~\cite[Section 1.4.3]{nc} to 2SRLSCs, and will allow us (using Lemma~\ref{lemma:min_bandwidth}) to reduce the number of variables of \textbf{OPT} by assuming that a flow is assigned the same 2SRC at every hop.
\begin{lemma}
\label{lemma:2srlsc_concat}
Consider flow~$i$ assigned token bucket $(r_i,b_i)$ 
and assigned on link~$j\in\mathcal{P}_i$ a 2SRLSC $\beta_{ij}$ with parameters $T_{ij}$, $R_{ij}$, $B_{ij}$, and $r_{ij} = r_{i}$.  The concatenation of the flow's 2SRLSCs is readily found to be another 2SRLSC of the form:
\begin{flalign*}
{\displaystyle \bigotimes_{j\in\mathcal{P}_i}\beta_{ij}} &\displaystyle{= \beta\Big\{T_{i}=\sum_{j\in\mathcal{P}_i}T_{ij}, R_{i}=\min_{j\in\mathcal{P}_i}R_{ij}, B_{i}=\min_{j\in\mathcal{P}_i}B_{ij}, r_i\Big\}}\\
&=\delta_{T_i} \otimes\sigma_{i},
\end{flalign*}
where $\delta_{T_i}$ is a delay element with parameter $T_i$ and $\sigma_{i}$ is a 2SRC with parameters $R_i,B_i,$ and $r_i$.
\end{lemma}
}

\highlight{\useColor}{The proof is in Appendix~\ref{app:concat}\onlineVersion{\useOnline}{~of~\cite{multihop22}}. We term $\sigma_{i}$ the flow's \emph{minimum reprofiler}, and}
from Section~\ref{sec:bounds}, the flow's end-to-end delay upper bound is then of the form:
\beq
\label{eq:delay_bound}
\Delta(\alpha_i, \bigotimes_{j\in\mathcal{P}_i}\beta_{ij}) = \sum_{j\in\mathcal{P}_i}T_{ij} + \Delta(\alpha_i, \sigma_{i}), \quad \forall\, 1\leq i \leq m,
\eeq
which is the sum of all the \emph{local deadlines} plus the \emph{reprofiling delay} from the minimum reprofiler $\sigma_{i}$\footnote{This is consistent with an earlier similar finding in~\cite{georgiadis96a}.}.

\highlight{\useColor}{
The next lemma formalizes that assigning flow~$i$ the same 2SRC as part of its 2SRLSC at each hop does not affect the optimal solution of \textbf{OPT}.
\begin{lemma}
    Given an optimal solution to \textbf{OPT}, the 2SRC of the 2SRLSC of every flow~$i, 1\leq i \leq m,$ on link~$j\in \mathcal{P}_i$, can be set to its minimum reprofiler~$\sigma_i$.
\end{lemma}
}

\highlight{\useColor}{This is consistent with results from~\cite[Section 1.5.3]{nc}.} The proof derives directly from \Eqref{eq:delay_bound}, which states that given flow $i$'s local deadline assignments $\mathbf{T}_{i},$ ensuring that its end-to-end latency target $d_i$ is met 
depends only on the values $R_i$ and $B_i$ of its minimum reprofiler.  Further, \highlight{\useColor}{as per Lemma~\ref{lemma:min_bandwidth},} 
the link bandwidth required to accommodate a set of 2SRLSCs is a non-decreasing function of the $R_{ij}$ and $B_{ij}$ of the flows sharing the link. Given an optimal solution to \textbf{OPT}, changing the parameters $R_{ij}$ and $B_{ij}$ of flow~$i$ on link~$j$ to $R_i$ and $B_i$ (their minima) can, therefore, never increase the required link bandwidth.  Hence an optimal solution exists where all flows are assigned their minimum reprofiler.

\subsubsection{Reprofiling Delay}
\label{sec:reprofiling_delay}

We next establish that for any given flow~$i,1\leq i\leq m,$ the variables $R_{i}$ and $B_{i}$ can be reduced to a single variable, namely, the flow's \emph{reprofiling delay} $D_i$, \highlight{\useColor}{using the optimal reprofiler $\sigma^{*}_{i}$ introduced in Lemma~\ref{lemma:optimal_reprofiler}}. 
\highlight{\useColor}{
\begin{lemma}
Given an optimal solution to \textbf{OPT}, the minimum reprofiler $\sigma_i$ of every flow $i, 1 \leq m$, can be set to its optimal reprofiler $\sigma^{*}_{i}$.
\end{lemma}
}

\highlight{\useColor}{According to Lemma~\ref{lemma:optimal_reprofiler}, the optimal reprofiler $\sigma^{*}_{i}$ ensures $\sigma_i(t) \geq \sigma^{*}_{i}(t), \forall t \geq 0$ 
among all reprofilers $\sigma_i$ with a reprofiling delay of $D_i$.} Hence, setting $R_{i} = b_{i}/D_{i}$ and $B_{i} = b_{i} - r_{i}D_{i}$ for all flows does not affect the optimality of the solution.  Consequently, \highlight{\useColor}{we have:}
\highlight{\useColor}{
\begin{proposition}
Solving \textbf{OPT} is equivalent to solving $\textbf{OPT}^-$
\beq\label{eq:opt-}
\text{\bf{OPT}}^-: \quad 
\min_{\substack{\mathbf{T}_i, D_i \\ 1\leq i \leq m }}
\sum_{j=1}^{n} C_j,
\eeq
subject to the constraints 
\beq
\label{eq:const1}
\begin{aligned}
\sum_{j \in \mathcal{P}_i}T_{ij} + D_i &\leq d_i, \quad \forall\, 1 \leq i \leq m,\\
D_i &\leq \frac{b_i}{r_i}, \quad \forall\, 1 \leq i \leq m,
\end{aligned}
\eeq
\end{proposition}
}
\noindent 
\highlight{\useColor}{where} the second set of constraints \highlight{\useColor}{ensure} $R_i \geq r_i$ for all minimum reprofilers.

\begin{sourcefigure}[!h]
\centering
\includegraphics[width=0.7\linewidth]{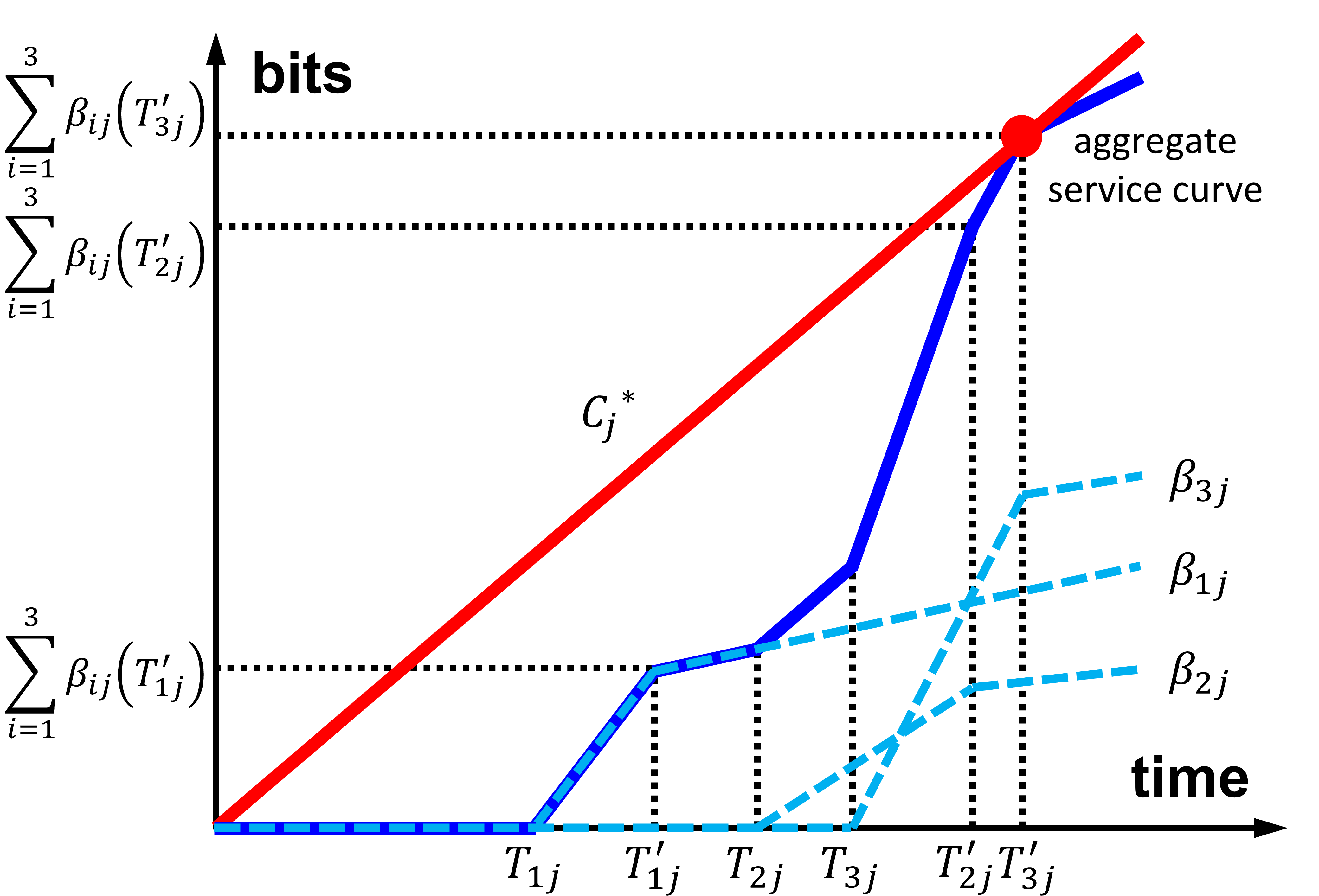}
\caption{Aggregate service curve and minimum required bandwidth.}
\label{fig:min_bandwidth}
\end{sourcefigure}
\subsection{Non-Linear Programs Formulation}
\label{sec:nlp}

\Eqref{eq:opt-} shows that solving $\textbf{OPT}^-$ calls for computing the link bandwidths $C_j, 1\leq j\leq n,$ given the variables $\mathbf{T}_i,D_i, 1\leq i\leq m$.  \fig{fig:min_bandwidth} illustrates for a given link~$j$ with three flows, the minimum bandwidth $C_j^*$ that satisfies the constraints imposed by the service curves $\beta_{ij}$'s of the flows traversing link~$j$.
The figure also hints at how $C_j^*$ can be determined\footnote{Note that $C_j^*$ can only be realized at points associated with bandwidth \emph{decreases}, and those correspond to the $T'_{ij}=T_{ij}+D_i$'s.} from the resulting aggregate service curve (the solid, dark-blue line in the figure). This aggregate service curve is a composite piece-wise linear function constructed from the three segments of all the individual $\beta_{ij}$'s, namely,
\begin{equation}
\label{eq:beta_ij}
    \beta_{ij}(t) = 
    \begin{cases}
    0 & 0 \leq t < T_{ij}, \\
    \frac{b_i}{D_i}(t - T_{ij}) & T_{ij} \leq t < \highlight{\useColor}{T'_{ij}}\strikenOut{clean}{T_{ij} + D_i},\\
    b_i + r_i(t - \highlight{\useColor}{T'_{ij}}\strikenOut{clean}{T_{ij} - D_i}) & t \geq \highlight{\useColor}{T'_{ij}}\strikenOut{clean}{T_{ij} + D_i},
    \end{cases}
\end{equation}
with the notation $T'_{ij}=T_{ij}+D_i$ used in \highlight{\useColor}{\Eqref{eq:beta_ij}} and \fig{fig:min_bandwidth}.

\highlight{\useColor}{Combining Lemma~\ref{lemma:min_bandwidth} with the expressions in \Eqref{eq:beta_ij} gives $C_j^* = \sup_{t \geq 0} \frac{\sum_{i \in \mathcal{F}_j} \beta_{ij}(t)}{t}$, which, as per the next lemma, can be further simplified into \Eqref{eq:min1hopbw}.
\begin{lemma}
Given a set of flows $\mathcal{F}_j$ sharing link~$j$ and assigned service curves in the form of 2SRLSCs as given in \Eqref{eq:beta_ij}, the minimum bandwidth $C_j^*$ required to meet the service curves is equal to:
\beq
\label{eq:min1hopbw}
C^*_j = \max_{k \in \mathcal{F}_j}\left\{\sum_{i \in \mathcal{F}_j} r_i, \frac{\sum_{i \in \mathcal{F}_j} \beta_{ij}(T'_{kj})}{T'_{kj}} \right\}.
\eeq
\end{lemma}
}
\highlight{\useColor}{\Eqref{eq:min1hopbw} derives directly from the proof of Proposition~$1$ in~\cite{onehop21}.} Solving $\textbf{OPT}^-$, therefore, conceptually calls for using \Eqrefs{eq:min1hopbw}{eq:beta_ij} to evaluate $\sum_{j=1}^n C_j$ for all combinations of the variables $T_{ij}$ and $D_i$ that meet the constraints of \Eqref{eq:const1}.  \highlight{\useColor}{As alluded to,} \fig{fig:min_bandwidth} suggests a more efficient approach.  

From \Eqref{eq:min1hopbw}, we see that $C_j^*$ has a closed-form functional expression \emph{as long as the set of linear segments making up the aggregate service curve remains unchanged} (condition \textbf{ORD}).  In other words, their slope and length can vary, but their number and respective positions should not. When this holds and given the functional expression of this aggregate service curve on each link, $\textbf{OPT}^-$ can then be formulated as a set of non-linear programs\footnote{The variables $D_i$'s and $T_{ij} + D_i$'s are in the denominator of~\Eqrefs{eq:beta_ij}{eq:min1hopbw}, respectively.} (NLPs). 

Those NLPs can be solved using standard solvers. Finalizing a solution to $\textbf{OPT}^-$ then calls for exploring how the problem's variables and constraints affect condition \textbf{ORD}.  Specifically, \textbf{ORD} depends on the relative positions (\emph{ordering}, hence the condition's name) of the inflection points of the link's aggregate service curve.  For link~$j$, these map to the variables $T_{ij}$ and $T'_{ij}=T_{ij} + D_i, i\in \mathcal{F}_j$.

The number of feasible orderings of these variables 
is combinatorial in nature (it grows super-exponentially in $m$ and $n$), and likely intractable even for a single hop. As a result, we rely on a standard randomized combinatorial search strategy to explore the space of feasible orderings\footnote{\highlight{\useColor}{We explore a logarithmic number of feasible orderings to achieve a reasonable trade-off between solution quality and computational efficiency.}}, and for each such ordering solve the associated NLP.  Appendix~\ref{app:nlp_example}\onlineVersion{\useOnline}{~of~\cite{multihop22}}
offers additional details on how we generate feasible orderings, and illustrates it with a simple three flows example.

\subsection{Buffer Bounds}
\label{sec:buffer_bound}

\highlight{\useColor}{Given a solution to $\textbf{OPT}^-$, \Eqref{eq:buffer} can be used to obtain buffer upper bounds. We do so for 
a non-work-conserving setting\footnote{\highlight{\useColor}{In a work-conserving setting buffer bounds are more challenging as  tight departure curves from the upstream SCED schedulers are difficult to derive.}}, where buffer requirements are split into (i) per flow ingress reprofiling buffers, (ii) per flow reprofiling buffers at each hop/link, and (iii) a per link scheduling buffer.}  

\highlight{\useColor}{Under non-work-conserving operation, a flow's reprofiling buffer, at both ingress and at each hop, receives data from the upstream link or source, stores it until it conforms to the flow's (re)profile, before releasing it to the link/scheduler for transmission.  The (link) scheduling buffer holds data from all flows waiting for transmission on the link.  Bounds can be obtained separately for each flow's reprofiling buffer and for the link scheduling buffer.  With those bounds in hand, an upper bound can be derived for a link's total buffer requirements simply by adding the reprofiling buffer bounds of the flows sharing the link and the link's scheduling buffer bound.  Expressions for those individual bounds are given next. }

\highlight{\useColor}{The scheduling buffer bound for link~$j$ is given by:
\begin{equation}
\label{eq:link_buffer}
\widehat{\Theta}_j = \sup_{t \geq 0 }\{\sum_{i\in\mathcal{F}_j}\sigma_i - C_jt\}.
\end{equation}
where $\sum_{i\in\mathcal{F}_j}\sigma_i$ represents the aggregate arrival curve of all flows sharing link $j$ as they leave their respective reprofiler, \ie the flow's 2SRC.  Conversely, $C_jt$ represents the link's service curve given its bandwidth $C_j$.
}

\highlight{\useColor}{
Similarly, we show in Appendix~\ref{app:reprofiler_buffer}\onlineVersion{\useOnline}{~of~\cite{multihop22}} that the reprofiling buffer of flow~$i$ at hop (link)~$j$ can be bounded based on the worst-case burst that can accumulate because of scheduling delays at the previous hop~$j'$ on its path $\mathcal{P}_i$:
\begin{equation}
\label{eq:reprofiler_buffer}
\widetilde{\Theta}_{ij} = \sup_{t \geq 0 }\{\sigma_i - \beta_{ij'}\},
\end{equation}
Finally, we note that flow~$i$'s ingress reprofiling buffer is readily bounded by its original token bucket burst size $b_i$.
}

\subsection{A Representative Example}
\label{sec:example}

To showcase the operation of our NLP approach and the type of solutions to $\textbf{OPT}^-$ it produces, we introduce next a few representative examples. In spite of their limited scope, they highlight the diversity of possible outcomes, and consequently the challenges in generating insight into how individual parameters affect solutions.

For clarity of exposition, we focus on the minimalist configuration of~\fig{fig:2h2f_network} with just $2$~hops and $2$~flows. Flow~$1$ traverses links~$1$ and~$2$, while flow~$2$ is limited to link~$2$.  \highlight{\useColor}{The optimal solution, therefore, only requires determining how to reprofile flow~$1$ and split its residual deadline (after subtracting its reprofiling delay) across the two links.  This is because flow~$1$ is reprofiled prior to reaching link~$2$, which, when combined with the optimality of EDF in the one-hop scenario~\cite{onehop21}, implies that minimizing the bandwidth of link~$2$ can be realized without reprofiling flow~$2$ (or further reprofiling flow~$1$).}

\begin{sourcefigure}[!h]
\centering
\includegraphics[width=0.8\linewidth]{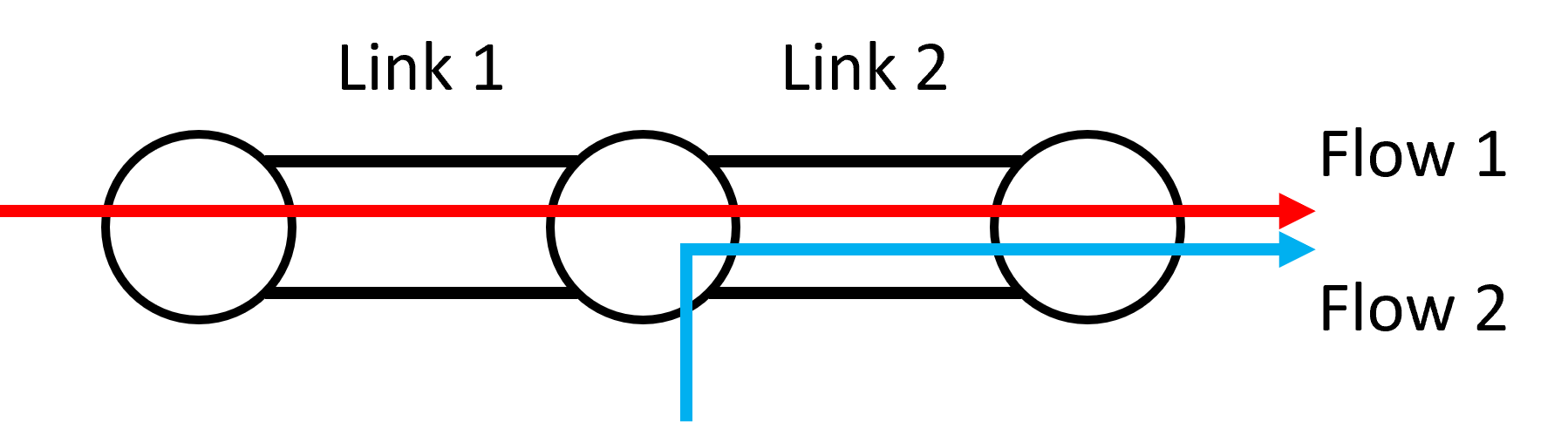}
\caption{Network with 2 hops and 2 flows.}
\label{fig:2h2f_network}
\end{sourcefigure}

\begin{table*}[ht]
\begin{center}
\caption{Optimal Solutions for the 2-hop, 2-flow Network of~\fig{fig:2h2f_network}.}
\label{table:solution_example}
\begin{tabular}{c|cc|cccc}
\hline
& \multicolumn{2}{c|}{\textbf{Profile $(r_i, b_i, d_i), i=1,2$}}
& \multicolumn{4}{c}{\textbf{Flow~$1$ Reprofiling and Deadlines}}\\\hline
& \multicolumn{1}{c|}{flow 1} & \multicolumn{1}{c|}{flow 2}
& \multicolumn{1}{c|}{$D_1$} & \multicolumn{1}{c|}{$T_{11}$} & \multicolumn{1}{c|}{$T_{12}$} & $D_1/\widehat{d}_1$\\\hline
\textbf{Expt 1}
& \multicolumn{1}{c|}{($98.75$, $88.18$, $0.20$)} & ($87.63$, $33.56$, $0.01$)
& \multicolumn{1}{c|}{$0.10$} & \multicolumn{1}{c|}{$0.09$} & \multicolumn{1}{c|}{$0.01$} & $50.68\%$\\ \hline
\textbf{Expt 2}
& \multicolumn{1}{c|}{($16.84$, $21.88$, $2.00$)} & ($57.37$, $70.14$, $1.00$)
& \multicolumn{1}{c|}{$1.27$} & \multicolumn{1}{c|}{$0$} & \multicolumn{1}{c|}{$0.73$} & $98.07\%$\\ \hline
\textbf{Expt 3}
& \multicolumn{1}{c|}{($28.26$, $71.05$, $2.00$)} & ($81.47$, $48.07$, $0.10$)
& \multicolumn{1}{c|}{$0.93$} & \multicolumn{1}{c|}{$0.97$} & \multicolumn{1}{c|}{$0.10$} & $46.29\%$\\ \hline
\textbf{Expt 4}
& \multicolumn{1}{c|}{($60.39$, $4.88$, $0.20$)} & ($86.24$, $61.55$, $0.10$)
& \multicolumn{1}{c|}{$0.05$} & \multicolumn{1}{c|}{$0.04$} & \multicolumn{1}{c|}{$0.11$} & $57.15\%$\\ \hline
\textbf{Expt 5}
& \multicolumn{1}{c|}{($33.11$, $6.19$, $0.20$)} & ($25.32$, $88.41$, $0.01$)
& \multicolumn{1}{c|}{$0.08$} & \multicolumn{1}{c|}{$0.11$} & \multicolumn{1}{c|}{$0.01$} & $44.56\%$\\ \hline\hline
\end{tabular}
\end{center}
\end{table*}
\add{Table \ref{table:solution_example} reports the structure (reprofiling delay and local deadline assignments) of optimal solutions for $5$ experiments associated with different combinations of flow profiles. Because the focus is on the \emph{structure} of the solution, we do not report the resulting bandwidth values, nor do we compare them to that of alternative approaches.  This discussion is deferred to the investigation of Section~\ref{sec:algorithm_comparison}.

Returning to Table \ref{table:solution_example}, the columns labeled ``flow~$1$'' and ``flow~$2$'' report the $(r_i,b_i,d_i)$, $i=1,2,$ profiles of the two flows, with rates, burst sizes, and deadlines taking different values in each of the $5$~experiments. The second group of columns labeled ``Flow~$1$ Reprofiling and Deadlines'' reports the reprofiling delay $D_1$ of flow~$1$ and its local deadlines $T_{11},T_{12},$ on links~$1$ and~$2$, respectively.  We also report $D_1$ relative to its maximum value\footnote{The reprofiling rate $R_1$ cannot be less than $r_1$.} $\widehat{d}_1=\min(d_1, b_1/r_1)$ to demonstrate how much reprofiling is applied to flow~$1$.
}

\add{As alluded to, no clear trends emerge from the data, \eg tighter deadlines imply neither smaller nor larger relative reprofiling delays.  Nevertheless, the optimal solutions 
share a few characteristics:  {\bf (1)}  Optimal reprofiling delays can be anywhere between full reprofiling $(D_i=\widehat{d}_i)$ and no reprofiling $(D_i=0)$, and {\bf (2)} Local deadlines can vary significantly across hops, \ie they are typically not equal.}

\section{A Greedy Reprofiling Algorithm}
\label{sec:greedy}

The NLP formulation's ability to find an ``optimal'' solution to $\textbf{OPT}^-$ notwithstanding, its complexity motivates the development of a heuristic that we describe next.  It relies on a simple greedy algorithm (Greedy) that, at least on the small topologies the NLP formulation can handle, performs nearly as well, but at a much lower computational cost.

As just mentioned, optimal solutions often have reprofiling delays $D_i$ anywhere in $[0,\widehat{d}_i]$, where $\widehat{d}_{i} = \min(d_i, b_i/r_i)$ is flow~$i$'s maximum reprofiling delay, and per hop deadlines that vary across hops.  In other words, those solutions select \emph{intermediate} reprofiling configurations, \ie strictly between ``no reprofiling'' and ``full reprofiling'', and distribute the remaining delay budget unevenly across hops.

\subsection{Overview}
\label{sec:greedy_oview}

\begin{figure}[!h]
\centering
\includegraphics[width=0.5\linewidth]{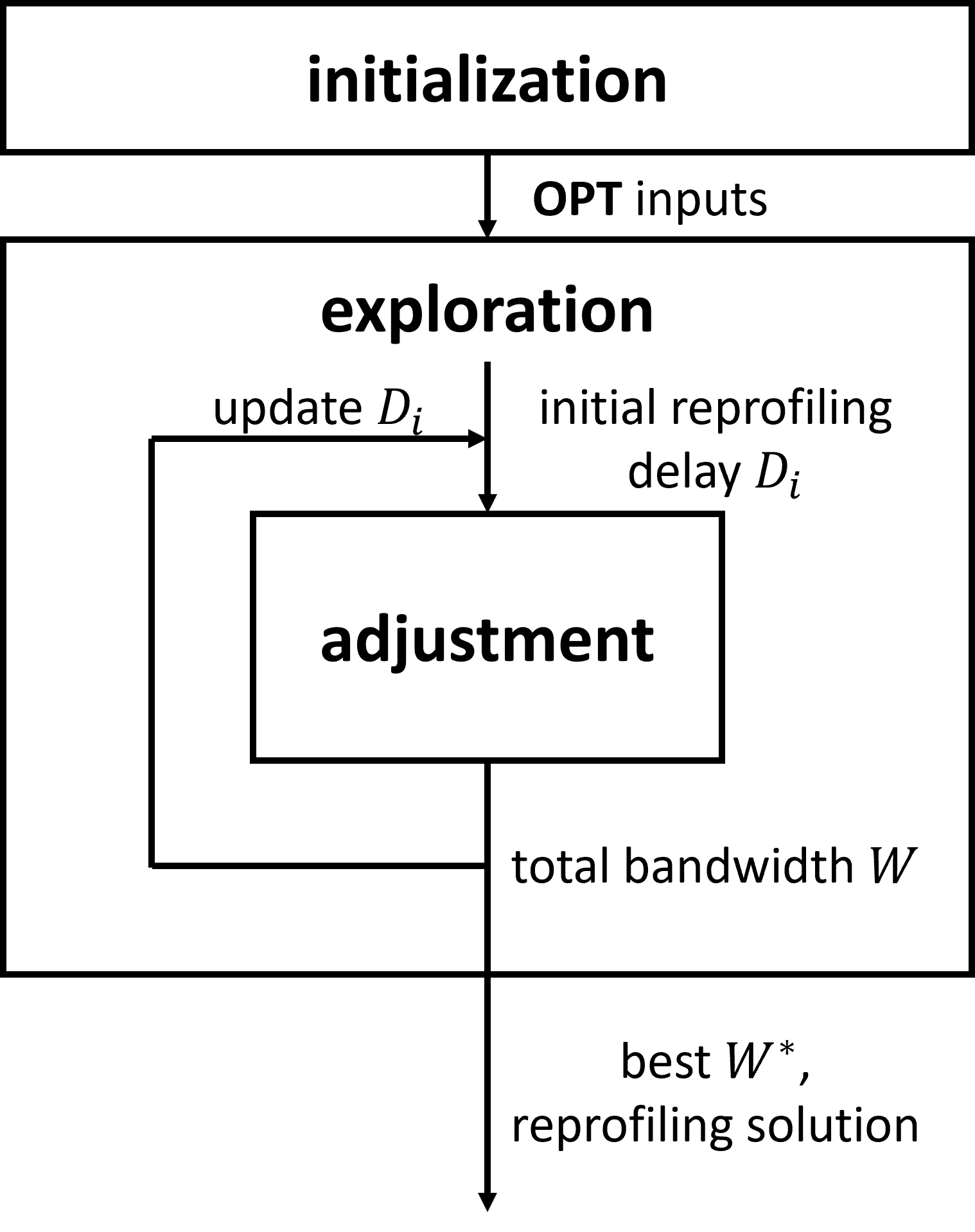}
\caption{Overview of Greedy.}
\label{fig:greedy_oview}
\end{figure}

\highlight{\useColor}{
The algorithm we present next captures both above aspects. It proceeds in two nested phases as shown in \fig{fig:greedy_oview}. 
\begin{enumerate}[(i),nosep]
\item  An iterative \emph{exploration} of reprofiling configurations, with initial flows' reprofiling delays spanning values between 0 and their maximum value $\widehat{d}_i$.
\item For each reprofiling configuration, \emph{adjustments} of local deadlines and reprofiling solutions at hops where flows' delay guarantees are strictly better than their target.
\end{enumerate}
}

\subsection{Exploration Phase}
\label{sec:explore}

\highlight{\useColor}{
As its name indicates, this phase \emph{explores} reprofiling configurations that span the range of possible reprofiling delays. The intent is to find the combination that yields the smallest overall bandwidth. It is detailed in Algorithm~\ref{algo:exploration}, where we refer to the line numbers on the left when describing the algorithm.}
\begin{algorithm}
\caption{Exploration}
\highlight{\useColor}{
\begin{algorithmic}[1]
\label{algo:exploration}
\renewcommand{\algorithmicrequire}{\textbf{Inputs:}}
\renewcommand{\algorithmicensure}{\textbf{Output:}}
\REQUIRE 
flow profiles $\mathbf{r} = (r_1, r_2, \ldots, r_m)$,\\ 
$\mathbf{b} = (b_1, b_2, \ldots, b_m)$, $\mathbf{d} = (d_1, d_2, \ldots, d_m)$\\
path matrix $\pmb{\mathcal{P}} = (\mathcal{P}_1, \mathcal{P}_2, \ldots, \mathcal{P}_m)$\\
exploration parameters $L, K$\\
adjustment threshold $\epsilon$
\ENSURE
minimum total bandwidth $W^*$\\
optimal reprofiling delays $\mathbf{D}^*$\\
local deadline assignments $\mathbf{T}^*$
\STATE $\widehat{d}_{i} = \min(d_i, b_i/r_i), \forall 1 \leq i \leq m$
\STATE $lr = 0, hr = 1$
\FOR {$l = 1$ to $L$}
\FOR {$k = 0$ to $K+1$}
\STATE $\gamma_k = lr + \frac{hr - lr}{K+1} \cdot k$
\STATE $D_i = \gamma_k \cdot \widehat{d}_{i}, \forall 1 \leq i \leq m$
\STATE $T_{ij} = (d_i - D_i) / |\mathcal{P}_i|, \forall 1 \leq i \leq m, j \in \mathcal{P}_i$
\STATE $W = \textmd{adjustment}((\mathbf{r}, \mathbf{b}, \mathbf{d}), \pmb{\mathcal{P}}, \mathbf{D}, \mathbf{T}, \epsilon)$
\IF {$W < W^*$}
\STATE $W^* = W, k^* = k$
\STATE $\mathbf{D}^* = \mathbf{D}, \mathbf{T}^* = \mathbf{T}$
\ENDIF
\ENDFOR
\STATE $lr = \gamma_{\max(k^*-1, 0)}, hr = \gamma_{\min(k^*+1, K+1)}$
\ENDFOR
\RETURN $W^*, \mathbf{D}^*, \mathbf{T}^*$
\end{algorithmic}
}
\end{algorithm}

\highlight{\useColor}{In considering possible reprofiling configurations, the choice of reprofiling delays can vary across flows, and each flow~$i$ boasts a different range $[0,\widehat{d}_i]$. Exploring all possible combinations is clearly intractable.  Our exploration seeks a trade-off between coverage and tractability.  It proceeds iteratively, and, in iteration~$x$, applies the \emph{same} reprofiling ratio $\gamma_x$ to all flows, and sets the reprofiling delay of flow~$i$ to $D_i(x)=\gamma_x\widehat{d}_i$.  This allows flows with different reprofiling ranges to be assigned different reprofiling delays, while enabling a systematic exploration of the underlying space, \ie from $D_i(x)=0$ when $\gamma_x=0$ to $D_i(x)=\widehat{d}_i$ when $\gamma_x=1$.}

\begin{figure}[!h]
\centering
\begin{subfigure}{0.49\linewidth}
  \centering
  \includegraphics[width=\linewidth]{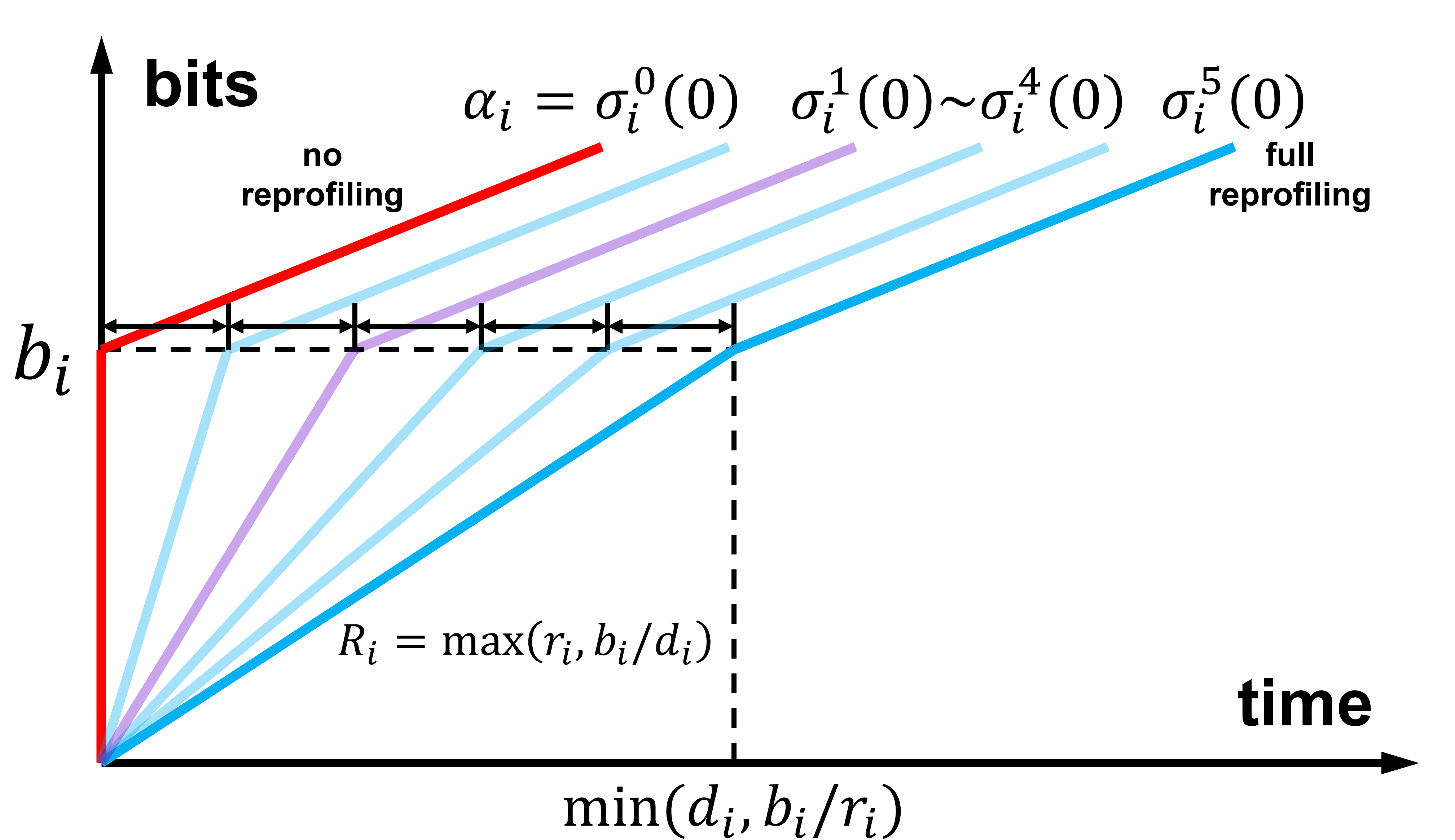}
  \caption{initial partition}
  \label{fig:uniform_sampling}
\end{subfigure}
\begin{subfigure}{0.49\linewidth}
  \centering
  \includegraphics[width=\linewidth]{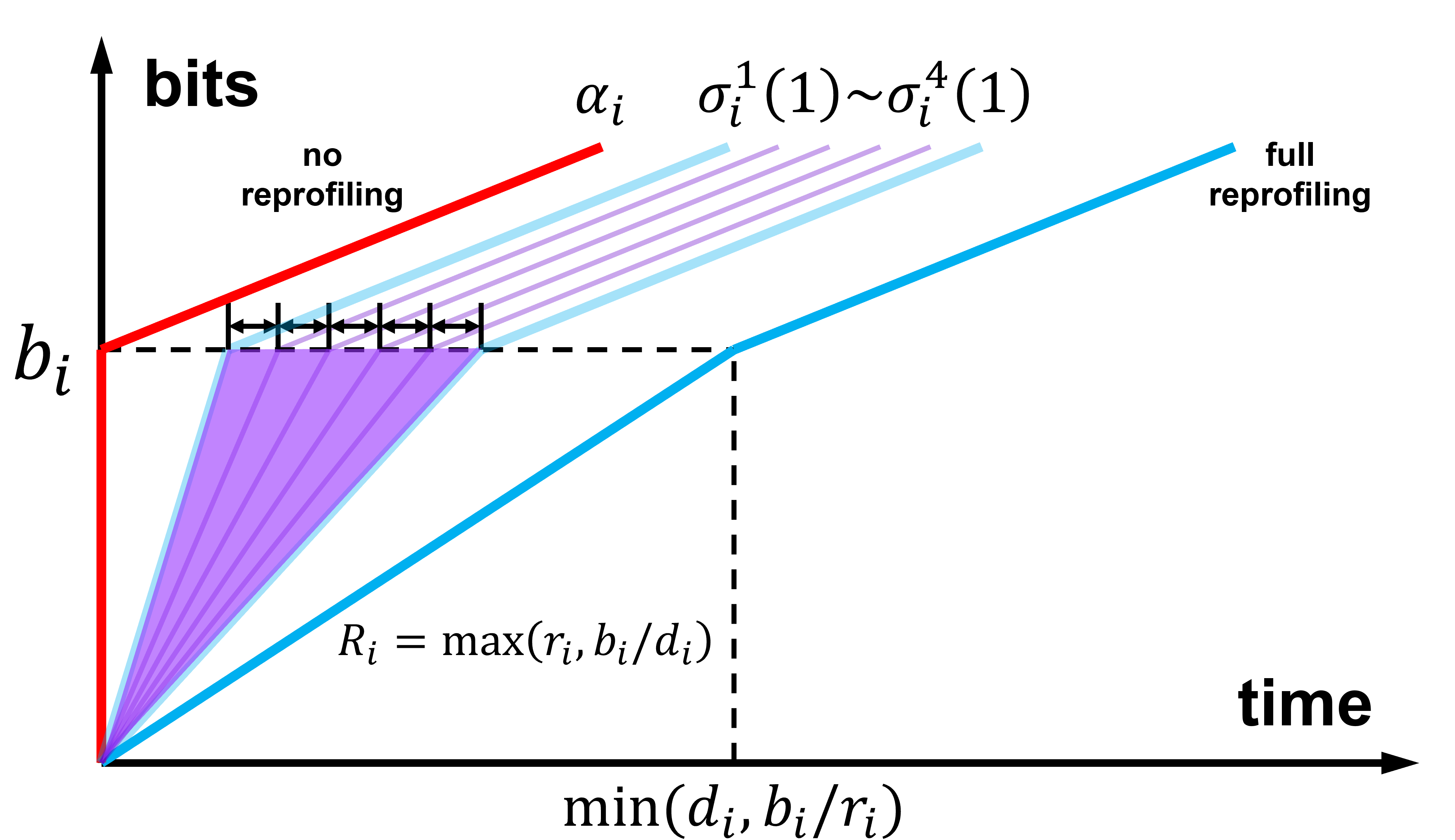}
  \caption{intermediate partition}
  \label{fig:progressive_update}
\end{subfigure}
\caption{Reprofiling exploration for $K=4$. $\sigma_i^k(l)$ specifies the $k^{th}$ 2SRC of flow $i$ at iteration $l$. $\sigma_i^2(0)$ is assumed to yield the best outcome in the first iteration.}
\label{fig:progressive_reprofiling}
\end{figure}

\highlight{\useColor}{The relative simplicity of this approach notwithstanding, its coverage depends on the number of distinct values $\gamma_x$ takes; with a fine-grain exploration requiring a large number of values.  To bound the resulting computational cost, we rely on an iterative approach that progressively refines the range of reprofiling ratios it explores. It is illustrated in~\fig{fig:progressive_reprofiling}.} 

\highlight{\useColor}{Turning next to the details of Algorithm~\ref{algo:exploration}, each iteration explores $K+2$ reprofiling ratios $\gamma_k=\frac{k}{K+1}, 0\leq k\leq K+1$, uniformly distributed in that iteration's range (line~$4$). The first iteration (line~$2$) is shown in~\fig{fig:uniform_sampling}, and spans the full range from $D_i(0)=0$ to $D_i(K+1)=\widehat{d}_i$. Subsequent iterations, for a total of~$L$, explore increasingly narrower ranges of reprofiling ratios to progressively refine and improve the best result.}

\highlight{\useColor}{Specifically, Algorithm~\ref{algo:exploration} iterates (line~$3$) by exploring new ranges centered at the reprofiling ratio that produced the smallest network bandwidth in the previous iteration (lines~$9-10$), and with lower and upper bounds set to the two adjacent values (line~$13$).  As discussed next, given a reprofiling ratio, the required network bandwidth is obtained (line~$8$) following an adjustment phase\footnote{\highlight{\useColor}{See Algorithm~\ref{algo:adjustment}.}} that, when feasible, shifts delay allocations between initial reprofiling delays and local deadlines.}

We note that implicit in the iterative approach of Algorithm~\ref{algo:exploration} is the assumption that, in any iteration, the region where the minimum of the objective function (network bandwidth) is located can be bracketed with few samples. \fig{fig:greedy_explore} of Appendix~\ref{sec:supplement}\onlineVersion{\useOnline}{~of~\cite{multihop22}} offers evidence in support of this assumption.

\subsection{Adjustment Phase}
\label{sec:adjust}

\begin{figure}[!h]
\centering
\begin{subfigure}{0.49\linewidth}
  \centering
  \includegraphics[width=\linewidth]{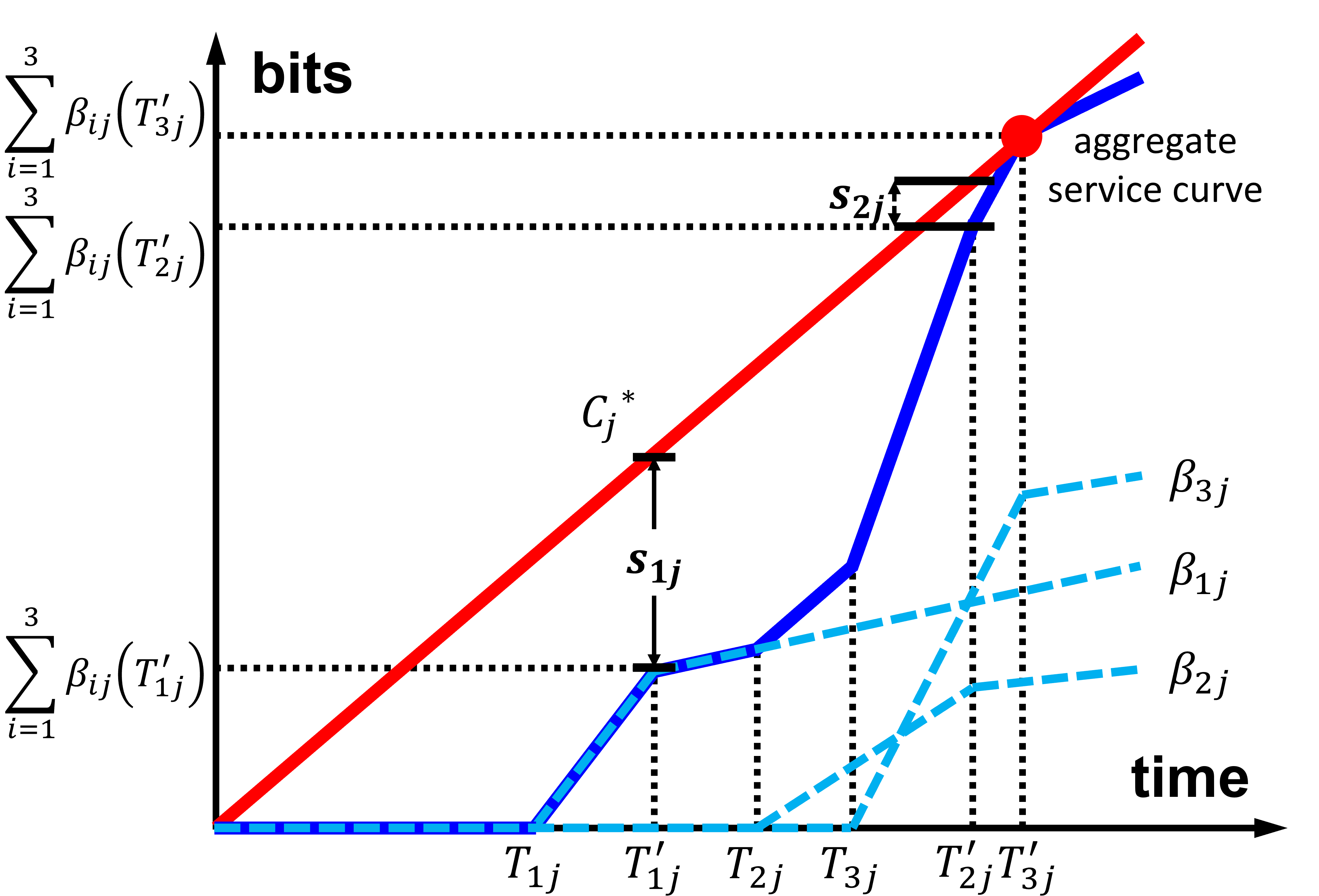}
  \caption{flow slack at $T'_{ij}$}
  \label{fig:margin}
\end{subfigure}
\begin{subfigure}{0.49\linewidth}
  \centering
  \includegraphics[width=\linewidth]{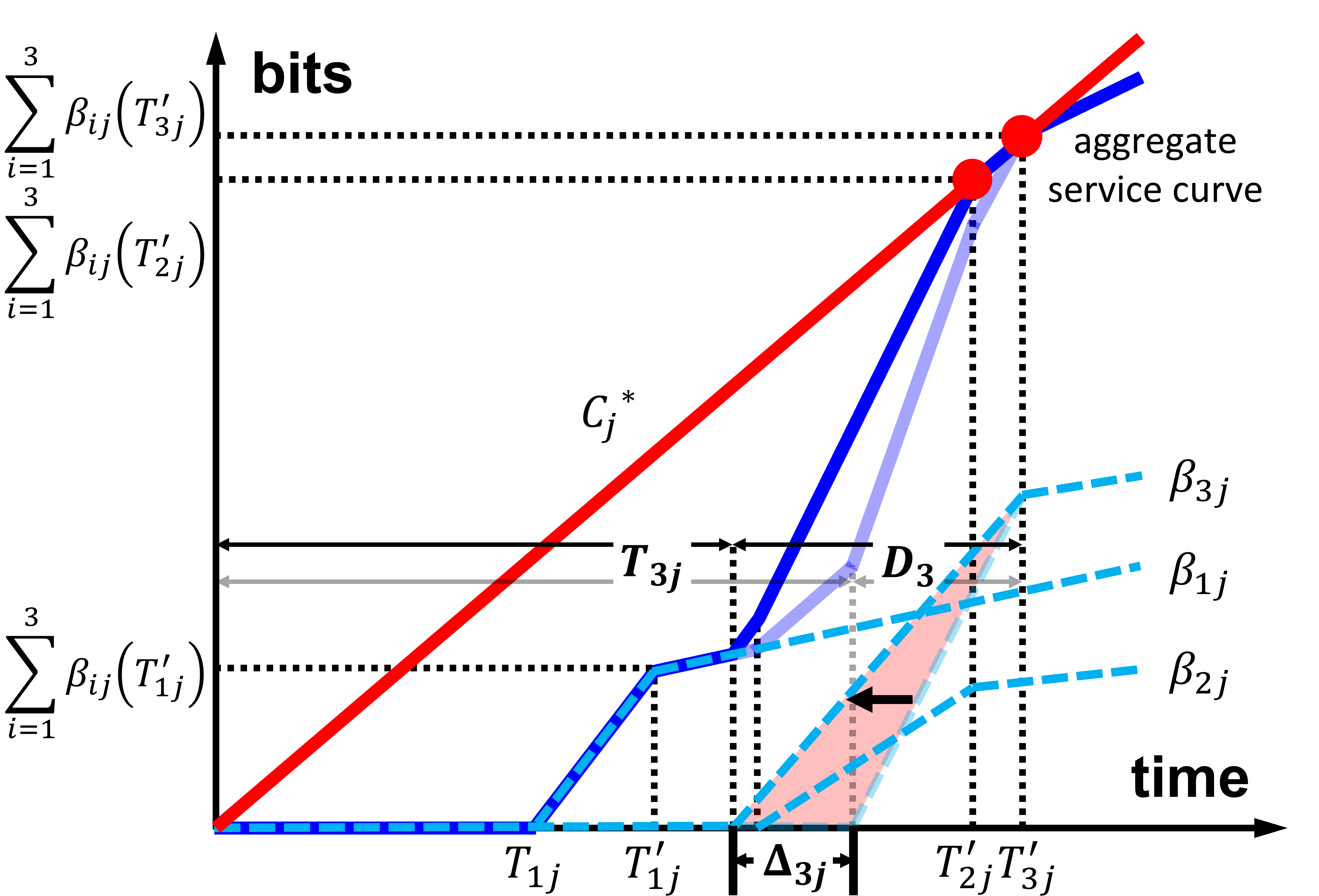}
  \caption{increase $D_i$ by $\Delta_{ij}$}
  \label{fig:smoothing}
\end{subfigure}
\caption{Greedy flow reprofiling adjustment.}
\label{fig:greedy_smoothing}
\end{figure}

\highlight{\useColor}{As shown in \fig{fig:greedy_oview}, the adjustment phase is invoked in each iteration of the exploration phase.  For an initial assignment of reprofiling delays and local deadlines, it checks every link to, when possible, greedily \emph{adjust} (increase) the reprofiling delays of flows with better deadlines than needed on that link.  The adjustment is bandwidth-neutral for the link, but, because increasing reprofiling makes flows smoother, it can reduce bandwidth requirements on other links. Algorithm~\ref{algo:adjustment} details those adjustments, with \fig{fig:greedy_smoothing} illustrating them on link~$j$.}
\begin{algorithm}
\caption{Adjustment}
\highlight{\useColor}{
\begin{algorithmic}[1]
\label{algo:adjustment}
\renewcommand{\algorithmicrequire}{\textbf{Input:}}
\renewcommand{\algorithmicensure}{\textbf{Output:}}
\REQUIRE flow profiles $\mathbf{r} = (r_1, r_2, \ldots, r_m)$,\\
$\mathbf{b} = (b_1, b_2, \ldots, b_m)$, $\mathbf{d} = (d_1, d_2, \ldots, d_m)$\\
path matrix $\pmb{\mathcal{P}} = (\mathcal{P}_1, \mathcal{P}_2, \ldots, \mathcal{P}_m)$\\
initial reprofiling delays $\mathbf{D} = (D_1, D_2, \ldots, D_m)$\\
initial local deadlines $\mathbf{T} = \{T_{ij}: \forall 1 \leq i \leq m, j \in \mathcal{P}_i\}$\\
improvement threshold $\epsilon$
\ENSURE total bandwidth $W$ after adjustment
\STATE $W = \infty, W' = \infty$
\STATE sort links in decreasing order of $|\bigcup_{i \in \mathcal{F}_j}\mathcal{P}_i|$
\WHILE {$(W' - W) / W' > \epsilon$}
\FOR {$j = 1$ to $n$}
\STATE $T'_{ij} = T_{ij} + D_i, \forall i \in \mathcal{F}_j$
\STATE compute $C^*_j$ according to~\Eqref{eq:min1hopbw}
\STATE sort flows in $\mathcal{F}_j$ in decreasing order of $T'_{ij}$
\FOR {$i \in \mathcal{F}_j$}
\STATE /* varies $T_{ij}$ in \Eqref{eq:slack} to find $T^*_{ij}$ that depletes slack at first preceding inflection point */
\STATE $T^*_{ij} = \max_{k \in \mathcal{F}_j, k > i}\{T_{ij}:s_{kj} = 0\}$
\STATE $T_{ij} = \max(T^*_{ij}, 0, T'_{ij} - b_i/r_i)$
\STATE $D_i = T'_{ij} - T_{ij}$
\ENDFOR
\ENDFOR
\STATE update $C^*_j, \forall 1 \leq j \leq n$ according to~\Eqref{eq:min1hopbw}
\STATE $W' = W, W = \sum_{1 \leq j \leq n}C^*_j$
\ENDWHILE
\RETURN $W$ 
\end{algorithmic}
}
\end{algorithm}

\highlight{\useColor}{
Recall that according to~\Eqref{eq:min1hopbw}, the minimum required bandwidth $C^*_j$ on link~$j$ depends on the values of the aggregate service curve at the inflection points $T'_{ij}$'s. As illustrated in \fig{fig:margin}, only one of those inflection points is typically involved in determining $C^*_j$. This indicates the presence of ``slack,'' \ie more bandwidth than necessary, at the other inflection points. We denote the slack at inflection point $T'_{ij}$ as $s_{ij}$, with:}
\begin{equation}
\label{eq:slack}
    s_{ij} = C^*_jT'_{ij} - \sum_{k \in \mathcal{F}_j} \beta_{kj}(T'_{ij}).
\end{equation}
\highlight{\useColor}{The purpose of the adjustment phase is to leverage this slack to increase the reprofiling delay of flows that can tolerate it.  For a given flow~$i$, this means increasing its reprofiling delay while correspondingly decreasing $T_{ij}$ (to keep $T'_{ij}$ constant). The resulting impact and how slack is consumed in the process, is shown in~\fig{fig:smoothing} for flow~$3$.  We note that increasing the reprofiling delay $D_3$ ``raises'' the aggregate service curve link~$j$ \emph{prior} to $T'_{3j}$.  Hence, the increase in $D_3$ can proceed until either one of the following three conditions is met (lines~$9-10$):
\begin{enumerate}[label=(\alph*),wide=0pt]
\item $s_{kj}$ is depleted\footnote{This is the case in \fig{fig:smoothing}, where $D_3$ is increased until $s_{2j}=0$.} at some $T'_{kj}, \, k \in \mathcal{F}_j$ and $T'_{kj} < T'_{ij}$.   
\item The flow's own local deadline $T_{ij}$ decreases to $0$.
\item The reprofiling delay $D_i$ reaches its maximum value $b_i/r_i$.
\end{enumerate}
Once adjustments have been finalized on a link, the updated profiles (2SRCs) are propagated to all the hops traversed by the affected flows. The resulting smoother profiles translate in potentially lower bandwidth requirements on those links.
}

We note that, except for the flow with the smallest $T'_{ij}$ that can always be fully reprofiled, reprofiling a flow affects slack at other $T'_{ij}$'s. Hence, the order in which flows are examined influences reprofiling.  We chose to reprofile flows in decreasing order of the $T'_{ij}$'s \highlight{\useColor}{(line~$7$)}. This is because a larger $T'_{ij}$ minimizes how many other flows are affected\footnote{Because reprofiling only affects slack at smaller $T'_{ij}$'s.}.

\highlight{\useColor}{
The other ordering affecting the adjustment phase is the one in which links are visited. We visit links
in decreasing order of $|\bigcup_{i \in \mathcal{F}_j}\mathcal{P}_i|$ (line~$2$) to prioritize links whose flows affect the most other links. This ordering notwithstanding, changes in a flow's profile affect slack on all the links it traverses, including previously visited links.  Hence, irrespective of the ordering chosen, new reprofiling opportunities may emerge.  Realizing them requires revisiting links.  To that end, we repeat the adjustment process (visiting all network links) until improvements fall below a threshold~$\epsilon$ (line~$3$). As the required network bandwidth is lower-bounded, this process converges.  Across all our experiments with a threshold of $\epsilon=0.1\%$, it terminated in~$4$ iterations or less $90\%$ of the time.
}

\subsection{Discussion}
\label{sec:greedy_overall}
\highlight{\useColor}{Recall that each iteration of the exploration phase starts with the \emph{same} reprofiling ratio $\gamma_x$ for all flows, and splits any remaining delay equally among links. We experimented with alternatives, including assigning different initial reprofiling delays to flows and heterogeneous initial local deadlines across hops. The results were almost statistically identical (slightly worse), and had a longer run time.}

Running Greedy calls for selecting $L$ (exploration iterations) and $K$ (reprofiling configurations).  The algorithm terminates after $L$ iterations or when an iteration yields an improvement below a configured threshold.  In practice, choosing $L=2$ and $K=4$ seems to realize a reasonable performance vs.~computations compromise. \fig{fig:greedy_config} of Appendix~\ref{sec:supplement}\onlineVersion{\useOnline}{~of~\cite{multihop22}} offers evidence in support of this conclusion.  We also note that because flows from the same traffic class (same end-to-end deadline) and following the same path can be aggregated, Greedy's run time complexity is $O(N_C\times N_{P})$, where $N_C$ and $N_{P}$ are the number of traffic classes and paths, respectively.

Finally, while Greedy, like the NLP formulation, combines reprofiling and unequal deadlines across hops, its primary driver is reprofiling, with unequal deadline assignments a side effect.  Hence, its close approximation of the NLP formulation, as shown next, hints that reprofiling is the dominant factor in decreasing bandwidth.  \fig{fig:greedy_init} of Appendix~\ref{sec:supplement}\onlineVersion{\useOnline}{~of~\cite{multihop22}} offers evidence in support of this claim.  \highlight{\useColor}{We also acknowledge that the evidence of Greedy's close approximation of NLP is only empirical.  Deriving tight approximation bounds remains elusive, in part because the ``optimal'' solution to $\textbf{OPT}^-$ relies on the combination of multiple locally optimal solutions for each NLP instance under consideration.  This makes applying traditional approximation techniques challenging.}

\section{NLP Formulation vs. Greedy}
\label{sec:algorithm_comparison}

\begin{figure}[!h]
\centering
  \includegraphics[width=\linewidth]{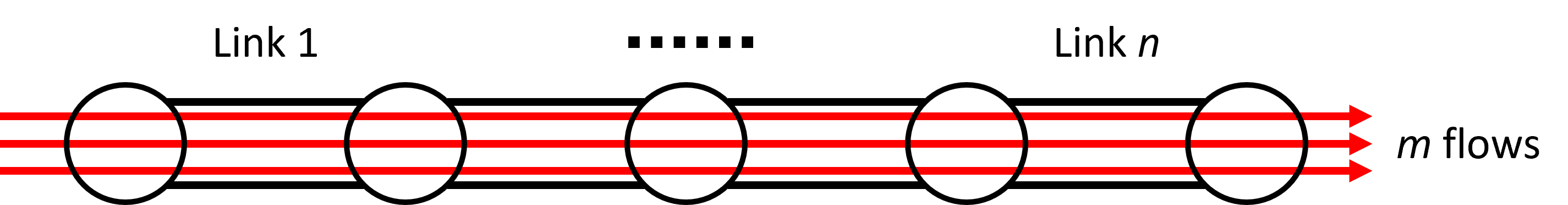}
  \caption{tandem topology}
  \label{fig:tandem}
\end{figure}

In this section, we compare the cost and performance (bandwidth reduction) of NLP and Greedy. As NLP's complexity grows rapidly with the number of links and flows, our comparison is limited to the simple tandem topology of \fig{fig:tandem}.

In comparing the two algorithms, we \highlight{\useColor}{vary the number $m$ of flows and the number $n$ of hops each flow traverses in the topology of \fig{fig:tandem}.  We also select from a range of traffic profiles, and use different values of \emph{target} per-hop deadlines to construct end-to-end deadlines.  The latter means that, as $n$ increases, so do end-to-end deadlines. \figs{fig:eval_fd_fr}{fig:eval_fd_nr} in Appendix~\ref{sec:synthetic}\onlineVersion{\useOnline}{~of~\cite{multihop22}} 
investigate an alternative where end-to-end deadlines are fixed, with target per-hop deadlines then decreasing as $n$ increases.  Results were qualitatively similar.} 

Specifically, we consider three per-hop deadlines with a dynamic range of~$100$, \ie $d_h\in\{0.01, 0.1,1\}$, \highlight{\useColor}{so that end-to-end deadlines are of the form $d=n\cdot d_h$, as the number of hops vary}.  The value $d_h$ of a flow is randomly chosen from these three values, and its traffic profile $(r,b)$ is set by randomly selecting values in the range $[1,100]$ for both $r$ and $b$.  

\begin{figure}[!h]
\centering
\begin{subfigure}{0.49\linewidth}
  \centering
  \includegraphics[width=\linewidth]{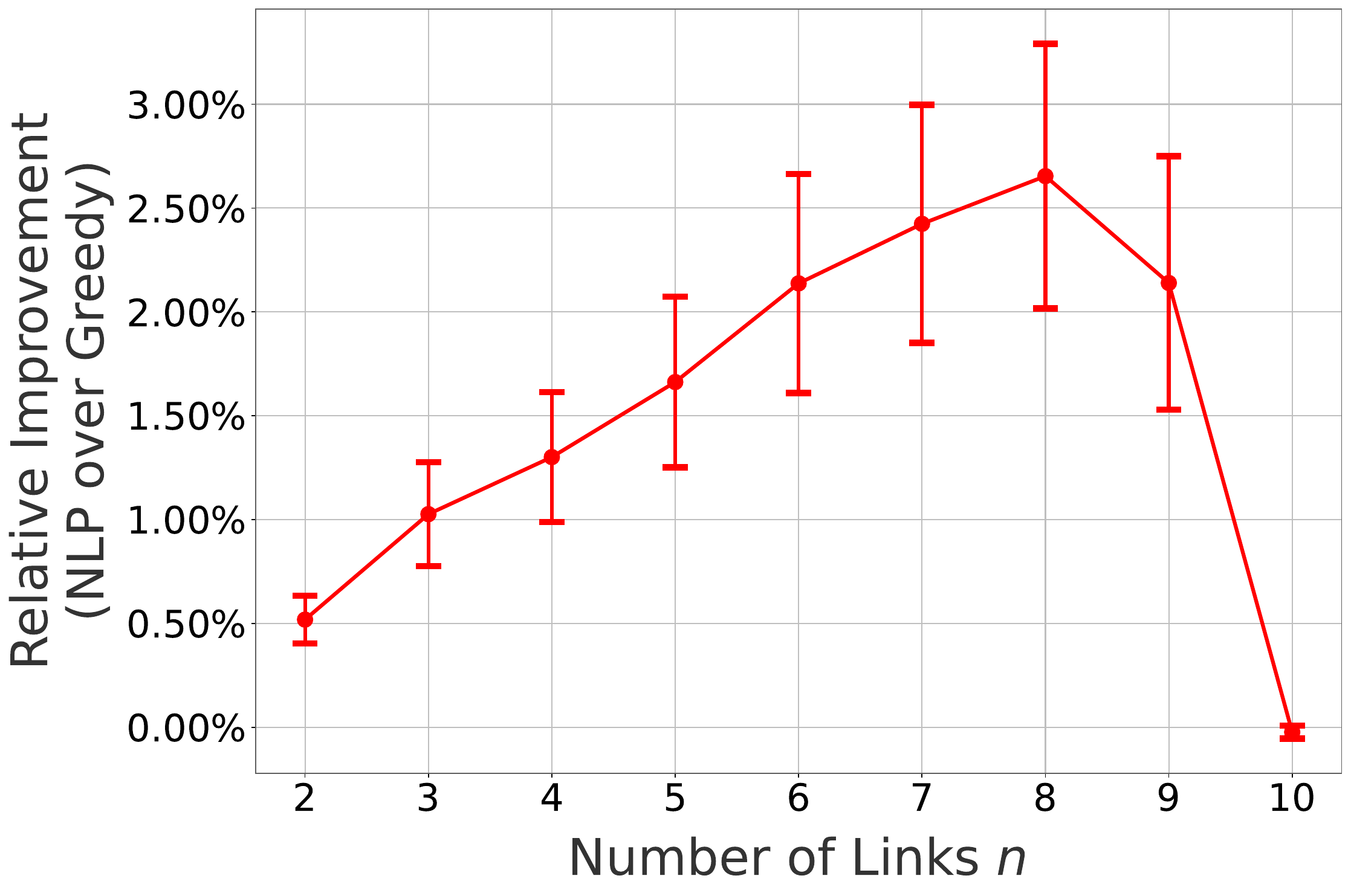}
  \caption{NLP over Greedy}
  \label{fig:solution1}
\end{subfigure}
\begin{subfigure}{0.49\linewidth}
  \centering
  \includegraphics[width=\linewidth]{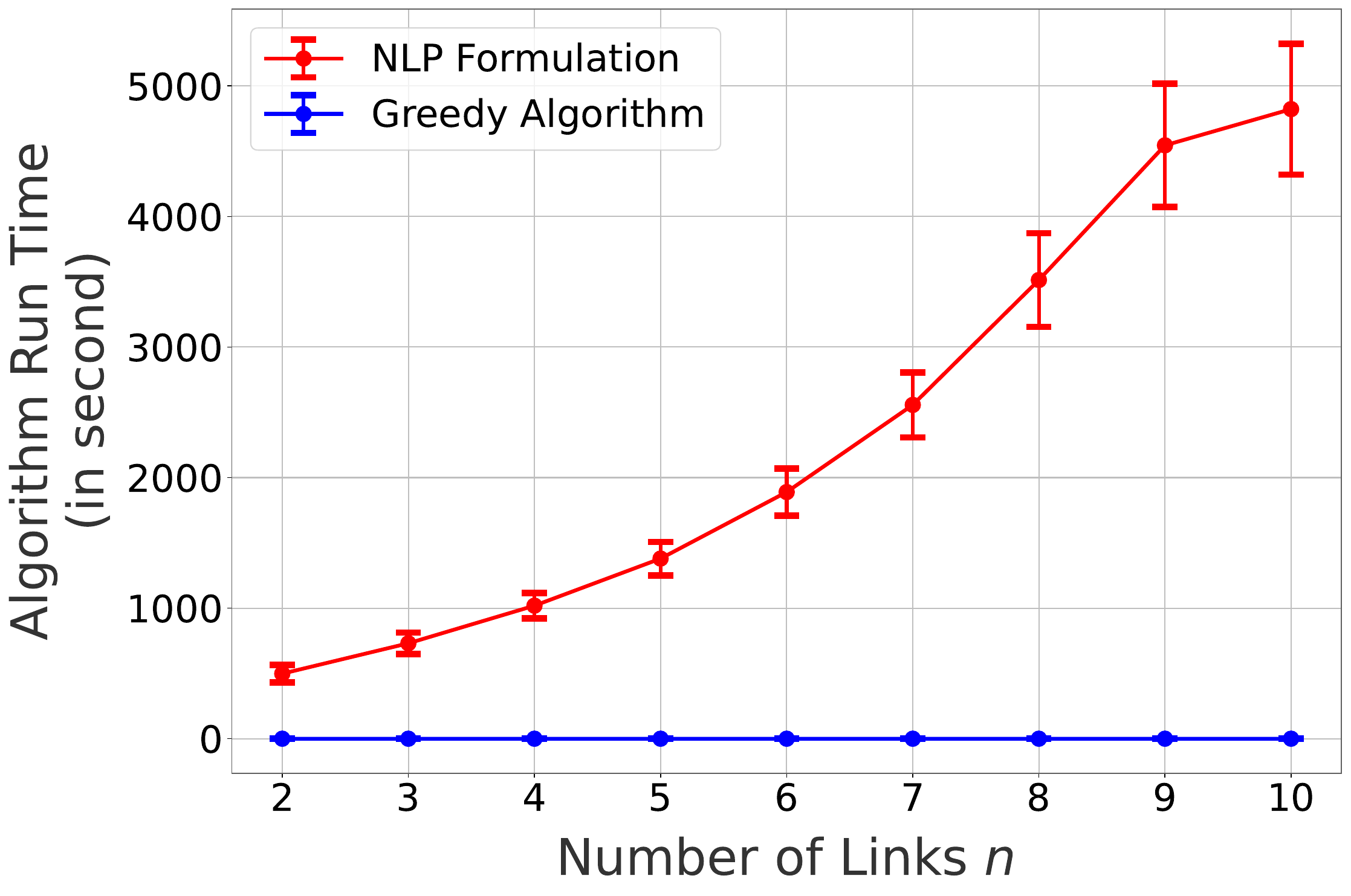}
  \caption{run time}
  \label{fig:time1}
\end{subfigure}
\caption{NLP vs.~Greedy ($m=3$ flows, \# links, $n$, varies).}
\label{fig:performance_hop}
\end{figure}

In our first set of experiments, we fix the number of flows to $m=3$ and vary the number of links $n$ from~$2$ to~$10$.  The results are shown in \fig{fig:performance_hop}, where each data point is based on \highlight{\useColor}{$150$} samples.  We use Octeract\footnote{Accessible at \url{https://octeract.gg/}.} as our NLP solver.  \fig{fig:solution1} reports the average relative reduction in overall bandwidth that the NLP formulation affords over Greedy, while \fig{fig:time1} gives the corresponding average run times on an Intel(R) Xeon(R) CPU E5-1620 v3 @ 3.50GHz server.  The error bars report the $95\%$ confidence interval for both \highlight{\useColor}{over a set of $150$ experiments}.  As expected from its ``optimality,'' the NLP formulation outperforms Greedy across all our experiments, but the gap was small (below $3\%$), and differences in run time quickly exceeded three orders of magnitude as $n$ increased.

\fig{fig:solution1} also highlights that while the gap between NLP and Greedy first increases with $n$, it eventually drops to~$0$.  This is expected and an artifact of keeping per-hop deadlines fixed.  Specifically, as $n$ grows, so do end-to-end deadlines until they reach a value for which reprofiling all flows to their average rate (a reprofiling delay of $D=b/r$) becomes feasible.  Both NLP and Greedy readily discover this solution.
\begin{figure}[!h]
\centering
\begin{subfigure}{0.49\linewidth}
  \centering
  \includegraphics[width=\linewidth]{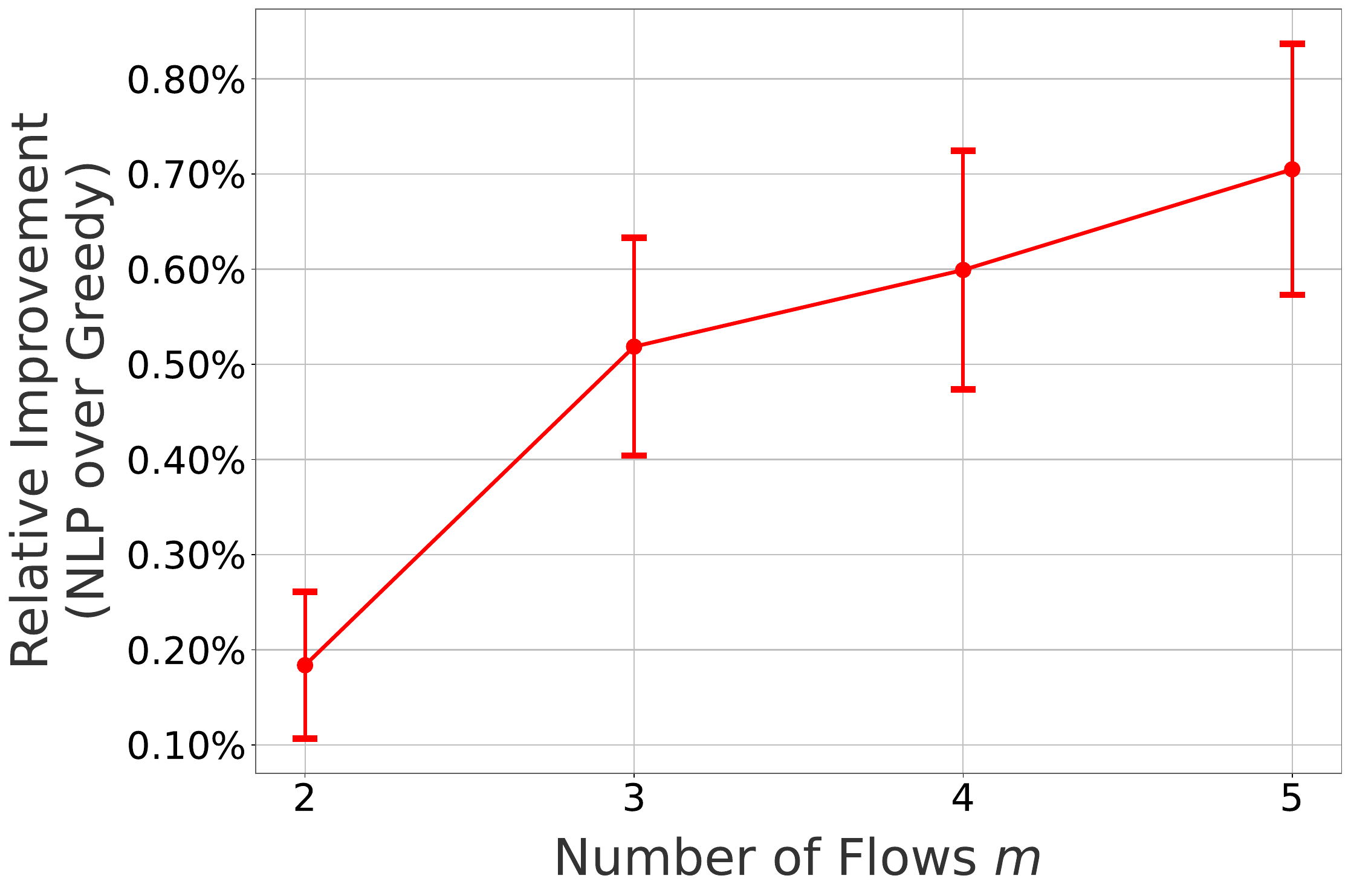}
  \caption{NLP over Greedy}
  \label{fig:solution2}
\end{subfigure}
\begin{subfigure}{0.49\linewidth}
  \centering
  \includegraphics[width=\linewidth]{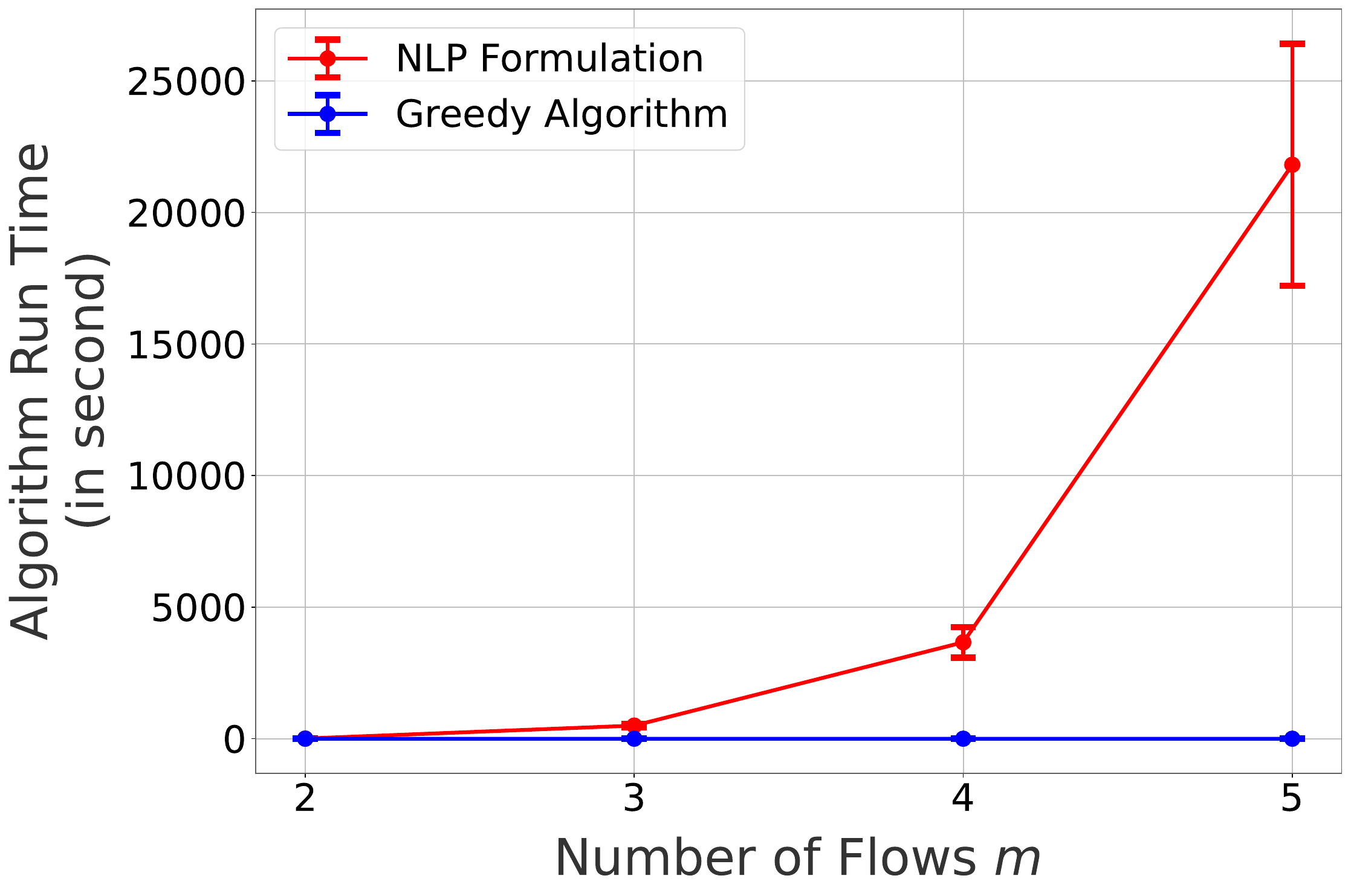}
  \caption{run time}
  \label{fig:time2}
\end{subfigure}
\caption{NLP vs.~Greedy ($n=2$ links, \# flows, $m$, varies).}
\label{fig:performance_flow}
\end{figure}

Our next set of experiments uses a similar setup as the first, but varies the number of flows~$m$ from~$2$ to~$5$ while keeping the number of hops at $n=2$.  The smaller range of the latter is because the complexity of the NLP formulation grows faster (factorial) in the number of flows than links.  The results are reported in \fig{fig:performance_flow} that parallels \fig{fig:performance_hop} with mostly similar conclusions.  Greedy remains within about $0.5\%$ of NLP on average across experiments, and it achieves this with even greater savings in run time, namely, improvements of more than six orders of magnitude over NLP when $m$ is just~$5$.

\highlight{\useColor}{For calibration purposes, 
we also compare in Appendix~\ref{sec:nlp_baseline}\onlineVersion{\useOnline}{~of~\cite{multihop22}} the NLP-based solution with two baselines. The first, ``Full Reprofiling'' or FR, reprofiles flows as much as possible with a reprofiling delay set to $D=\widehat{d}=\min(d, b/r)$. When the deadline~$d$ is small, \ie $d \leq b/r$, it is fully dedicated to reprofiling and the network delay is zero\footnote{Recall that we are assuming a fluid model.}.  When the deadline is larger, \ie $d > b/r$, the residual delay budget, $d - b/r$, is split evenly across hops.  The second, ``No Reprofiling'' or NR, forfeits reprofiling altogether, \ie sets $D=0$, and splits the full~$d$ evenly across hops.  As Greedy includes FR and NR among its initial solutions, it never performs worse than either, and often does better because of its ability to explore intermediate reprofiling configurations.}

The conclusion from our two sets of experiments is that while the NLP formulation generates better solutions, its computational complexity makes it impractical for anything but very small configurations.  In contrast, Greedy scales well and produces solution that perform nearly as well.  From examining the solutions produced by the two approaches, we note though that they usually differ significantly in the configurations they produce.  As pointed out in Section~\ref{sec:greedy_overall}, this is not unexpected given their structural differences, and hints at a relatively ``flat'' solution space around optimal solutions.

\section{Evaluation}
\label{sec:evaluation}

In Appendix~\ref{sec:synthetic}\onlineVersion{\useOnline}{~of~\cite{multihop22}} we conduct a comprehensive evaluation of the benefits of reprofiling using synthetic topologies and flow profiles whose parameters we vary systematically.  The investigation shows that reprofiling  yields consistent improvements, often in the double digits in relative bandwidth decreases.  However, except for a few intuitive findings, \eg that full reprofiling is increasingly the best solution as the number of hops a flow traverses increases\footnote{As discussed in Appendix~\ref{sec:synthetic}\onlineVersion{\useOnline}{~of~\cite{multihop22}} this is true for both fixed hop-by-hop and end-to-end deadlines, albeit for different reasons.}, it does not offer much insight into the structure of optimal reprofiling decisions.  As a result, we concentrate on reporting the benefits of reprofiling (Greedy) for two more realistic configurations.

The first configuration is a vehicle network operating according to the TSN standard.  The second mimics a network connecting multiple datacenters and the traffic flows between them.  For both, the benefits of Greedy's reprofiling decisions are compared to the two baselines, FR and NR.

\highlight{\useColor}{In reporting results,} the focus is on the bandwidth improvements that Greedy yields over both baselines, with, as before, error bars giving the $95\%$ confidence intervals.  \highlight{\useColor}{Of interest is that Greedy's outcome often falls between the two extremes of NR and FR.  This demonstrates not only the benefits of reprofiling, but more importantly that they involve a trade-off between making flows smoother and preserving scheduling flexibility in the network.  Because of space constraints, results on buffer bounds are relegated to Appendices~\ref{app:buffer_tot} and~\ref{app:buffer_distribution}\onlineVersion{\useOnline}{~of~\cite{multihop22}}.} 

\subsection{Application-Derived Topologies and Flow Profiles}
\label{sec:realistic}

Exploring how reprofiling performs in a more realistic setting calls for identifying representative network topologies and creating a traffic mix composed of flows with profiles derived from actual application traffic.  We consider two application scenarios; one reflective of a TSN deployment,
and the other that arises in an inter-datacenter network setting.

\subsubsection{Time Sensitive Networking (TSN)}
\label{sec:eval_tsn}
\paragraph{Representative Network Topology}
\label{sec:net_topo_tsn}

We first consider a TSN-inspired network setting. The Orion Crew Exploration Vehicle (CEV) network~\cite{paulitsch2011time} is the instantiation we use for our evaluation.  Fig.~$3$ of~\cite{tuamacs2014optimization} illustrates the network that consists of $31$ end devices and $13$ switches connected by $47$ bidirectional links, each viewed as two independent unidirectional links for bandwidth accounting.  Every end device is a possible source or destination for traffic flows.  We note that SCED is not a scheduler included in the TSN standard, so that our focus is on evaluating the benefits of reprofiling (for SCED) in a setting where hard delay bounds are relevant.

Paths between sources and destinations (S-D pairs) are minimum hop paths.  When multiple choices exist, one is randomly selected as the default.  In addition, when generating a set of paths, we omit all instances of single hop paths.  This is to avoid bias and noise when comparing Greedy to FR and NR, respectively\footnote{Recall that the EDF policy of SCED is optimal on a single link.}.

\paragraph{Flow Profiles}
\label{sec:traffic_tsn}

The TSN standard specifies several traffic classes, including a Control-Data Traffic (CDT) class with stringent latency requirements, and classes A and B with looser latency targets.  The traffic characteristics of those three classes are given in Table~$2$ of~\cite{thangamuthu2015analysis} through their maximum frame size, minimum frame inter-arrival time and maximum end-to-end delay target. We use that information to create traffic profiles $(r,b,d)$ for flows from all three classes.

End-to-end deadlines $d$ for classes CDT, A, and B are set to $0.1, 2$, and $50$ms, respectively, directly from Table~$2$ of~\cite{thangamuthu2015analysis}.

Specifying the tuples $(r,b)$ requires some intermediate steps.  We first specify average flow rates for all three classes. We do so by relying on the frame sizes of Table~$2$ of~\cite{thangamuthu2015analysis} ($128,256,$  and $256$~bytes for classes CDT, A, and B respectively), and selecting frame inter-arrival times randomly from among a discrete set of values that range from the minimum values of~\cite{thangamuthu2015analysis} ($500\mu$s, $125\mu$s and $250\mu$s for classes CDT, A, and B respectively) up to maximum values a couple of orders of magnitudes larger. Once a flow's average rate is set, we rely on the approach of~\cite{srtm21} to determine suitable token bucket parameters $(r,b)$. We assume that frames arrive according to a Poisson process, and choose the flow's token rate $r$ to be $10\%$ higher than the average flow rate, with $b$ then set equal to $25$ frame sizes.  This combination ensures a $99^{th}$ percentile of the token bucket access delay of approximately zero.

\paragraph{Scenarios}
\label{sec:scenario_tsn}

Following again the setup of~\cite{thangamuthu2015analysis}, we sample applications and their associated flow profiles from a mixture of the three traffic classes, CDT, A, and B, using a sampling ratio of 1:4:4.  Further, because in a TSN setting flows can be either unicast, multicast, or broadcast, we select each transmission mode equally for each flow. A flow's source is first chosen randomly from the 31 end devices in the Orion CEV network.  In the case of unicast and multicast, the destinations are also selected randomly from the remaining nodes, with, in the multicast case, the number of distinct destinations itself random between $2$ and $29$.  Hence, an application maps to, on average, $15.5$ traffic flows or distinct S-D pairs. 

Each experiment consists of $1000$ instances of a given scenario.  Results are reported across experiments involving an increasing number of applications (from $10$ to $200$ in steps of $10$), and therefore a correspondingly increasing expected number of flows (S-D pairs).  For scalability, traffic flows from the same class and S-D pair are aggregated.

\paragraph{Results}
\label{sec:realistic_results_tsn}

\begin{figure}[!h]
\centering
\begin{subfigure}{0.49\linewidth}
  \centering
  \includegraphics[width=\linewidth]{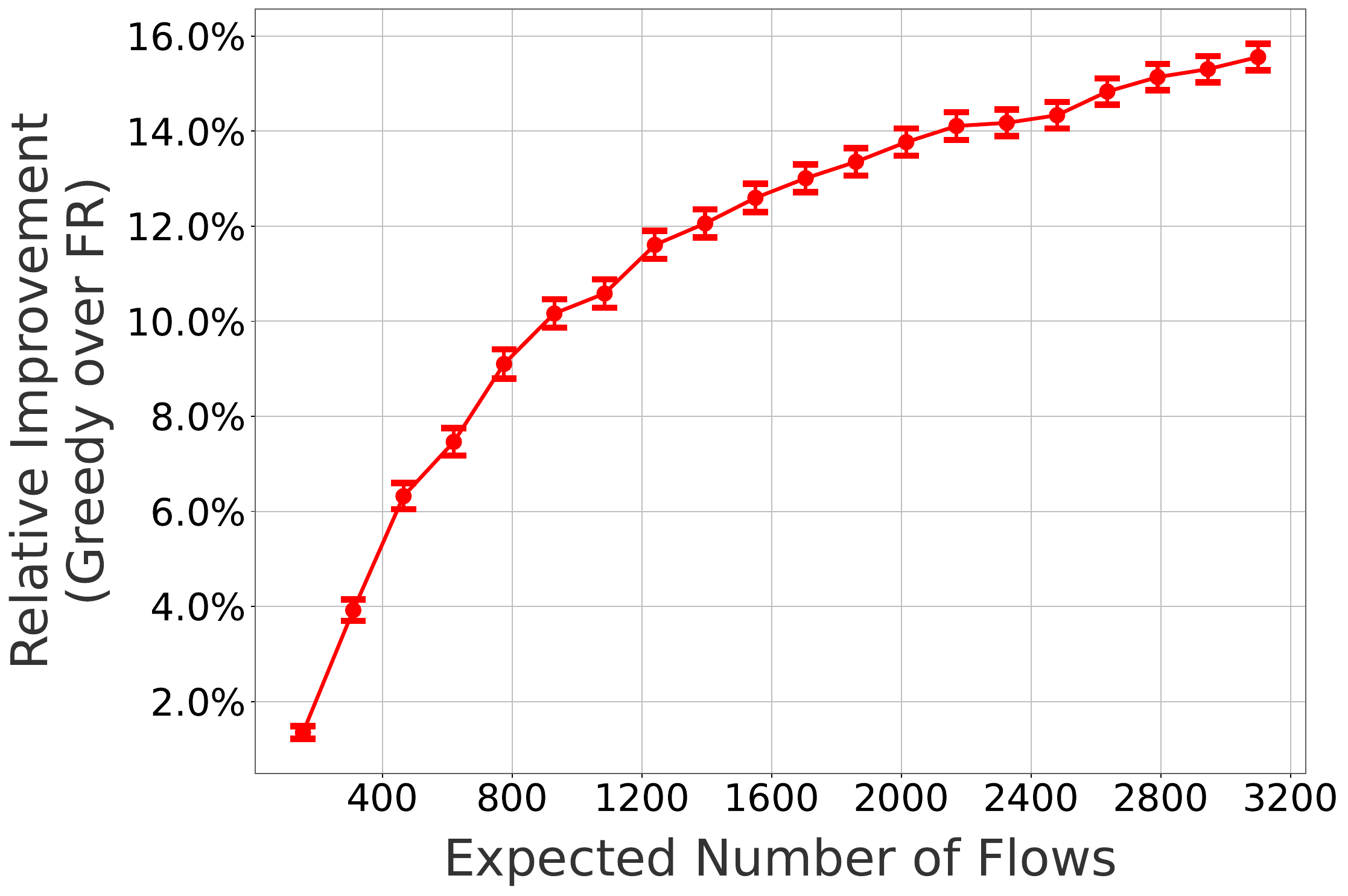}
  \caption{Greedy over FR}
  \label{fig:real_fr_tsn}
\end{subfigure}
\begin{subfigure}{0.49\linewidth}
  \centering
  \includegraphics[width=\linewidth]{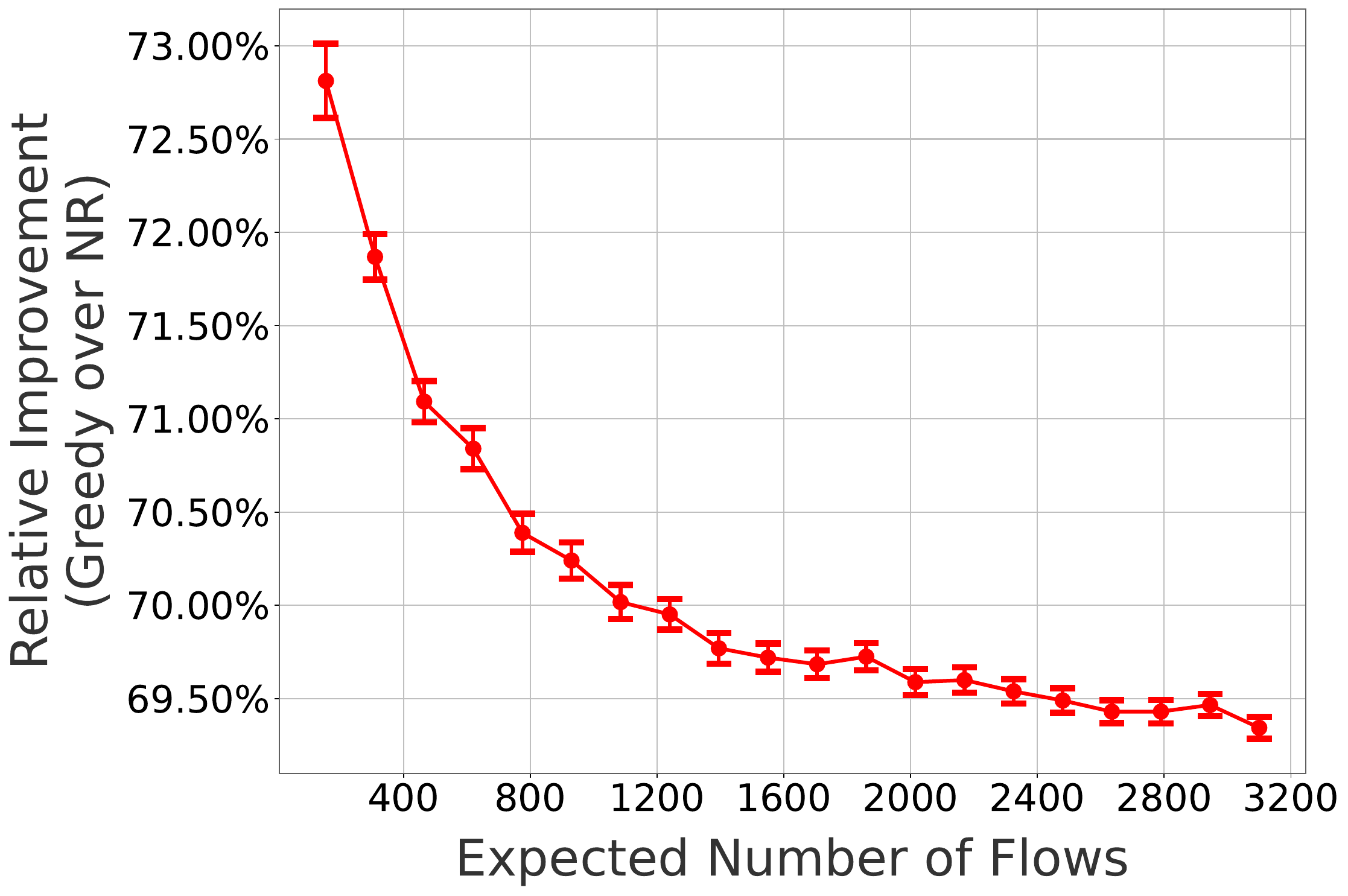}
  \caption{Greedy over NR}
  \label{fig:real_nr_tsn}
\end{subfigure}
\caption{Greedy's bandwidth improvements for Orion CEV.}
\label{fig:improvement_realistic_tsn}
\end{figure}

\figs{fig:real_fr_tsn}{fig:real_nr_tsn} report, as a function of the expected number of flows, the bandwidth improvements of Greedy over \highlight{\useColor}{the two baselines,} FR and NR, respectively. The figures show that on the Orion CEV network, Greedy yields bandwidth improvements of up to $16\%$ over FR and $73\%$ over NR.

Having significantly larger improvements over NR than FR is consistent with our intuition from small configurations.  Consider a simple scenario with a single flow crossing two hops.  In this basic configuration, both FR and Greedy produce the same solution, while NR requires twice as much bandwidth\footnote{Each hop receives only half the total delay budget, and the original flow profile is assumed at both.  Note that avoiding this penalty was the motivation behind the RCSD policies of~\cite{georgiadis96a} that essentially emulate FR.}.  Hence, in this single flow setting, Greedy generates an improvement of $0\%$ over FR and of $50\%$ over NR.

We also note that Greedy's improvements over FR increase with the number of flows.  This is again intuitive, as more flows means more deadline combinations, and correspondingly scheduling flexibility that Greedy can leverage as it seeks to balance the benefits of smoother flows and scheduling flexibility. In contrast, FR largely forfeits this option by allocating the full delay budget to making flows smoother. Hence, Greedy outperforms FR as its ability to leverage scheduling flexibility increases with the number flows.

Analyzing the improvements of Greedy over NR as the number of flows varies is more challenging. It involves a complex trade-off between how the number of flows affects the ``error'' of NR's end-to-end delay bounds and its ability to leverage greater scheduling flexibility, all of which depend on the exact combinations of flow profiles.  Hence, the main conclusion of \fig{fig:real_nr_tsn} is that Greedy consistently outperforms NR, irrespective of the number of flows.

Finally, as the number of flows increases, the improvements of Greedy over both FR and NR eventually stabilize.  This is because flows associated with the same S-D pair (and therefore path) and from the deadline class can be aggregated.  Hence, given that the number of S-D pairs is bounded, once the number of flows is large enough, traffic mixes on all links converge to a steady-state configuration that in turn yields a certain level of improvements.

\begin{figure}[!h]
\centering
\begin{subfigure}{0.49\linewidth}
  \centering
  \includegraphics[width=\linewidth]{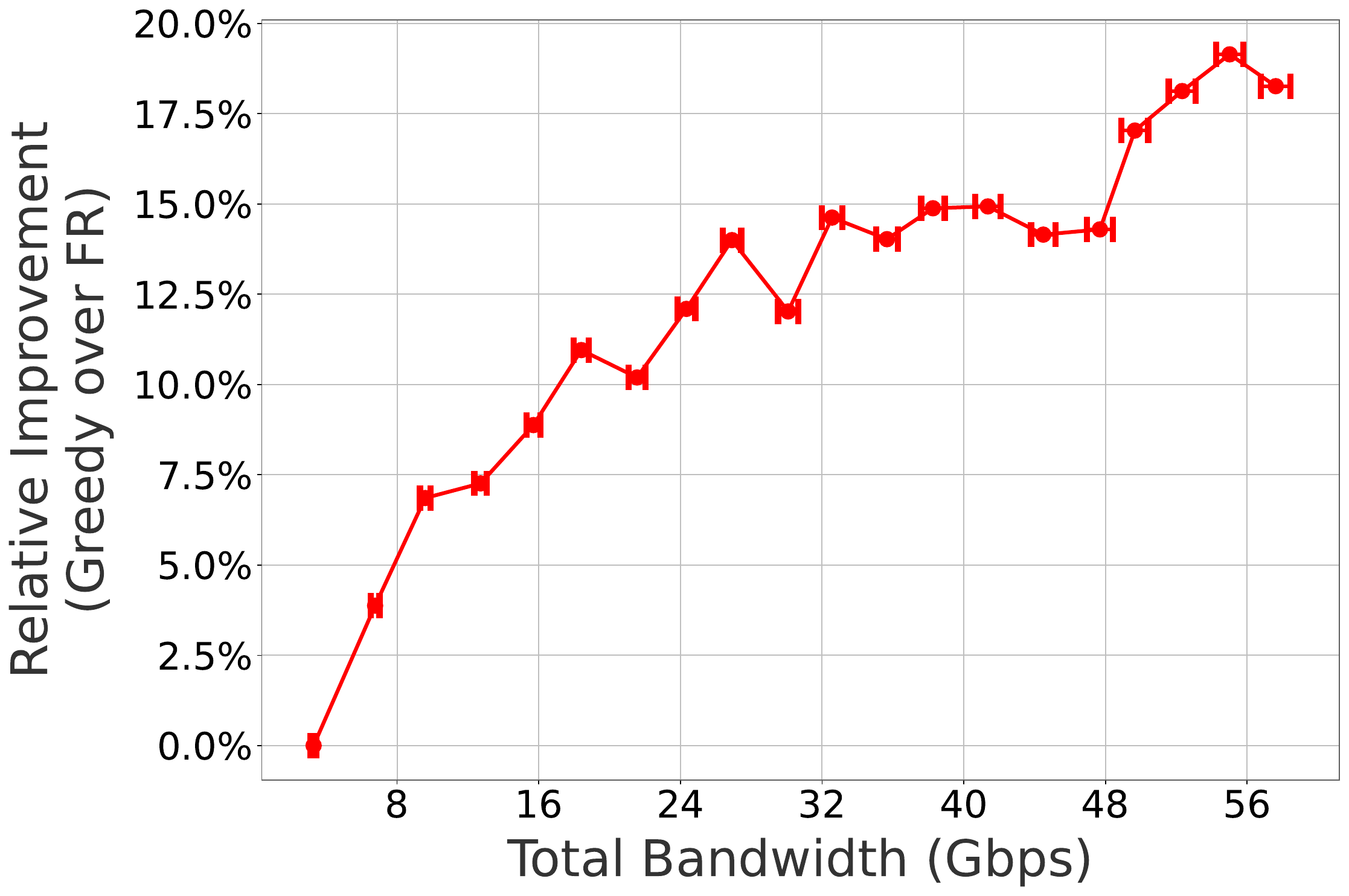}
  \caption{Greedy over FR}
  \label{fig:real_fr_flow_tsn}
\end{subfigure}
\begin{subfigure}{0.49\linewidth}
  \centering
  \includegraphics[width=\linewidth]{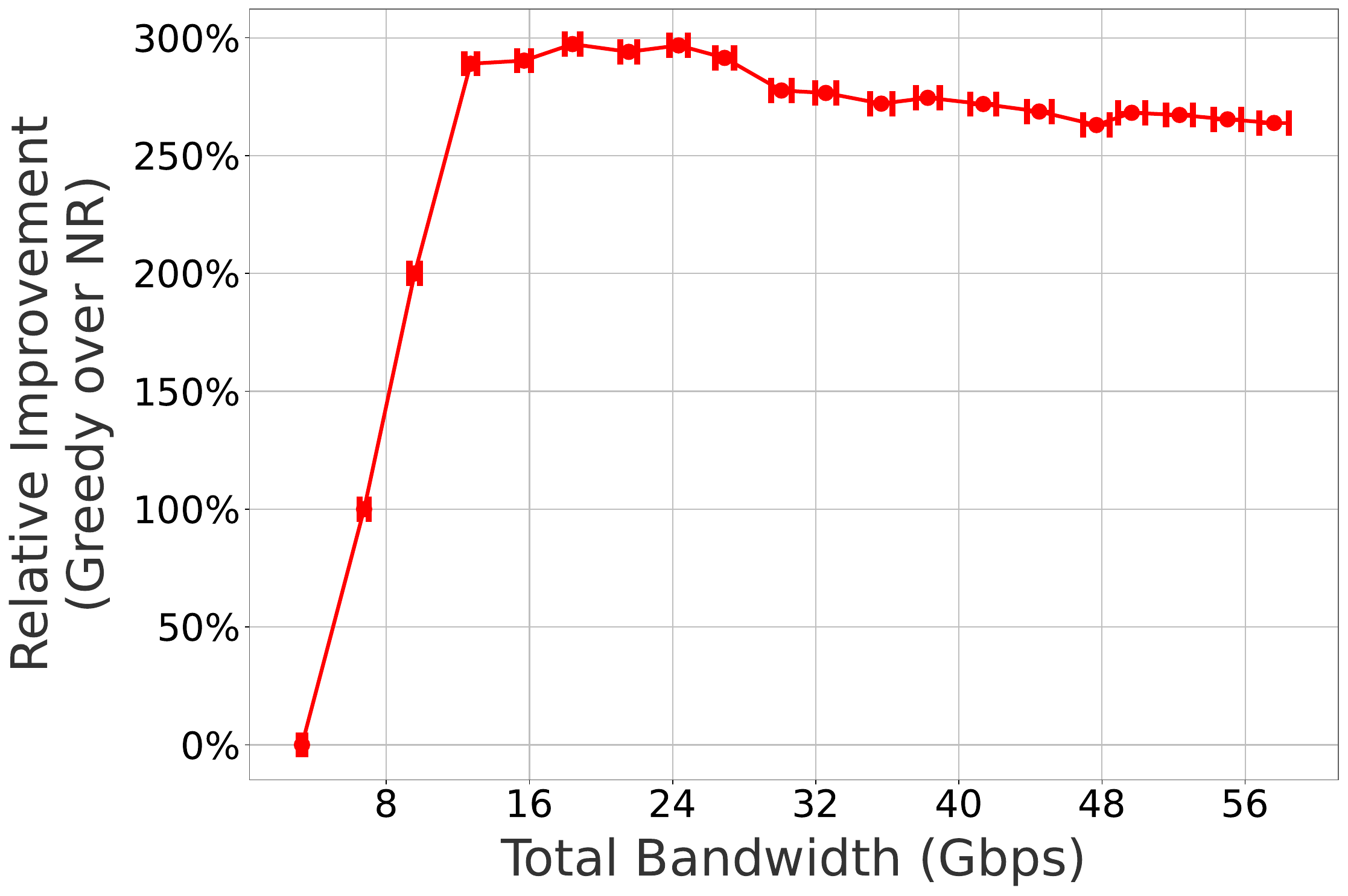}
  \caption{Greedy over NR}
  \label{fig:real_nr_flow_tsn}
\end{subfigure}
\caption{Greedy's improvements in number of flows accommodated for Orion CEV.}
\label{fig:improvement_realistic_flow_tsn}
\end{figure}

\fig{fig:improvement_realistic_flow_tsn} demonstrates the benefit of reprofiling from a different perspective.  It shows Greedy's improvements in number of flows accommodated compared to FR and NR, \emph{assuming that the network bandwidth is fixed}. In other words, it captures the additional traffic (number of flows) a \emph{given} network can carry from relying on Greedy's reprofiling solution. The figure shows that reprofiling allows about $20\%$ more flows than FR and $300\%$ more flows than NR, which is consistent with the bandwidth improvements results from \fig{fig:improvement_realistic_tsn}.  Specifically, when the number of flows across classes is large enough, the bandwidth required by an ``average'' flow can be approximated by dividing the total network bandwidth by the number of flows it carries. Under such an assumption, a relative bandwidth improvement of $x$ can then be shown\footnote{See Appendix~\ref{sec:linear_assumption}\onlineVersion{\useOnline}{~of~\cite{multihop22}}
for details.} to yield a relative improvement of $\frac{x}{1-x}$ in the number of flows carried.

\begin{figure}[!h]
\centering
\includegraphics[width=0.6\linewidth]{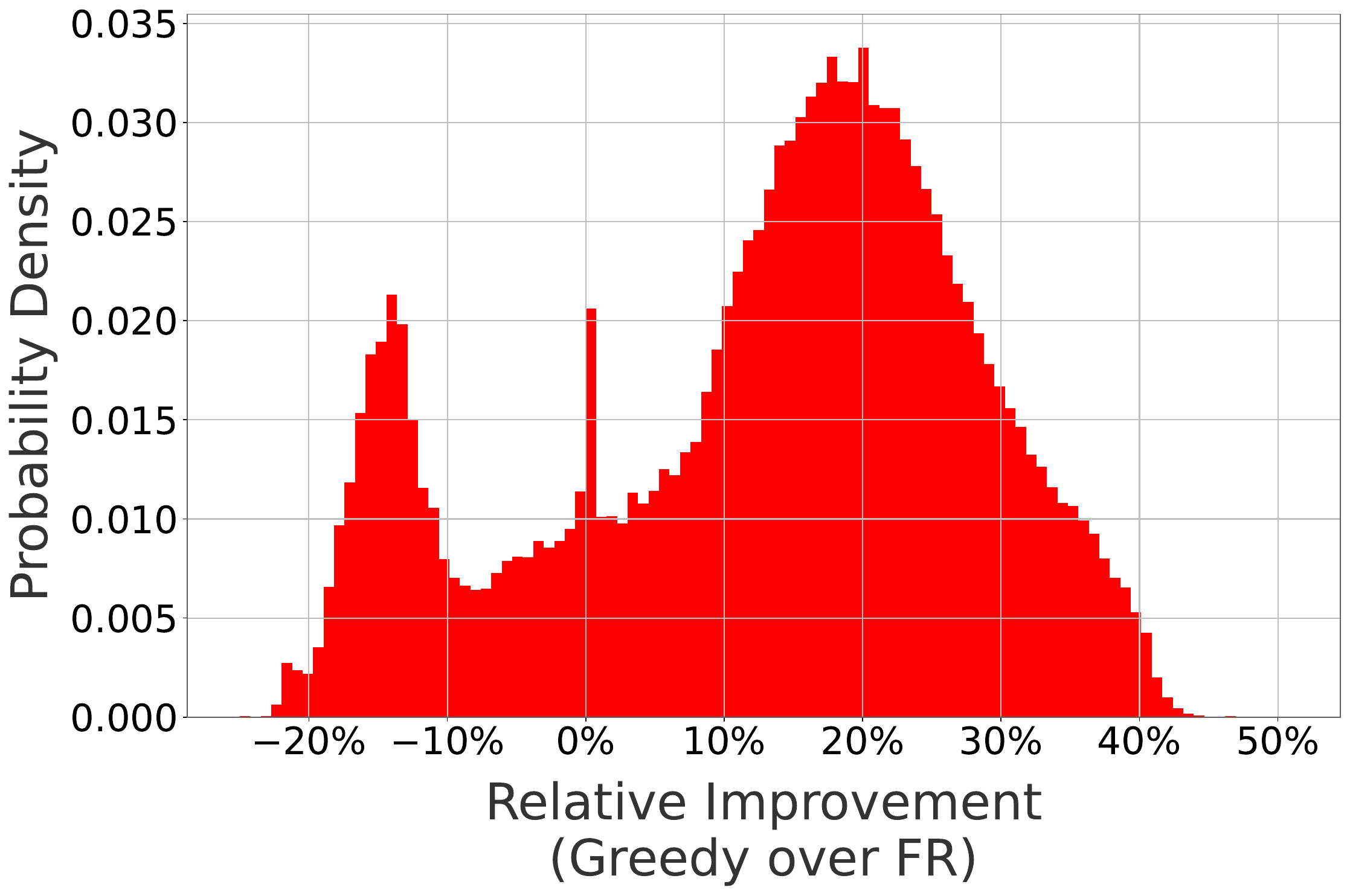}
\caption{Distribution of Greedy's bandwidth improvements over FR across the links of Orion CEV ($3100$~flows).}
\label{fig:improvement_realistic_dist_tsn}
\end{figure}

\fig{fig:improvement_realistic_dist_tsn} offers additional insight into how Greedy's improvements are realized.  It plots for a $3100$~flows configuration, the distribution (over the 1000 instances of the experiment) of the relative \emph{per link} improvements of Greedy over FR across the $47$ directional links of the Orion CEV network.  It shows that while most links have a lower bandwidth than under FR, this is not true for all links.  This is because 
differences in traffic mix at each hop can affect how much the relative benefits of greater scheduling flexibility matter.  In some cases, a (locally) sub-optimal performance on a given link affords larger savings on other links, or savings on multiple links.

\begin{figure}[!h]
\centering
\includegraphics[width=0.6\linewidth]{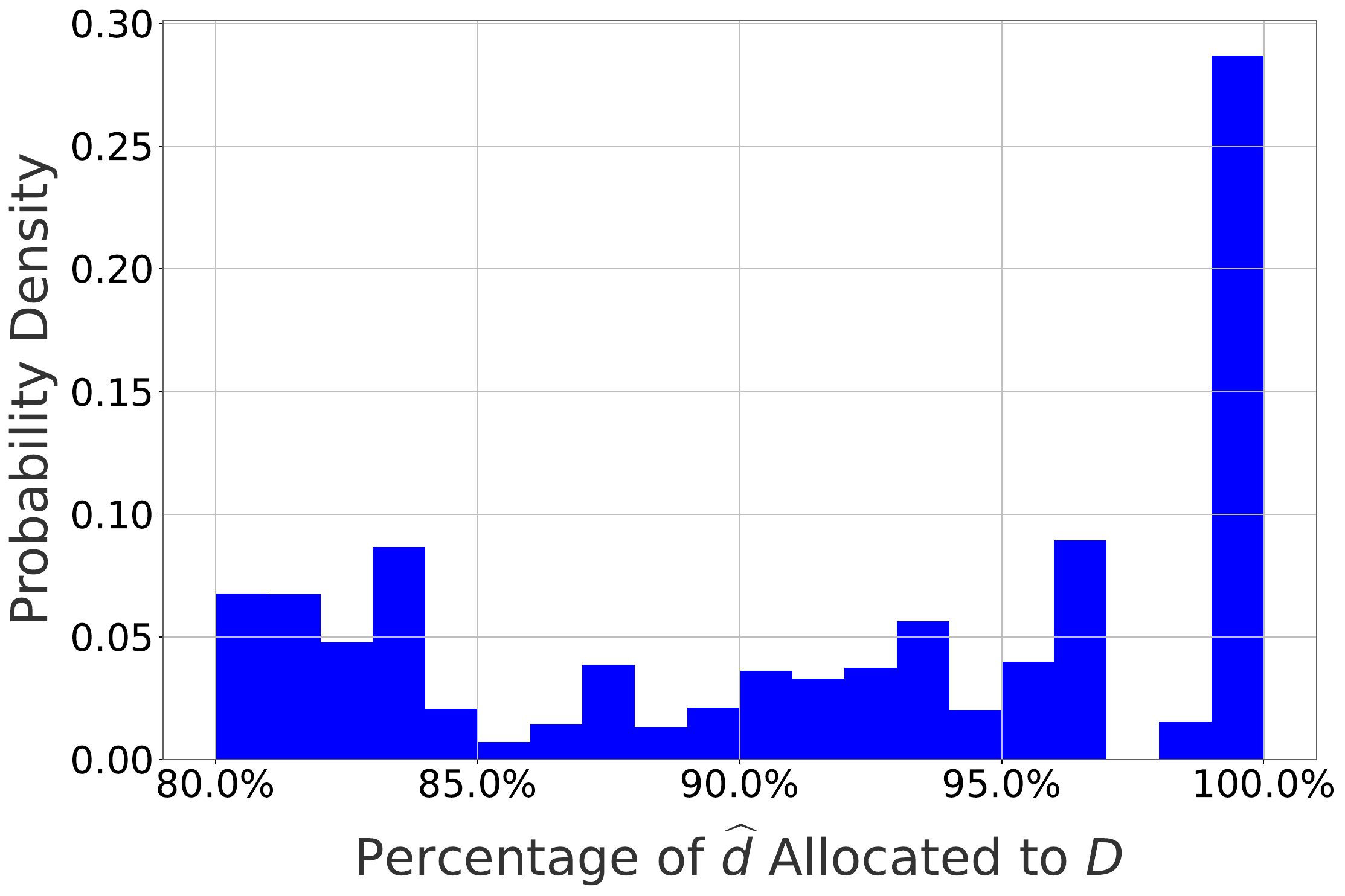}
\caption{Distribution of Greedy's reprofiling ratio $D/\widehat{d}$ on Orion CEV ($3100$ flows).}
\label{fig:real_dist_tsn}
\end{figure}

\fig{fig:real_dist_tsn} offers another perspective. It reports 
the distribution across flows of the fraction of the maximum reprofiling delay $\widehat{d}$ used. The figure shows that, while some flows are fully reprofiled (the mode at $100\%$), Greedy also uses intermediate solutions that preserve some scheduling flexibility, which helps it outperform FR. This confirms the existence of a ``sweet spot'' in the trade-off between tighter network delays (because of the added reprofiling delay), and the benefits of smoother traffic.

\begin{figure}[!h]
\centering
\begin{subfigure}{0.49\linewidth}
  \centering
  \includegraphics[width=\linewidth]{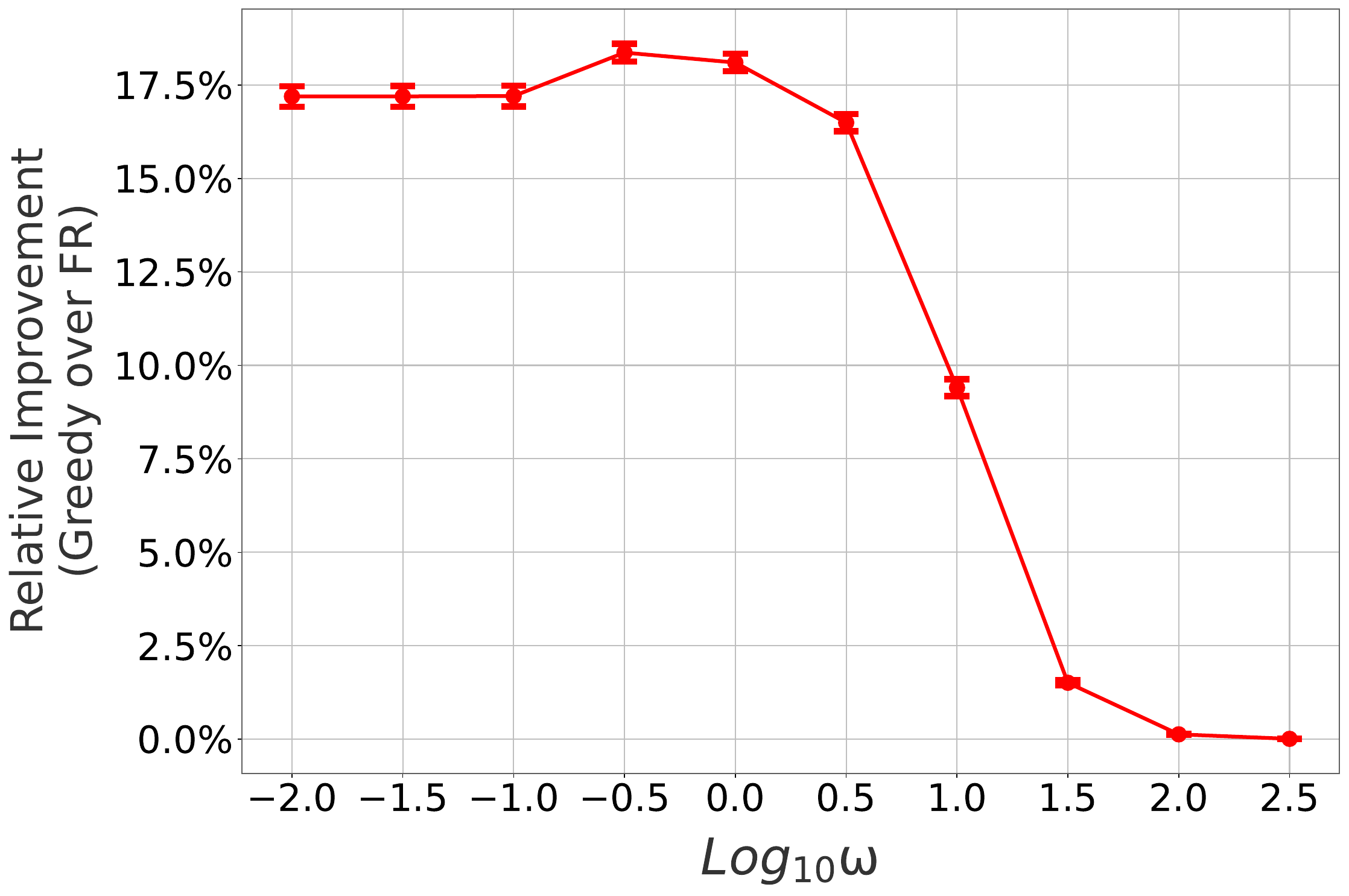}
  \caption{Greedy over FR}
  \label{fig:deadline_improvement_tsn}
\end{subfigure}
\begin{subfigure}{0.48\linewidth}
  \centering
  \includegraphics[width=\linewidth]{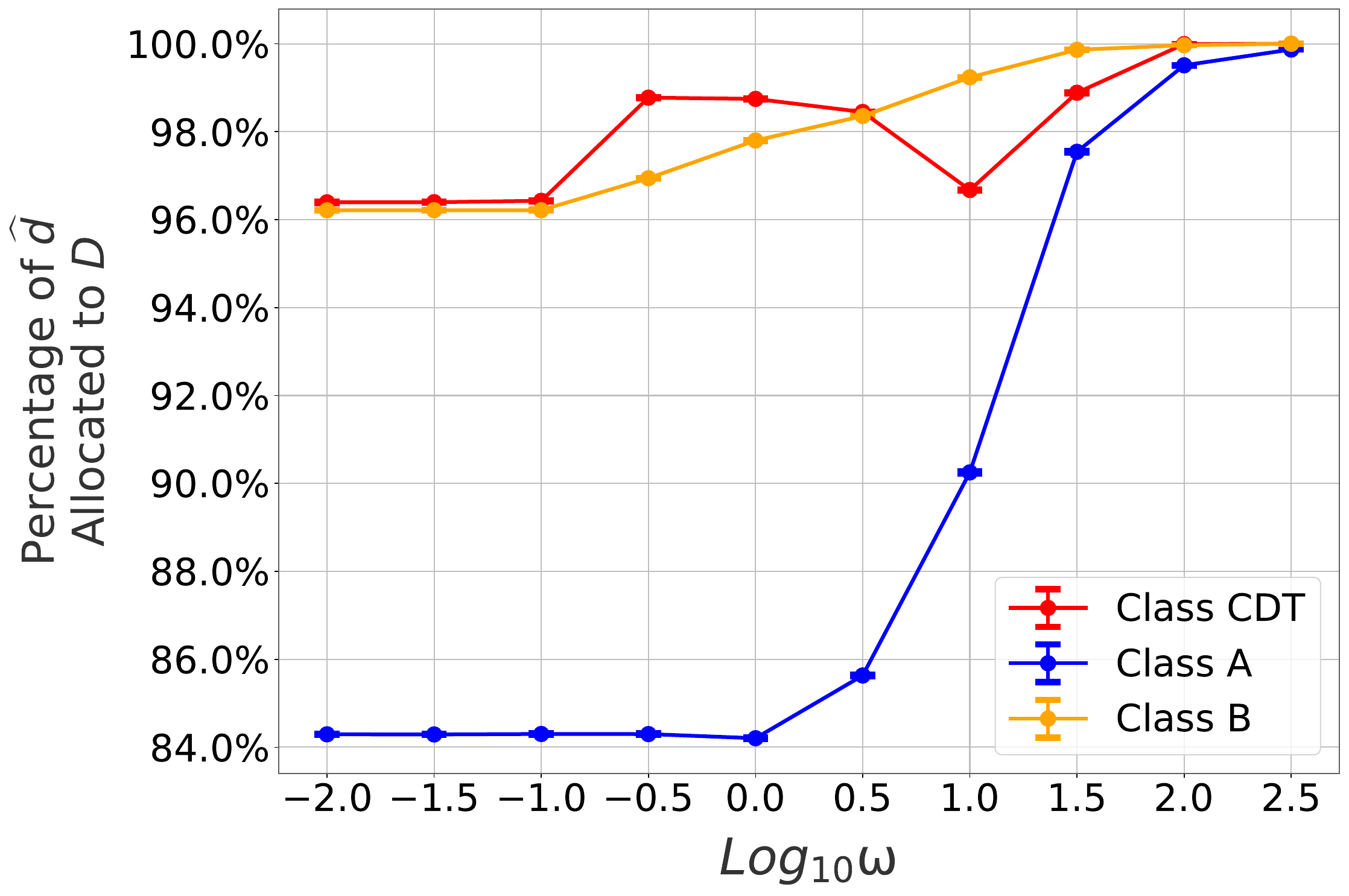}
  \caption{reprofiling ratio $D/\widehat{d}$}
  \label{fig:deadline_reprofiling_tsn}
\end{subfigure}
\caption{Reprofiling's behavior on Orion CEV ($3100$~flows) as a function of deadline scaling ($\omega$).}
\label{fig:deadline_realistic_tsn}
\end{figure}

Although our choice of deadlines was inspired by Table~$2$ of~\cite{thangamuthu2015analysis}, it is of interest to explore the sensitivity of the results to different choices. 
We do so using again the $3100$~flows configuration but scaling the deadlines of the three traffic classes by a common factor $\omega\in [10^{-2}, 10^{2.5}]$.

\fig{fig:deadline_improvement_tsn} reports the relative improvements in bandwidth of Greedy over FR as deadlines vary (the comparison to NR is omitted as it consistently under-performs FR).  The $y$-axis is the relative bandwidth reduction of Greedy over FR, while the $x$-axis is the $\log_{10}$ of the deadlines' scaling factor $\omega$. 
The figure shows that, as deadlines increase, Greedy's improvements over FR eventually decrease. This is expected since large deadlines allow all flows to be reprofiled at their token rate, \ie $R=r$,
in which case Greedy and FR converge to the same solution. 
In contrast, tighter deadlines result in increasing improvements that, for the Orion CEV network,
stabilize\footnote{Note though that once deadlines reach~$0$, the required bandwidth is infinite, so that no improvements are then possible.} at about $17\%$.

\fig{fig:deadline_reprofiling_tsn} investigates possible causes for those improvements.
It plots the reprofiling ratio $D/\widehat{d}$ for the three traffic classes, CDT, A, and B, as $\omega$ varies.
As expected, class CDT has a consistently high reprofiling ratio.  This is because it has the smallest deadline target
and, as mentioned in Section~\ref{sec:adjust}, the flow with the smallest $T'_{ij}$ (sum of local deadline and reprofiling delay) can always be fully reprofiled during Greedy's adjustment phase.  A similar phenomenon is observed for class B, albeit for a different reason.  Since class CDT and class~A have smaller deadlines and burst sizes no larger than those of class~B, class~B can keep trading local deadlines for reprofiling without depleting the slack of the other two classes in the adjustment phase\footnote{Recall that the adjustment phase considers the largest deadline flows first.}.

The situation is different for class A.  When $\omega$ is large, CDT's reprofiling rate is set to $R=r$, which eliminates all burstiness.  This changes as $\omega$ decreases. $R$ gets larger and CDT's burstiness starts to increase. It then becomes possible for class~A to use CDT's bandwidth once CDT's initial burst subsides.  Realizing this opportunity calls for a smaller reprofiling ratio to reintroduce local deadlines that allow class A to wait for CDT's additional bandwidth to become available.

\begin{figure}[!h]
\centering
\includegraphics[width=0.6\linewidth]{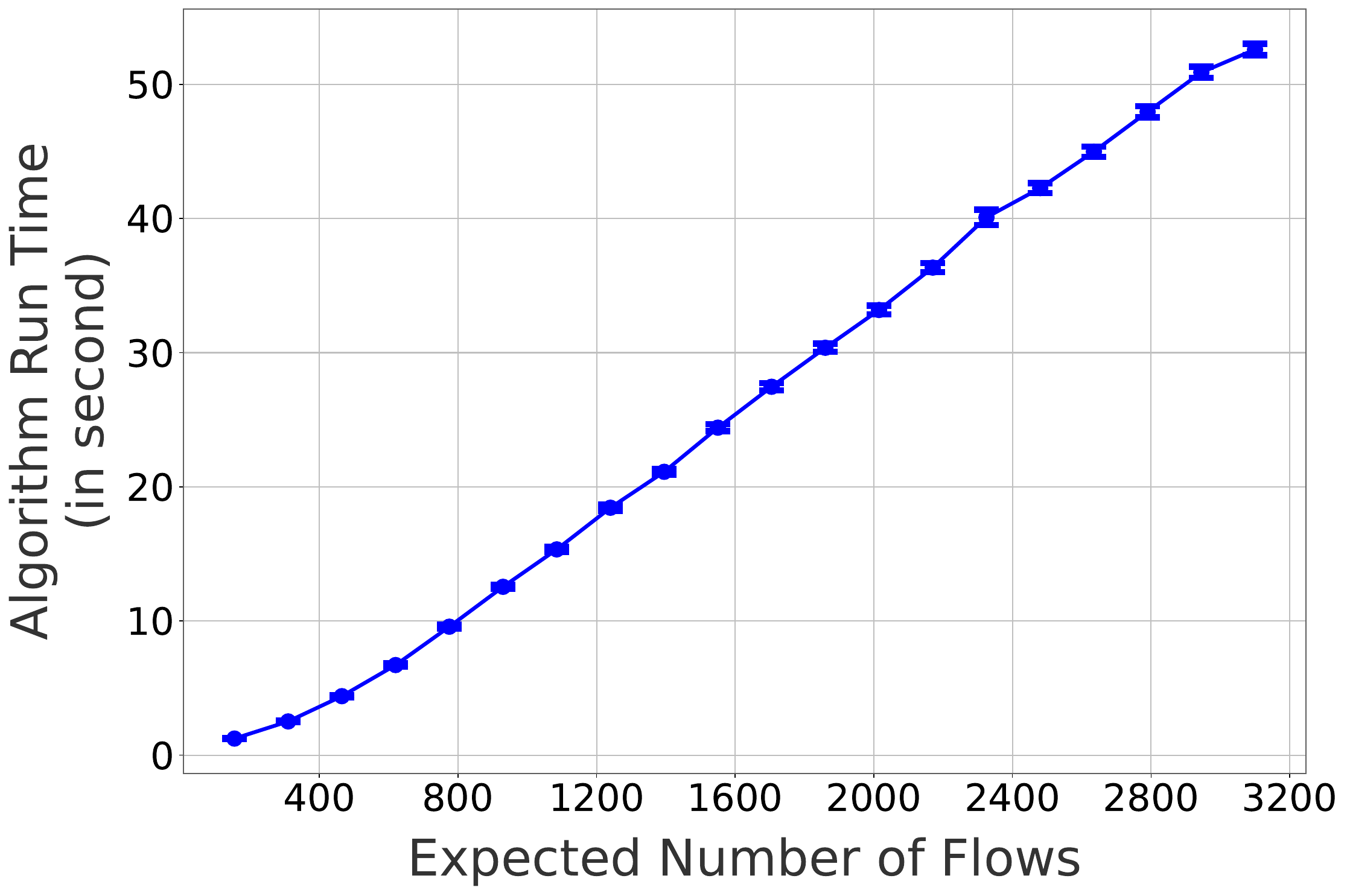}
\caption{Greedy's run time on Orion CEV.}
\label{fig:runtime_realistic_tsn}
\end{figure}

Finally, \fig{fig:runtime_realistic_tsn} is intended to demonstrate the scalability of Greedy.  Recall that, as mentioned in Section~\ref{sec:scenario_tsn}, flows sharing the same path and traffic class are aggregated.  As a result, the number of (aggregate) flow combinations that Greedy needs to consider is upper-bounded by the number of distinct paths in the network times the number of traffic classes (end-to-end deadline classes).
Although this upper-bound is not reached with the largest $(3100)$ number of flows we consider, the corresponding run time of about $50$~secs obtained for the Orion CEV network is reasonable especially in light of the low-end server on which the computations run.

\subsubsection{Inter-Datacenter Network}
\label{sec:eval_inter_dc}
\paragraph{Representative Network Topology}
\label{sec:net_topo}

\begin{figure}[!h]
\centering
\includegraphics[width=0.6\linewidth]{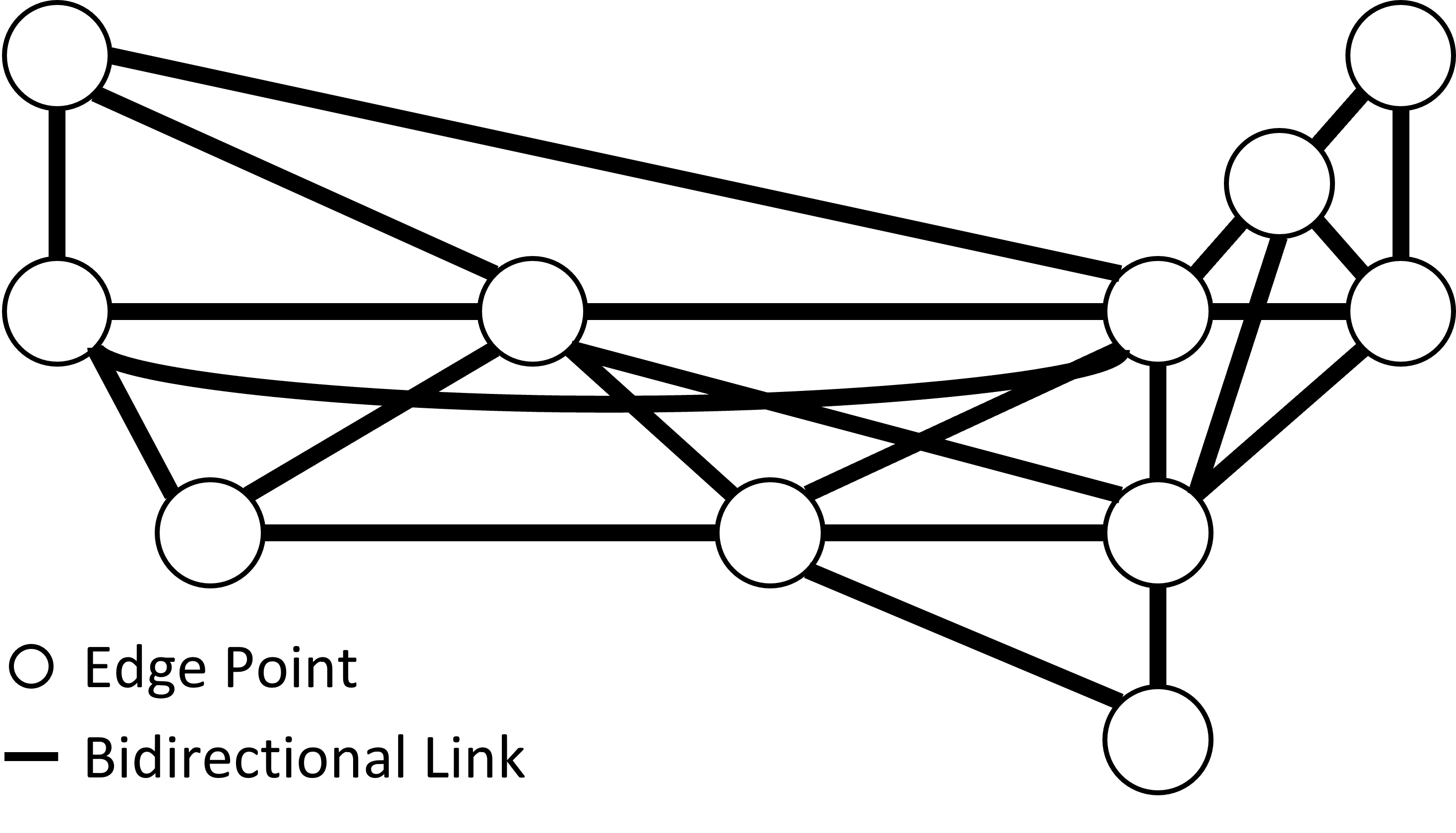}
\caption{US-Topo network topology (from~\cite{gcp-loc}).}
\label{fig:google_network}
\end{figure}

This scenario targets a topology representative of networks connecting large datacenters, and mimics the North-American network connecting the (edge) datacenters of a major cloud provider~\cite{gcp-loc}.  The resulting topology (US-Topo) is shown in \fig{fig:google_network} and consists of $11$ nodes and $23$ bidirectional links.  Sources and destinations are selected from the $11$ nodes and, as with the Orion CEV network, we use minimum hop routing and only multi-hop paths are considered.

\paragraph{Flow Profiles}
\label{sec:traffic}

Creating realistic traffic mixes is challenging and few studies provide the necessary information.  Other works targeting datacenter network optimization~\cite{qjump15,silo15,chameleon20} have faced a similar problem, and~\cite{chameleon20} introduced a set of candidate flow profiles in its own evaluation.  Those profiles unfortunately focus on \emph{intra} datacenter traffic, so that most deadlines can be met by guaranteeing a flow its long-term rate, \ie flows can be reprofiled down to their average rate. This does not affect the evaluation of~\cite{chameleon20} that is focuses on the benefits of fine-grained routing, but is a poor fit for our investigation of the benefits of reprofiling.  As a result, we opt to extract profiles from a study of the traffic traversing segments of a large network connecting datacenters~\cite{roy15}.

\begin{figure}[!h]
\centering
\includegraphics[width=0.6\linewidth]{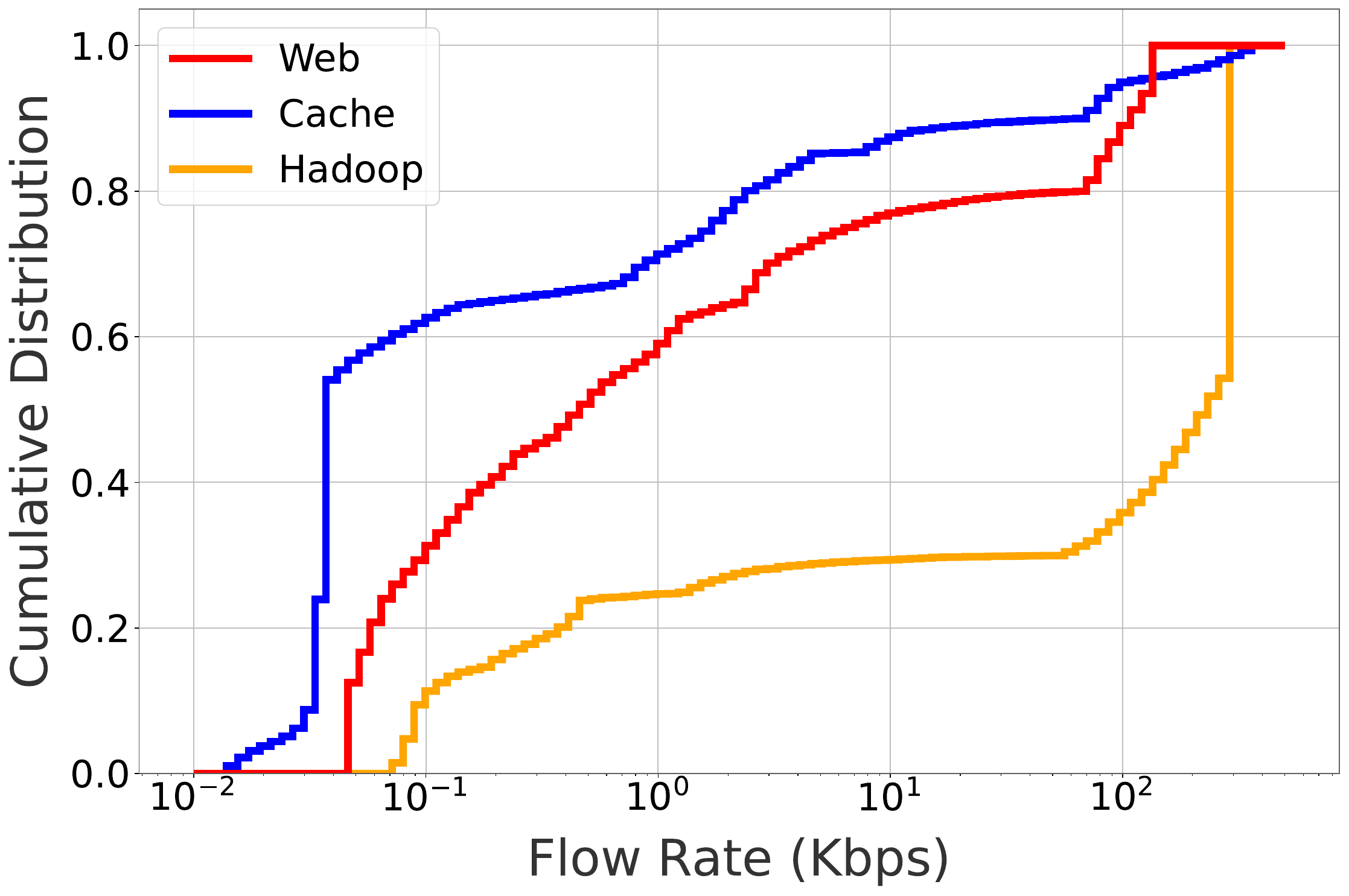}
\caption{CDF of flow rate $r$ for Web, Cache, and Hadoop applications derived from~\cite{roy15}.}
\label{fig:rate_cdf}
\end{figure}

We selected three applications from~\cite{roy15}: Web (W), Cache read and replacement (C), and Hadoop (H). Fig.~$4$ of~\cite{roy15} reports their traffic contributions to inter-datacenter links, with Figs.~$6$ and~$7$ reporting the distributions of their flow sizes and durations.  Because those distributions are reported separately, additional assumptions are needed to create complete profiles.  Specifically, we assumed perfectly positively correlated flow sizes and durations, \ie larger flows last longer than smaller flows, and sample flow size and duration pairs jointly from their respective distributions. This then readily yields cumulative distribution functions (CDFs) for the average (token) rates $r$ of the three applications (see \fig{fig:rate_cdf}).

In choosing bucket sizes, we assume that the smallest flow sizes of each application, $S_W=0.15$Kbytes, $S_C=0.4$Kbytes, and $S_H=0.3$Kbytes, correspond to the transmission of an isolated burst.  In determining overall burstiness, we rely on the statement from~\cite{roy15} that flows from these applications tend to be internally bursty, with Cache significantly burstier than Hadoop, and Web somewhere in between.  To limit access delay, token bucket sizes are configured to accommodate multiple successive bursts.  Specifically, $b$ is drawn randomly from $[0,20S_C]$, $[0,10S_W]$, and $[0,2S_H]$, for Cache, Web, and Hadoop, respectively.

Finally, we set the end-to-end deadlines of Web, Cache, and Hadoop to $10$ms, $50$ms, and $200$ms, respectively. This is motivated by discussions in~\cite{roy15} regarding the services relying on them, \ie Web search, user data query, and offline analysis (\eg data mining).

\paragraph{Scenarios}
\label{sec:scenario}
We specify possible traffic matrices by selecting a traffic mix of Web, Cache, and Hadoop in the ratio 3:9:1, which seeks to capture their relative usage across services\footnote{Loosely inspired from Table~$4$ of~\cite{roy15}.}.  The number of flows is then varied from~$50$ to~$3000$ in steps of~$50$, with again, for a given experiment (number of flows), $1000$ random instances generated for each result.  Traffic from the same application and S-D pair is again aggregated for scalability.

\paragraph{Results}
\label{sec:realistic_results}

\begin{figure}[!h]
\centering
\begin{subfigure}{0.49\linewidth}
  \centering
  \includegraphics[width=\linewidth]{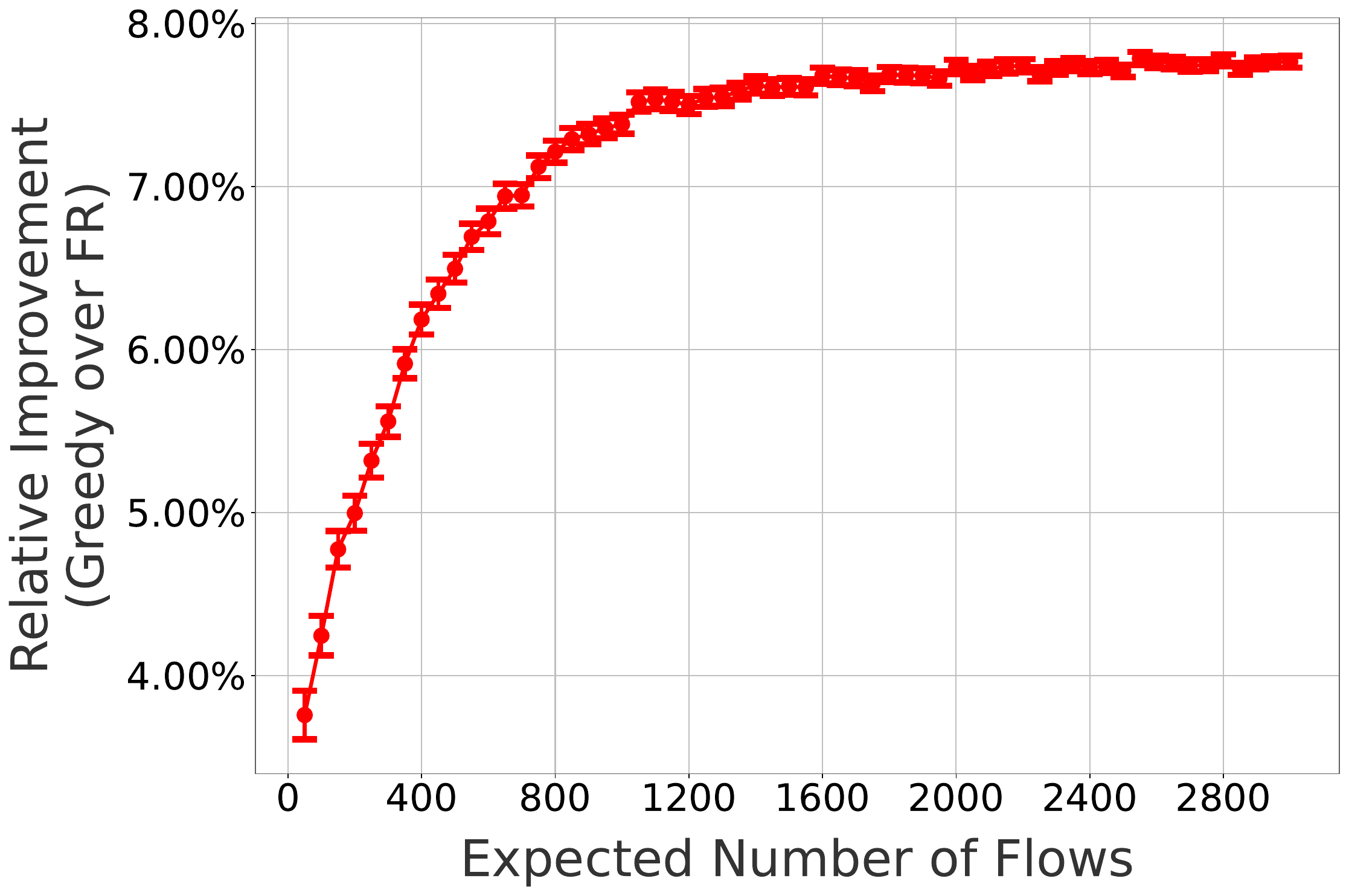}
  \caption{Greedy over FR}
  \label{fig:real_fr}
\end{subfigure}
\begin{subfigure}{0.49\linewidth}
  \centering
  \includegraphics[width=\linewidth]{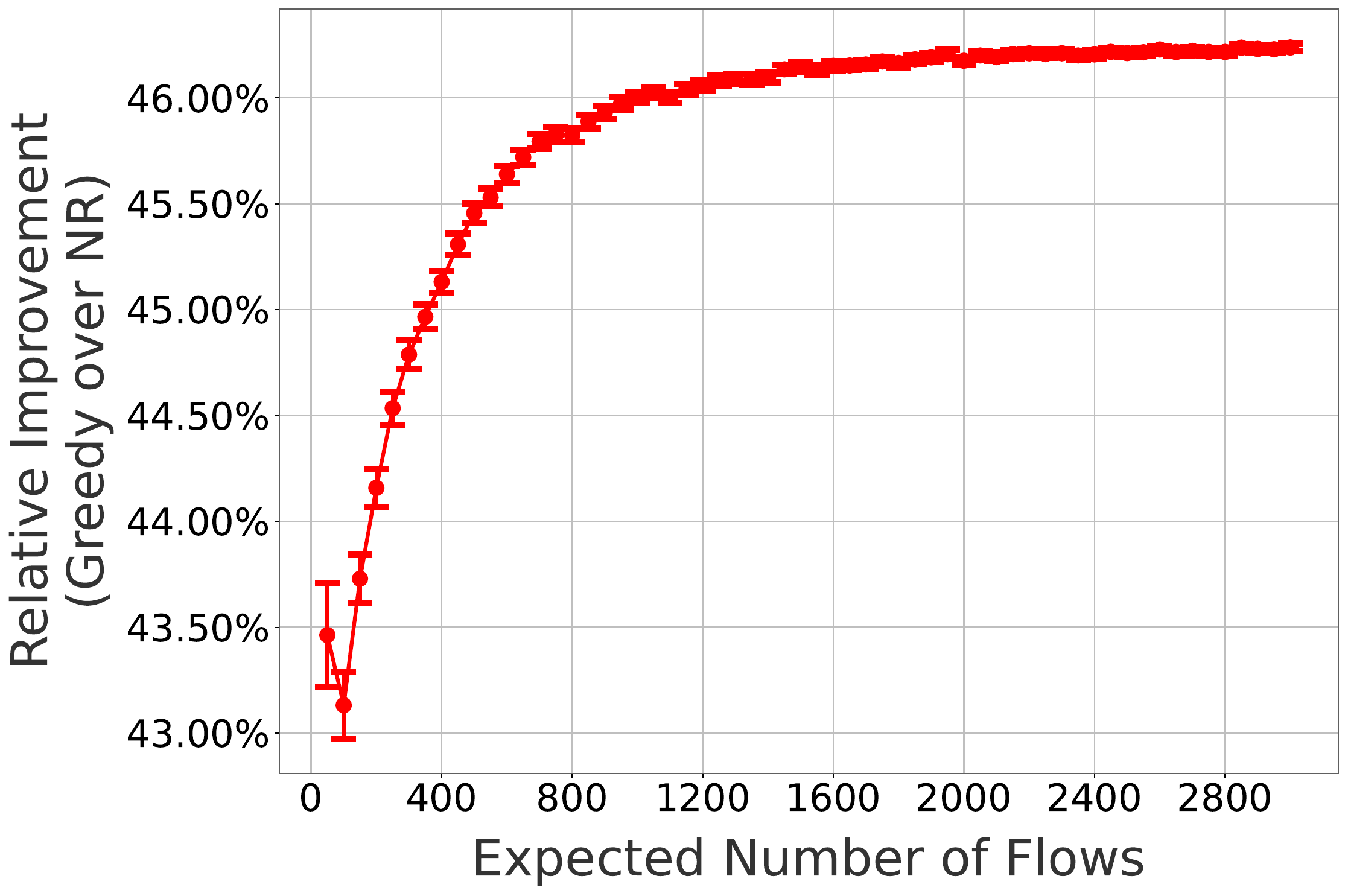}
  \caption{Greedy over NR}
  \label{fig:real_nr}
\end{subfigure}
\caption{Greedy's bandwidth improvements for US-Topo.}
\label{fig:improvement_realistic}
\end{figure}

\begin{figure}[!h]
\centering
\begin{subfigure}{0.49\linewidth}
  \centering
  \includegraphics[width=\linewidth]{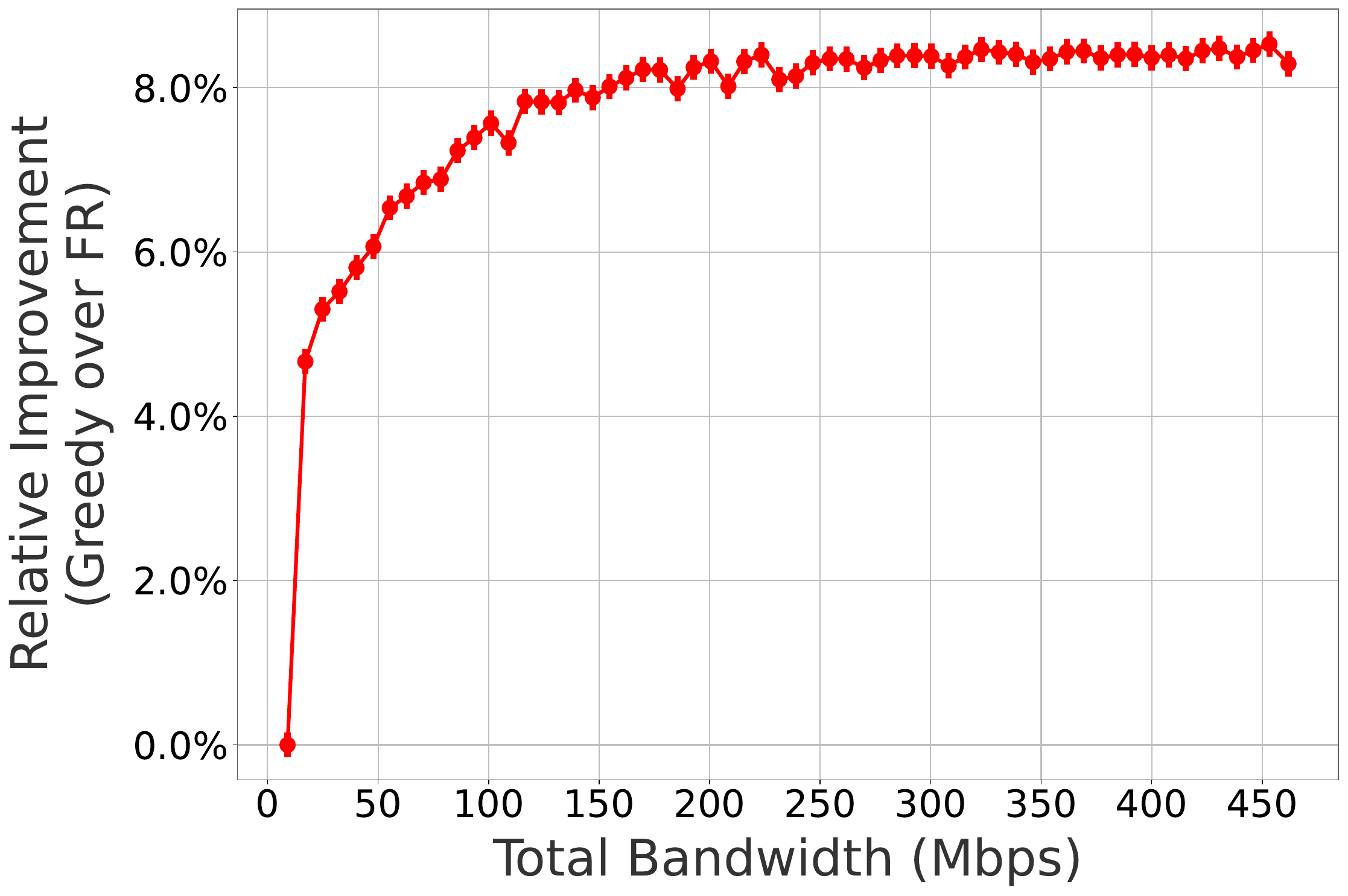}
  \caption{Greedy over FR}
  \label{fig:real_fr_flow}
\end{subfigure}
\begin{subfigure}{0.49\linewidth}
  \centering
  \includegraphics[width=\linewidth]{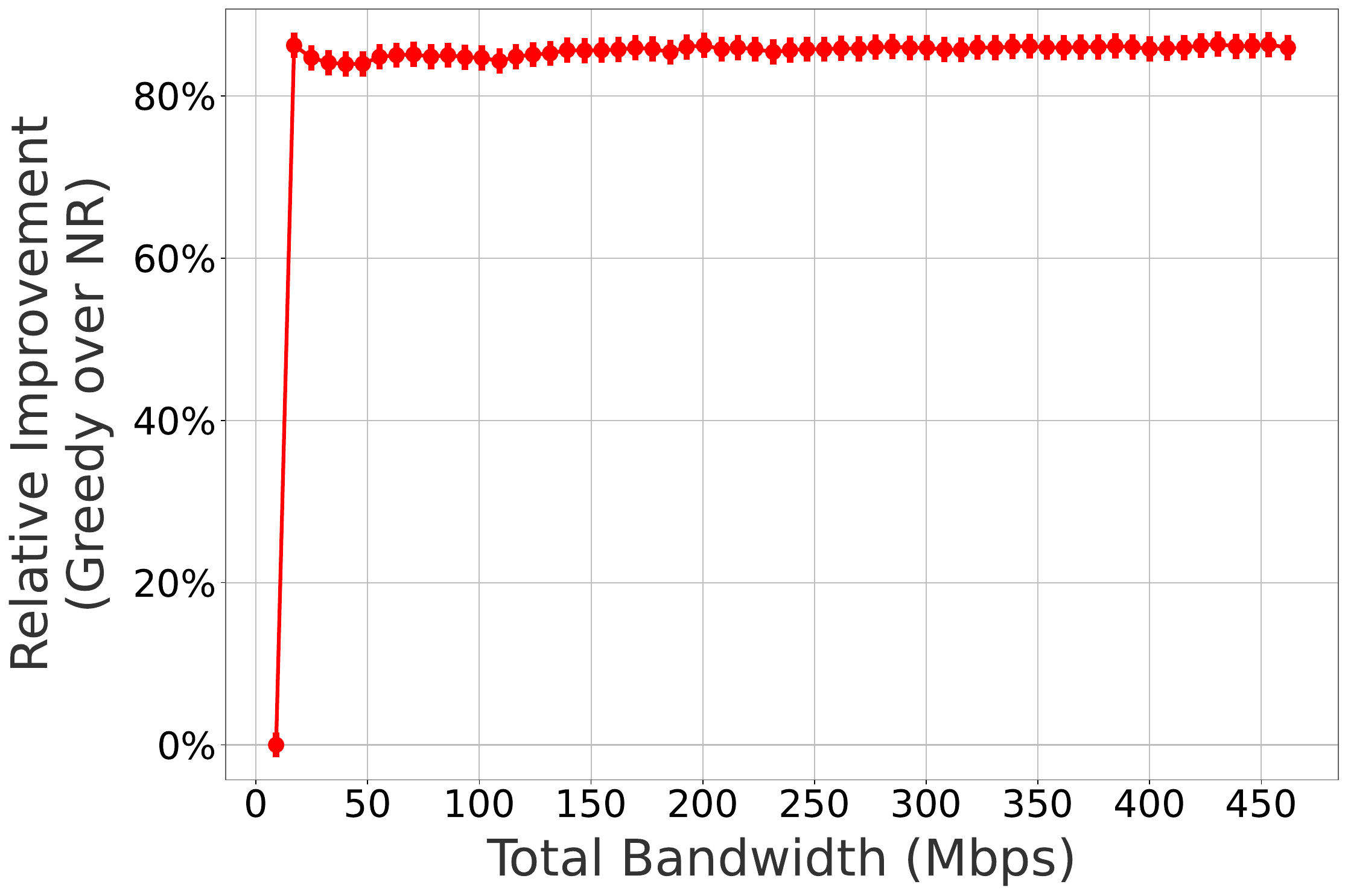}
  \caption{Greedy over NR}
  \label{fig:real_nr_flow}
\end{subfigure}
\caption{Greedy's improvements in number of flows accommodated for US-Topo.}
\label{fig:improvement_realistic_flow}
\end{figure}

\fig{fig:improvement_realistic} mirrors \fig{fig:improvement_realistic_tsn} and reports the relative bandwidth improvements of Greedy over FR and NR.  The results are qualitatively similar to those of the Orion CEV network.  Bandwidth improvements are a little smaller but still meaningful (about~$8\%$ over FR and~$46\%$ over NR).  Similarly, \fig{fig:improvement_realistic_flow} evaluates the number of flows the network can accommodate, and Greedy can allows $8\%$ and $90\%$ more flows than FR and NR, respectively.
\begin{figure}[!h]
\centering
\includegraphics[width=0.6\linewidth]{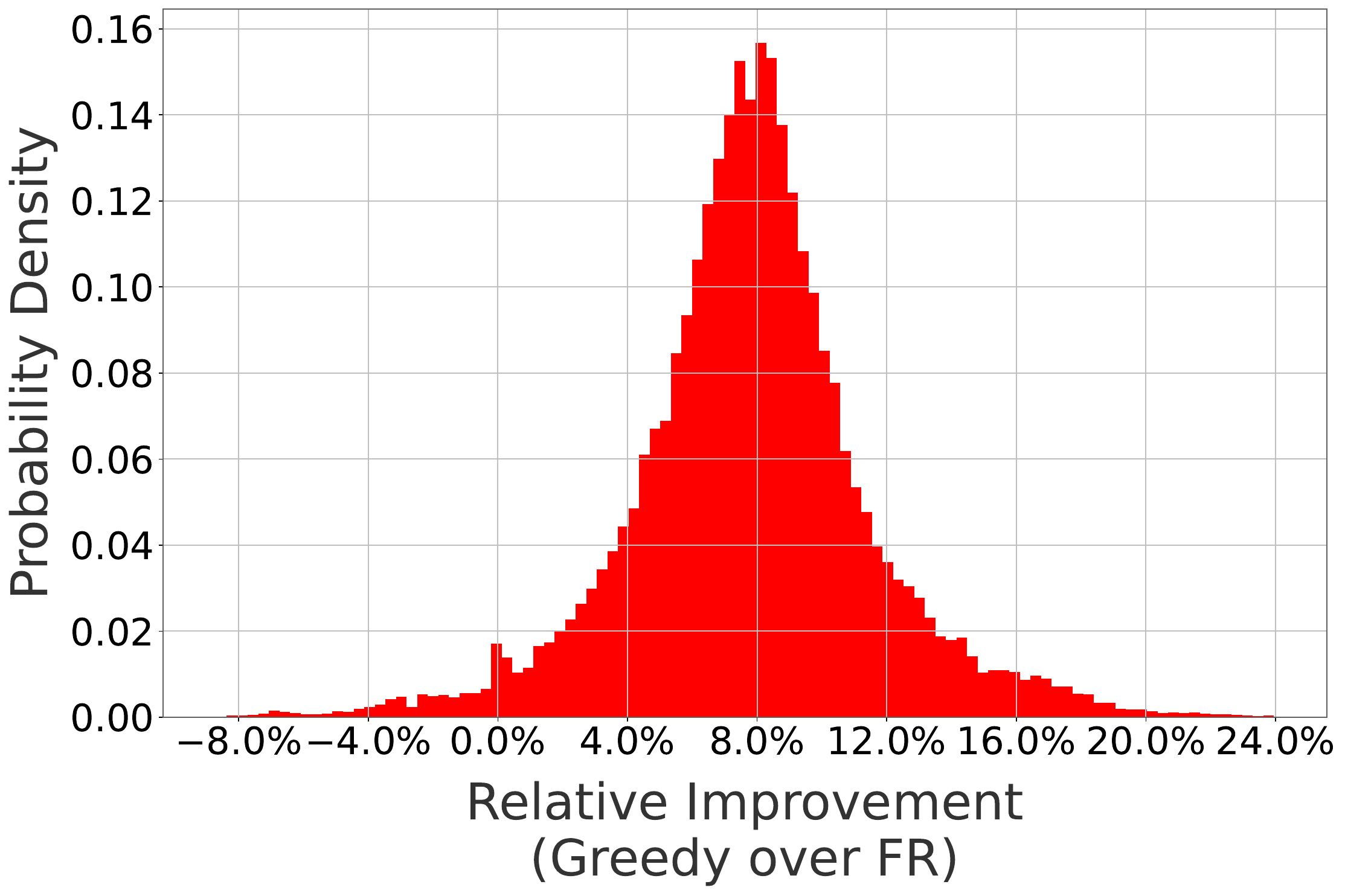}
\caption{Distribution of Greedy's bandwidth improvements over FR across the links of US-Topo ($3000$~flows).}
\label{fig:improvement_realistic_dist}
\end{figure}

\begin{figure}[!h]
\centering
\includegraphics[width=0.6\linewidth]{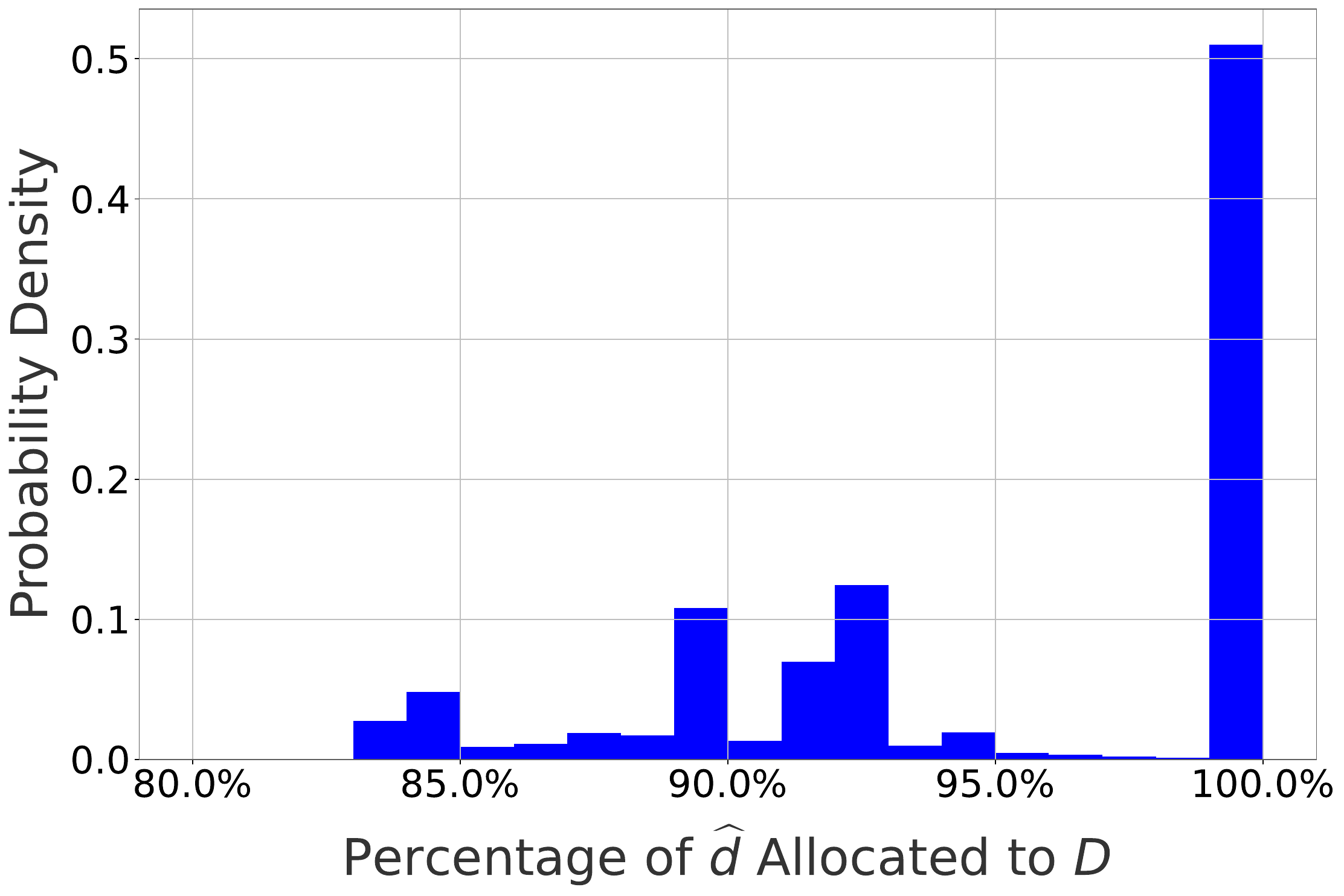}
\caption{Distribution of Greedy's reprofiling ratio $D/\widehat{d}$ on US-Topo ($3000$ flows).}
\label{fig:real_dist}
\end{figure}

\fig{fig:improvement_realistic_dist} duplicates \fig{fig:improvement_realistic_dist_tsn} for US-Topo and plots the distribution of link bandwidth improvements of Greedy over FR for a $3000$~flows configuration.  The shape of the distribution differs from that of the Orion CEV network, but the results confirm that some links see bandwidth increases (negative improvements) in exchange for larger decreases on other links.  How those improvements are realized is reported in \fig{fig:real_dist} that parallels \fig{fig:real_dist_tsn} and plots the distribution of reprofiling ratios for the $3000$~flows configuration. As with the Orion CEV network, intermediate reprofiling solutions are common.

\begin{figure}[!h]
\centering
\begin{subfigure}{0.49\linewidth}
  \centering
  \includegraphics[width=\linewidth]{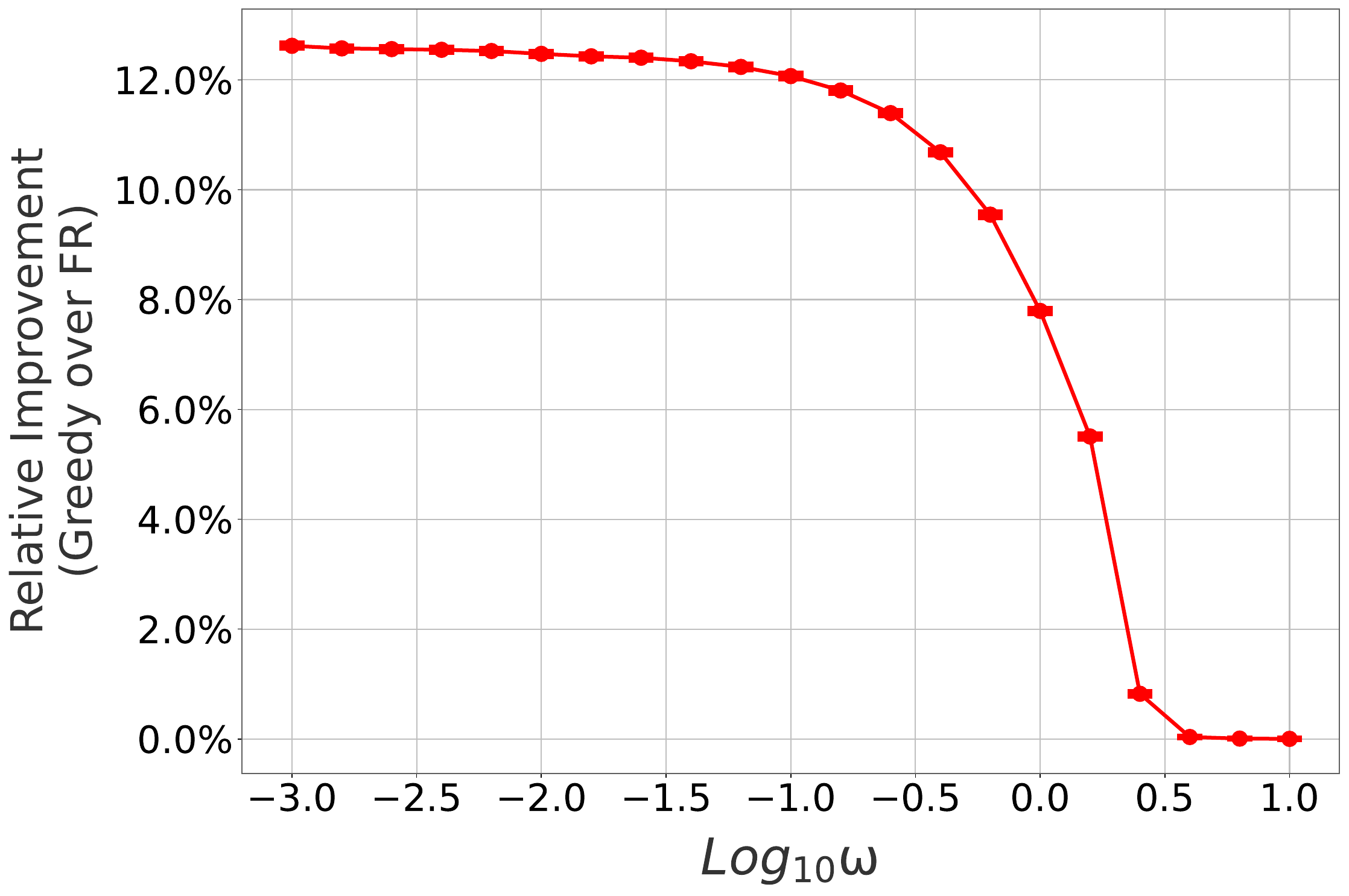}
  \caption{Greedy over FR}
  \label{fig:deadline_improvement}
\end{subfigure}
\begin{subfigure}{0.48\linewidth}
  \centering
  \includegraphics[width=\linewidth]{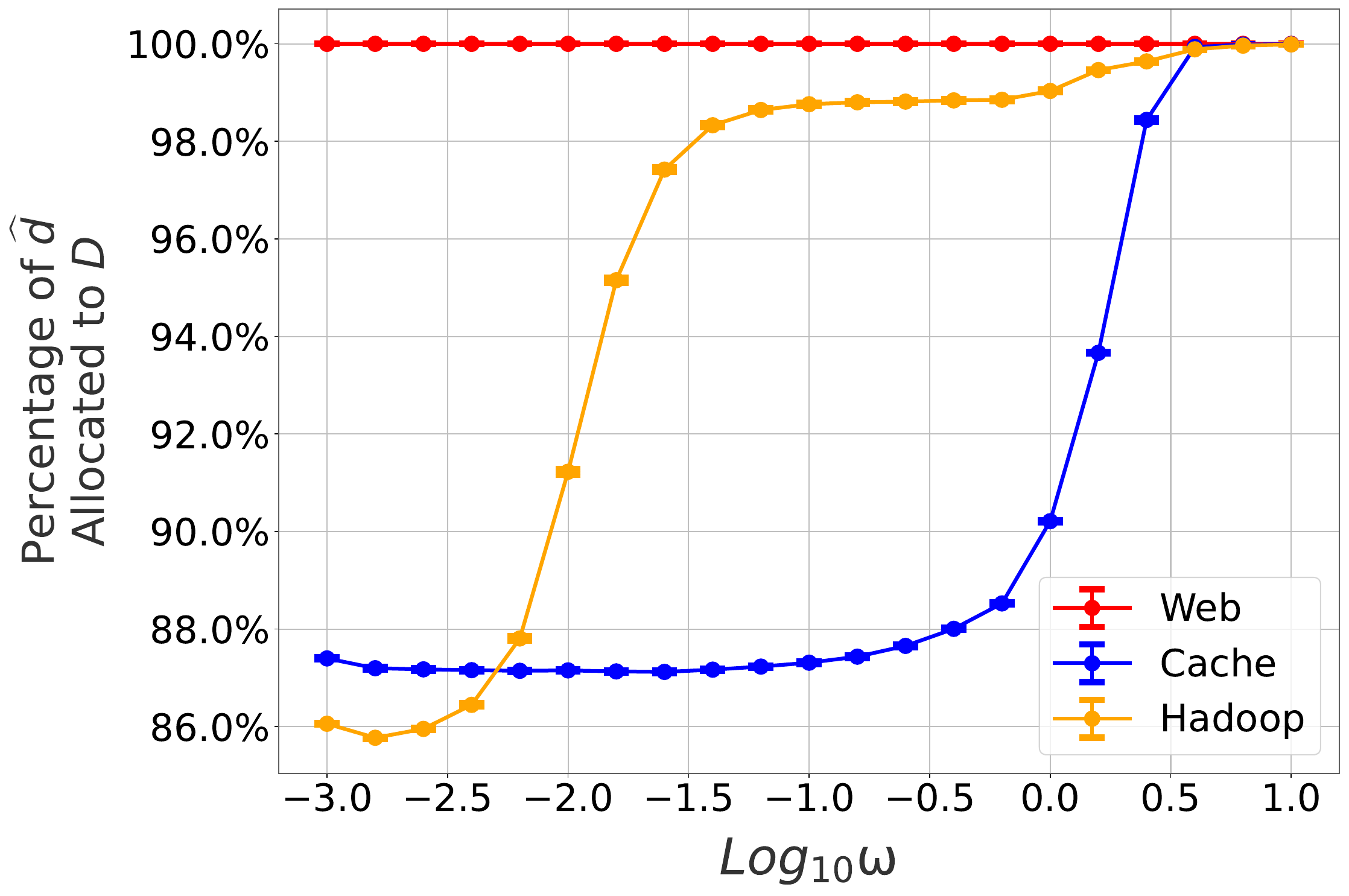}
  \caption{reprofiling ratio $D/\widehat{d}$}
  \label{fig:deadline_reprofiling}
\end{subfigure}
\caption{Reprofiling's behavior on US-Topo ($3000$~flows) as a function of deadline scaling ($\omega$).}
\label{fig:deadline_realistic}
\end{figure}

\fig{fig:deadline_realistic} explores the impact of reprofiling as deadlines vary. As before the deadlines from each class are scaled by a factor $\omega$ now spanning the range $[0.001, 10]$. The relative bandwidth improvement of Greedy over FR and the reprofiling ratios of the three applications are reported in \figs{fig:deadline_improvement}{fig:deadline_reprofiling}, respectively, with results largely consistent with those of \fig{fig:deadline_realistic_tsn}.  A minor difference is that, unlike Clas~B in Orion CEV, the reprofiling ratio of the largest deadline application, Hadoop, drops to about $86\%$ as $\omega$ decreases.  This is because, unlike Orion CEV classes~CDT and~A, Web and Cache have larger burst sizes relative to Hadoop. Hence, as $\omega$ decreases so that Hadoop full reprofiling rate exceeds its token rate $r$, it can start exploiting scheduling flexibility to lower its required bandwidth by taking advantage of bandwidth that becomes available when the initial bursts of Web and Cache clear.  This is similar to what happens to class A in Orion CEV.

\begin{figure}[!h]
\centering
\includegraphics[width=0.6\linewidth]{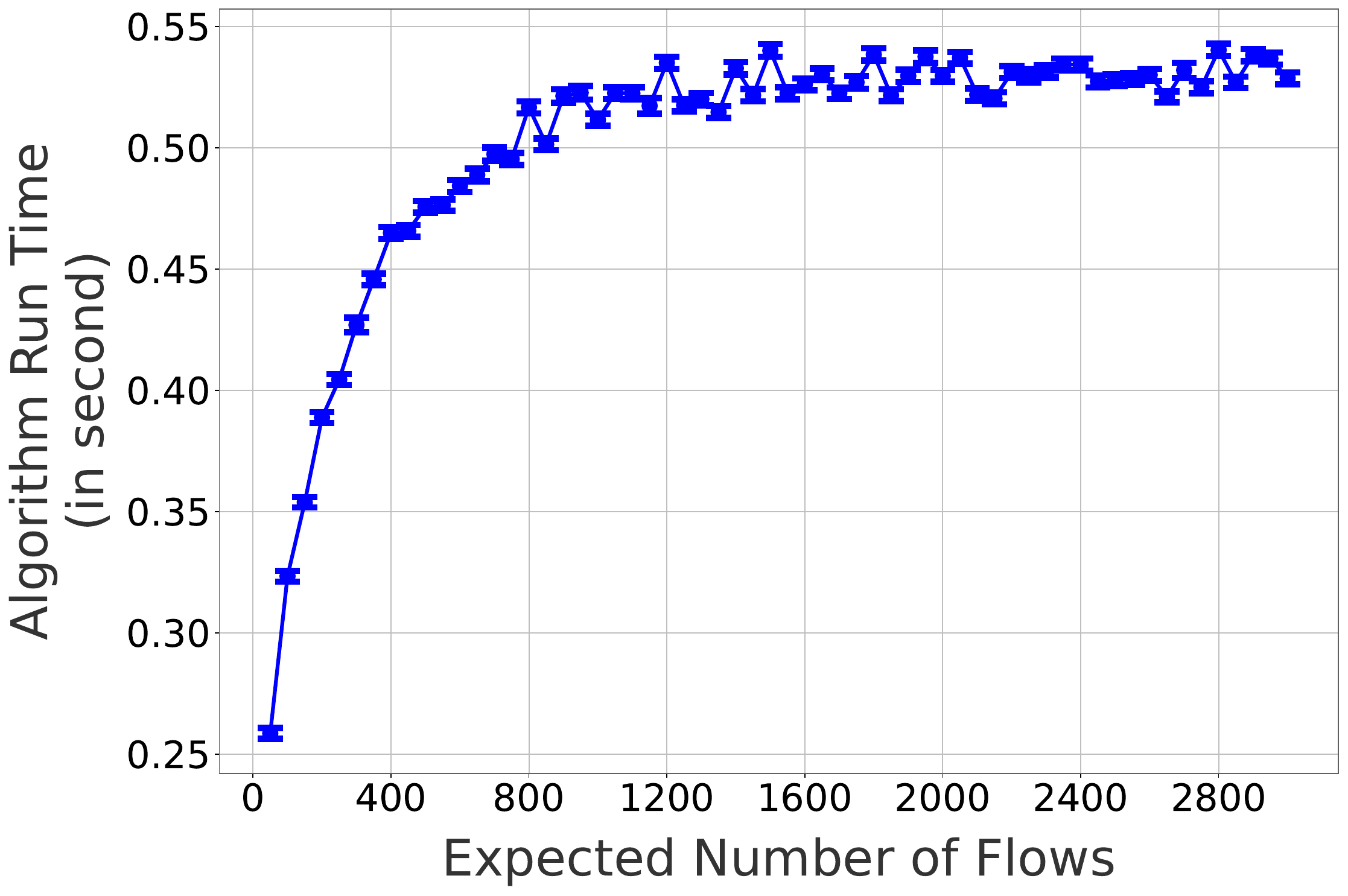}
\caption{Greedy's run time on US-Topo.}
\label{fig:runtime_realistic}
\end{figure}

Finally, \fig{fig:runtime_realistic} duplicates \fig{fig:runtime_realistic_tsn} in reporting Greedy's average run time as we increase the expected number of flows on the US-Topo network.  As with Orion CEV, since flows are aggregated the run time converges to an upper-bound once all paths have been sampled with a stable mix of applications. This happens quickly in the small configuration of US-Topo.

\section{Related Work}
\label{sec:related}

\subsection{Time Sensitive Networking (TSN)}
\label{sec:tsn}

As alluded to, the TSN (and DetNet) standard targets hard delay guarantees for traffic with deterministic traffic profiles.  Recent surveys~\cite{tsn_survey22,walrand23} offer comprehensive reviews of both the standard and various scheduling and shaping solutions developed under its guise.  

The optimization framework that underlies many of those studies have connections to the problem we address.  Their objective function is typically in the form of either minimizing delay or maximizing the amount of traffic that can be accommodated given network resources.  In contrast, we primarily target the dual problem of minimizing network resources given a set of flows.  More importantly, like most prior scheduling work, traffic profiles are assumed fixed and the impact of reprofiling is not considered.

\subsection{Packet-level shaping and scheduling}

Delay guarantees in packet networks have received much attention, and we only sample a few directly relevant works.

An early work~\cite{schmitt02} introduced latency-two-rates (L2R) service curves similar to the 2SRLSC of this paper. However, it focused on minimizing the resources allocated to \emph{individual} flows~\cite[Section IV.A]{schmitt02}, rather than the network-wide, multi-flow perspective of this paper. More importantly, the aspect of flow reprofiling was altogether absent.

Another early example closer to this paper is~\cite{georgiadis96a,georgiadis96b} that relied on \emph{rate-controlled service disciplines} (RCSDs)~\cite{zhang95}.  The papers demonstrated that by selecting shaping parameters derived from the service curves of \emph{rate proportional processor sharing} (RPPS) disciplines~\cite{parekh93,parekh94}, RCSDs could outperform those disciplines and afford a larger schedulability region. The papers, however, explored neither the reprofiling (reshaping) of flows, nor how to allocate per-hop deadlines.

\cite{giacomazzi06} investigated splitting a delay budget between shaping and network delays to minimize bandwidth.  The focus was, however, on a single flow and single link.

Finally,~\cite{bouillard2010tight}, and~\cite{bouillard2014exact} relied on linear programming (LP) and mixed-integer linear programming (MILP) to compute tight worst-case delay bounds under blind and FIFO scheduling, respectively.  Flow reprofiling was again absent.

\subsection{Datacenter solutions}
\label{sec:related_perf}

Several papers~\cite{qjump15,silo15,compactor17,chameleon20} proposed solutions for predictable and differentiated performance in datacenters. They 
vary in their type of schedulers and traffic profile models, but typically ignore the potential impact of reprofiling.

WorkloadCompactor~\cite{compactor17} is the exception, and while it relies on static priority schedulers, it comes closest to this paper in exploring reprofiling. It investigates how to adjust token bucket parameters to guarantee tail latency while minimizing the number of servers required by a set of workloads.  Its reprofiling is, however, limited to a family of feasible regulators\footnote{Regulators above the \emph{r-b} curve using the terminology of~\cite{compactor17}.} that do not delay workload requests.  In contrast, the reprofiling options we consider can result in added access delays that must then be offset by lower network delays. 

\subsection{(Re)shaping bulk transfers} 

Bulk data transfers
must often complete by a deadline, and various approaches~\cite{marcon11,thyaga12,amoeba15,pretium16,trafficshaper18,butler19,cascara21} have been proposed to meet this goal while minimizing bandwidth costs.  
The main difference with this paper is the focus on \emph{transfer} rather than \emph{packet} deadlines.  The latter are relevant to real-time or interactive applications.  The two perspectives are, however, complementary, as tight packet deadlines often result in relatively low link bandwidth utilization that can then accommodate the coarser bandwidth shaping schedules of bulk data transfers.

\subsection{Deadline assignments}

Another set of relevant works~\cite{li02,marinca04,hong11} considered how to split end-to-end deadlines across hops, albeit under very different models from ours. In~\cite{li02}, local deadlines are dynamically adjusted based on actual delays incurred at upstream hops. Statically partitioning end-to-end deadlines across hops, as in this paper, is explored in~\cite{marinca04,hong11}. However, as the notion of arrival curves is absent from both, their solutions are neither applicable nor do they address reprofiling.

\section{Conclusion}
\label{sec:conclusion}

The paper demonstrated that reprofiling flows can help a network offer delay guarantees with less bandwidth. This highlighted the existence of a trade-off between the added delay this reprofiling entails and the benefits it affords the network from smoother traffic.  This trade-off exists in part because the reprofiling delay penalty is incurred once, while its benefits accrue at every hops the flows traverse.

There are many directions in which to extend the work. Simpler schedulers, \eg static priority or FIFO, are clearly of interest, as they are known to benefit from reprofiling even in the single hop case~\cite{onehop21}.  Such an extension would, however, likely also include reliance on TSN's inter-leaved shapers~\cite{leboudec18} rather than per-flow greedy shapers to minimize in-network complexity.  Another natural extension is to consider statistical guarantees.  Hard delay bounds are arguably extreme statistics that may not be suitable for all environments.  In particular, while they are appropriate in the context of the Orion CEV network, the inter-datacenter networks of Section~\ref{sec:eval_inter_dc} may benefit from instead considering lower order statistics.  Exploring the extent to which reprofiling can still be beneficial when focusing on such metrics is, therefore, also of interest.


%

\ifCLASSOPTIONcaptionsoff
  \newpage
\fi



\bibliographystyle{IEEEtran}
\bibliography{IEEEabrv,ref}

\begin{thebibliography}{10}
\providecommand{\url}[1]{#1}
\csname url@samestyle\endcsname
\providecommand{\newblock}{\relax}
\providecommand{\bibinfo}[2]{#2}
\providecommand{\BIBentrySTDinterwordspacing}{\spaceskip=0pt\relax}
\providecommand{\BIBentryALTinterwordstretchfactor}{4}
\providecommand{\BIBentryALTinterwordspacing}{\spaceskip=\fontdimen2\font plus
\BIBentryALTinterwordstretchfactor\fontdimen3\font minus
  \fontdimen4\font\relax}
\providecommand{\BIBforeignlanguage}[2]{{%
\expandafter\ifx\csname l@#1\endcsname\relax
\typeout{** WARNING: IEEEtran.bst: No hyphenation pattern has been}%
\typeout{** loaded for the language `#1'. Using the pattern for}%
\typeout{** the default language instead.}%
\else
\language=\csname l@#1\endcsname
\fi
#2}}
\providecommand{\BIBdecl}{\relax}
\BIBdecl

\bibitem{automotive21}
\BIBentryALTinterwordspacing
M.~Ashjaei, L.~L. Bello, M.~Daneshtalab, G.~Patti, S.~Saponara, and S.~Mubeen,
  ``Time-sensitive networking in automotive embedded systems: State of the art
  and research opportunities,'' \emph{Journal of Systems Architecture}, vol.
  117, p. 102137, 2021. [Online]. Available:
  \url{https://www.sciencedirect.com/science/article/pii/S1383762121001028}
\BIBentrySTDinterwordspacing

\bibitem{afdx}
\emph{Avionics Full Duplex Switched Ethernet ({AFDX}) Network}, Airlines
  Electronic Engineering Committee, Aircraft Data Network Part 7, ARINC
  Specification 664, Aeronautical Radio, Annapolis, MD, USA, 2002.

\bibitem{factory20}
\BIBentryALTinterwordspacing
C.~Zunino, A.~Valenzano, R.~Obermaisser, and S.~Petersen, ``Factory
  communications at the dawn of the fourth industrial revolution,''
  \emph{Computer Standards \& Interfaces}, vol.~71, p. 103433, 2020. [Online].
  Available:
  \url{https://www.sciencedirect.com/science/article/pii/S0920548919300868}
\BIBentrySTDinterwordspacing

\bibitem{smartgrid23}
\BIBentryALTinterwordspacing
T.~Docquier, Y.-Q. Song, V.~Chevrier, L.~Pontnau, and A.~Ahmed-Nacer,
  ``Performance evaluation methodologies for smart grid substation
  communication networks: A survey,'' \emph{Computer Communications}, vol. 198,
  pp. 228--246, 2023. [Online]. Available:
  \url{https://www.sciencedirect.com/science/article/pii/S0140366422004285}
\BIBentrySTDinterwordspacing

\bibitem{aws22}
\BIBentryALTinterwordspacing
(2022) {AWS} global network. [Online]. Available:
  \url{https://aws.amazon.com/about-aws/global-infrastructure/global_network/}
\BIBentrySTDinterwordspacing

\bibitem{google-netw21}
\BIBentryALTinterwordspacing
(2022) Google cloud networking overview. [Online]. Available:
  \url{https://cloud.google.com/blog/topics/developers-practitioners/google-cloud-networking-overview}
\BIBentrySTDinterwordspacing

\bibitem{msft21}
\BIBentryALTinterwordspacing
(2021, January) Microsoft global network. [Online]. Available:
  \url{https://docs.microsoft.com/en-us/azure/networking/microsoft-global-network}
\BIBentrySTDinterwordspacing

\bibitem{tsn18}
J.~Farkas, L.~L. Bello, and C.~Gunther, ``Time-sensitive networking
  standards,'' \emph{IEEE Communications Standards Magazine}, vol.~2, no.~2,
  2018.

\bibitem{parsons22}
G.~Parsons, ``The rise of time-sensitive networking ({TSN}) in automobiles,
  industrial automation, and aviation,'' In Compliance - Electronic Design,
  Testing \& Standards, January 2022,
  https://incompliancemag.com/article/the-rise-of-time-sensitive-networking-tsn-in-automobiles-industrial-automation-and-aviation/.

\bibitem{seol21}
Y.~Seol, D.~Hyeon, J.~Min, M.~Kim, and J.~Paek, ``Timely survey of
  time-sensitive networking: Past and future directions,'' \emph{IEEE Access},
  vol.~9, pp. 142\,506--142\,527, 2021.

\bibitem{detnet}
\BIBentryALTinterwordspacing
N.~Finn, P.~Thubert, B.~Varga, and J.~Farkas, ``{Deterministic Networking
  Architecture},'' RFC 8655, October 2019. [Online]. Available:
  \url{https://www.rfc-editor.org/info/rfc8655}
\BIBentrySTDinterwordspacing

\bibitem{leboudec18}
J.-Y. {Le Boudec}, ``A theory of traffic regulators for deterministic networks
  with application to interleaved regulators,'' \emph{IEEE/ACM Transactions on
  Networking}, vol.~26, no.~6, 2018.

\bibitem{onehop21}
J.~Song, J.~Qiu, R.~Gu{\'e}rin, and H.~Sariowan, ``On the benefits of traffic
  ``reprofiling'' the single hop case,'' \emph{IEEE/ACM Transactions on
  Networking}, vol. TBD, no. TBD, Feburary 2024, extended version available at
  \url{https://arxiv.org/abs/2104.02222}.

\bibitem{Georgiadis97}
L.~Georgiadis, R.~Gu\'{e}rin, and A.~K. Parekh, ``Optimal multiplexing on a
  single link: delay and buffer requirements,'' \emph{IEEE Transactions on
  Information Theory}, vol.~43, no.~5, 1997.

\bibitem{liebeherr96}
J.~Liebeherr, D.~E. Wrege, and D.~Ferrari, ``Exact admission control for
  networks with a bounded delay service,'' \emph{IEEE/ACM Transactions on
  Networking}, vol.~4, no.~6, 1996.

\bibitem{sced}
H.~Sariowan, R.~L. Cruz, and G.~C. Polyzos, ``{SCED}: A generalized scheduling
  policy for guaranteeing quality-of-service,'' \emph{IEEE/ACM Transactions on
  Networking}, vol.~7, no.~5, 1999.

\bibitem{nc}
J.-Y. {Le Boudec} and P.~Thiran, \emph{Network Calculus: A Theory of
  Deterministic Queuing Systems for the Internet}.\hskip 1em plus 0.5em minus
  0.4em\relax Springer, 2001, available at
  \url{https://leboudec.github.io/netcal/}.

\bibitem{bouillard16}
\BIBentryALTinterwordspacing
A.~Bouillard and G.~Stea, ``Worst-case analysis of tandem queueing systems
  using network calculus,'' in \emph{Quantitative Assessments of Distributed
  Systems}, ser. Book Series on Performability Engineering, D.~Bruneo and
  S.~Distefano, Eds.\hskip 1em plus 0.5em minus 0.4em\relax John Wiley \& Sons,
  2016, ch.~15. [Online]. Available: \url{https://hal.inria.fr/hal-01272090}
\BIBentrySTDinterwordspacing

\bibitem{sharma20}
N.~K. Sharma, C.~Zhao, M.~Liu, P.~G. Kannan, C.~Kim, A.~Krishnamurthy, and
  A.~Sivaraman, ``Programmable calendar queues for high-speed packet
  scheduling,'' in \emph{Proc. USENIX NSDI}, 2020.

\bibitem{sivaraman16}
A.~Sivaraman, S.~Subramanian, M.~Alizadeh, S.~Chole, S.-T. Chuang, A.~Agrawal,
  H.~Balakrishnan, T.~Edsall, S.~Katti, and N.~McKeown, ``Programmable packet
  scheduling at line rate,'' in \emph{Proc. ACM SIGCOMM}, 2016.

\bibitem{georgiadis96a}
L.~Georgiadis, R.~Gu\'{e}rin, V.~Peris, and K.~N. Sivarajan, ``Efficient
  network {QoS} provisioning based on per node traffic shaping,''
  \emph{IEEE/ACM Transactions on Networking}, vol.~4, no.~4, 1996.

\bibitem{paulitsch2011time}
M.~Paulitsch, E.~Schmidt, B.~Gst{\"o}ttenbauer, C.~Scherrer, and H.~Kantz,
  ``Time-triggered communication (industrial applications),''
  \emph{Time-Triggered Communication}, 2011.

\bibitem{tuamacs2014optimization}
D.~T{\u{a}}ma{\c{s}}-Selicean and P.~Pop, ``Optimization of {TTEthernet}
  networks to support best-effort traffic,'' in \emph{Proc. IEEE ETFA}, 2014.

\bibitem{thangamuthu2015analysis}
S.~Thangamuthu, N.~Concer, P.~J. Cuijpers, and J.~J. Lukkien, ``Analysis of
  ethernet-switch traffic shapers for in-vehicle networking applications,'' in
  \emph{2015 Design, Automation \& Test in Europe Conference \& Exhibition
  (DATE)}, 2015.

\bibitem{srtm21}
C.~Li, J.~Liu, C.~Lu, R.~Gu\'{e}rin, and C.~D. Gill, ``Impact of distributed
  rate limiting on load distribution in a latency-sensitive messaging
  service,'' in \emph{Proc. CLOUD}, 2021.

\bibitem{gcp-loc}
\BIBentryALTinterwordspacing
(2022) Google cloud global locations. [Online]. Available:
  \url{https://cloud.google.com/about/locations#network}
\BIBentrySTDinterwordspacing

\bibitem{qjump15}
M.~P. Grosvenor, M.~Schwarzkopf, I.~Gog, R.~N.~M. Watson, A.~W. Moore, S.~Hand,
  and J.~Crowcroft, ``Queues {Don{\textquoteright}t} matter when you can {JUMP}
  them!'' in \emph{Proc. USENIX NSDI}, 2015.

\bibitem{silo15}
K.~Jang, J.~Sherry, H.~Ballani, and T.~Moncaster, ``Silo: Predictable message
  latency in the cloud,'' in \emph{Proc. ACM SIGCOMM}, 2015.

\bibitem{chameleon20}
A.~Van~Bemten, N.~Deri\'{c}, A.~Varasteh, S.~Schmid, C.~Mas-Machuca, A.~Blenk,
  and W.~Kellerer, ``Chameleon: Predictable latency and high utilization with
  queue-aware and adaptive source routing,'' in \emph{Proc. ACM CoNEXT}, 2020.

\bibitem{roy15}
A.~Roy, H.~Zeng, J.~Bagga, G.~Porter, and A.~C. Snoeren, ``Inside the social
  network's (datacenter) network,'' in \emph{Proc. ACM SIGCOMM}, 2015.

\bibitem{tsn_survey22}
\BIBentryALTinterwordspacing
T.~St\"{u}ber, L.~Osswald, S.~Lindner, and M.~Menth, ``A survey of scheduling
  in time-sensitive networking ({TSN}),'' 2022. [Online]. Available:
  \url{https://arxiv.org/abs/2211.10954}
\BIBentrySTDinterwordspacing

\bibitem{walrand23}
J.~Walrand, ``A concise tutorial on traffic shaping and scheduling in
  time-sensitive networks,'' \emph{IEEE Communications Surveys \& Tutorials},
  pp. 1--1, May 2023.

\bibitem{schmitt02}
J.~B. {Schmitt}, ``On the allocation of network service curves for
  bandwidth/delay-decoupled scheduling disciplines,'' in \emph{Proc. IEEE
  GLOBECOM}, 2002.

\bibitem{georgiadis96b}
L.~Georgiadis, R.~Gu\'{e}rin, V.~Peris, and K.~N. Sivarajan, ``The effect of
  traffic shaping in efficiently providing end-to-end performance guarantees,''
  \emph{Telecommunications Systems}, vol.~5, no.~1, 1996.

\bibitem{zhang95}
H.~Zhang, ``Providing end-to-end performance guarantees using
  non-work-conserving disciplines,'' \emph{Computer Communications}, vol.~18,
  no.~10, 1995.

\bibitem{parekh93}
A.~K. {Parekh} and R.~G. {Gallager}, ``A generalized processor sharing approach
  to flow control in integrated services networks: the single-node case,''
  \emph{IEEE/ACM Transactions on Networking}, vol.~1, no.~3, 1993.

\bibitem{parekh94}
------, ``A generalized processor sharing approach to flow control in
  integrated services networks: the multiple node case,'' \emph{IEEE/ACM
  Transactions on Networking}, vol.~2, no.~2, 1994.

\bibitem{giacomazzi06}
P.~Giacomazzi, L.~Musumeci, G.~Saddemi, and G.~Verticale, ``Optimal selection
  of token bucket parameters for the admission of aggregate flows in {IP}
  networks.'' in \emph{Proc. IEEE GLOBECOM}, 2006.

\bibitem{bouillard2010tight}
A.~Bouillard, L.~Jouhet, and E.~Thierry, ``Tight performance bounds in the
  worst-case analysis of feed-forward networks,'' in \emph{Proc. IEEE
  INFOCOM}.\hskip 1em plus 0.5em minus 0.4em\relax IEEE, 2010, pp. 1--9.

\bibitem{bouillard2014exact}
A.~Bouillard and G.~Stea, ``Exact worst-case delay in fifo-multiplexing
  feed-forward networks,'' \emph{IEEE/ACM Transactions on Networking}, vol.~23,
  no.~5, pp. 1387--1400, 2014.

\bibitem{compactor17}
T.~Zhu, M.~A. Kozuch, and M.~Harchol-Balter, ``{WorkloadCompactor}: Reducing
  datacenter cost while providing tail latency {SLO} guarantees,'' in
  \emph{Proc. SoCC}, 2017.

\bibitem{marcon11}
M.~Marcon, M.~Dischinger, K.~P. Gummadi, and A.~Vahdat, ``The local and global
  effects of traffic shaping in the internet,'' in \emph{Proc. COMSNETS}, 2011.

\bibitem{thyaga12}
T.~Nandagopal and K.~P. Puttaswamy, ``Lowering inter-datacenter bandwidth costs
  via bulk data scheduling,'' in \emph{Proc. CCGRID}, 2012.

\bibitem{amoeba15}
H.~Zhang, K.~Chen, W.~Bai, D.~Han, C.~Tian, H.~Wang, H.~Guan, and M.~Zhang,
  ``Guaranteeing deadlines for inter-datacenter transfers,'' in \emph{Proc.
  EuroSys}, 2015.

\bibitem{pretium16}
V.~Jalaparti, I.~Bliznets, S.~Kandula, B.~Lucier, and I.~Menache, ``Dynamic
  pricing and traffic engineering for timely inter-datacenter transfers,'' in
  \emph{Proc. ACM SIGCOMM}, 2016.

\bibitem{trafficshaper18}
W.~Li, X.~Zhou, K.~Li, H.~Qi, and D.~Guo, ``{TrafficShaper}: Shaping
  inter-datacenter traffic to reduce the transmission cost,'' \emph{IEEE/ACM
  Transactions on Networking}, vol.~26, no.~3, 2018.

\bibitem{butler19}
Z.~Yang, Y.~Cui, X.~Wang, Y.~Liu, M.~Li, S.~Xiao, and C.~Li, ``Cost-efficient
  scheduling of bulk transfers in inter-datacenter {WANs},'' \emph{IEEE/ACM
  Transactions on Networking}, vol.~27, no.~5, 2019.

\bibitem{cascara21}
R.~Singh, S.~Agarwal, M.~Calder, and P.~Bahl, ``Cost-effective cloud edge
  traffic engineering with cascara,'' in \emph{Proc. USENIX NSDI}, 2021.

\bibitem{li02}
C.~Li and E.~W. Knightly, ``Coordinated multihop scheduling: A framework for
  end-to-end services,'' \emph{IEEE/ACM Transactions on Networking}, vol.~10,
  no.~6, 2002.

\bibitem{marinca04}
D.~Marinca, P.~Minet, and L.~George, ``Analysis of deadline assignment methods
  in distributed real-time systems,'' \emph{Computer Communications}, vol.~27,
  no.~15, 2004.

\bibitem{hong11}
S.~Hong, T.~Chantem, and X.~S. Hu, ``Meeting end-to-end deadlines through
  distributed local deadline assignments,'' in \emph{Proc. IEEE RTSS}, 2011.

\end{thebibliography}
%

%

\vspace{-1.1cm}
\begin{IEEEbiography}[{\includegraphics[width=1in,height=
1.25in,clip,keepaspectratio]{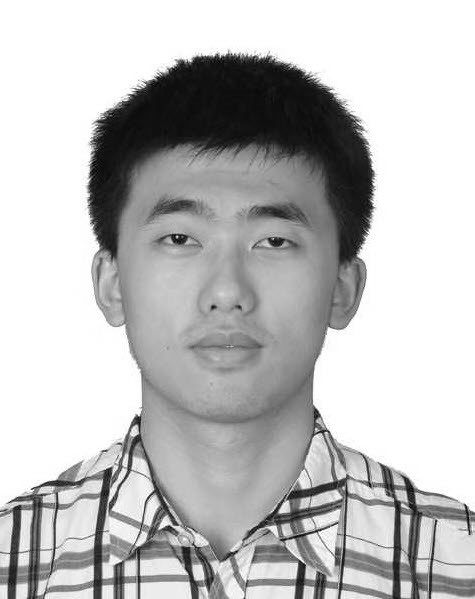}}]{Jiaming Qiu}
received his B.S. in 2020 and is currently a Ph.D candidate, both in Computer Science from Washington University in St. Louis. His research interests include network performance evaluation and optimization, edge computing, and machine learning.
\end{IEEEbiography}

\vspace{-1.0cm}

\begin{IEEEbiography}[{\includegraphics[width=1in,height=
1.25in,clip,keepaspectratio]{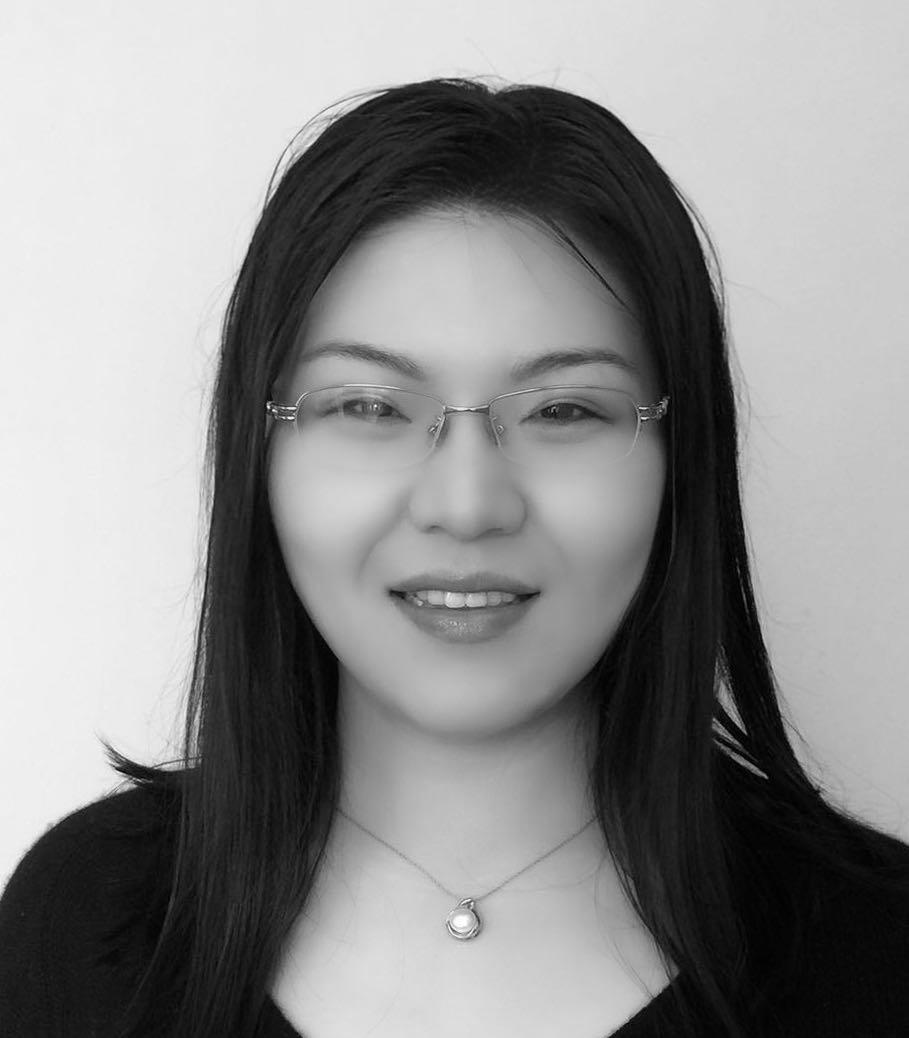}}]{Jiayi Song}
received her Ph.D. in computer science from Washington University in St. Louis. Prior to her doctoral studies, she obtained her bachelor's degree from Wuhan University, majoring in both applied mathematics and economics. Currently, she is with Bytedance Inc.’s Technical Infrastructures team.
\end{IEEEbiography}

\vspace{-1.0cm}

\begin{IEEEbiography}[{\includegraphics[width=1in,height=
1.25in,clip,keepaspectratio]{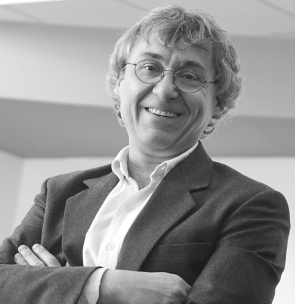}}]{Roch Gu\'{e}rin} (F IEEE '01 / F ACM '06) received his Ph.D. from Caltech, worked at the IBM T.~J.~Watson Research Center and then the University of Pennsylvania.  He is currently with the Department of Comp. Sci. \& Eng. of Washington University in St. Louis, as the Harold B.~and Adelaide G.~Welge Professor and department chair.  Dr. Gu\'erin was the Editor-in-Chief for the IEEE/ACM Transactions on Networking from 2009 till 2013 and served as chair of ACM SIGCOMM.  In 2010 he received the IEEE INFOCOM Achievement Award for ``Pioneering Contributions to the Theory and Practice of QoS in Networks''.
\end{IEEEbiography}

\vspace{-1.0cm}
\begin{IEEEbiography}[{\includegraphics[width=1in,height=
1.25in,clip,keepaspectratio]{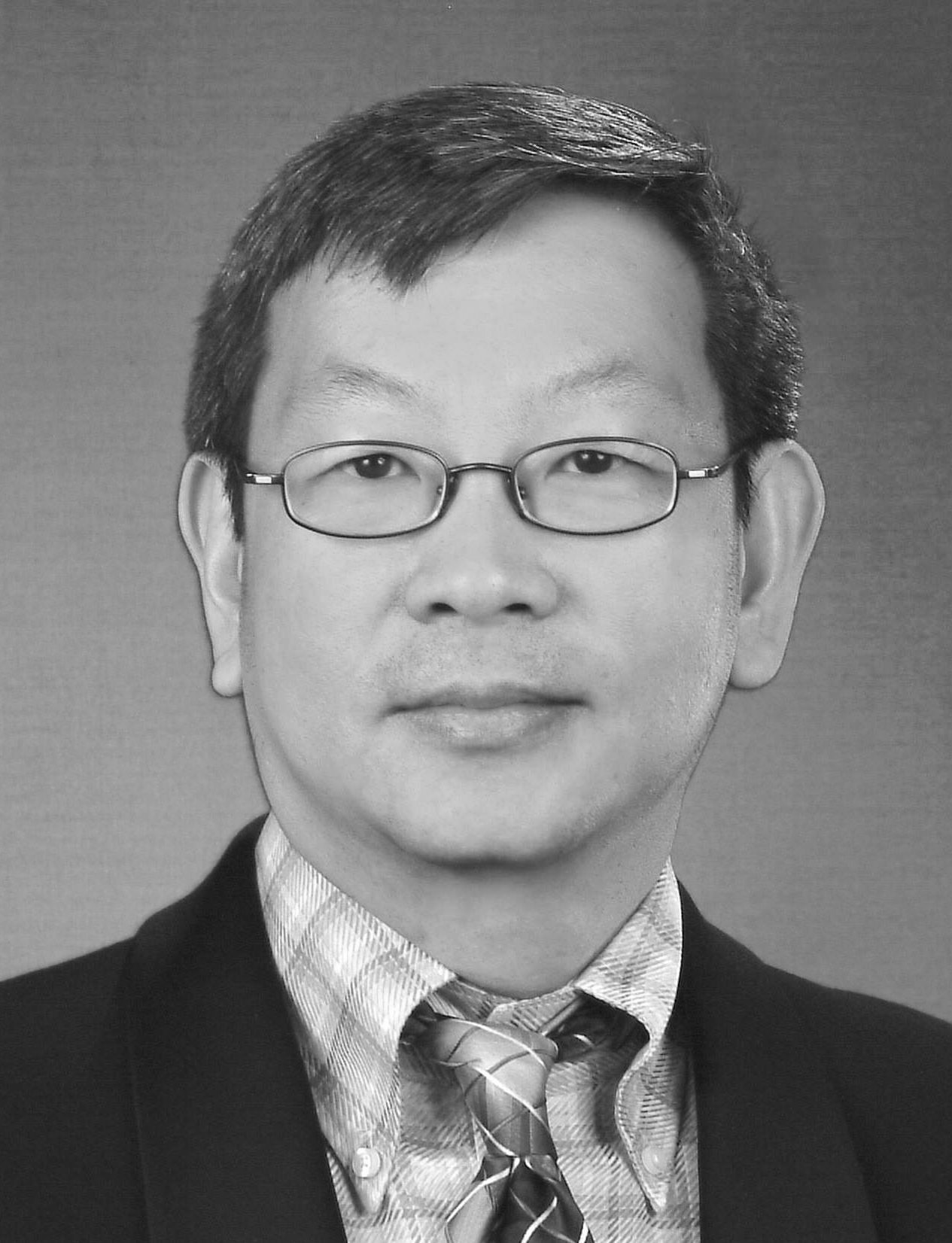}}]{Henry Sariowan}
(SM IEEE, '03) received his Ph.D. from the University of California, San Diego, M.S. from Columbia University, New York, and B.S. from Sepuluh Nopember Institute of Technology, Surabaya, Indonesia, all in electrical engineering.  He is currently with Google's Global Networking team, part of Google Cloud.  Dr. Sariowan previously worked with several technology companies in San Diego, CA and published papers in the areas of network quality of service and video transmission.
\end{IEEEbiography}




\vfill


\appendices
\clearpage

\section{Summary of Notation and Acronyms}

\begin{center}
\begin{supertabular}{ c | c }
\hline
{\textbf{Acronyms}} & {\textbf{Definition}} \\ \hline
2SRC & Two-Slope Reprofiling Curve \\ \hline
2SRLSC & Two-Slope Rate-Latency Service Curve \\ \hline
DESC & Delay Element Service Curve \\ \hline
EDF & Earliest Deadline First scheduler \\ \hline
PBOO & Pay Burst Only Once \\ \hline
RCSD & Rate-Controlled Service Disciplines \\ \hline
RPPS & Rate Proportional Processor Sharing \\ \hline
SCED & Service Curve Earliest Deadline first \\ \hline\hline
{\textbf{Notation}} & {\textbf{Definition}} \\ \hline
$\alpha$ & (token bucket) arrival curve  \\ \hline
$\beta$ & (two-slope rate-latency) service curve \\ \hline
$\delta$ & delay element service curve \\ \hline
$\sigma$ & two-slope reprofiling curve \\ \hline
\multirow{2}{*}{$\Delta(\alpha,\beta)$} & delay upper bound for \\ & arrival curve $\alpha$ under service curve $\beta$ \\ \hline
\multirow{2}{*}{\highlight{\useColor}{$\Theta(\alpha,\beta)$}} & \highlight{\useColor}{buffer upper bound for} \\ & \highlight{\useColor}{arrival curve $\alpha$ under service curve $\beta$} \\ \hline
$\sigma_i$ & minimum reprofiler of flow $i$ \\ \hline
$\sigma^{*}_i$ & \highlight{\useColor}{optimal} reprofiler of flow $i$ \\ \hline
$m$ & number of flows in the network \\ \hline
$n$ & number of links in the network \\ \hline
$i$ & flow index \\ \hline
$j$ & link (hop) index \\ \hline
$r$ & flow long-term rate \\ \hline
$b$ & flow burst size \\ \hline
$d$ & flow end-to-end latency target \\ \hline
\multirow{2}{*}{$\widehat{d}$} & \add{flow maximum reprofiling delay} \\ & \add{$\min(d, b/r)$} \\ \hline
$\mathcal{P}_i$ & set of links on flow $i$'s path (route) \\ \hline
$\mathcal{F}_j$ & set of flows on link~$j$ \\ \hline
$(r_i, b_i, d_i)$ & profile of flow~$i$ \\ \hline
$(\mathbf{r}, \mathbf{b}, \mathbf{d})$ & vector of flow profiles \\ \hline
$\pmb{\mathcal{P}}$ & flow path matrix \\ \hline
$t$ & time \\ \hline
$R$ & 2SRLSC short-term rate \\ \hline
$B$ & 2SRLSC burst size \\ \hline
$T$ & 2SRLSC local deadline \\ \hline
$T'$ & shortcut for $T + B/(R-r)$ \\ \hline
$D$ & reprofiling delay \\ \hline
$C$ & transmission link bandwidth capacity \\ \hline
$C^{*}$ & minimum required bandwidth capacity \\ \hline
\highlight{\useColor}{$\widehat{\Theta}$} & \highlight{\useColor}{scheduling buffer bound} \\ \hline
\highlight{\useColor}{$\widetilde{\Theta}$} & \highlight{\useColor}{reprofiling buffer bound} \\ \hline
$s$ & flow slack \\ \hline
\multirow{2}{*}{$L$} & \add{number of exploration iterations} \\
    & \add{of the greedy algorithm} \\ \hline
\multirow{2}{*}{$K$} & \add{number of initial solutions per iteration} \\
    & \add{of the greedy algorithm} \\ \hline
\end{supertabular}
\end{center}

\section{Proofs and Details}
\subsection{Packet Model Extension}
\label{app:packet_model}
\highlight{\useColor}{The reprofiling algorithms developed in the paper rely on Network Calculus~\cite{nc} with a fluid model, \ie data arrives and is transmitted a bit at the time. The Network Calculus model can, however, readily be extended to account for a packet model. Specifically, compared to a fluid model, a packet model introduces two additional delay components in the expression of a flow's end-to-end delay: (1) a packetization delay, and (2) a scheduling delay that accounts for the scheduler's non-preemptive operation in a packet model, \ie it must complete serving the current packet irrespective of its ``priority''.}

\highlight{\useColor}{Theorem~1.7.1 of~\cite{nc} characterizes the packetization delay using the ``packetizer'' model of Definition~1.7.3, with \cite[Section 1.7.2]{nc} presenting an illustrative example for the resulting end-to-end delay when flows are provided with a guaranteed rate service curve by Generalized Processor Sharing (GPS) schedulers. }

\highlight{\useColor}{The impact of non-preemption can be derived from Theorem~$1$ of~\cite{Georgiadis97} that offered an early account of Theorem~1.7.1 of~\cite{nc} in addition to also considering non-preemptive scheduling. More specifically, the theorem considers flow~$i$ with maximum packet size $l_{\max, i}$,  traversing $m$ hops and guaranteed rate $R_i$ at each hop, \eg through a reprofiling service curve as in Section~\ref{sec:2slope}. The theorem then states that the increase in the end-to-end delay of flow $i$ between a packet and a fluid model is upper-bounded by:
}
\highlight{\useColor}{
\begin{equation*}
    (m-1)\frac{l_{\max, i}}{R_i} + \sum_{j=1}^m \frac{l_{\max}}{C_j},
\end{equation*}
where $l_{\max}$ is the maximum packet size among all the flows sharing the $m$ hops, and $C_j$ is the link bandwidth at hop $j$.  The first term in the above expression is the previously mentioned packetization delay at each hop, while the second term represents the non-preemption delay. 
}

\highlight{\useColor}{Extending our approach to a packet model, therefore, requires accounting for these additional delay terms.  This can be realized by subtracting them from the end-to-end latency target of each flow before proceeding to compute the optimal reprofiling solution.}

\subsection{Minimum Link Bandwidth}
\label{app:min_bandwidth}
\highlight{\useColor}{
Recalling Lemma~\ref{lemma:min_bandwidth}, we have
\begin{flushleft}
\textbf{Lemma}~\ref{lemma:min_bandwidth}. \textit{
Given a set of service curves $\beta_i(t), i=1,\ldots,m$, any scheduling mechanism requires a link bandwidth of at least:
\begin{equation*}
    C^* = \sup_{t \geq 0} \frac{\sum_i^m \beta_{i}(t)}{t},
\end{equation*}
to guarantee those service curves. SCED realizes those service curves with a link bandwidth of exactly $C^*$.}
\end{flushleft}
}

\highlight{\useColor}{We first show that $C^*$ is a lower bound on the link bandwidth required by any scheduling mechanism to guarantee the service curve assignments.}

\begin{proof}
\highlight{\useColor}{
Define the cumulative amount of traffic from flow $i$ arrived and serviced by time $t$ as $A_{i}(t)$ and $S_{i}(t)$, respectively.  According to the definition of a service curve, to guarantee all service curves $\beta_i, i=1,\ldots,m,$ the scheduler needs to ensure that $\forall i\in\{1,\ldots,m\}, \, \forall t\geq 0, S_{i}(t) \geq (A_{i} \otimes \beta_{i})(t)$.
}
\highlight{\useColor}{
Recalling the definition of min-plus convolution, this implies
}
\highlight{\useColor}{
\bequn
\begin{aligned}
(A_{i} \otimes \beta_{i})(t) &= \inf_{0\leq\tau\leq t}\big\{A_{i}(\tau) + \beta_{i}(t - \tau)\big\}\\
&\leq A_{i}(0) + \beta_{i}(t) = \beta_{i}(t).
\end{aligned}
\eequn
In other words, at any time~$t$, the maximum number of bits that the scheduler needs to have transmitted is upper-bounded by $\sum_i^m (A_{i} \otimes \beta_{i})(t) \leq \sum_i^m \beta_{i}(t)$.  Hence, ensuring $\inf_{t\geq0}\{Ct - \sum_i^m \beta_{i}(t)\} \geq 0$ is sufficient to accommodate all the service curves $\beta_{i}, i=1,\ldots,m$.
}

\highlight{\useColor}{
Next, we use a counter example to prove that satisfying $\inf_{t\geq0}\{Ct - \sum_i^m \beta_{i}(t)\} \geq 0$ is necessary to ensure that the scheduler can guarantee all service curves $\beta_{i}, i=1,\ldots,m$. Assume that a link bandwidth of $C'$ is sufficient, where $C'$ is such that $\inf_{t\geq0}\{C't - \sum_i^m \beta_{i}(t)\} < 0$.  Consider then a scenario where each flow $i=1,\ldots,m,$ sends an infinite number of bits at $t_0=0^{+}$, \ie $A_{i}(t) = \infty, i=1,\ldots m, t > 0$.  From the definition of service curves, we then get
}
\highlight{\useColor}{
\bequn
S_{i}(t) \geq (\infty \otimes \beta_{i})(t) = \beta_{i}(t).
\eequn
}
\highlight{\useColor}{
This implies that $\forall t\geq 0, \sum_i^m S_{i}(t) = \sum_i^m \beta_{i}(t)$, which is the minimum number of bits that the scheduler must be able to send by any time~$t\geq 0$ to guarantee that the service curves are met. However, this conflicts with the fact that $\inf_{t\geq0}\{C't - \sum_i^m \beta_{i}(t)\} < 0$ , which implies\footnote{\highlight{\useColor}{Note that $\widetilde{t}>0$ since $A_i(0)=\beta_i(0)=0, i=1,\ldots,m.$}} that $\exists\, \widetilde{t} > 0, s.t. \, C'\tilde{t} < \sum_i^m \beta_{i}(\tilde{t})$.  In other words, a bandwidth of $C'$ results in the scheduler violating the service curve guarantees at $t=\widetilde{t}$. 
}
\highlight{\useColor}{
Combining the above necessary and sufficient conditions, the minimum bandwidth $C^*$ required to allow any scheduling mechanism to provide a service curve guarantee of $\beta_{i}$ to each flow $i$ satisfies:
}
\highlight{\useColor}{
\beq
\label{eq:min1hopbw1}
C^* = \sup_{t \geq 0} \frac{\sum_i^m \beta_{i}(t)}{t}.
\eeq
}

\highlight{\useColor}{
Next, we show that SCED can provide the service curve guarantees with a link bandwidth of just $C^*$. Recall that SCED prioritizes bits according to the latest possible time $t'$ that meets their service curve guarantees:
\begin{equation*}
    A_i(t) = (A_i \otimes \beta_{i})(t').
\end{equation*}
Therefore, SCED's ability to guarantee the service curves with a link bandwidth of just $C^*$ stems from the fact that, when the schedulability condition of \Eqref{eq:min1hopbw1} is met, SCED can transmit any bit arriving at $t$ no later than the deadline $t'$. This latter result is a consequence of the optimality of the EDF scheduler, which is shown in Proposition~$2$ of~\cite{onehop21}.
}
\end{proof}

\subsection{Optimal Reprofiler}
\label{app:reprofiler}

\highlight{\useColor}{
Recalling Lemma~\ref{lemma:optimal_reprofiler}, we have
}

\begin{flushleft}
\highlight{\useColor}{
\textbf{Lemma}~\ref{lemma:optimal_reprofiler}. \textit{
Consider a flow with token bucket profile $\alpha=(r, b)$ that is reprofiled using $\sigma=\bigotimes_{j=1}^n\widetilde{\alpha}_j$, where $\widetilde{\alpha}_j=(r_j,b_j), j=1,\ldots,n,$ are two-parameters token buckets, and $\sigma$ is such that $\Delta(\alpha,\sigma)=D$.  Let $\sigma^{*}=\alpha_1^*\otimes\alpha_2^*$, where $\alpha_1^*$ and $\alpha_2^*$ are token buckets with profiles $\alpha_1^*=(R^{*}= b/D,0)$ and $\alpha_2^*=(r,B^{*}=b-rD)$, then $\Delta(\alpha,\sigma^*)=D$ and $\sigma(t) \geq \sigma^{*}(t), \forall t \geq 0$.
}
}
\end{flushleft}

\highlight{\useColor}{
Before proceeding with the proof of Lemma~\ref{lemma:optimal_reprofiler}, we first establish that the reprofiler resulting from the concatenation of token buckets is a concave, piece-wise linear function.
\begin{lemma}
\label{lemma:token_bucket_concat}
Given a set of $n$ token bucket shapers $\widetilde{\alpha}_j = (r_j, b_j), 1 \leq j \leq n$, $\bigotimes_{j=1}^{n}\widetilde{\alpha}_j = \widetilde{\alpha}_1 \otimes \widetilde{\alpha}_2 \otimes \ldots \otimes \widetilde{\alpha}_n = \min(\widetilde{\alpha}_1, \widetilde{\alpha}_2, \ldots, \widetilde{\alpha}_n)$.
\end{lemma}
\begin{proof}
The proof is a direct consequence of Theorem~$3.1.3$ and~$3.1.6$ of~\cite{nc}, where a token bucket shaper $\widetilde{\alpha}$ is shown to fall under the umbrella of the so-called  ``star-shaped function'' of Definition~$3.1.9$ of~\cite{nc}. 
\end{proof}
}

\begin{figure}[!h]
\centering
\begin{subfigure}{0.45\linewidth}
  \centering
  \includegraphics[width=\linewidth]{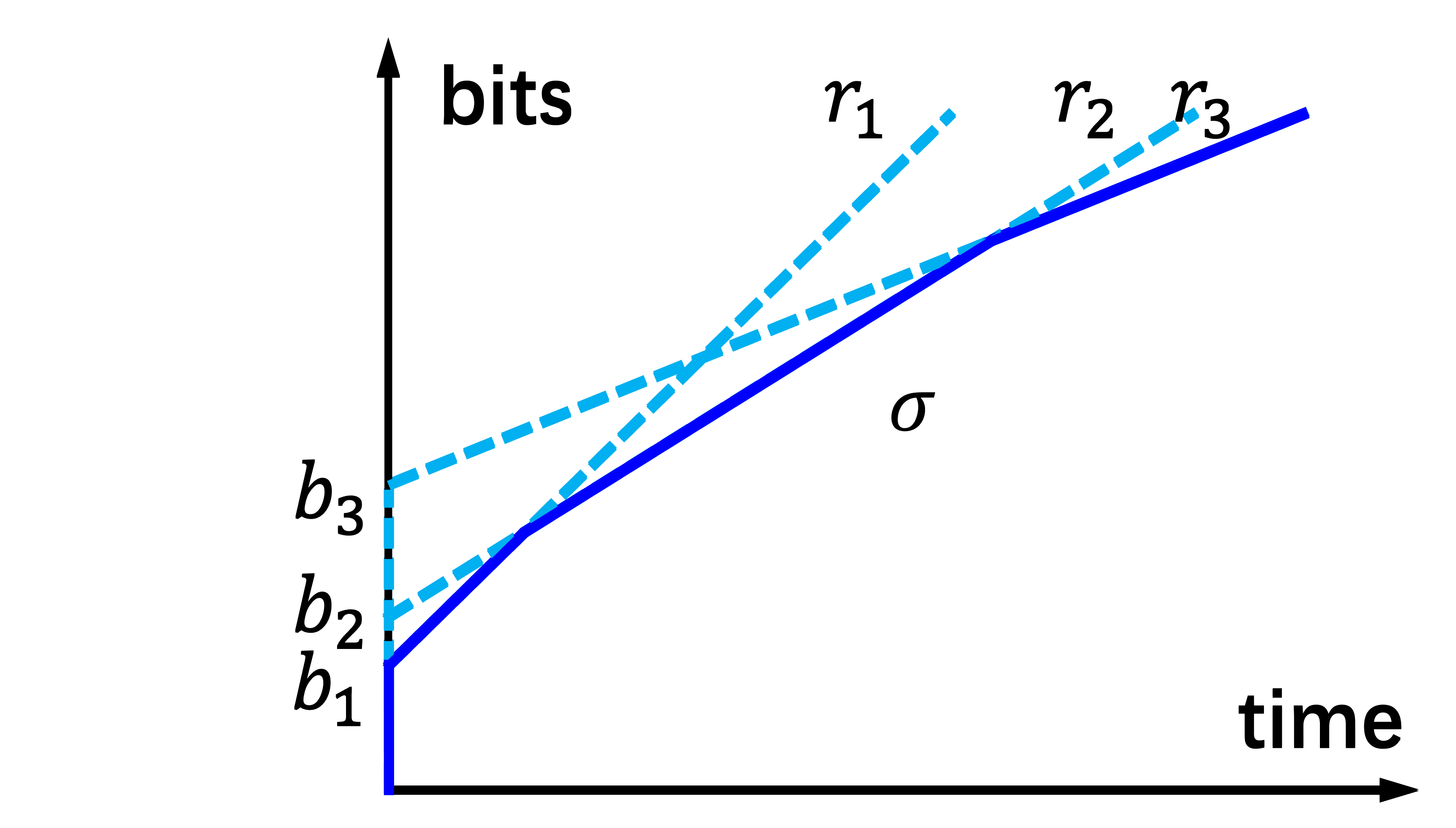}
  \caption{}
  \label{fig:opt1}
\end{subfigure}
\begin{subfigure}{0.45\linewidth}
  \centering
  \includegraphics[width=\linewidth]{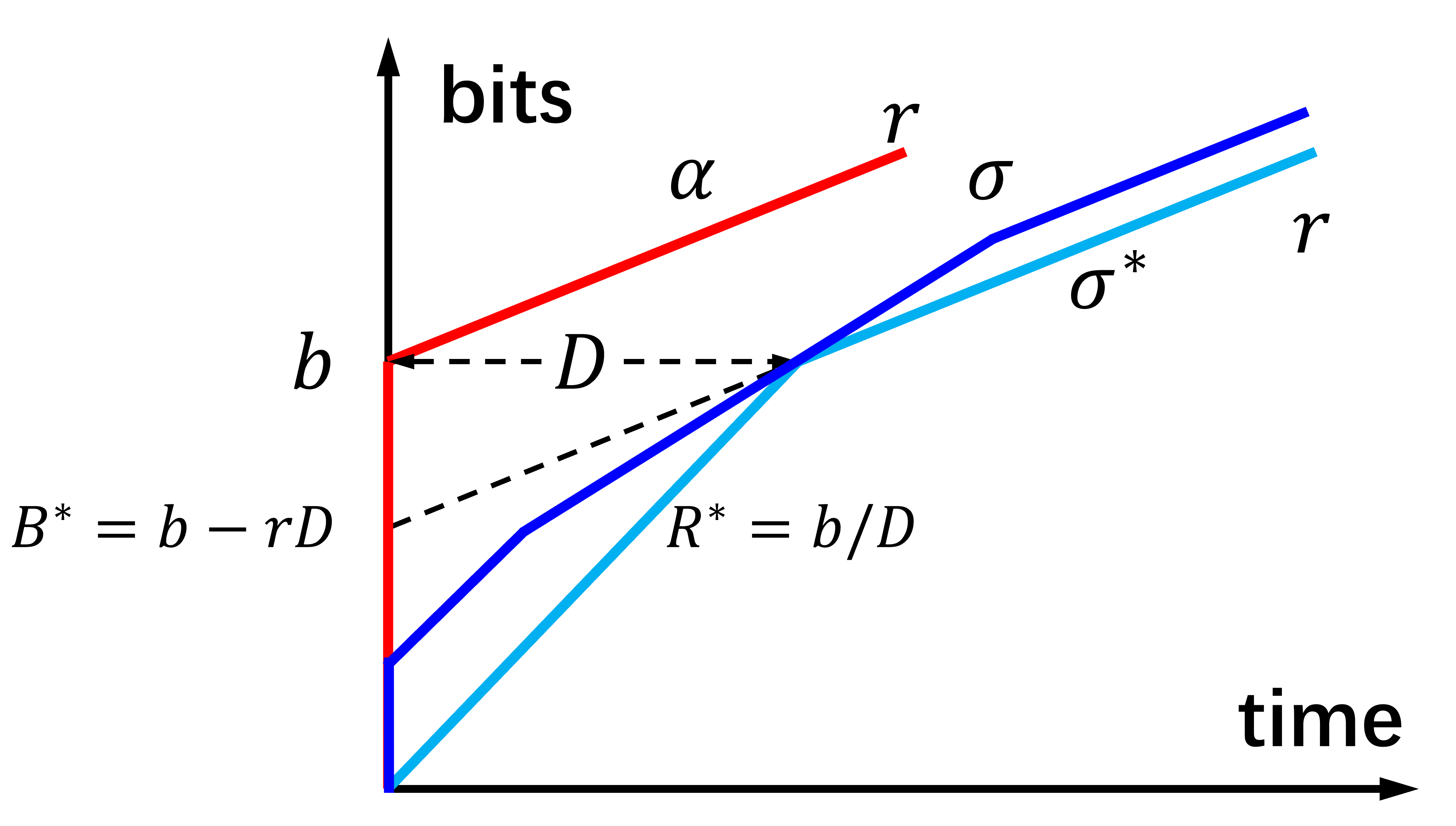}
  \caption{}
  \label{fig:opt2}
\end{subfigure}
\caption{Token bucket concatenation and the optimal reprofiler.}
\label{fig:opt_curve}
\end{figure}

\highlight{\useColor}{
\fig{fig:opt1} illustrates a concatenation of three token buckets, and the fact that it results in a concave, piece-wise linear function.  The latter derives directly from the fact that concatenation is simply the minimum across the three token buckets.
}

\highlight{\useColor}{
We are now ready to proceed with the proof of Lemma~\ref{lemma:optimal_reprofiler}.}
\begin{proof}
\highlight{\useColor}{
Following Lemma~\ref{lemma:token_bucket_concat}, we assume that the reprofiler $\sigma$ is a concave, piece-wise linear function consisting of $n$ segments:
\begin{equation*}
    l_j(t) = b_j + r_jt, \quad t \in [t_j, t_{j+1}), \quad 1 \leq j \leq n,
\end{equation*}
where $0^+ = t_0 < t_1 < \ldots < t_n < t_{n+1} = \infty$ and $\infty > r_0 > r_1 > \ldots > r_n = r$. Note that if $b_i\leq b_k, k< i$, then we can ignore the corresponding token bucket $\widetilde{\alpha}_i$ since it is dominated by $\widetilde{\alpha}_k$. We assume $r_j$ is lower bounded by $r$, the rate of the flow's original token bucket profile, to ensure a finite reprofiling delay. WLOG, we further assume $r_n = r$ since according to Theorem 1.5.3 of~\cite{nc}, a flow is still constrained by its original token bucket after reprofiling.
}

\highlight{\useColor}{
From Section~\ref{sec:bounds}, the reprofiling delay $D$ is given by the maximum horizontal distance between $\alpha$ and $\sigma$:
\begin{equation*}
\begin{aligned}
    D &= \Delta(\alpha, \sigma) = \sup_{t \geq 0 }\{h(t)\}\\
    h(t) &= \inf\{\tau \geq 0: \alpha(t) \leq \sigma(t + \tau)\}.
\end{aligned}
\end{equation*}
We next show that this maximum distance is achieved at $t=0^+$, \ie $h(0^+) = \inf\{\tau \geq 0: \alpha(0^+) \leq \sigma(\tau)\}$. First we note that $h(0) = 0$ since $\alpha(0) = \sigma(0) = 0$ by definition. Second, $h(t)$ is a decreasing function for $t > 0$. Specifically, if $\sigma(t + \tau) = l_j(t + \tau)$, $h(t) = \inf\{\tau \geq 0: \alpha(t) \leq l_j(t + \tau)\}$, which implies:
\begin{equation*}
\begin{aligned}
    \alpha(t) &= l_j(t + h(t))\\
    b + rt &= b_j + r_j \cdot (t + h(t))\\
    h(t) &= \frac{b - b_j + (r - r_j)t}{r_j}.
\end{aligned}
\end{equation*}
Since $r \leq r_j, 1 \leq j \leq n$, $h(t)$ is a decreasing function of $t$ regardless of the segment on which $h(t)$ is achieved. Consequently, the reprofiling delay $D = \sup_{t \geq 0 }\{h(t)\} = h(0^+)$.
}

\highlight{\useColor}{
We define the optimal reprofiler $\sigma^{*}$ as a 2SRC with $R^{*} = b/D$ and $B^{*} = b - rD$, such that $\sigma$ is tangential to $\sigma^{*}$ at $t = D$ as~\fig{fig:opt2} illustrates. Since $\sigma$ is a concave function, we readily find that $\sigma^{*}(t) \leq \sigma(t), \forall t \geq 0$.
}
\end{proof}

\subsection{2SRLSC Concatenation}
\label{app:concat}

\highlight{\useColor}{
Recalling Lemma~\ref{lemma:2srlsc_concat}, we have
}
\begin{flushleft}
\highlight{\useColor}{
\textbf{Lemma}~\ref{lemma:2srlsc_concat}. \textit{
Consider flow~$i$ assigned token bucket $(r_i,b_i)$ 
and assigned on link~$j\in\mathcal{P}_i$ a 2SRLSC $\beta_{ij}$ with parameters $T_{ij}$, $R_{ij}$, $B_{ij}$, and $r_{ij} = r_{i}$.  The concatenation of the flow's 2SRLSCs is readily found to be another 2SRLSC of the form:
\begin{flalign*}
\bigotimes_{j\in\mathcal{P}_i}\beta_{ij} &= \beta\Big\{T_{i}=\sum_{j\in\mathcal{P}_i}T_{ij}, R_{i}=\min_{j\in\mathcal{P}_i}R_{ij}, B_{i}=\min_{j\in\mathcal{P}_i}B_{ij}, r_i\Big\}\\
&=\delta_{T_i} \otimes\sigma_{i},
\end{flalign*}
where $\delta_{T_i}$ is a delay element with parameter $T_i$ and $\sigma_{i}$ is a 2SRC with parameters $R_i,B_i,$ and $r_i$.}
}
\end{flushleft}
\begin{proof}
Recalling the notation of Section~\ref{sec:background}, $\alpha_{r,B}$ is associated with a token bucket controlled arrival curve with rate parameter $r$ and burst size $B$, $\delta_T$ is a delay element service curve with a delay value of $T$, $\sigma_{R,B,r}$ is a a two-slope reprofiling curve (2SRC) with short-term rate $R$, a long-term rate $r$, and a duration of transmissions at the short-term rate determined by $B$, and finally $\beta_{T,R,B,r}$ is a two-slope rate-latency service curve (2SRLSC) obtained from concatenating $\delta_T$ and $\sigma_{R,B,r}$.
We start with an example consisting of two 2SRLSCs and then generalize the result to any finite number of concatenations.  
The identity of each service curve is indicated in the subscript.
\begin{equation*}
\begin{aligned}
    &\beta_{T_1, R_1, B_1, r} \otimes \beta_{T_2, R_2, B_2, r}\\
    = &(\delta_{T_1} \otimes \sigma_{R_1, B_1, r}) \otimes (\delta_{T_2} \otimes \sigma_{R_2, B_2, r})\\
    = &(\delta_{T_1} \otimes (\alpha_{R_1, 0} \otimes \alpha_{r, B_1})) \otimes (\delta_{T_2} \otimes (\alpha_{R_2, 0} \otimes \alpha_{r, B_2}))\\
    = &(\delta_{T_1} \otimes \delta_{T_2}) \otimes (\alpha_{R_1, 0} \otimes \alpha_{R_2, 0}) \otimes (\alpha_{r, B_1} \otimes \alpha_{r, B_2})\\
    = &\delta_{T_1 + T_2} \otimes \alpha_{\min(R_1, R_2), 0} \otimes \alpha_{r, \min(B_1, B_2)}\\
    = &\delta_{T_1 + T_2} \otimes \sigma_{\min(R_1, R_2), \min(B_1, B_2), r}\\
    = &\beta_{T_1 + T_2, \min(R_1, R_2), \min(B_1, B_2), r}
\end{aligned}
\end{equation*}
where we have relied on the commutativity and associativity of min-plus convolution, as well as the concatenation results of token buckets and delay element service curves.

Since the concatenation is a 2SRLSC with the same long-term rate $r$, we can readily prove by induction that concatenating multiple 2SRLSCs yields:
\begin{equation*}
    \beta_{\sum_j T_j, \min_j(R_j), \min_j(B_j), r}
    \label{eqt:end2end_service}
\end{equation*}
where $j$ is used to index network hops.
\end{proof}

\subsection{Searching for Feasible NLP Orderings}
\label{app:nlp_example}

Repeating for convenience \fig{fig:min_bandwidth} that illustrates the piece-wise linear form of the aggregate service curve of flows sharing a given link~$j$, as well as the optimization $\textbf{OPT}^-$ we seek to solve
\beq\tag{\ref{eq:opt-}}
\text{\bf{OPT}}^-: \quad 
\min_{\substack{\mathbf{T}_i, D_i \\ 1\leq i \leq m }}
\sum_{j=1}^{n} C_j,
\eeq
subject to the constraints 
\beq
\tag{\ref{eq:const1}}
\begin{aligned}
\sum_{j \in \mathcal{P}_i}T_{ij} + D_i &\leq d_i, \quad \forall\, 1 \leq i \leq m,\\
D_i &\leq \frac{b_i}{r_i}, \quad \forall\, 1 \leq i \leq m,
\end{aligned}
\eeq
we recall that our NLP formulation relies on preserving the ordering of the variables $T_{ij}$ and $T'_{ij}$ to ensure a consistent closed-form expression for this aggregate service curve.
\reusefigure[!h]{fig:min_bandwidth}

As a result, problem $\textbf{OPT}^-$ calls for solving a set of NLPs across all possible such orderings, with enumerating such orderings combinatorial in nature.  As an exhaustive exploration of possible orderings is impractical, we instead rely on a standard randomized search heuristic.  A key aspect of this search is to generate what we call \emph{feasible} orderings of the variables $T_{ij}$ and $T'_{ij}, i\in \mathcal{F}_j$ for all links $j, 1\leq j\leq n$.  

Specifically, although on a given link~$j$ we can choose any ordering of the variables $T_{ij}$ and $T'_{ij}$ that satisfies\footnote{Recall that $T'_{ij}=T_{ij}+D_i$ and $D_i\geq 0$.} $T'_{ij}\geq T_{ij}, i\in \mathcal{F}_j$, the fact that the reprofiling delay $D_i$ of flow~$i$ is the same on all links $j\in \mathcal{P}_i$, introduces complex dependencies across links.  To address this issue, we first generate a random ordering of the $D_i, 1\leq i\leq m$, that applies to all the links the flows traverse, and then for each link $j, 1\leq j\leq n,$ generate a random ordering of the $T_{ij}, i\in \mathcal{F}_j.$ This combination of orderings induces a number of pair-wise relationships between the $T_{ij}$ and $T'_{ij}, i\in \mathcal{F}_j$.  Our last step is to generate, again for each link $j, 1\leq j\leq n$, an ordering of $T'_{ij}$ (and $T_{ij}$), $i\in \mathcal{F}_j,$ that is consistent with those relationships, \ie is a feasible ordering.  We do so by relying on a variation of topological sorting.

With a feasible ordering in place for all links, we have access to closed-form expressions for the aggregate service curve of flows on all links~$j, 1\leq j\leq n$.  As stated earlier, those can be used to determine the minimum link bandwidths $C_j^*, 1\leq j\leq n,$ using \Eqref{eq:min1hopbw}. In the next phase of our solution, the different terms in \Eqref{eq:min1hopbw} are mapped to a set of additional constraints of the link bandwidth $C_j$ that are then fed to the NLP formulation together with those of \Eqref{eq:const1} and the objective function of \Eqref{eq:opt-}.  Of note is that the NLP solver is the mechanism we rely on to verify that the constraints of \Eqref{eq:const1} can be satisfied by the orderings we select, \ie by having the NLP return a feasible solution. We illustrate this process next using the simple $3$~flows example of \fig{fig:min_bandwidth}.

\subsubsection{Generating a Feasible Ordering}

Assuming a network with only $3$~flows, our first step is to generate a random ordering of the flows' reprofiling delays $D_i, 1\leq i\leq m$, say,
\begin{equation*}
    D_3 \leq D_1 \leq D_2.
\end{equation*}
Considering next a given link~$j$ shared by the three flows, \ie $\mathcal{F}_j = \{1, 2, 3\}$, we specify a random ordering of the $T_{ij}$'s, say,
\begin{equation*}
    T_{1j} \leq T_{2j} \leq T_{3j}.
\end{equation*}
Together, the orderings of the $D_i$'s and $T_{ij}$'s imply several pairwise relationships among the $T'_{ij}$'s, namely,
\begin{equation*}
    T_{1j} \leq T_{2j} \text{ and } D_1 \leq D_2 \Rightarrow T'_{1j} \leq T'_{2j}.
\end{equation*}
Finally, we generate an ordering of the $T_{ij}$'s and $T'_{ij}$'s that is consistent with those pairwise relationships.  As mentioned earlier, this is done using a variation of topological sorting \highlight{\useColor}{that applies a stochastic rather than deterministic strategy to randomly generate a feasible ordering~\footnote{\highlight{\useColor}{See~\url{https://github.com/qiujiaming315/traffic-reprofiling} for details on the randomized strategy that our implementation relies on in its variation of topological sort.}}. For purpose of illustration, one such possible ordering would then look like:}
\begin{equation*}
    T_{1j} \leq T'_{1j} \leq T_{2j} \leq T_{3j} \leq T'_{2j} \leq T'_{3j}.
\end{equation*}
that ensures that condition \textbf{ORD} of Section~\ref{sec:nlp} is met.

\begin{figure*}[!h]
\centering
\captionsetup{justification=centering}
\begin{subfigure}{0.35\linewidth}
  \centering
  \includegraphics[width=\linewidth]{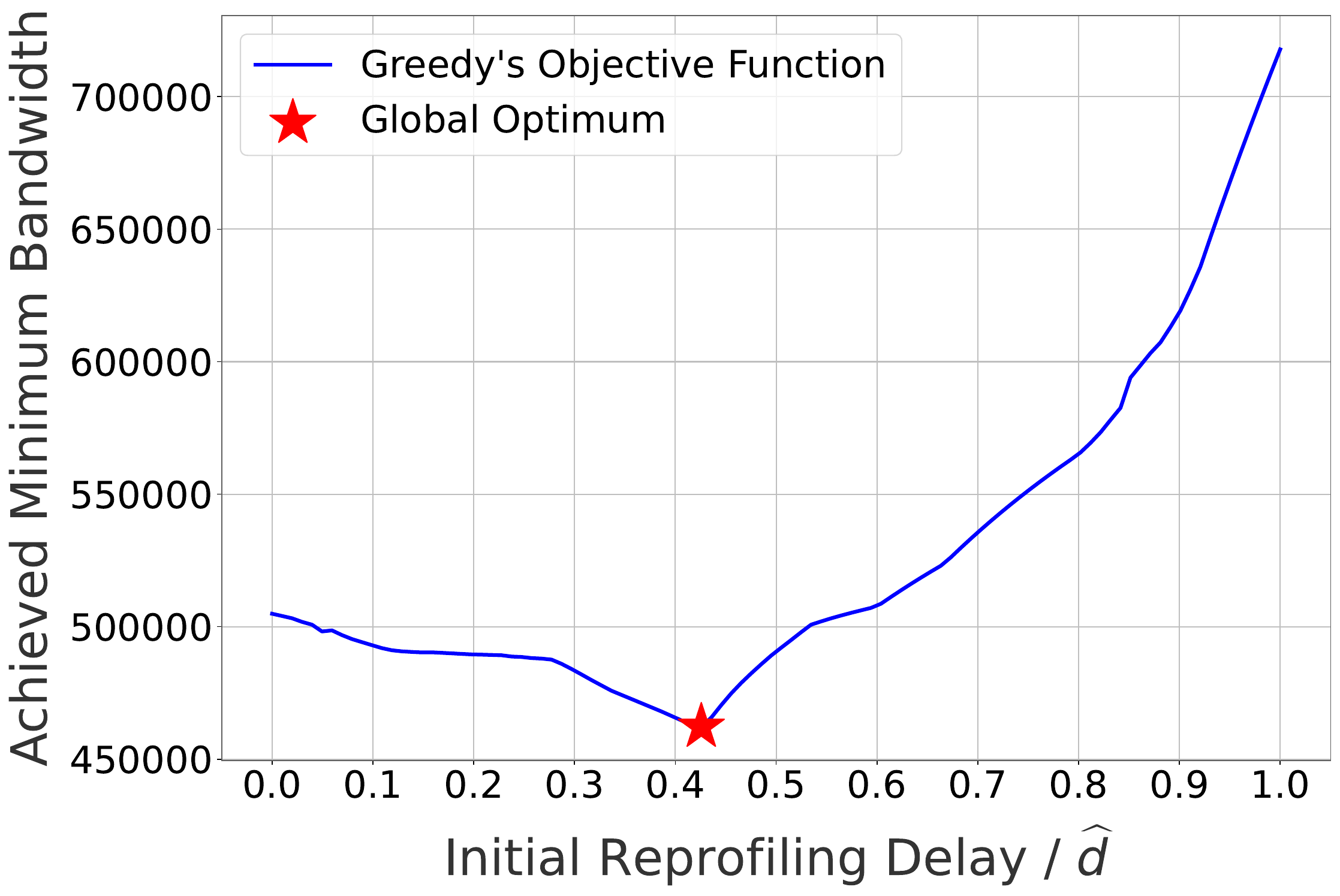}
  \caption{target objective function}
  \label{fig:greedy_obj1}
\end{subfigure}
\begin{subfigure}{0.35\linewidth}
  \centering
  \includegraphics[width=\linewidth]{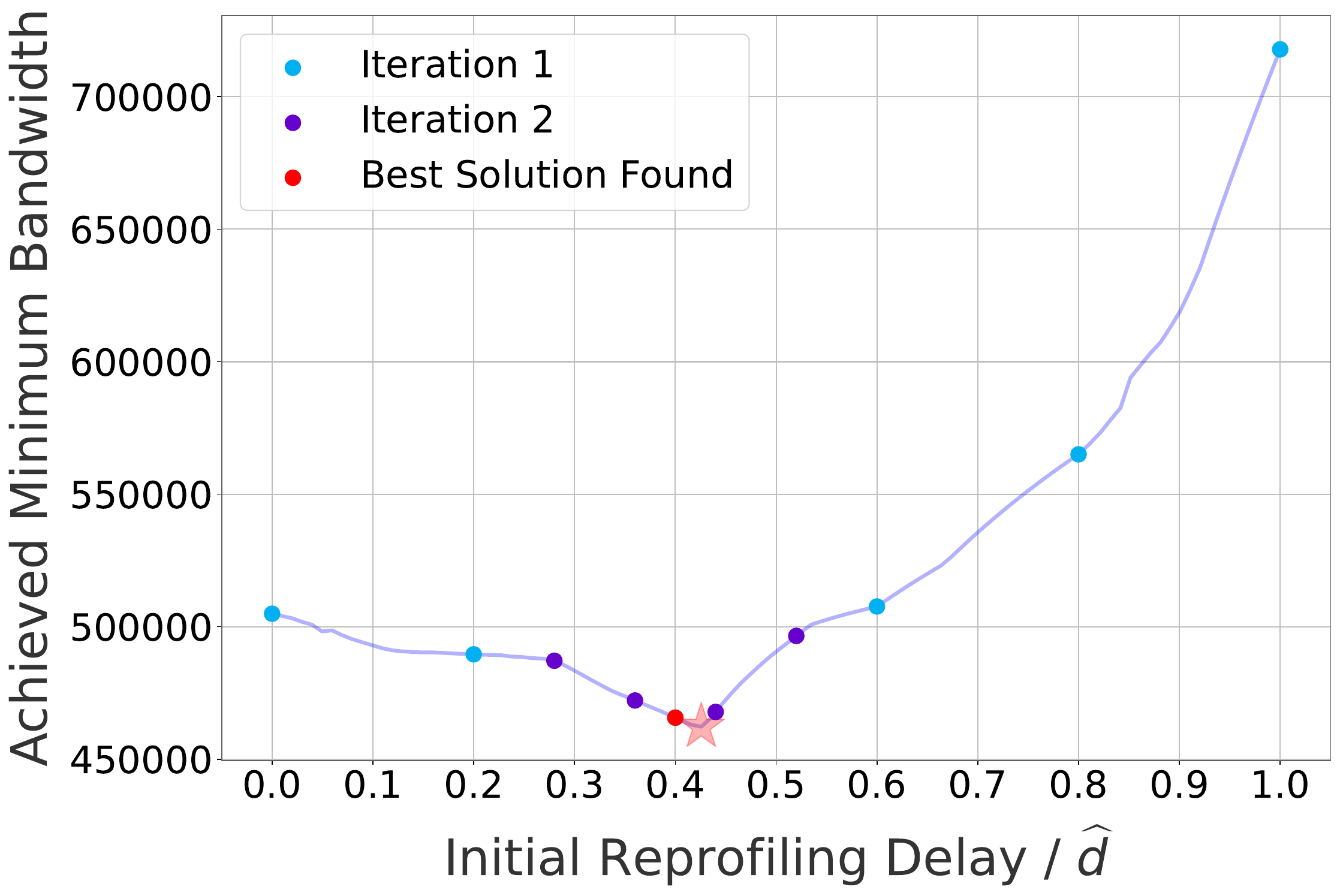}
  \caption{Greedy's exploration}
  \label{fig:greedy_obj2}
\end{subfigure}
\caption{Identifying the optimal reprofiling delay with Greedy.\\($L=2$, $K=4$, topology~$2$, configuration~$2$, $m=128$, $n=2$)}
\label{fig:greedy_explore}
\end{figure*}

\subsubsection{From Ordering to Constraints}

The above ordering is then input as constraints to our NLP formulation.  Recalling for convenience \Eqref{eq:min1hopbw},
\beq
\tag{\ref{eq:min1hopbw}}
C^*_j = \max_{k \in \mathcal{F}_j}\left\{\sum_{i \in \mathcal{F}_j} r_i, \frac{\sum_{i \in \mathcal{F}_j} \beta_{ij}(T'_{kj})}{T'_{kj}} \right\}.
\eeq
We see that with those constraints in place, the expression of \Eqref{eq:min1hopbw} remains unchanged as the variables $D_i$ and $T_{ij}$ vary across their feasible ranges.  It can then be used to determine the minimum bandwidth $C_j^*$ of link~$j$.  Next the link bandwidth $C_j$ needs to be included as another variable in the NLP, which can be realized by introducing additional constraints that pertain to $C_j$. 

These start with the stability constraint (first term of \Eqref{eq:min1hopbw}):
\begin{equation*}
    C_j \geq \sum_{i \in \mathcal{F}_j} r_i = r_1 + r_2 + r_3.
\end{equation*}
and then proceed with additional constraints, one for each of the terms in the summation of \Eqref{eq:min1hopbw}. Returning to our $3$~flows example, this corresponds to:
{\small
\begin{equation*}
\begin{aligned}
    C_j &\geq \frac{R_1(T'_{1j} - T_{1j})}{T'_{1j}} = \frac{b_1}{T_{1j} + D_1}\\
    C_j &\geq \frac{B_1 + r_1(T'_{2j}-T_{1j}) + R_2(T'_{2j} - T_{2j}) + R_3(T'_{2j} - T_{3j})}{T'_{2j}}\\
    &= \frac{b_1 + r_1(T_{2j} + D_2 - T_{1j} - D_1) + b_2 + \frac{b_3}{D_3}(T_{2j} + D_2 - T_{3j})}{T_{2j} + D_2}\\
    C_j &\geq \frac{B_1 + r_1(T'_{3j}-T_{1j}) + B_2 + r_2(T'_{3j}-T_{2j}) + R_3(T'_{3j} - T_{3j})}{T'_{3j}}\\
    &= \frac{\sum_{i=1}^2 b_i + r_i(T_{3j} + D_3 - T_{ij} - D_i) + b_3}{T_{3j} + D_3}
\end{aligned}
\end{equation*}
}
Solving the corresponding NLP instance across all network links produces an optimal solution for this particular ordering.  As mentioned before, computing a \emph{global} optimal solution to $\textbf{OPT}^-$ calls for exploring all feasible orderings, which we approximate by relying on a randomized search procedure. 

\highlight{\useColor}{In our experiments we limited the number of orderings we explore to $\log N$, where $N$ corresponds to the total number of possible orderings for that experiment.  This is because $N$ is often very large.  We assessed the extent to which this limit affected the quality of our solution, by checking how often the optimization terminated because we had reached it rather than because of our termination criterion (\ie no better solution had been found for a given number of orderings) had been met. In most cases, computations stopped because of the termination criterion.  This offers empirical support that the limit we impose on the number of orderings we explore does not significantly affect the quality of the solution we produce.}
\begin{figure*}[!h]
\centering
\captionsetup{justification=centering}
\begin{subfigure}{0.35\linewidth}
  \centering
  \includegraphics[width=\linewidth]{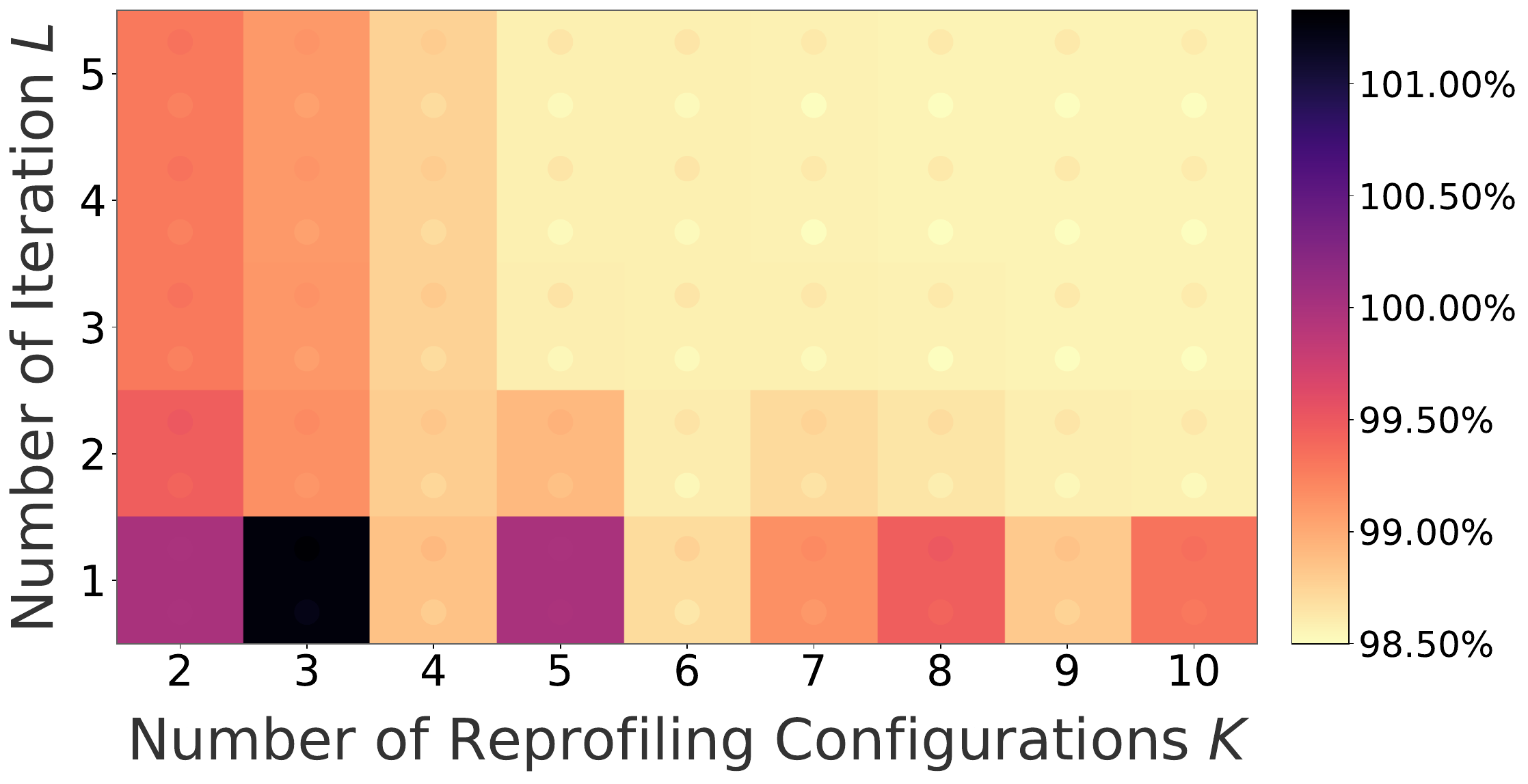}
  \caption{bandwidth performance vs.~$L=1,K=2$}
  \label{fig:greedy_config_solution}
\end{subfigure}
\begin{subfigure}{0.35\linewidth}
  \centering
  \includegraphics[width=\linewidth]{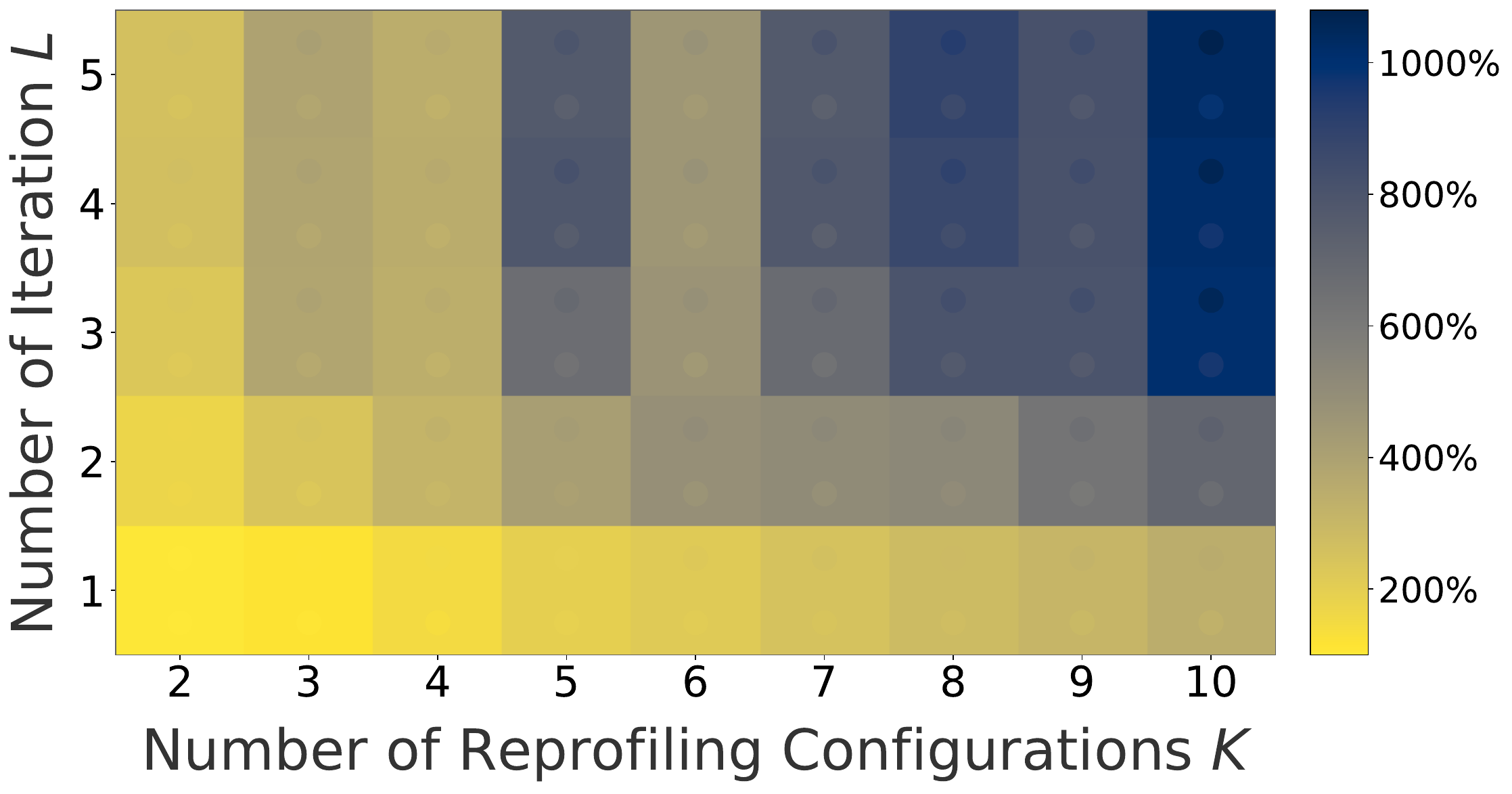}
  \caption{run time vs.~$L=1,K=2$}
  \label{fig:greedy_config_time}
\end{subfigure}
\caption{Greedy's computational cost vs.~bandwidth performance as a function of $K$ and $L$.\\(topology~$2$, configuration~$2$, $m=128$, $n=2$)}
\label{fig:greedy_config}
\end{figure*}

\subsection{Reprofiling Buffer Bound}
\label{app:reprofiler_buffer}

\highlight{\useColor}{We show that, when reprofiling operates in a non-work-conserving mode, \ie as a shaper, the reprofiling backlog of flow~$i$ at hop~$j$ is upper bounded by the maximum burst size $\widetilde{X}_{ij'}$ that flow~$i$ can accumulate in the scheduler at the previous hop~$j'$, namely,
\beq\tag{\ref{eq:reprofiler_buffer}}
\widetilde{X}_{ij'} = \sup_{t \geq 0 }\{\sigma_i - \beta_{ij'}\},
\eeq
where $\sigma_i$ is the 2SRC of flow~$i$ and $\beta_{ij'}$ is the 2SRLSC of flow~$i$ at hop~$j'$, upstream of~$j$ on flow~$i$'s path $\mathcal{P}_i$.
}

\highlight{\useColor}{As $\widetilde{X}_{ij'}$ is the maximum burst size flow~$i$ can generate leaving hop~$j'$, the arrival curve of flow~$i$ at hop~$j$ conforms to token bucket~\cite[Theorem~$1.4.3$]{nc} $\alpha_{ij} = (r_i, \widetilde{X}_{ij'})$. From~\Eqref{eq:buffer}, the reprofiling buffer of flow~$i$ at hop~$j$ is then upper-bounded by $\widetilde{\Theta}_{ij}$:
\begin{equation*}
    \widetilde{\Theta}_{ij}=\sup_{t \geq 0 }\{\alpha_{ij} - \sigma_i\} = \widetilde{X}_{ij'}= \sup_{t \geq 0 }\{\sigma_i - \beta_{ij'}\}.
\end{equation*}
}

\subsection{Exploring Greedy's Behavior}
\label{sec:supplement}

This appendix provides results in support of assumptions on which we rely in the design of the exploration phase of Greedy.  Specifically, we offer empirical evidence that the shape of the objective function that Greedy seeks to optimize  is consistent with the assumption on which its exploration phase is based.  We also investigate the trade-off between the quality of the solution that Greedy produces, and its computational cost as we vary Greedy's parameters, \ie $K+2$ (number of delay samples) and $L$ (number of iterations).  The results offer support for the choice made in the investigation of Section~\ref{sec:evaluation}. Finally, we demonstrate the bandwidth improvement from the adjustment phase of Greedy compared to the initial reprofiling solutions.  The results confirm that flow reprofiling offers more significant contributions to bandwidth minimization compared uneven network deadline allocation.

\subsubsection{Greedy's Exploration Phase}
Recall that Greedy aims to identify the optimal reprofiling delay to be applied to each flow to minimize overall network bandwidth (the objective function).  It relies on an iterative process that progressively refines the range from which it selects a fixed number $(K+2)$ of delay values at which to evaluate its objective function.  As iterations proceed, the range decreases with the expectation that the $K+2$ estimates get progressively closer to the optimal value.  The selection of a new range after each iteration is essentially greedy, and centered around the best result of the current iteration.  Convergence to the global optimum, therefore, requires that, at each iteration, the best result be associated with a steady progression towards the global optimum.  This is readily ensured if the objective function has a ``nice'' form, \eg, convex, but needs not necessarily hold.

\fig{fig:greedy_explore} shows a representative example of the network bandwidth (objective function) whose minimum value Greedy seeks to find.  The results are collected on a network with $m=128$ flows and $n=2$ links, where the network topology and flow profiles are according to topology~$2$ and configuration~$2$ of Section~\ref{sec:algorithm_comparison}.
The $x$-axis represents the range of reprofiling delays (relative to the maximum reprofiling delay $\widehat{d}$) allocated to each flow when creating initial solutions for Greedy.
The $y$-axis gives the corresponding network bandwidth values.

In this example, the objective function can be seen to be approximately convex, with a minimum achieved with a reprofiling delay of slightly more than $40\%$ of the maximum reprofiling delay $\widehat{d}$ of each flow.  The progression of the results produced by Greedy during its exploration phase is shown in~\fig{fig:greedy_obj2}, when Greedy is configured with $L=2$ and $K=4$.  The iterations that Greedy goes through are represented using dots of different colors. 

In its first iteration, Greedy evaluates the network bandwidth for $K+2=6$ equidistant reprofiling delay values ranging from $0$ to $\widehat{d}$ in steps of $0.2\widehat{d}$.  A reprofiling delay of $0.4\widehat{d}$ produces the best result. The second iteration, therefore, selects a new range of the form $[0.2\widehat{d},0.6\widehat{d}]$, \ie centered around that minimum value. It then proceeds to compute $K=4$ additional bandwidth values for reprofiling delays in the interior of the new range.  Since Greedy has explored for the maximum number of $L=2$ iterations, and none of the solutions obtained in this second iteration improve on the best solution of the previous iteration, Greedy terminates and reports the result achieved with a reprofiling delay of $0.4\widehat{d}$ as its best solution.

\subsubsection{Greedy's Accuracy vs.~Computational Cost}
The accuracy of Greedy depends on both $K$ and $L$.  $K$ affects Greedy's ability to converge to the correct range that contains the global minimum of its objective function\footnote{This is not a concern when the objective function is approximately convex, as the previous section suggested is common, but is relevant when local minima are present.}.  Conversely, $L$ determines Greedy's ability to ultimately identify a delay value close to the true optimum.  In general, larger $K$ and $L$ values correspond to better solutions.
\begin{figure*}[!h]
\centering
\begin{subfigure}{0.35\linewidth}
  \centering
  \includegraphics[width=\linewidth]{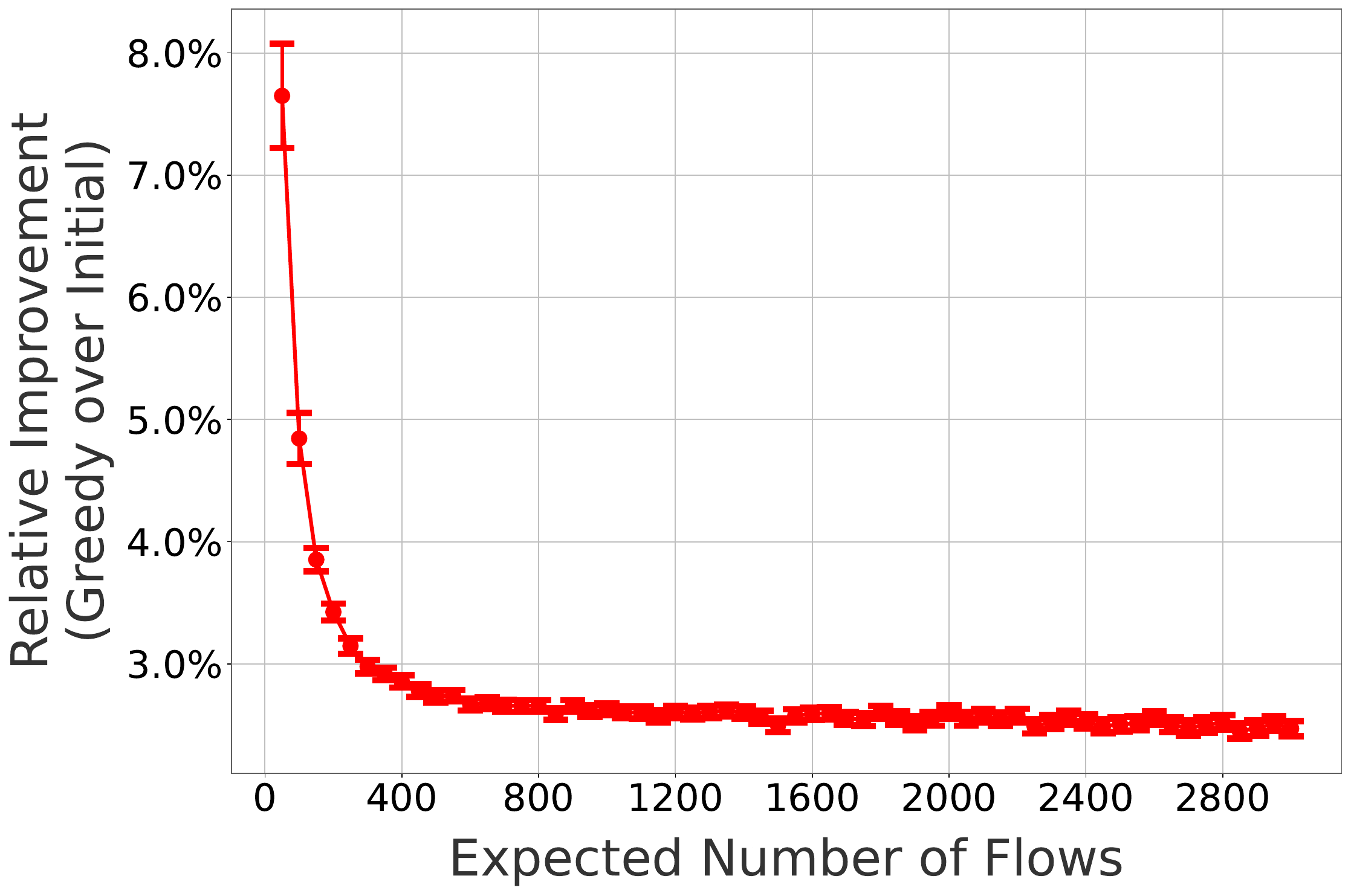}
  \caption{Greedy over initial reprofiling solutions}
  \label{fig:greedy_init_improvement}
\end{subfigure}
\begin{subfigure}{0.35\linewidth}
  \centering
  \includegraphics[width=\linewidth]{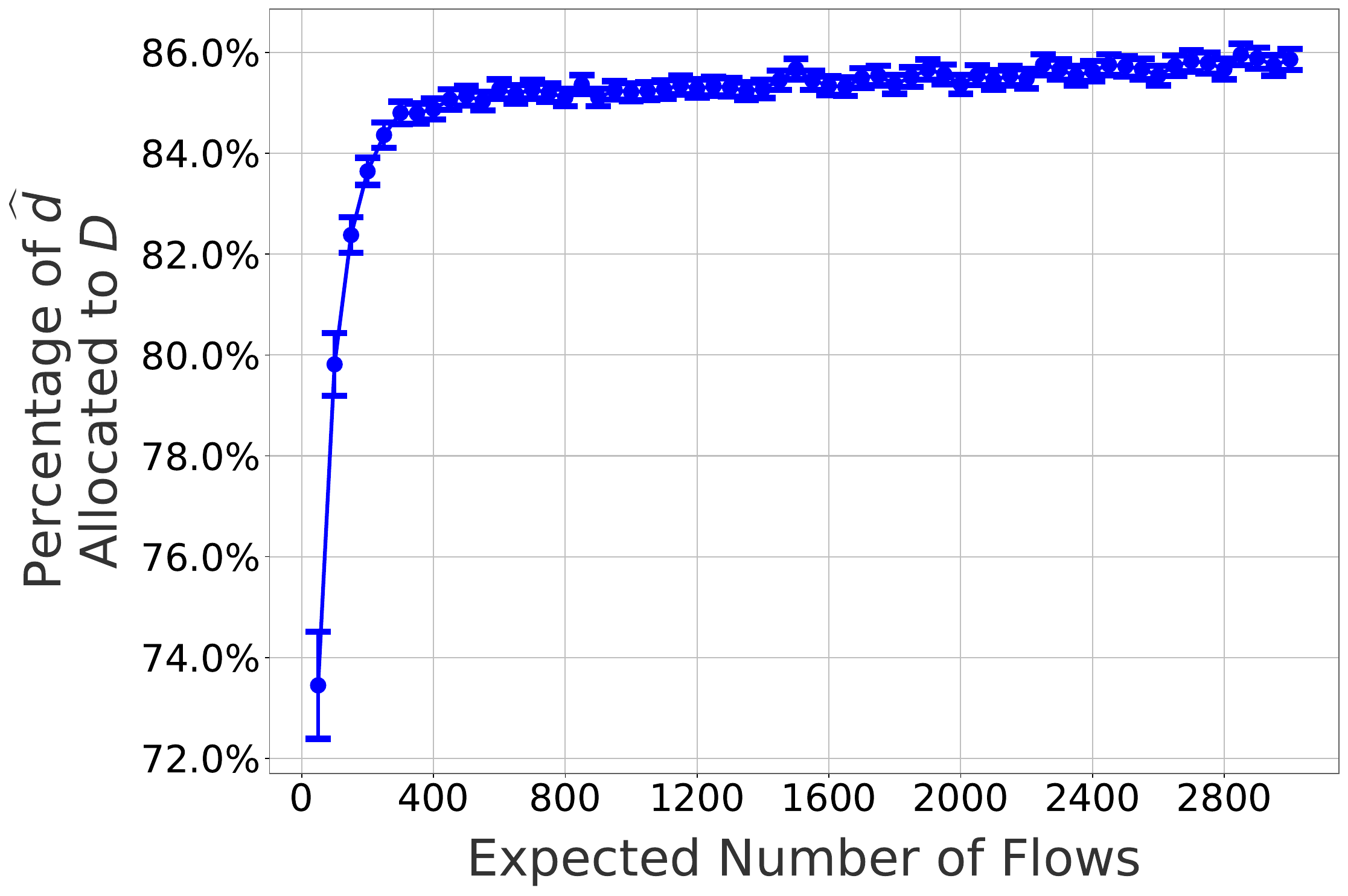}
  \caption{initial reprofiling ratio $D/\widehat{d}$}
  \label{fig:greedy_init_reprofiling}
\end{subfigure}
\caption{Greedy's bandwidth improvements and reprofiling strategy in the adjustment phase for US-Topo.}
\label{fig:greedy_init}
\end{figure*}

\fig{fig:greedy_config} reports the performance of Greedy as a function of $K$ and $L$ using again network topology~$2$ and flow profile configuration~$2$ with $m=128$ flows and $n=2$ links.  Similar results were observed for other network configurations. The results are again reported in the form of heat maps with lighter color corresponding to ``better'' outcomes, \ie lower bandwidth or lower run time.

\fig{fig:greedy_config_solution} shows that a single iteration $(L=1)$ is typically insufficient unless $K$ is quite large, but that similar performance can be realized with a smaller $K$ by increasing $L$.  The resulting gain in run time is displayed in \fig{fig:greedy_config_time}, which illustrates that intermediate configurations can offer an effective compromise between accuracy and computational cost, \eg the choice of $K=4$ and $L=2$ used in our evaluations.

\subsubsection{Improvement from Greedy's Adjustment Phase}
As we discussed in Section~\ref{sec:greedy}, Greedy approximates the NLP formulation through its ability to (1) explore different reprofiling configurations in its exploration phase, and (2) adjust local deadline assignments in its adjustment phase.  In this section, we investigate which phase contributes more to the bandwidth improvement.

\fig{fig:greedy_init} shows how Greedy improves the initial reprofiling solutions in its adjustment phase.  To create network configurations with heterogeneous traffic mix at each hop that magnify the importance of local deadline allocation, we consider the US-Topo introduced in Section~\ref{sec:net_topo} motivated by a real-world network topology.  We randomly select source and destination nodes to create ``S-D pairs'', and for each S-D pair sample flow profiles according to the three applications introduced in Section~\ref{sec:traffic}.  \fig{fig:greedy_init_improvement} shows the relative improvement in bandwidth that Greedy achieves in its adjustment phase compared to the initial reprofiling solution it starts from.  \fig{fig:greedy_init_reprofiling} reports the reprofiling ratio $D/\widehat{d}$ that the initial solution applies to every flow.  Both figures explore the number of S-D pairs that ranges from $10$ to $400$ with a step of $10$, with the y-axis showing the average and the $95^{th}$ percentile confidence interval across $1000$ random sets of flow profiles for each number of S-D pair.

As~\fig{fig:greedy_init_improvement} demonstrates, the bandwidth improvement that Greedy offers through its local deadline adjustments is relatively small.  As the number of S-D pairs increases, the average improvement quickly drops to about $2\%$ compared to the initial reprofiling solution.  The ability of Greedy to keep improving solution in the adjustment phase, however, is determined by the initial reprofiling configuration that would yield the best solution.  \fig{fig:greedy_init_reprofiling} shows that, the optimal solution is usually achieved at about $86\%$ initial reprofiling as the number of S-D pair increases, which only leaves a room of about $14\%$ for Greedy to make adjustments.  The small relative improvements in~\fig{fig:greedy_init_improvement} is, therefore, not surprising.  However, as we discussed in Section~\ref{sec:discuss}, it should be noted that, one of the reasons that Greedy usually favors a large initial reprofiling ratio is that the delay bounds at smaller reprofiling ratios tend to be looser.  Consequently, Greedy's ability to make bandwidth improvements through adjustments is limited by this potential looseness.

\subsection{NLP vs. FR and NR}
\label{sec:nlp_baseline}
\begin{figure*}[!h]
\centering
\begin{subfigure}{0.35\linewidth}
  \centering
  \includegraphics[width=\linewidth]{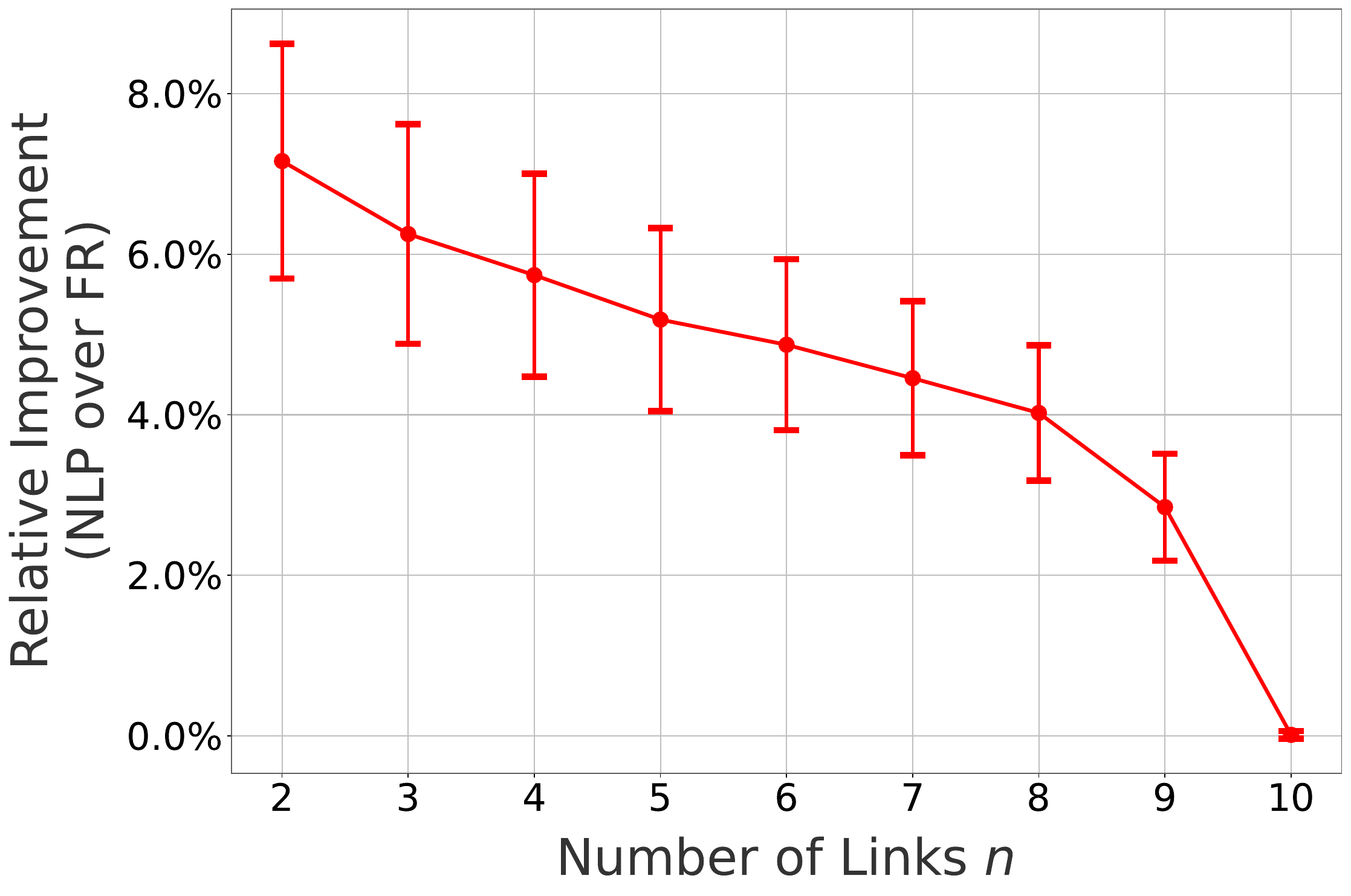}
  \caption{NLP vs.~FR}
  \label{fig:nlp_11}
\end{subfigure}
\begin{subfigure}{0.35\linewidth}
  \centering
  \includegraphics[width=\linewidth]{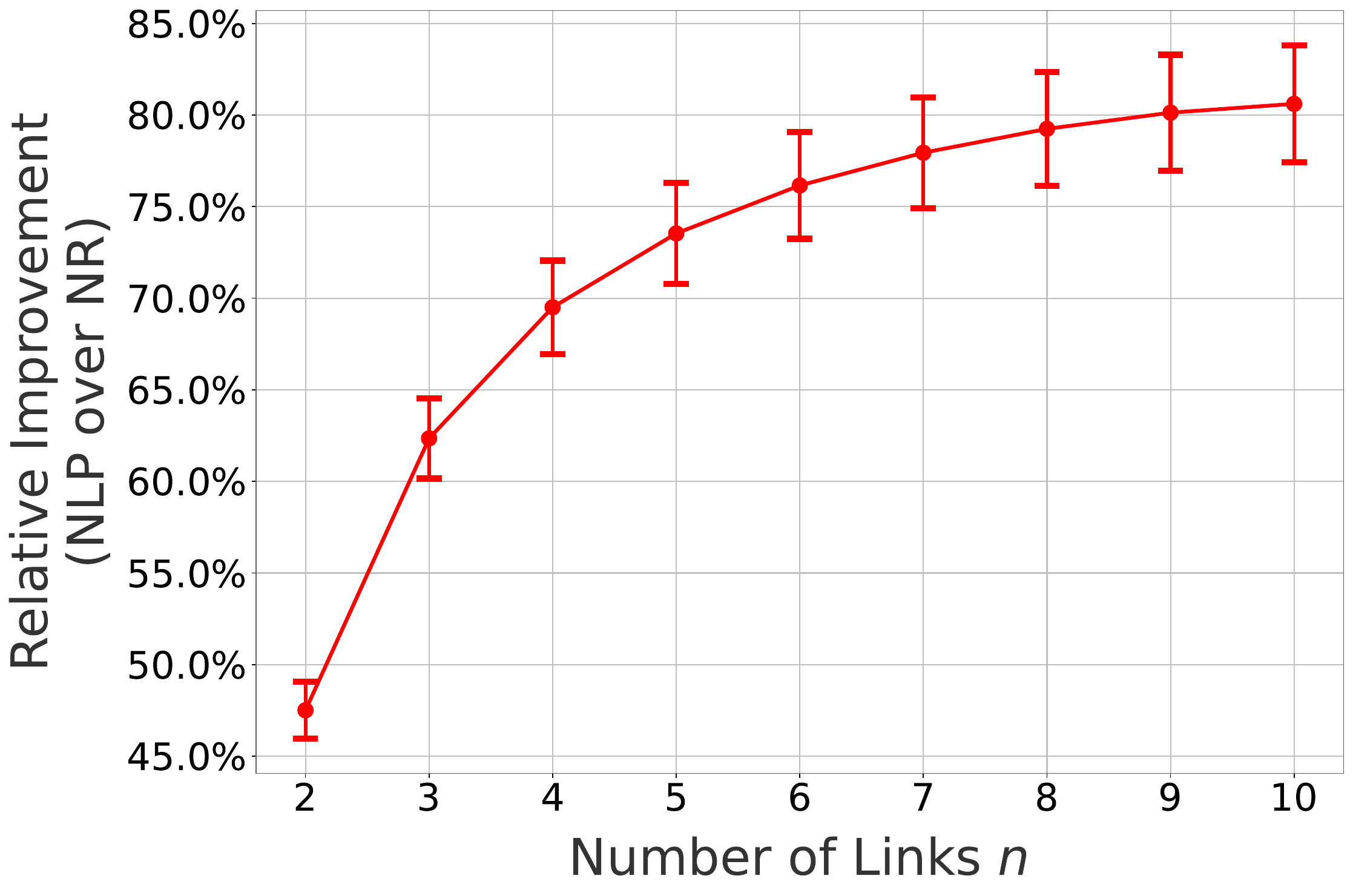}
  \caption{NLP vs.~NR}
  \label{fig:nlp_12}
\end{subfigure}
\caption{Relative improvement of NLP over FR and NR ($m=3$ flows, \# links, $n$, varies).}
\label{fig:eval_nlp1}
\end{figure*}

\highlight{\useColor}{To provide some perspective on Greedy's ability to closely approximate the NLP-based solution, we also compared NLP to the two baselines of FR and NR.  Greedy's ability to approximate NLP is meaningful mostly when NLP's own ability to improve over both baselines is substantial.  We carry out this comparison for the two sets of evaluations of Section~\ref{sec:algorithm_comparison}, with \figs{fig:eval_nlp1}{fig:eval_nlp2} reporting NLP's relative improvement for the configurations of~\figs{fig:performance_hop}{fig:performance_flow}, respectively.}

\highlight{\useColor}{Not surprisingly, NLP (and therefore Greedy) achieves a significant improvement over NR, typically above $50\%$. According to~\fig{fig:nlp_12}, this grows with the number of links $n$.  This is because the benefits of reprofiling accrue at each hop that flows traverse. In contrast, as illustrated by~\fig{fig:nlp_11}, the improvement of NLP over FR decreases with $n$, eventually dropping down to about $0$ for $n=10$.  This is also not surprising, again because a larger number of hops favors the consistent benefits that reprofiling accrues at each hop over the occasional benefit afforded by preserving some scheduling flexibility through non-zero local deadlines.  Hence, FR becomes the solution of choice.}

\begin{figure*}[!h]
\centering
\begin{subfigure}{0.35\linewidth}
  \centering
  \includegraphics[width=\linewidth]{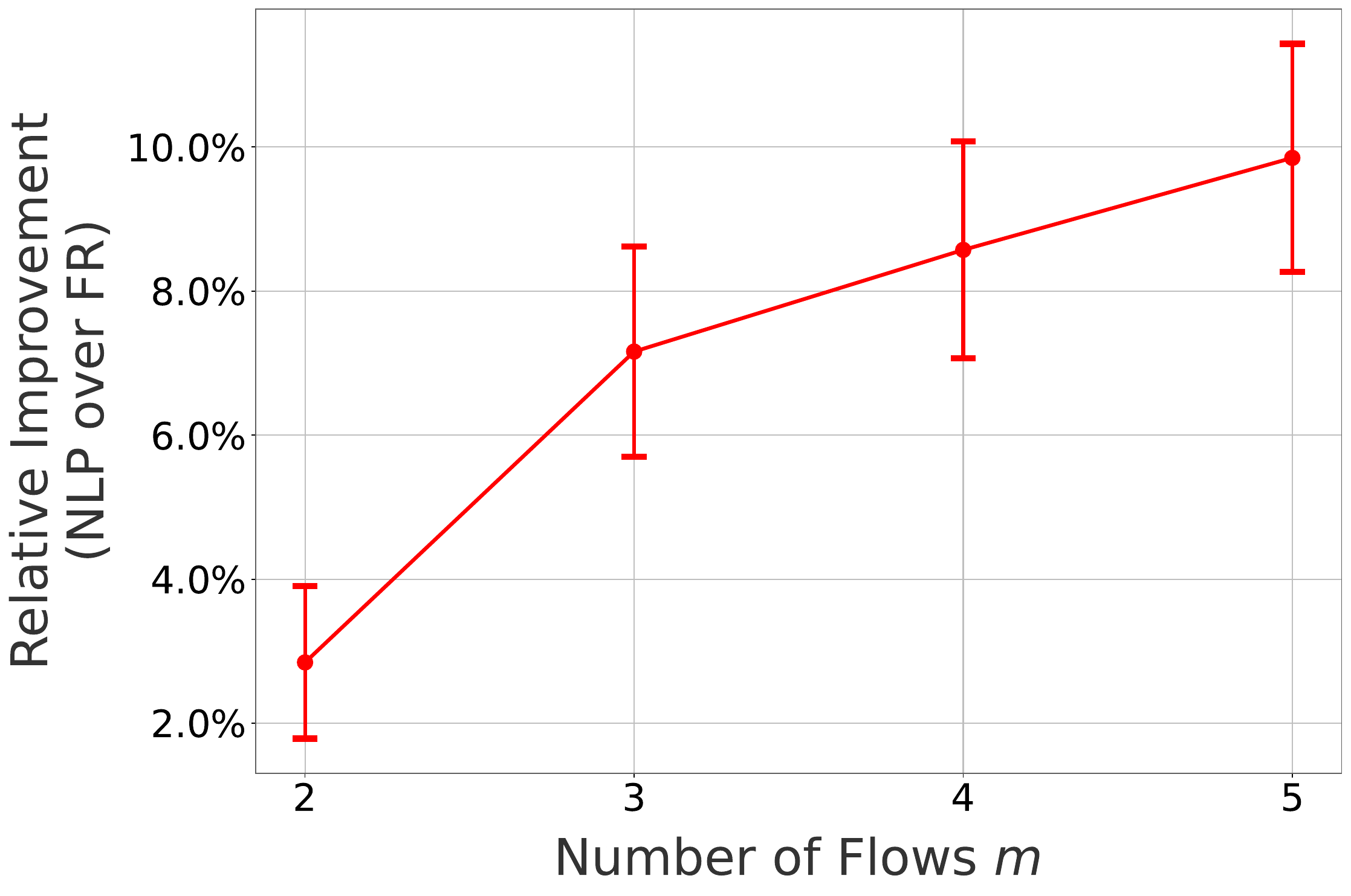}
  \caption{NLP vs.~FR}
  \label{fig:nlp_21}
\end{subfigure}
\begin{subfigure}{0.35\linewidth}
  \centering
  \includegraphics[width=\linewidth]{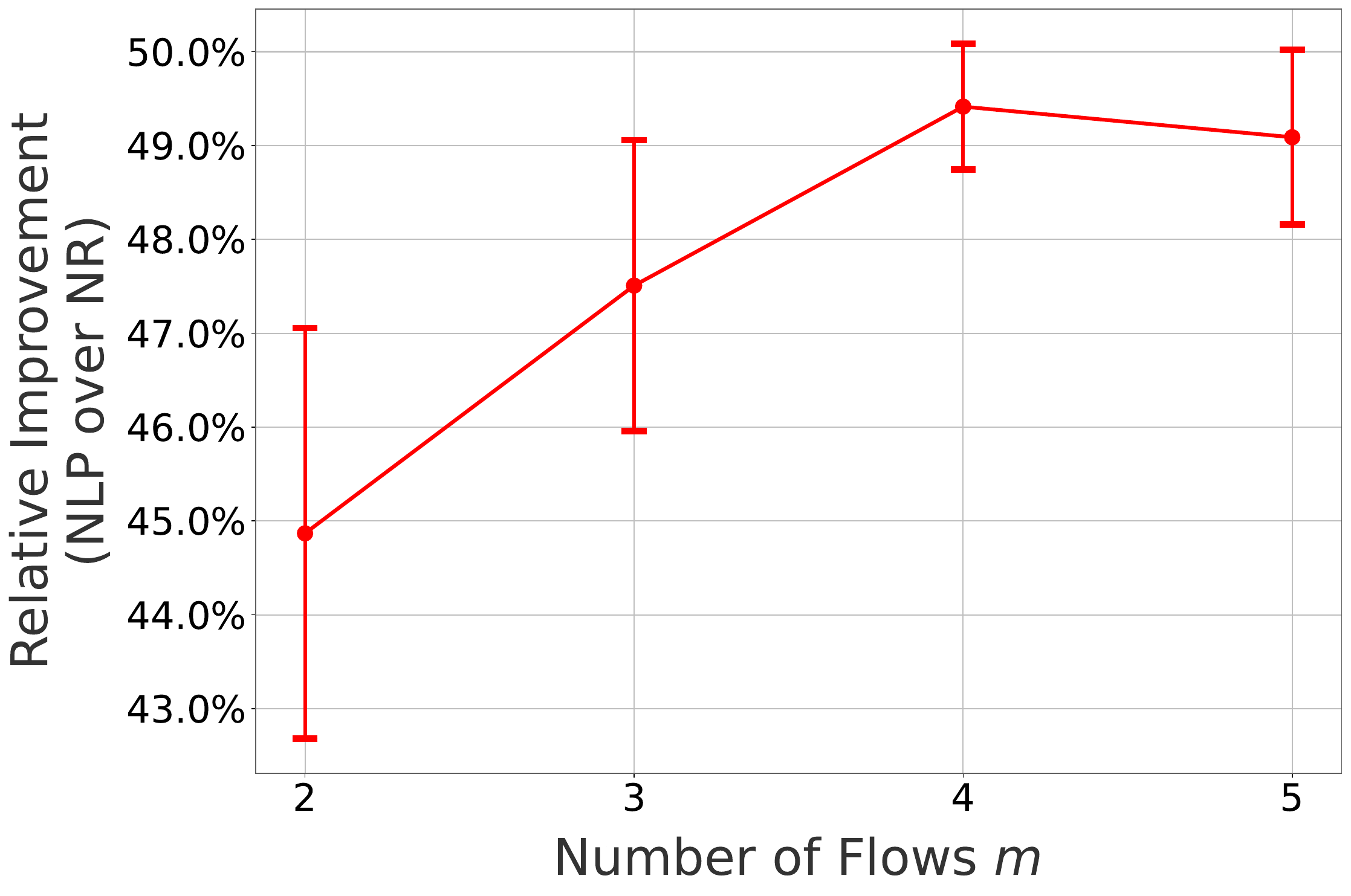}
  \caption{NLP vs.~NR}
  \label{fig:nlp_22}
\end{subfigure}
\caption{Relative improvement of NLP over FR and NR ($n=2$ links, \# flows, $m$, varies).}
\label{fig:eval_nlp2}
\end{figure*}

\highlight{\useColor}{Conversely and as illustrated in \fig{fig:nlp_21}, when the number of hops is fixed, increasing the number of flows creates more opportunities to leverage scheduling flexibility.  This then leads NLP (and Greedy) to select intermediate solutions that combine reprofiling with non-zero local deadlines to take advantage of those opportunities.  The latter allows NLP and Greedy\footnote{\highlight{\useColor}{\fig{fig:performance_flow} shows that Greedy is within about $1\%$ of the NLP-based solution in those cases.}} to outperform FR by more than $10\%$ as illustrated in~\fig{fig:nlp_21} for $m=5$.  This makes Greedy particularly valuable since NLP becomes quickly intractable as the number of flows increases, while Greedy scales well as shown in~\fig{fig:time2}.}

\subsection{Evaluation Results for Synthetic Networks}
\label{sec:synthetic}
\begin{figure}[!h]
\centering
\begin{subfigure}{0.8\linewidth}
  \centering
  \includegraphics[width=\linewidth]{Figure/parking_lot1.png}
  \caption{topology~$1$: tandem}
  \label{fig:tandem1}
\end{subfigure}
\begin{subfigure}{0.8\linewidth}
  \centering
  \includegraphics[width=\linewidth]{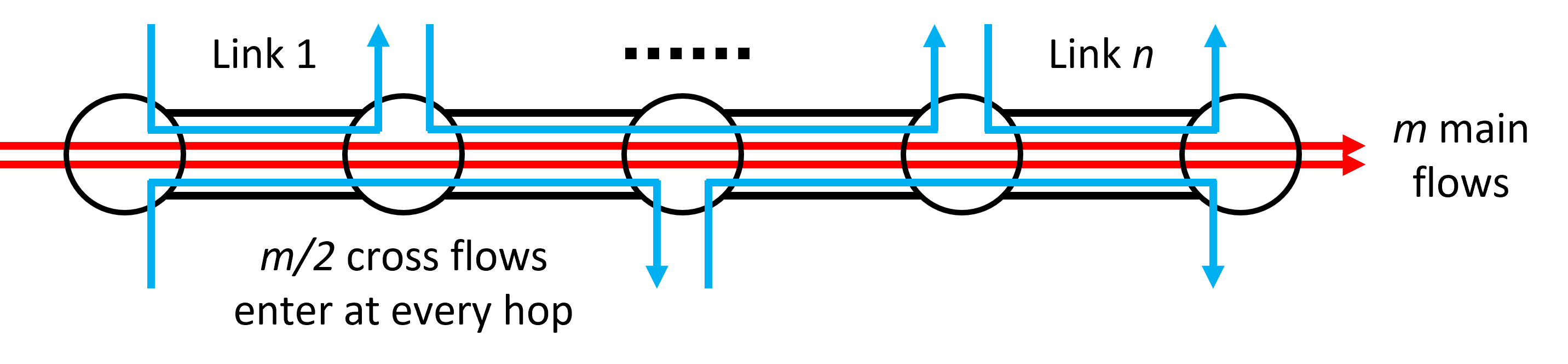}
  \caption{topology~$2$: parking lot}
  \label{fig:tandem2}
\end{subfigure}
\caption{Synthetic network topologies.}
\label{fig:tandem_network}
\end{figure}

Our synthetic topologies include both the tandem topology (topology~$1$) investigated in Section~\ref{sec:algorithm_comparison}, and a parking lot topology (topology~$2$) of \fig{fig:tandem2}.
In the parking lot topology, $m/2$ ``cross flows'' enter the network at every hop and overlap with the $m$ main traffic flows for~$2$ hops, thereby creating $m$ cross flows on every link. In addition,
the investigation is expanded to include additional flow profiles, namely, both configurations~$1$ (investigated in Section~\ref{sec:algorithm_comparison}) and~$2$ of Table~\ref{tab:profile}.  Configuration~$2$ has~$4$ per-hop deadlines values, $\{0.01, 0.025, 0.05, 0.1\}$, with a dynamic range of~$10$ vs.~$100$ for configuration~$1$.
\begin{table}[!h]
\begin{center}
\caption{Synthetic flow profiles}
\begin{tabular}{|l|c|c|}
\hline
\textbf{Configurations} & \textbf{1} & \textbf{2} \\\hline
\textbf{Rate} & \multicolumn{2}{c|}{uniformly distributed over $[1, 100]$}\\\hline
\textbf{Burst size} & \multicolumn{2}{c|}{uniformly distributed over $[1, 100]$}\\\hline
\textbf{Deadline Classes} & $\{0.01, 0.1, 1\}$ &  $\{0.01, 0.025, 0.05, 0.1\}$\\\hline
\end{tabular}
\label{tab:profile}
\end{center}
\end{table}
\begin{figure*}[!h]
\centering
\begin{subfigure}{0.35\linewidth}
  \centering
  \includegraphics[width=\linewidth]{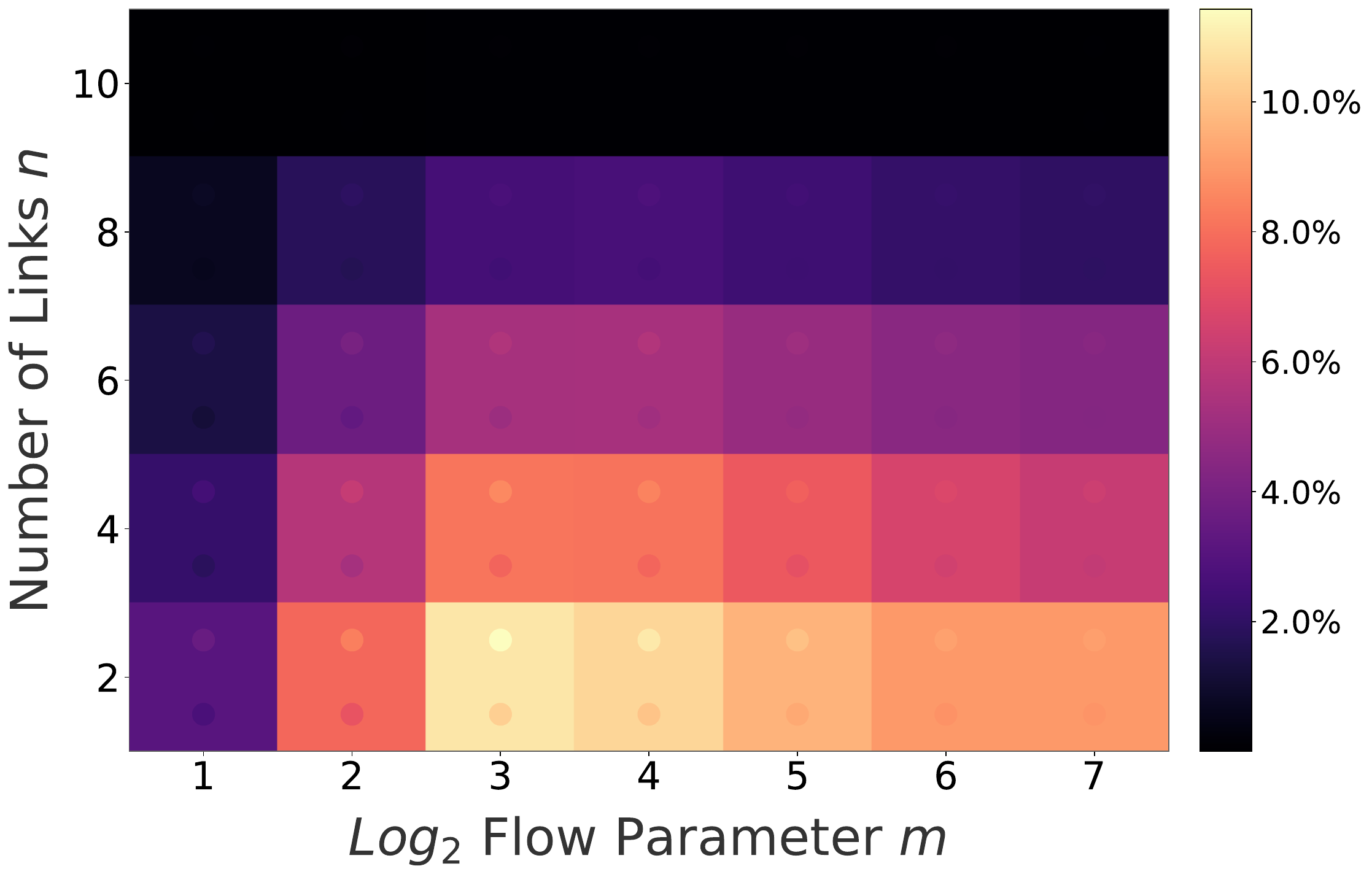}
  \caption{topology 1, configuration 1}
  \label{fig:fr_11}
\end{subfigure}
\begin{subfigure}{0.35\linewidth}
  \centering
  \includegraphics[width=\linewidth]{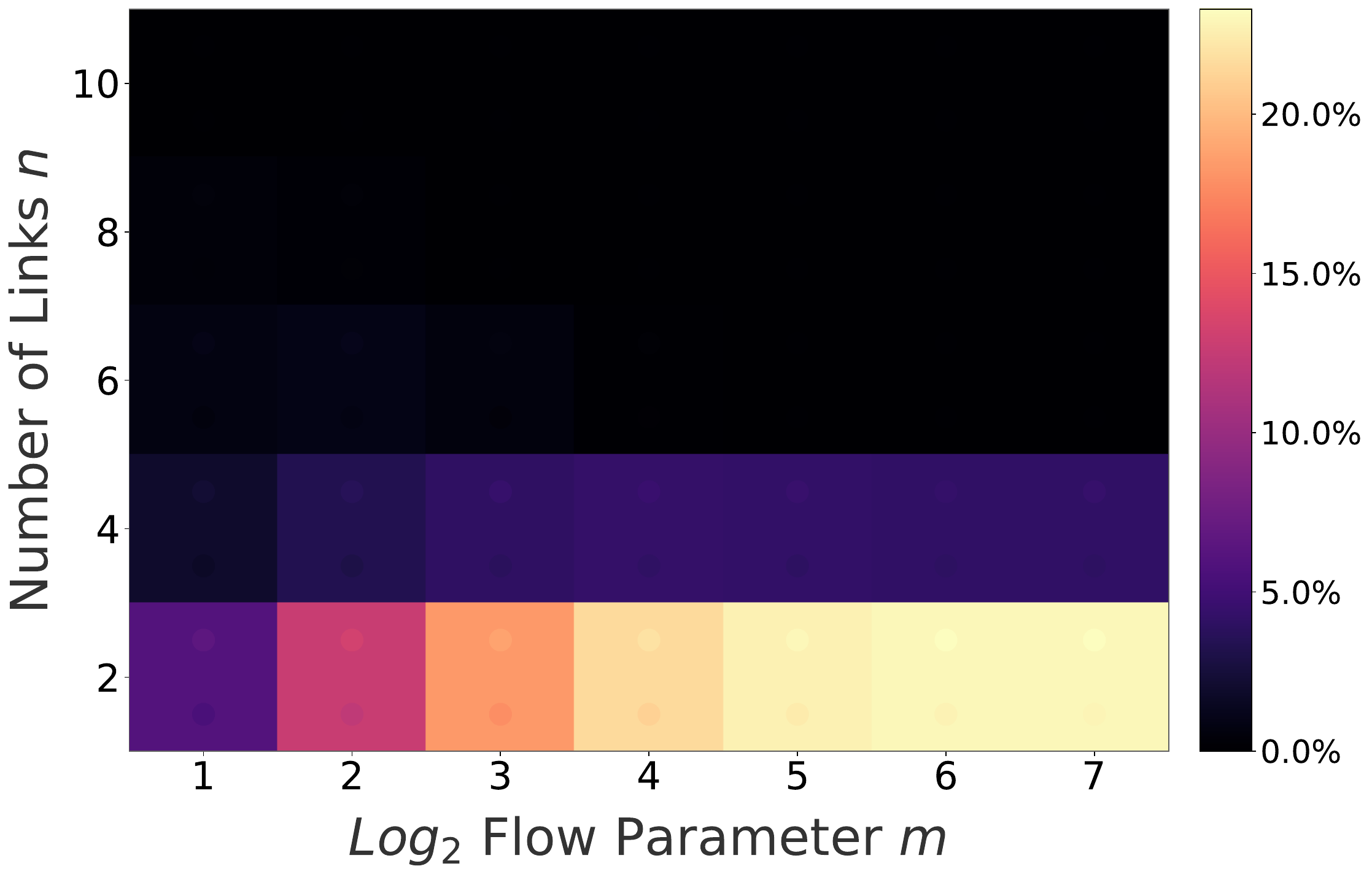}
  \caption{topology 1, configuration 2}
  \label{fig:fr_12}
\end{subfigure}
\begin{subfigure}{0.35\linewidth}
  \centering
  \includegraphics[width=\linewidth]{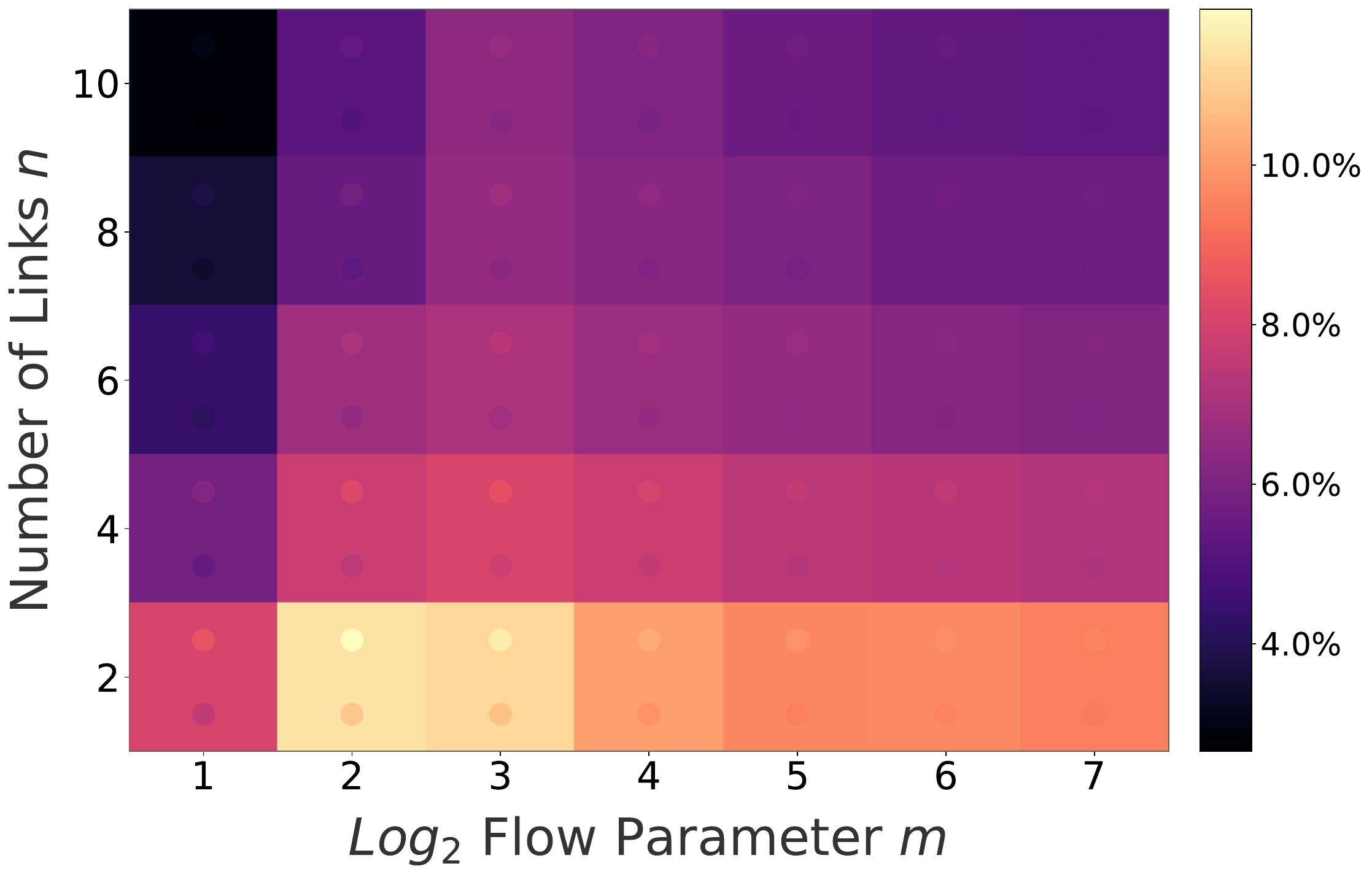}
  \caption{topology 2, configuration 1}
  \label{fig:fr_21}
\end{subfigure}
\begin{subfigure}{0.35\linewidth}
  \centering
  \includegraphics[width=\linewidth]{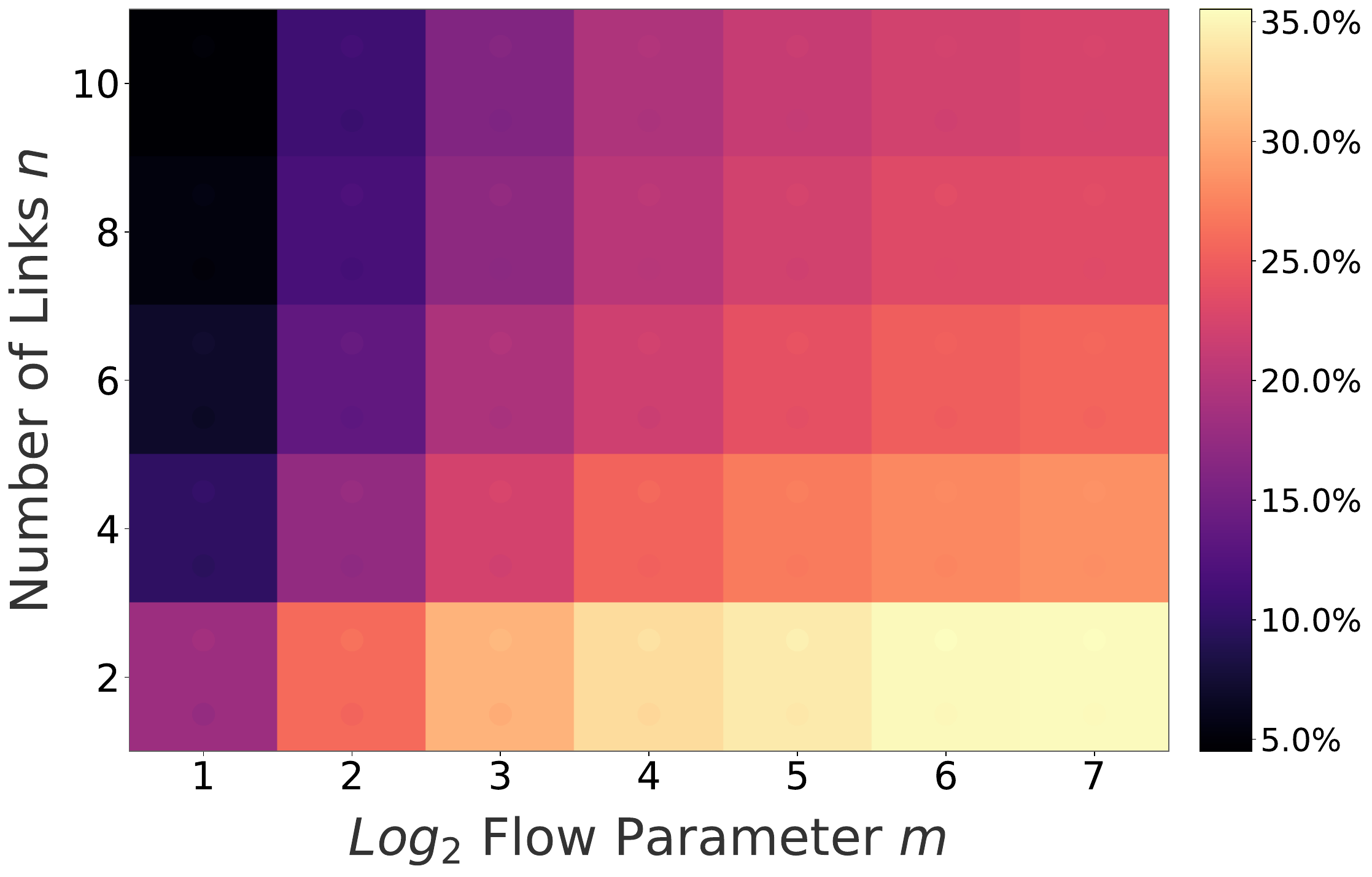}
  \caption{topology 2, configuration 2}
  \label{fig:fr_22}
\end{subfigure}
\caption{Relative improvement of Greedy over FR (fixed hop-by-hop deadlines).}
\label{fig:eval_fr}
\end{figure*}
\linebreak

\begin{figure*}[!h]
\centering
\begin{subfigure}{0.35\linewidth}
  \centering
  \includegraphics[width=\linewidth]{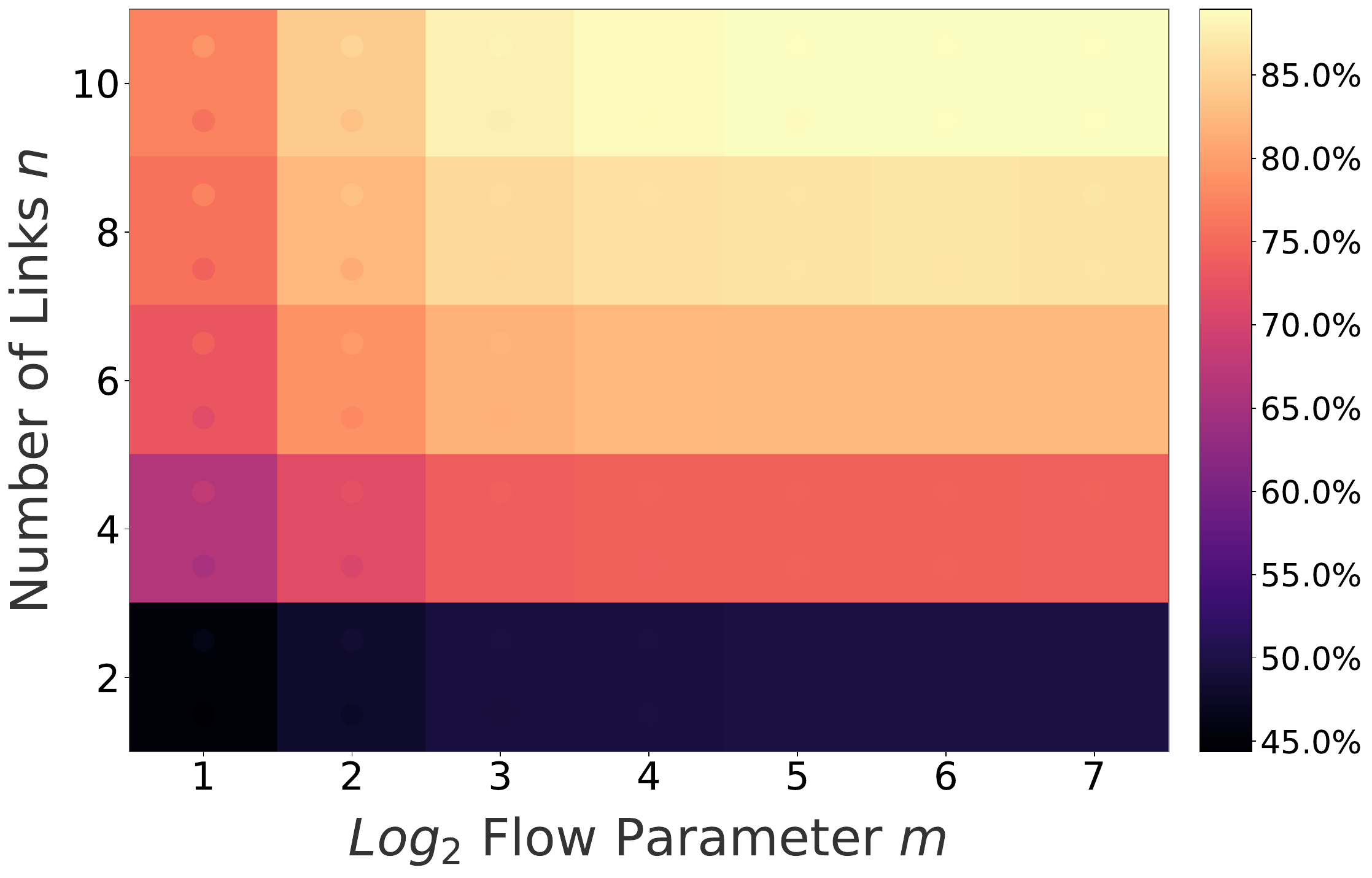}
  \caption{topology 1 configuration 1}
  \label{fig:nr_11}
\end{subfigure}
\begin{subfigure}{0.35\linewidth}
  \centering
  \includegraphics[width=\linewidth]{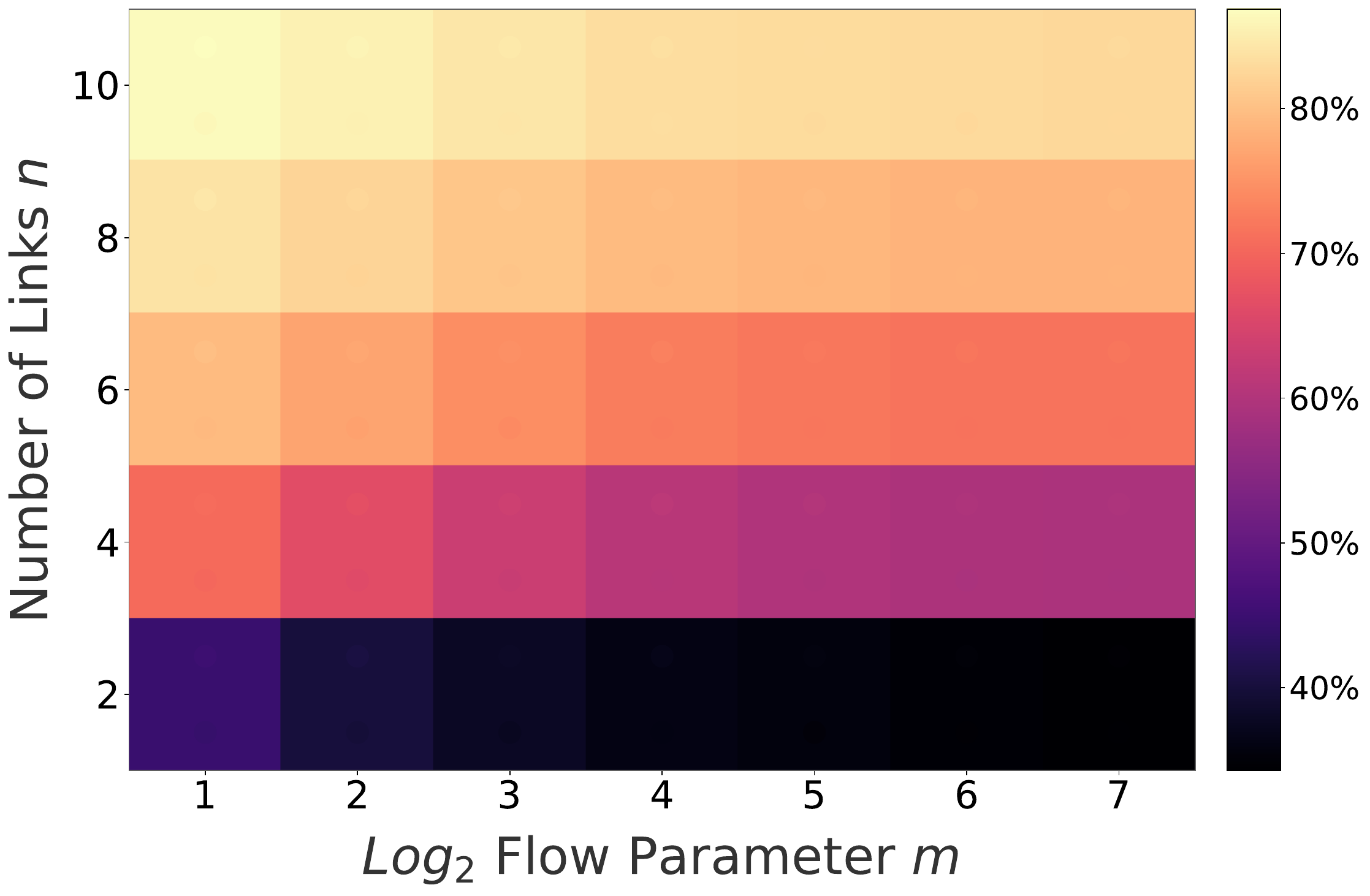}
  \caption{topology 1 configuration 2}
  \label{fig:nr_12}
\end{subfigure}
\begin{subfigure}{0.35\linewidth}
  \centering
  \includegraphics[width=\linewidth]{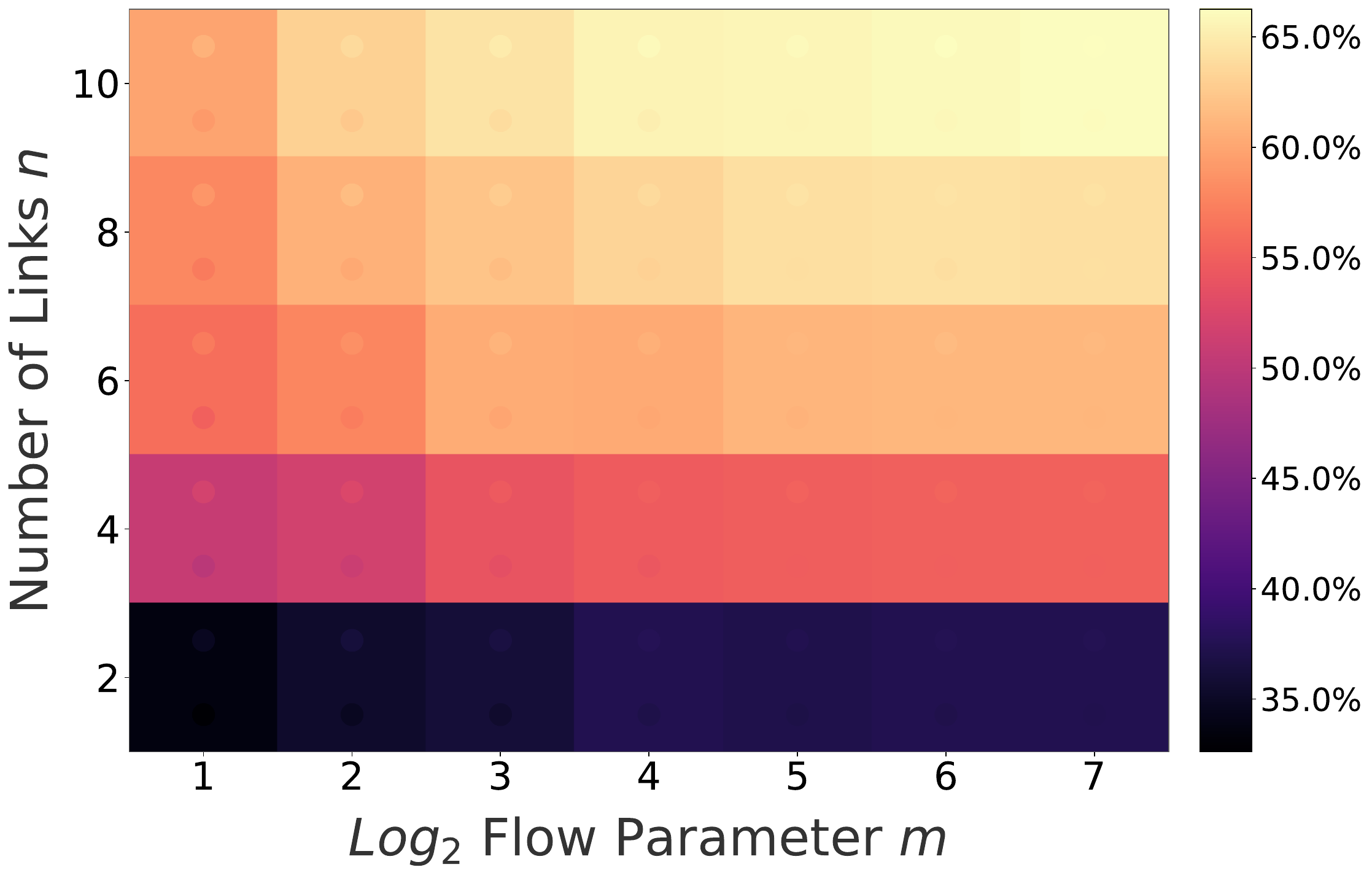}
  \caption{topology 2 configuration 1}
  \label{fig:nr_21}
\end{subfigure}
\begin{subfigure}{0.35\linewidth}
  \centering
  \includegraphics[width=\linewidth]{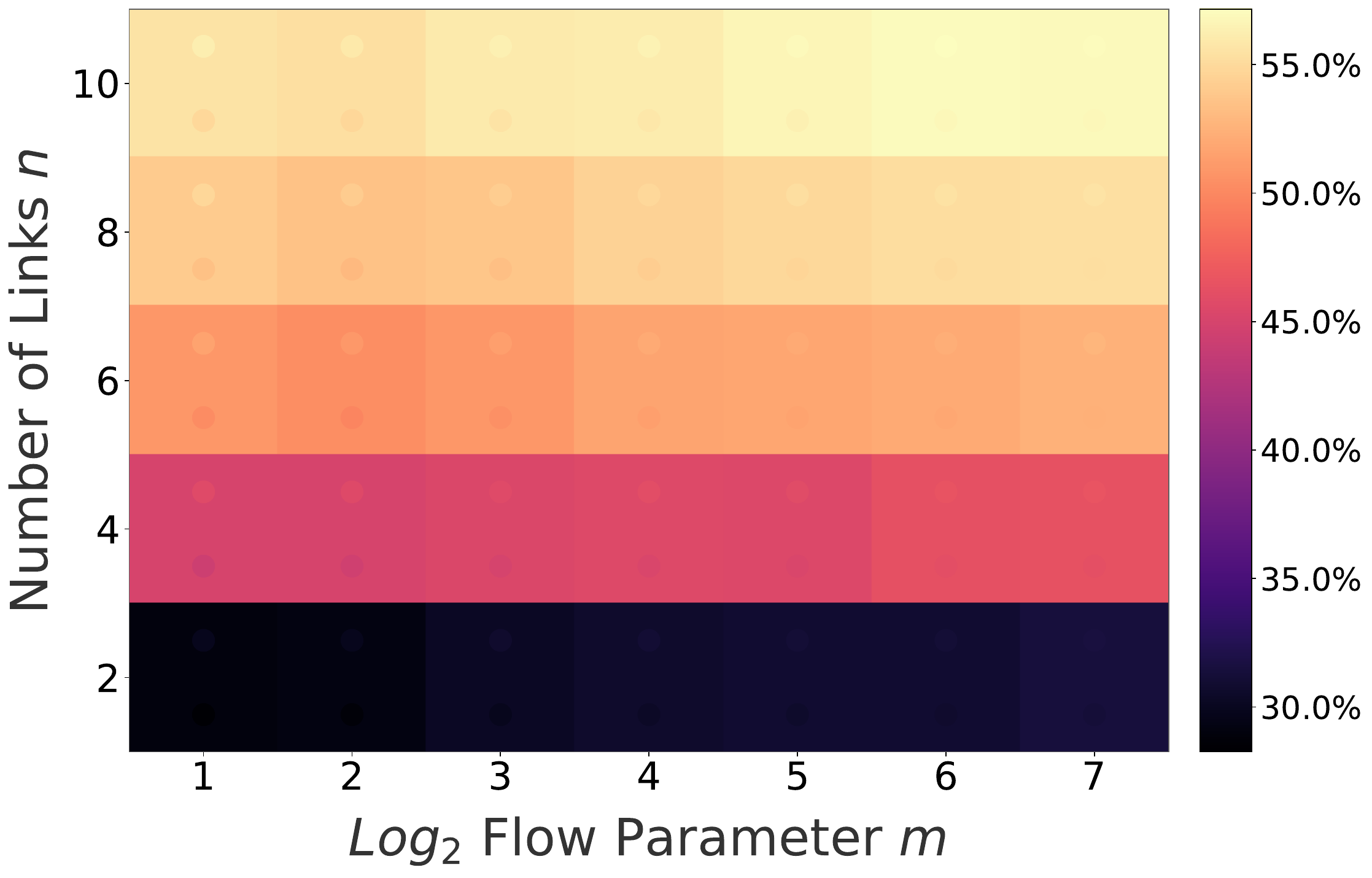}
  \caption{topology 2 configuration 2}
  \label{fig:nr_22}
\end{subfigure}
\caption{Relative Improvement of Greedy over NR (fixed hop-by-hop deadlines).}
\label{fig:eval_nr}
\end{figure*}

We report the relative average improvement of Greedy over FR and NR for four combinations spanning topologies $1$ and $2$ and flow profile configurations $1$ and $2$. We note that Greedy's potential improvements come from both reprofiling and its unequal assignment of deadlines across hop even if, as mentioned at the end of Section~\ref{sec:greedy_overall}, the latter is only a consequence of Greedy's reprofiling adustment phase. For each combination, we perform a ``grid search'' varying $m$ from $2^1$ to $2^7$, and $n$ from $2$ to $10$ using a step size of~$2$, with $1000$ flow profiles sampled for each $(m,n)$ pair.  The results are reported in~\figs{fig:eval_fr}{fig:eval_nr} as color-coded heat maps of Greedy's relative average improvement with lighter colors corresponding to larger improvements.  The $x$-axis of each heat map shows the $\log_2$ of the number of flows, $m$, while the $y$-axis shows the number of links (hops), $n$.  Each entry in the heat maps also includes two vertically aligned, color-coded dots\footnote{Unfortunately, a common color scheme makes those dots difficult to see in configurations where their value is close to that of the mean.} that represent the $95^{th}$ percentile confidence intervals across the $1000$ instances of each experiment.  

We note that, for readability, the color scales vary across figures. Hence, comparisons between configurations across figures cannot be ``color-based.''  Instead, they should rely on the actual values reported next to each color scale.

We first compare Greedy to FR in \fig{fig:eval_fr}.  The figure illustrates that although the relative benefits of reprofiling vary as a function of topology, flow profiles, number of flows, and number of hops/links, they are consistently present and even in double digits in many scenarios.

\begin{figure*}[!h]
\centering
\begin{subfigure}{0.35\linewidth}
  \centering
  \includegraphics[width=\linewidth]{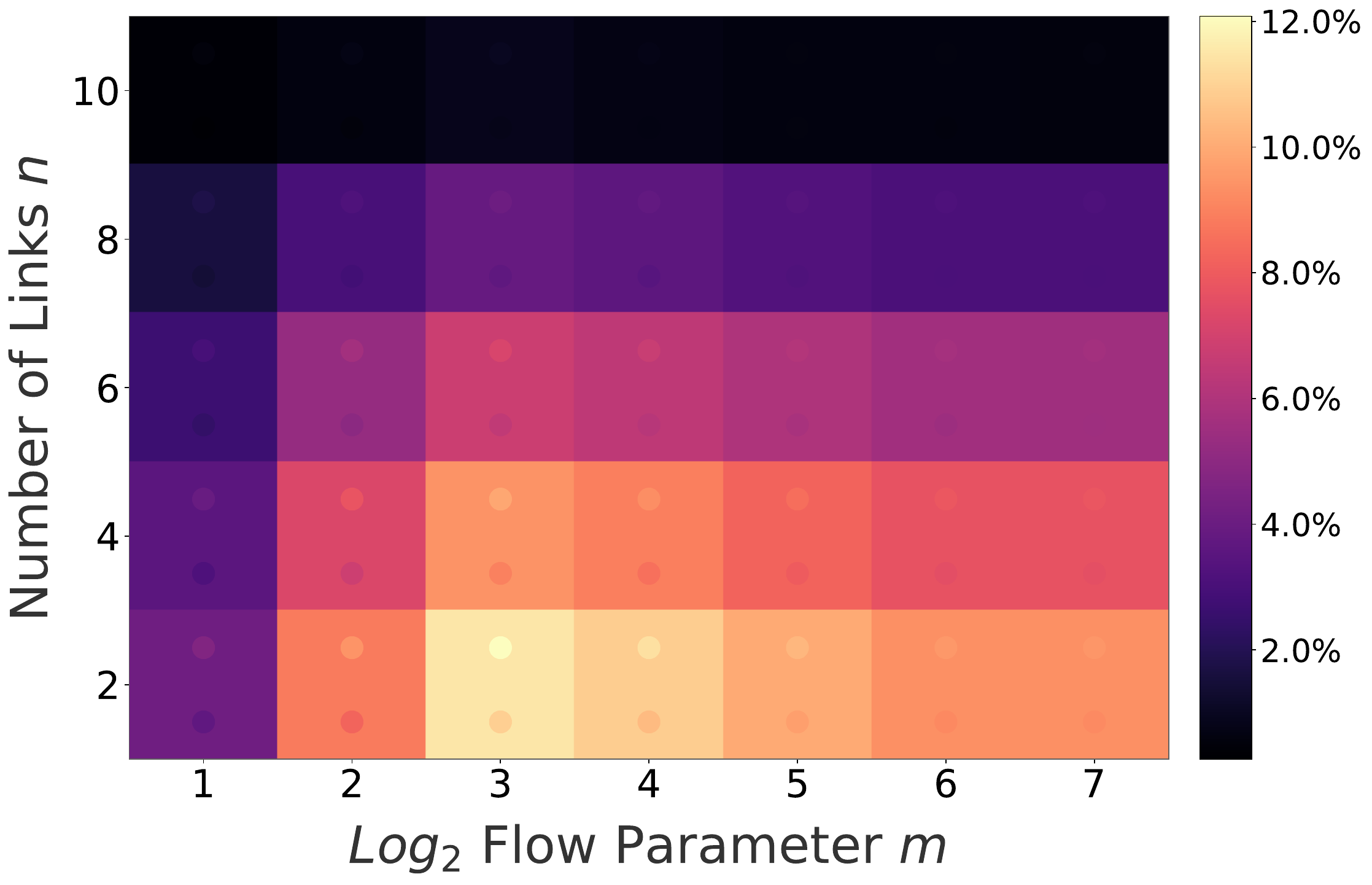}
  \caption{topology 1 configuration 1}
  \label{fig:fd_fr_11}
\end{subfigure}
\begin{subfigure}{0.35\linewidth}
  \centering
  \includegraphics[width=\linewidth]{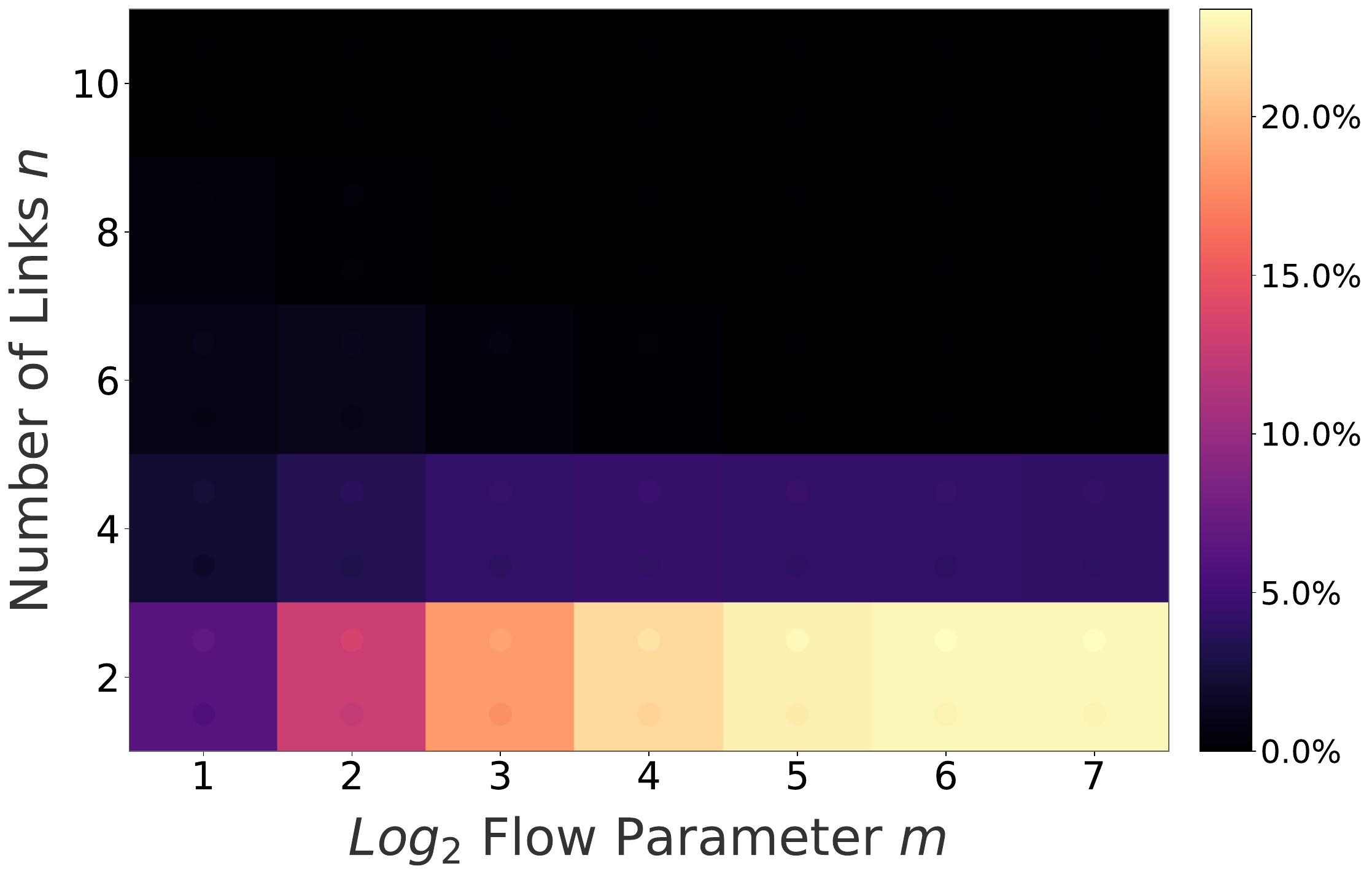}
  \caption{topology 1 configuration 2}
  \label{fig:fd_fr_12}
\end{subfigure}
\begin{subfigure}{0.35\linewidth}
  \centering
  \includegraphics[width=\linewidth]{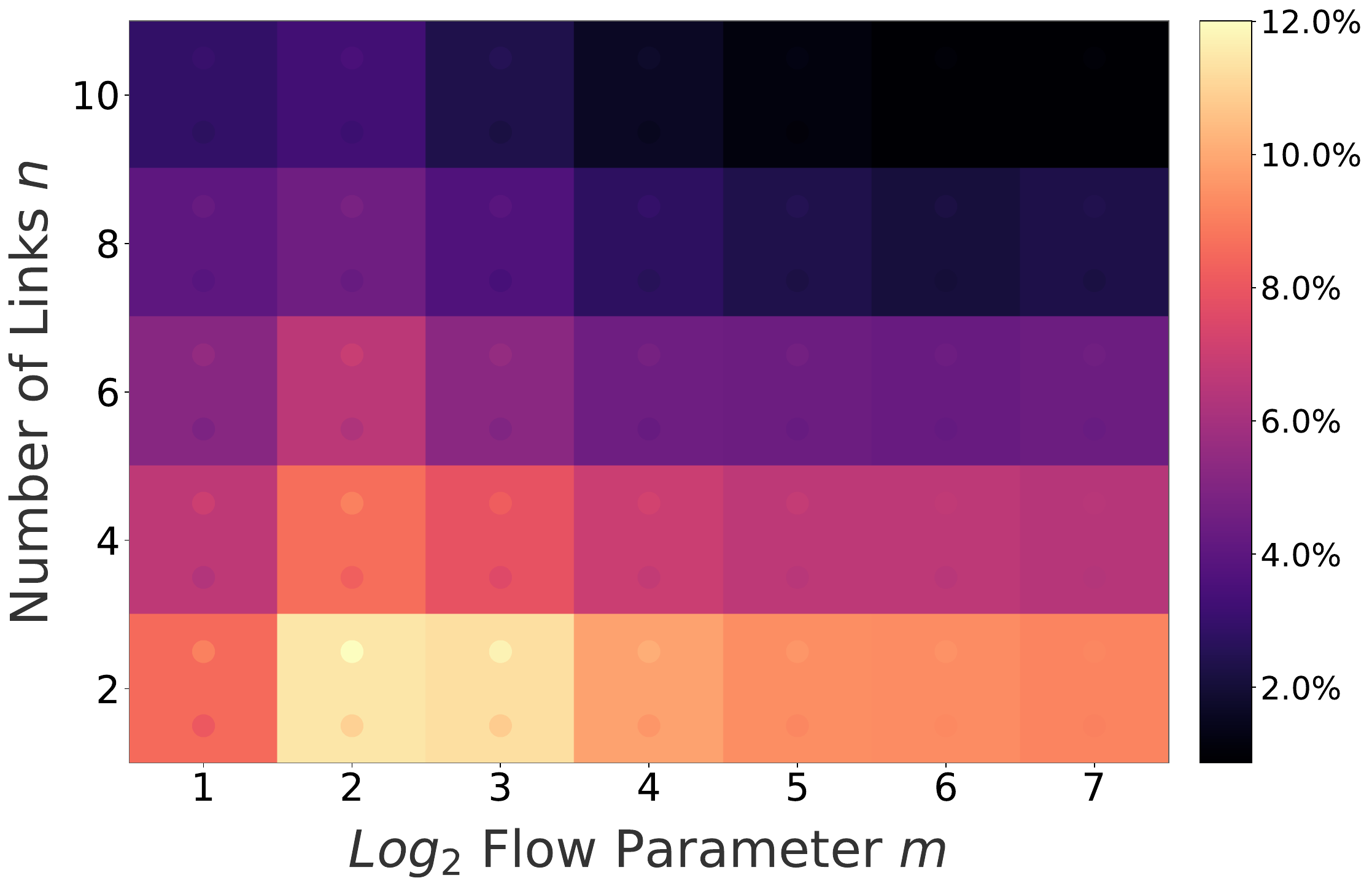}
  \caption{topology 2 configuration 1}
  \label{fig:fd_fr_21}
\end{subfigure}
\begin{subfigure}{0.35\linewidth}
  \centering
  \includegraphics[width=\linewidth]{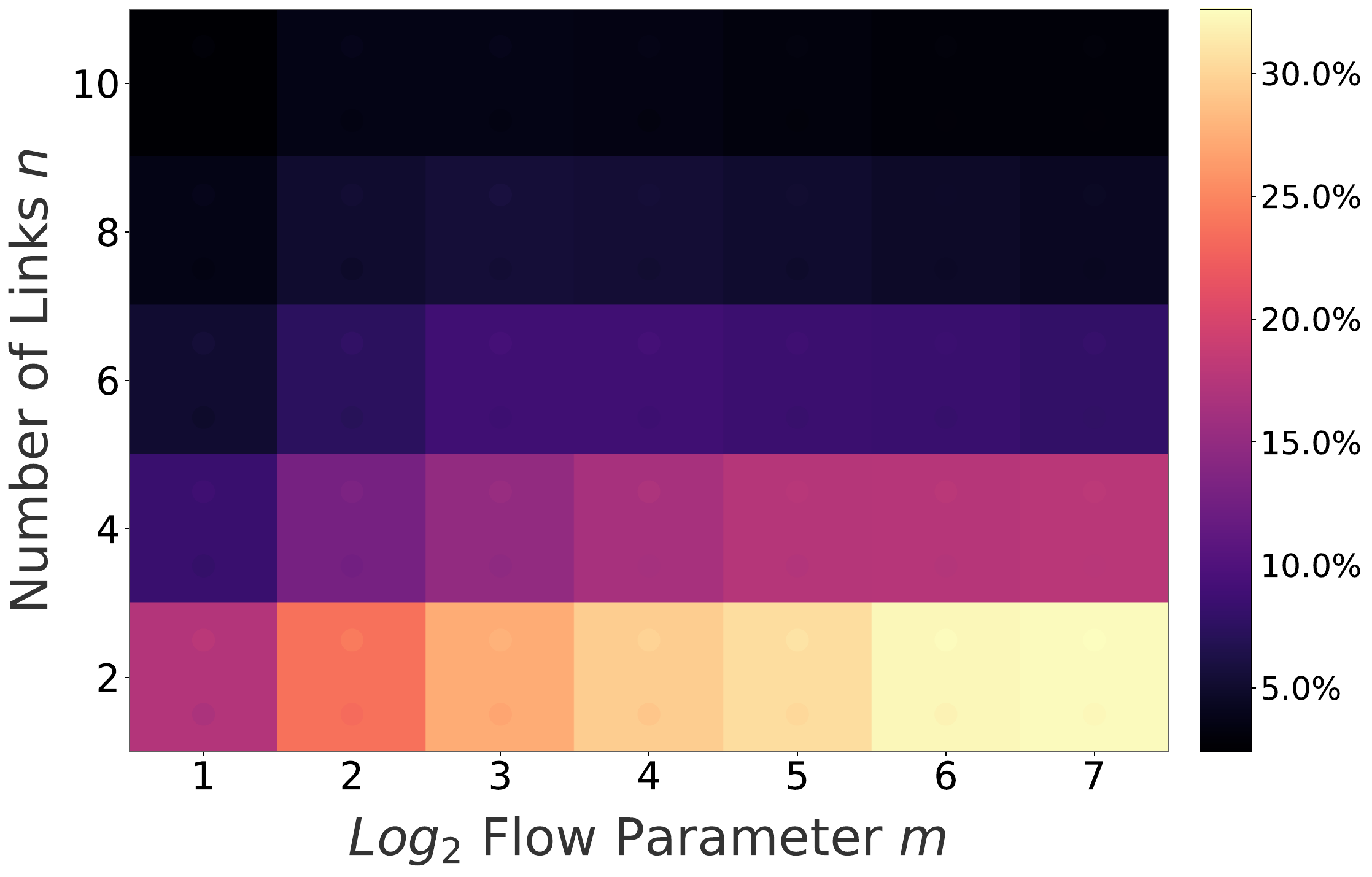}
  \caption{topology 2 configuration 2}
  \label{fig:fd_fr_22}
\end{subfigure}
\caption{Relative Improvement of Greedy over FR (fixed end-to-end deadlines).}
\label{fig:eval_fd_fr}
\end{figure*}

The one behavior that emerges somewhat consistently is that, as alluded to earlier, the benefits of Greedy over FR decrease as the number of hops, $n$, increases.  Recall that this is because we keep per-hop deadlines constant so that a larger $n$ means a larger end-to-end deadline that affords more opportunities for reprofiling, and ultimately convergence to \add{reprofiling} at the flow's long-term rate, \ie FR. As previously discussed, the alternative strategy of keeping end-to-end rather than per-hop deadlines fixed has a similar limiting behavior even if for a different reason (the per-hop deadlines converge to~$0$, which ultimately forces the reprofiling policy to mimic FR). As a confirmation, we performed a similar set of experiments as those of \figs{fig:eval_fr}{fig:eval_nr} but using fixed end-to-end deadlines.  The results are available in \figs{fig:eval_fd_fr}{fig:eval_fd_nr}.

%

\begin{figure*}[!h]
\centering
\begin{subfigure}{0.35\linewidth}
  \centering
  \includegraphics[width=\linewidth]{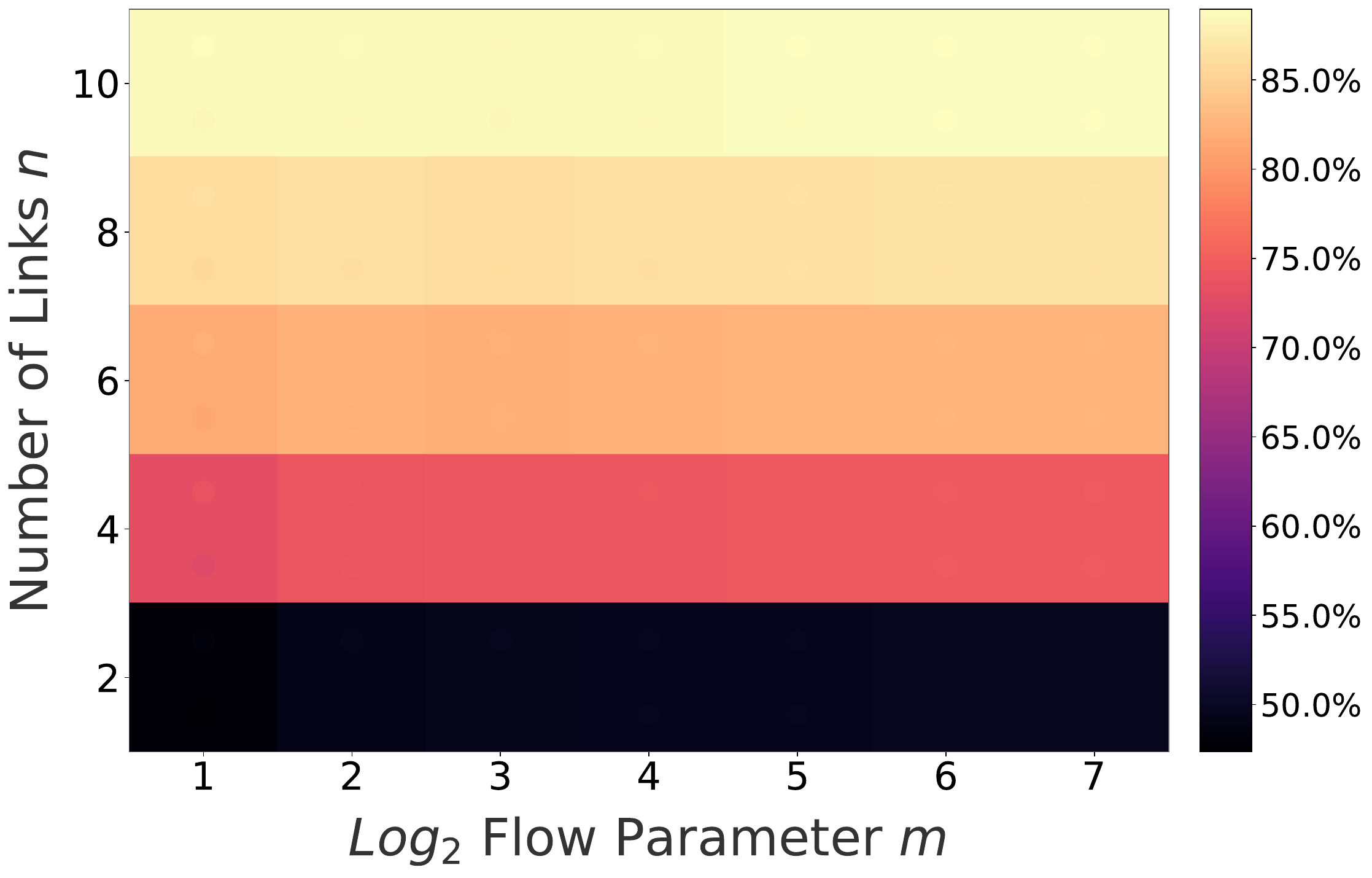}
  \caption{topology 1 configuration 1}
  \label{fig:fd_nr_11}
\end{subfigure}
\begin{subfigure}{0.35\linewidth}
  \centering
  \includegraphics[width=\linewidth]{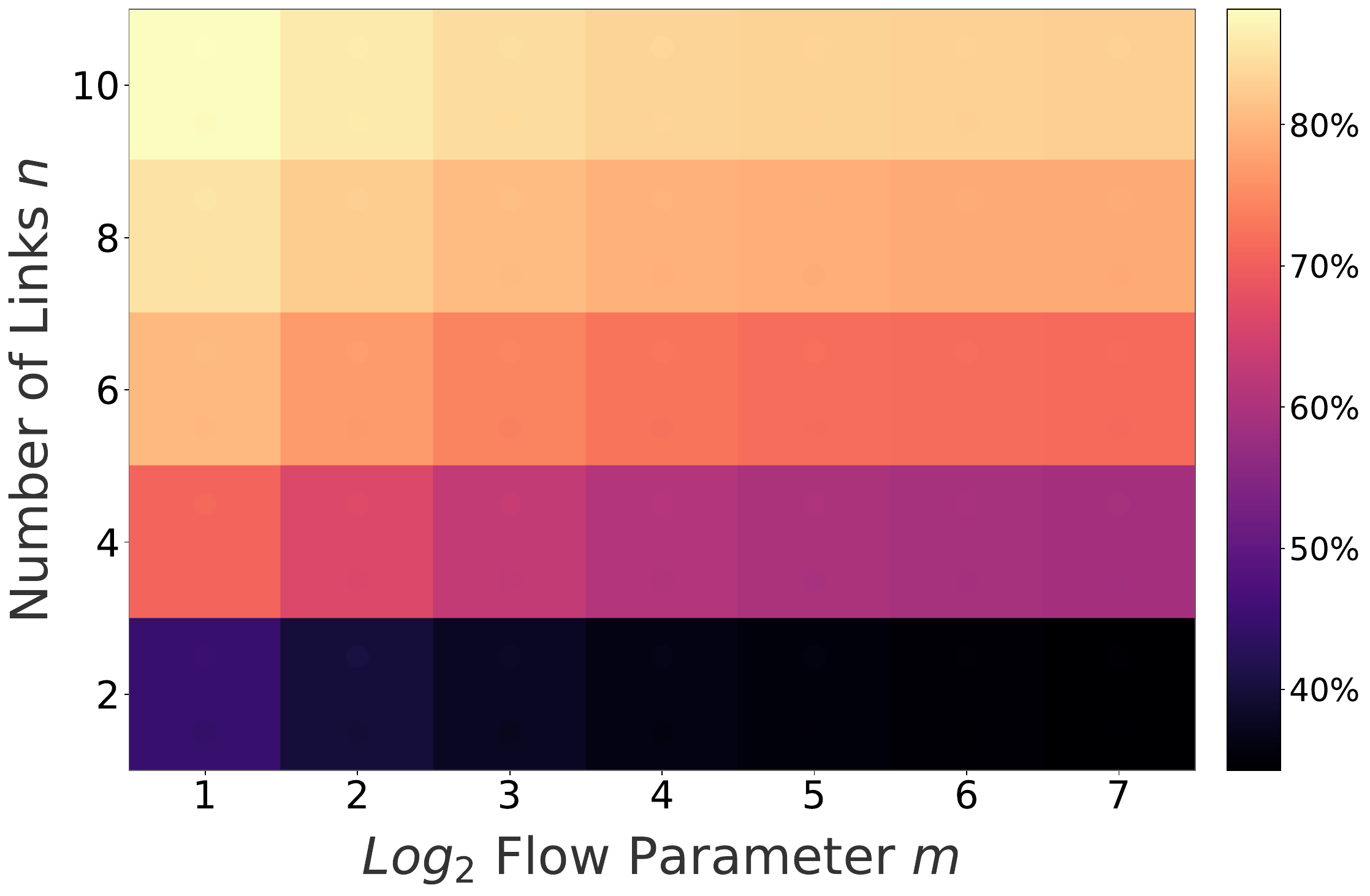}
  \caption{topology 1 configuration 2}
  \label{fig:fd_nr_12}
\end{subfigure}
\begin{subfigure}{0.35\linewidth}
  \centering
  \includegraphics[width=\linewidth]{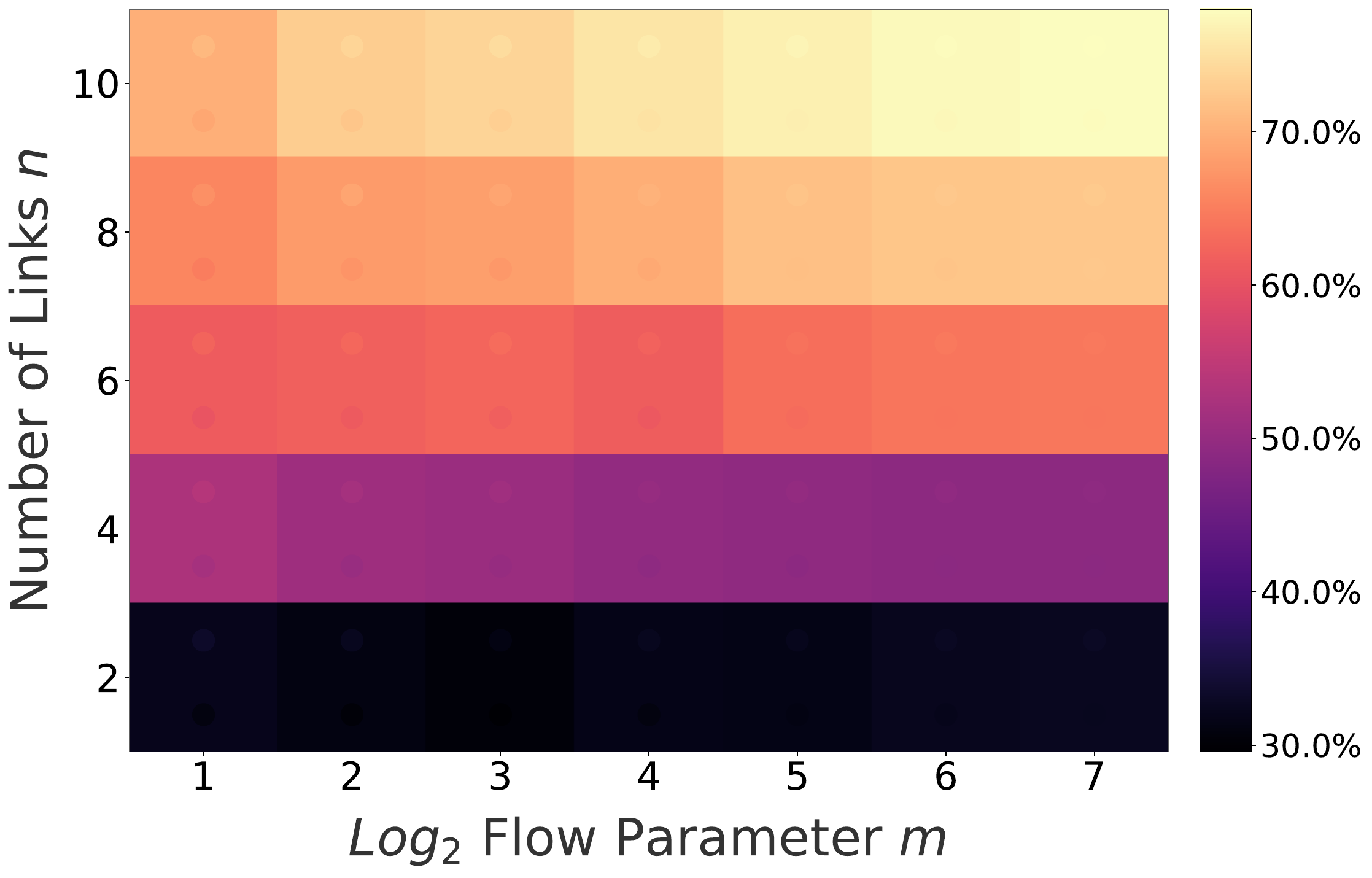}
  \caption{topology 2 configuration 1}
  \label{fig:fd_nr_21}
\end{subfigure}
\begin{subfigure}{0.35\linewidth}
  \centering
  \includegraphics[width=\linewidth]{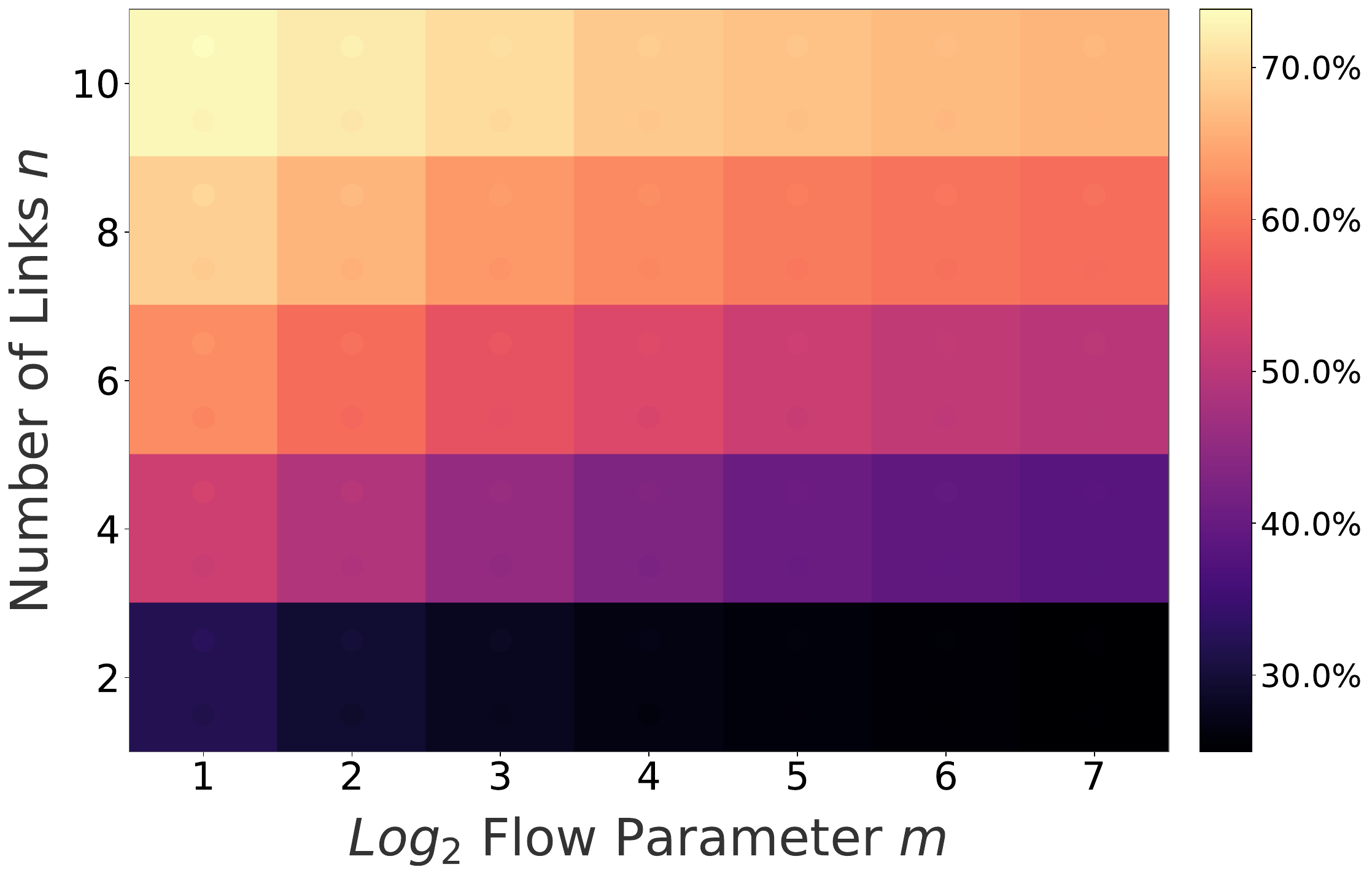}
  \caption{topology 2 configuration 2}
  \label{fig:fd_nr_22}
\end{subfigure}
\caption{Relative Improvement of Greedy over NR (fixed end-to-end deadlines).}
\label{fig:eval_fd_nr}
\end{figure*}

We demonstrate next in \fig{fig:eval_nr} the relative improvement of Greedy over NR.  The benefits of reprofiling are significantly larger than with FR, nearing $90\%$ in some cases.  \add{As alluded to in Section~\ref{sec:discuss}}, this is partly due to reliance on loose delay bounds.  Greedy's bound are also loose, but the reprofiling it preemptively applies to flows mitigates the magnitude of the resulting ``errors.'' In addition, the unequal deadline assignments that Greedy produces, even if they are a consequence of its adjustment phase, reflect the need to accommodate heterogeneity at each hop, which the equal deadline assignment strategy of NR fails to account for. 

We note that the combination of ``maximal'' smoothing and a simple (linear) service (and therefore departure) curve of FR facilitates tighter bounds.  This makes the improvements of Greedy over FR that much more remarkable and highlights the trade-off that exists between making traffic smoother and preserving scheduling flexibility in the network, \ie avoid consuming the entire end-to-end delay budget to make the flow smoother, as FR does.

In summary, the synthetic topologies and flow profiles of \fig{fig:tandem_network} and Table~\ref{tab:profile} are artificial, but they allowed a systematic investigation of the benefits of judicious reprofiling in meeting end-to-end deadlines with less network bandwidth.

\subsection{Overall Buffer Requirements}
\label{app:buffer_tot}

\highlight{\useColor}{We use the buffer bounds of Section~\ref{sec:buffer_bound} to report, for different numbers of flows, the total (across all ingress and network links) buffer sizes to ensure zero packet loss. The figures report (i) ingress buffers, if any, (ii) reprofiling buffers, and (iii) scheduling buffers. Ingress and reprofiling buffers are per flow and used to enforce flows' profiles on both ingress (except under NR) and at each link. Scheduling buffers are per link and shared by all flows on the link. We chose to report total buffer sizes as it offers a compact representation of the respective buffer requirements of each scheme. Finer grain information is available in Appendix~\ref{app:buffer_distribution} that reports distributions across links of the relative differences in reprofiling and scheduling buffer sizes between FR, Greedy and NR.}

\subsubsection{Orion CEV Network}

\begin{figure}[!h]
\centering
\includegraphics[width=0.6\linewidth]{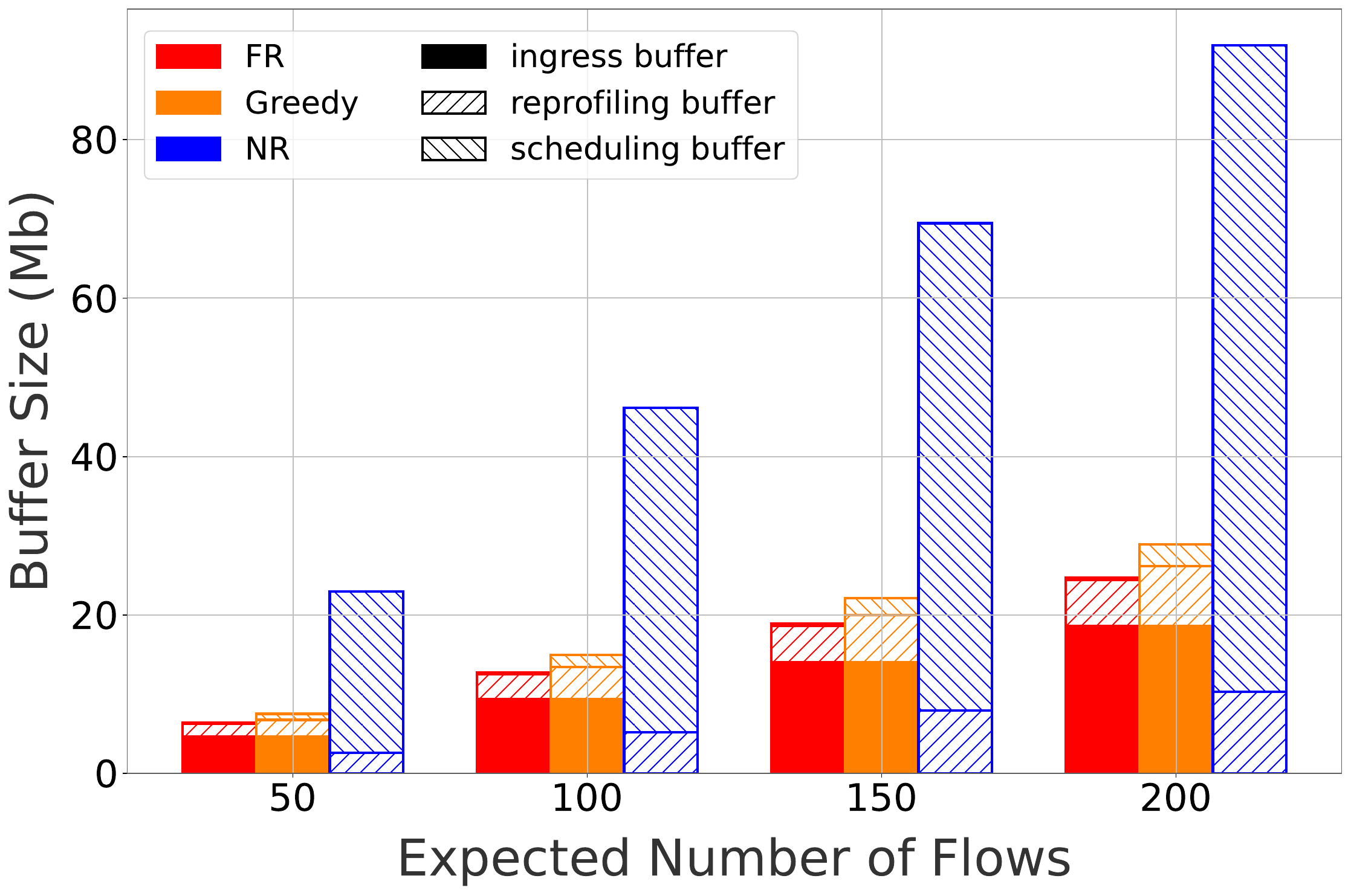}
\caption{Buffer requirements (Mb) of FR, Greedy, and NR on Orion CEV.}
\label{fig:buffer_tsn}
\end{figure}

\highlight{\useColor}{\fig{fig:buffer_tsn} reports total buffer requirements on the Orion CEV network based on averages across $1000$ experiments. The relative requirements of the three schemes are as expected,  with total buffer sizes growing with the number of flows, and with FR and NR having the smallest and largest values respectively, and Greedy close to FR. Both FR and Greedy incur an ingress reprofiling cost that NR avoids.  However, FR and Greedy then benefit from smoother flows and require both smaller reprofiling buffers, as well as much smaller scheduling buffers on every link. Also of note is the fact that Greedy's buffer requirements are close to those of FR, while Greedy affords meaningful reductions in bandwidth as previously shown.}

\subsubsection{US-Topo Network}

\begin{figure}[!h]
\centering
\includegraphics[width=0.6\linewidth]{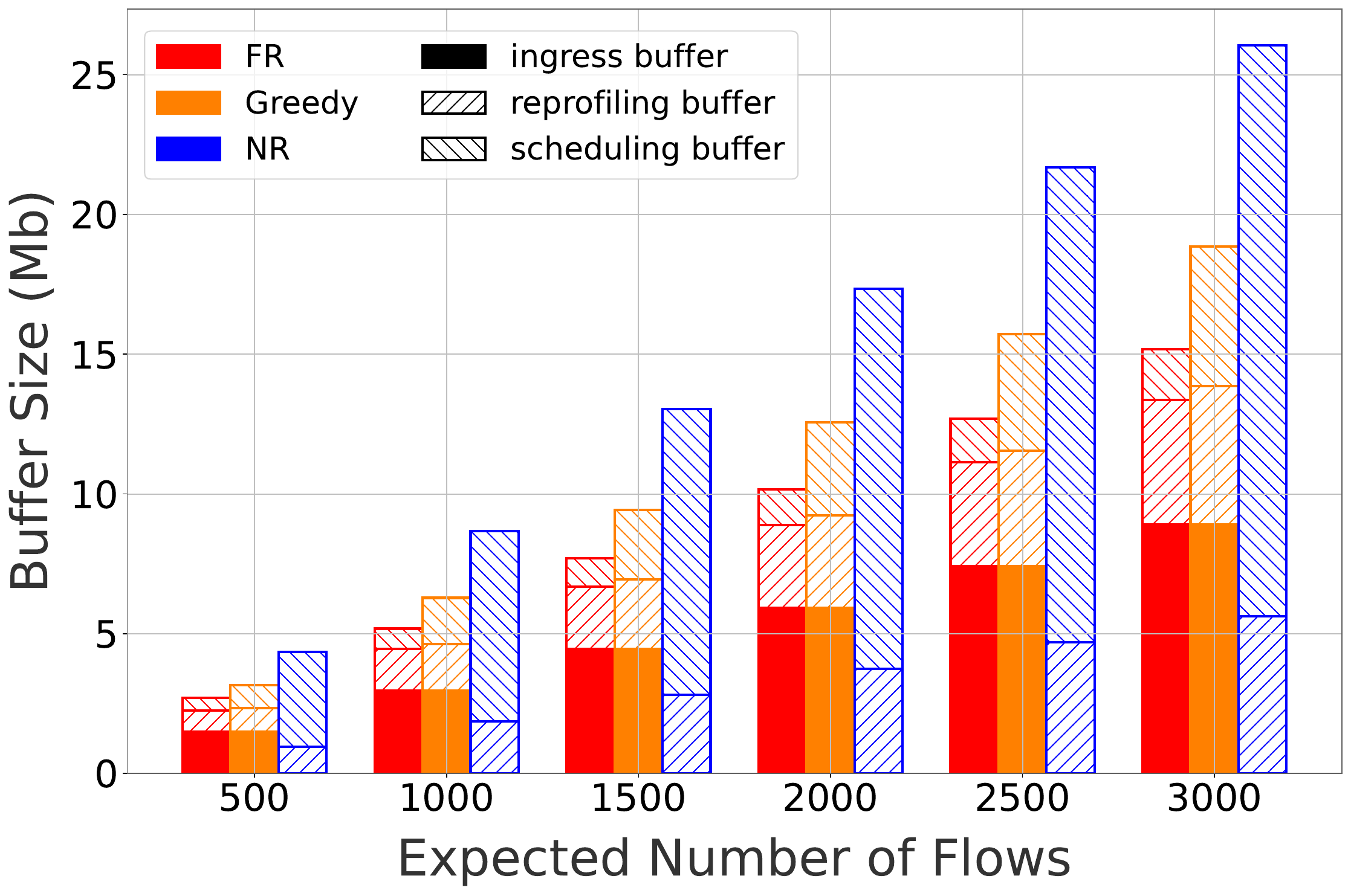}
\caption{Buffer requirements (Mb) of FR, Greedy, and NR on US-Topo.}
\label{fig:buffer}
\end{figure}

\highlight{\useColor}{\fig{fig:buffer} offers a similar perspective as \fig{fig:buffer_tsn} and reports the total required buffer sizes in US-Topo for different number of flows. The results are consistent across the two networks, with \fig{fig:buffer} exhibiting slightly smaller differences across schemes, mainly because of the smaller number of hop that flows traverse in US-Topo.}

\begin{figure*}[!h]
\centering
\begin{subfigure}{0.35\linewidth}
  \centering
  \includegraphics[width=\linewidth]{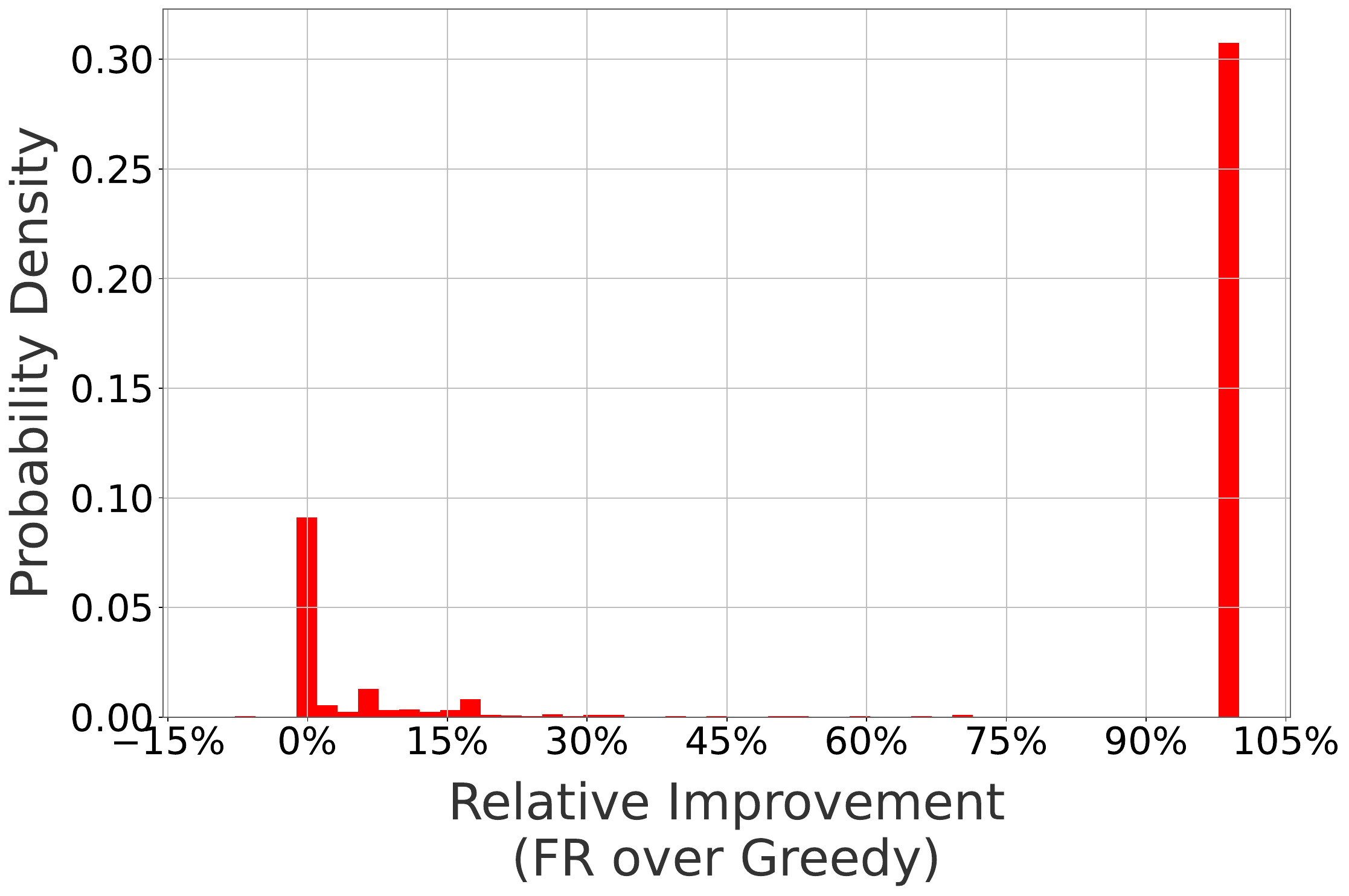}
  \caption{FR over Greedy}
  \label{fig:reprofiling_buffer_tsn_1}
\end{subfigure}
\begin{subfigure}{0.35\linewidth}
  \centering
  \includegraphics[width=\linewidth]{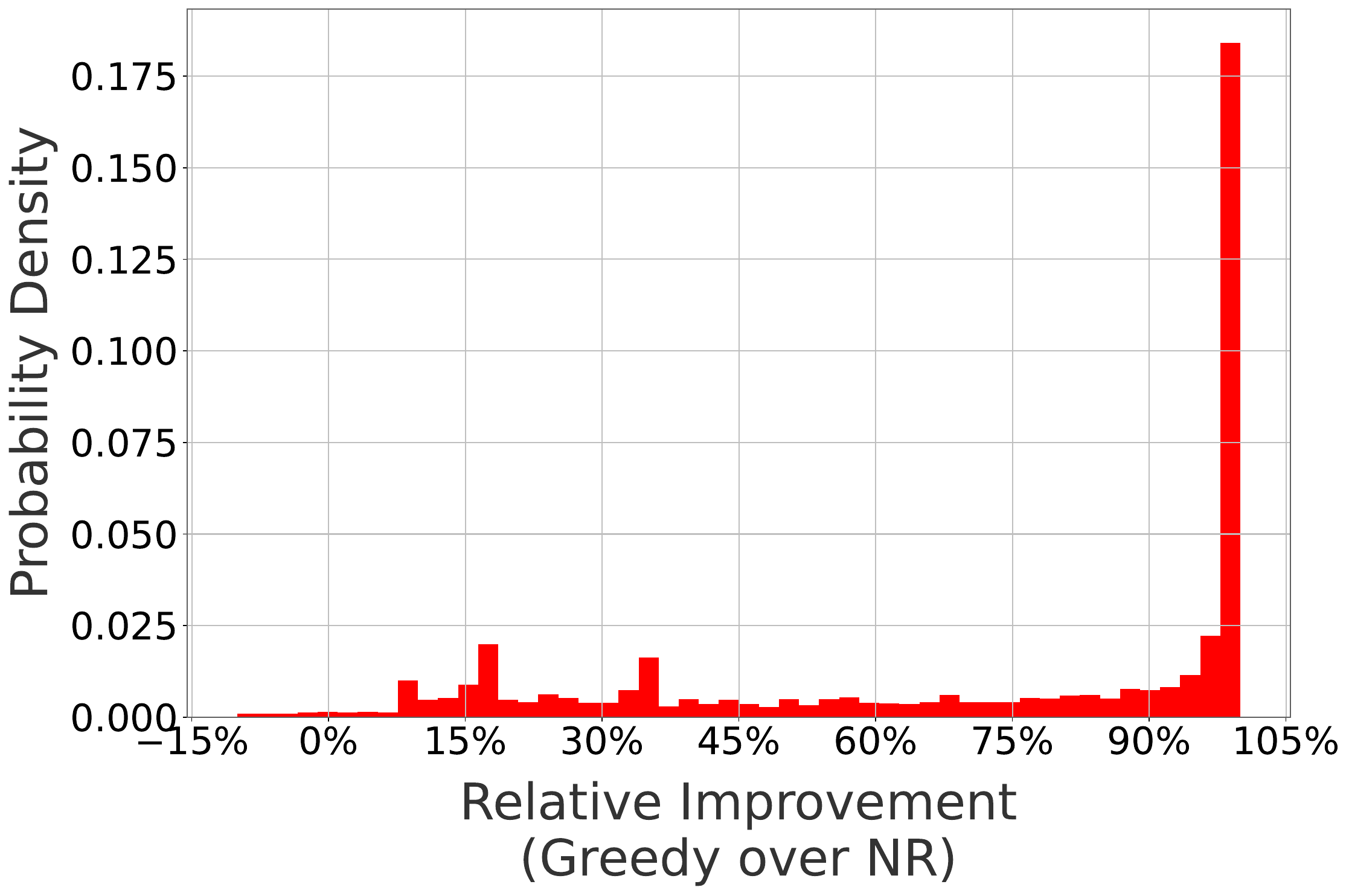}
  \caption{Greedy over NR}
  \label{fig:reprofiling_buffer_tsn_2}
\end{subfigure}
\caption{Distribution of Relative Improvement in Reprofiling Buffer Sizes (Orion CEV).}
\label{fig:reprofiling_buffer_tsn}
\end{figure*}

\begin{figure*}[!h]
\centering
\begin{subfigure}{0.35\linewidth}
  \centering
  \includegraphics[width=\linewidth]{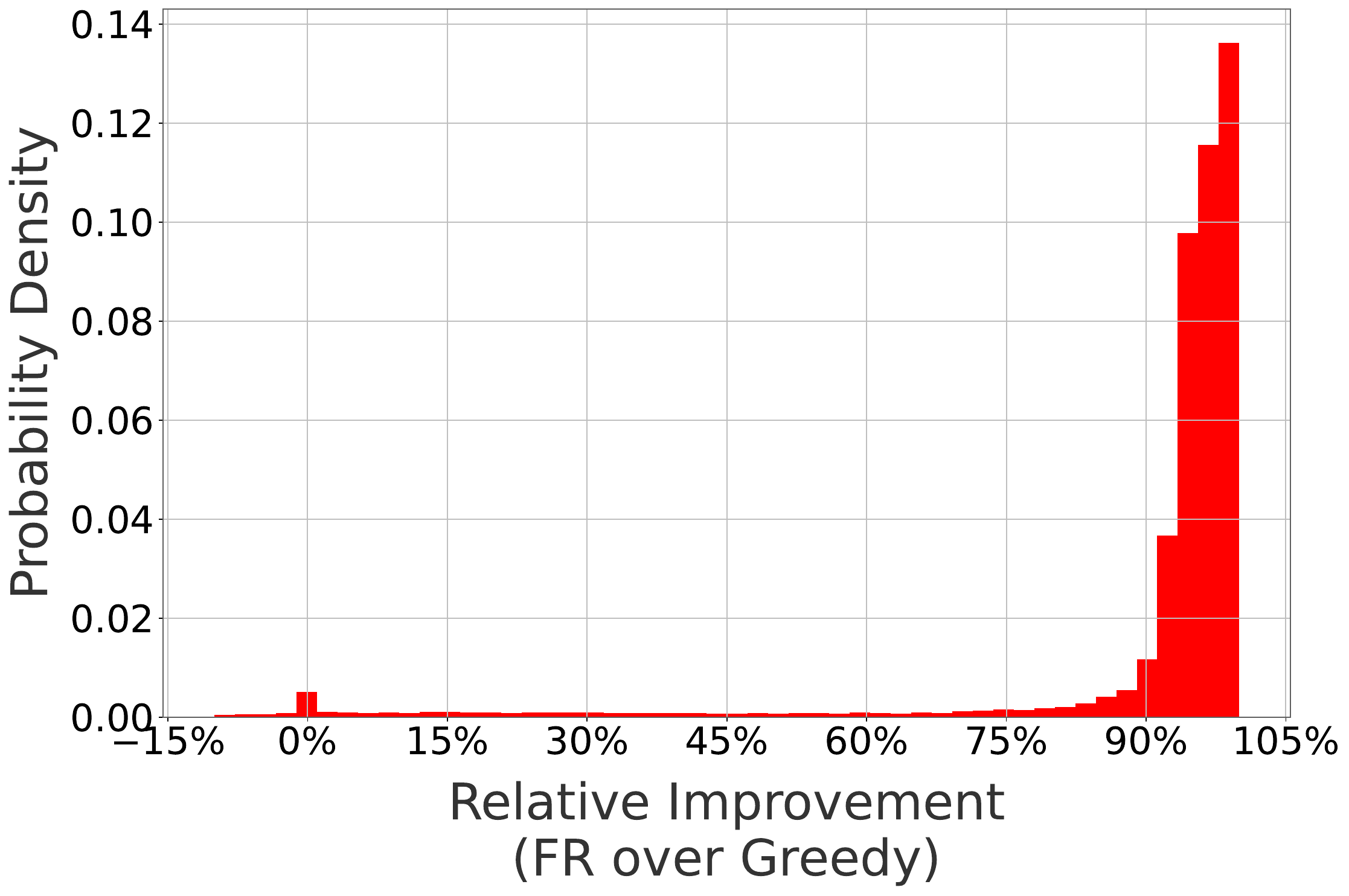}
  \caption{FR over Greedy}
  \label{fig:link_buffer_tsn_1}
\end{subfigure}
\begin{subfigure}{0.35\linewidth}
  \centering
  \includegraphics[width=\linewidth]{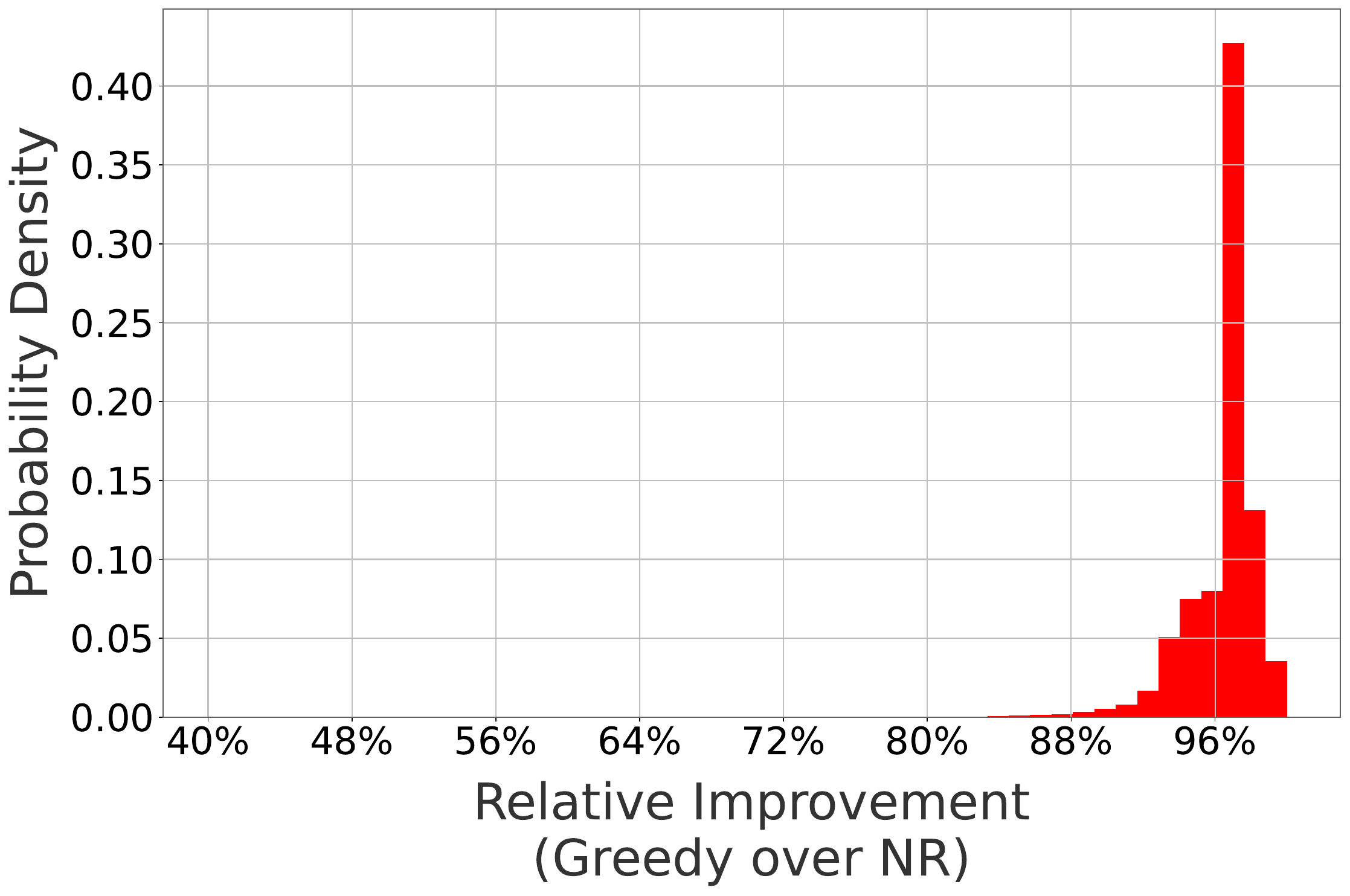}
  \caption{Greedy over NR}
  \label{fig:link_buffer_tsn_2}
\end{subfigure}
\caption{Distribution of Relative Improvement in Scheduling Buffer Sizes (Orion CEV).}
\label{fig:link_buffer_tsn}
\end{figure*}

\subsection{Distribution of Reprofiling and Scheduling Buffers}
\label{app:buffer_distribution}

\highlight{\useColor}{We present next more detailed results on buffer bounds by not only separating reprofiling and scheduling buffers, but more importantly by reporting the distributions of the relative improvements in those buffer sizes for (1) FR over Greedy and (2) Greedy over NR across the same $1000$ experiments.  As before, we first do so for the Orion CEV network before providing similar results for the US-Topo network.  The results are consistent with those for total buffer requirements, with FR (Greedy) improving over Greedy (NR) on most links. }

\highlight{\useColor}{Starting with the Orion CEV network, \figs{fig:reprofiling_buffer_tsn}{fig:link_buffer_tsn} report relative improvements in scheduling and reprofiling buffer sizes respectively, for a $3100$ flows scenario. As alluded to, the results are broadly consistent with \fig{fig:buffer_tsn}, with FR (Greedy) improving over Greedy (NR). }

\highlight{\useColor}{Focusing on reprofiling buffers first, \fig{fig:reprofiling_buffer_tsn_1} indicates that in most cases FR results in improvements of either $0\%$ or $100\%$ over Greedy. The presence of a mode at $0\%$ is intuitive as \fig{fig:real_dist_tsn} shows that Greedy fully reprofiles a fair number of flows.  In those cases it is indistinguishable from FR, at least when it comes to reprofiling buffers.   Conversely, for flows for which Greedy departs from FR and for which FR fully exhausts the flow's delay budget, the resulting lack of local deadlines under FR ensures that bursts never form in the network. Hence, reprofiling and the associated buffering is not required.  This results in an improvement of $100\%$ of FR over Greedy for those flows, and consequently the second mode.  In general, the combination of smaller local deadlines and smaller arrival curves under FR implies that FR's reprofiling buffer requirements never exceed those of Greedy.}

\highlight{\useColor}{This latter observation notwithstanding, \fig{fig:link_buffer_tsn_1} identifies a few instances of ``negative'' improvements of FR relative to Greedy when it comes to scheduling buffers.  In other words, while FR typically requires smaller scheduling buffers than Greedy, the inverse is possible in a few cases. This is in part because the comparison is not entirely an ``apple-to-apple'' comparison, as link bandwidths can differ between the two schemes.  Specifically, as shown in~\fig{fig:improvement_realistic_dist_tsn}, even though Greedy requires less overall bandwidth than FR, this is not necessarily true on all links.  On the few links where Greedy boasts a higher bandwidth than FR, this translates into a faster buffer clearing rate.  Such a faster clearing rate can then in turn imply smaller buffer requirements. }

\begin{figure*}[!h]
\centering
\begin{subfigure}{0.35\linewidth}
  \centering
  \includegraphics[width=\linewidth]{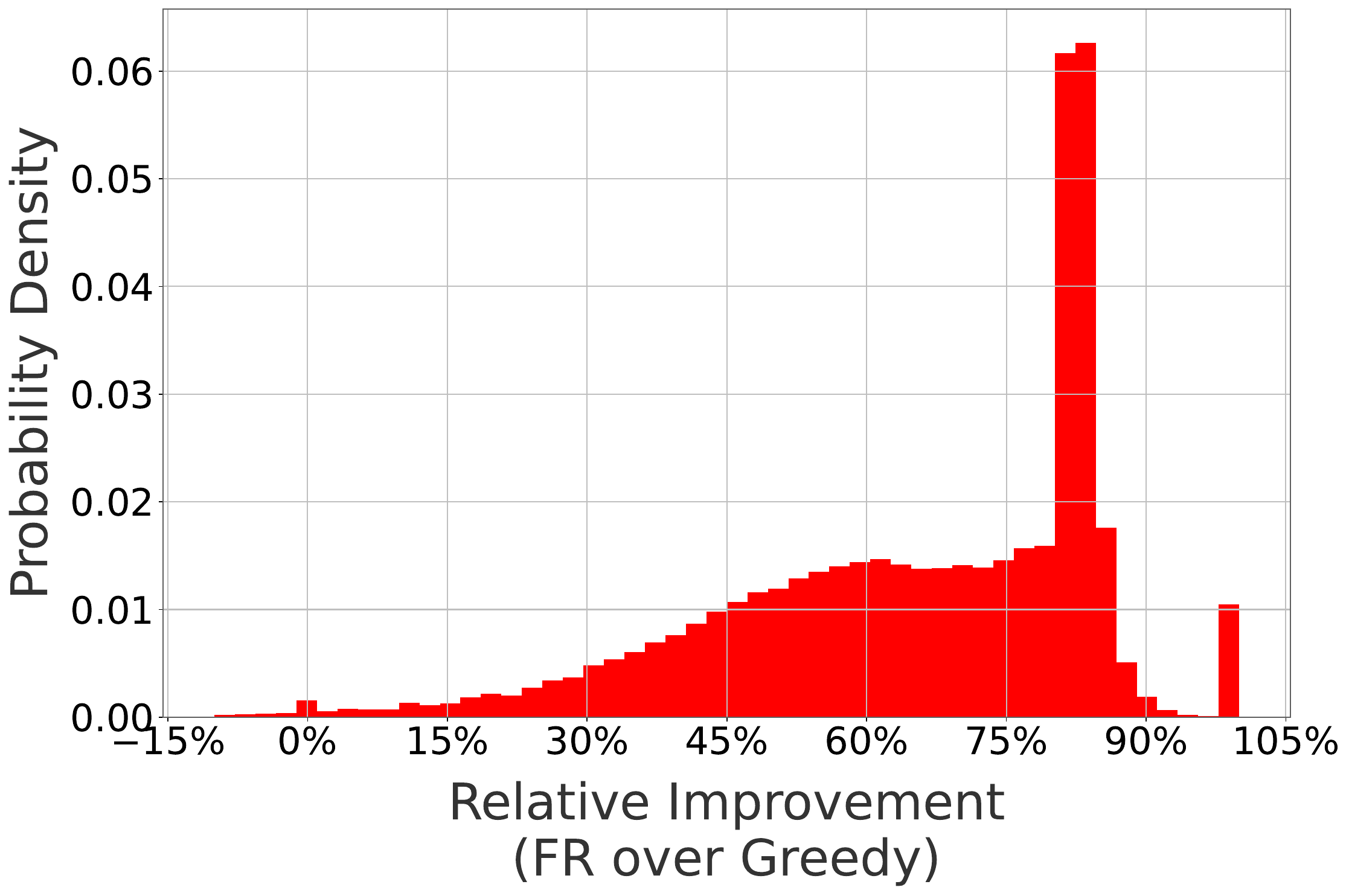}
  \caption{FR over Greedy}
  \label{fig:link_buffer_1}
\end{subfigure}
\begin{subfigure}{0.35\linewidth}
  \centering
  \includegraphics[width=\linewidth]{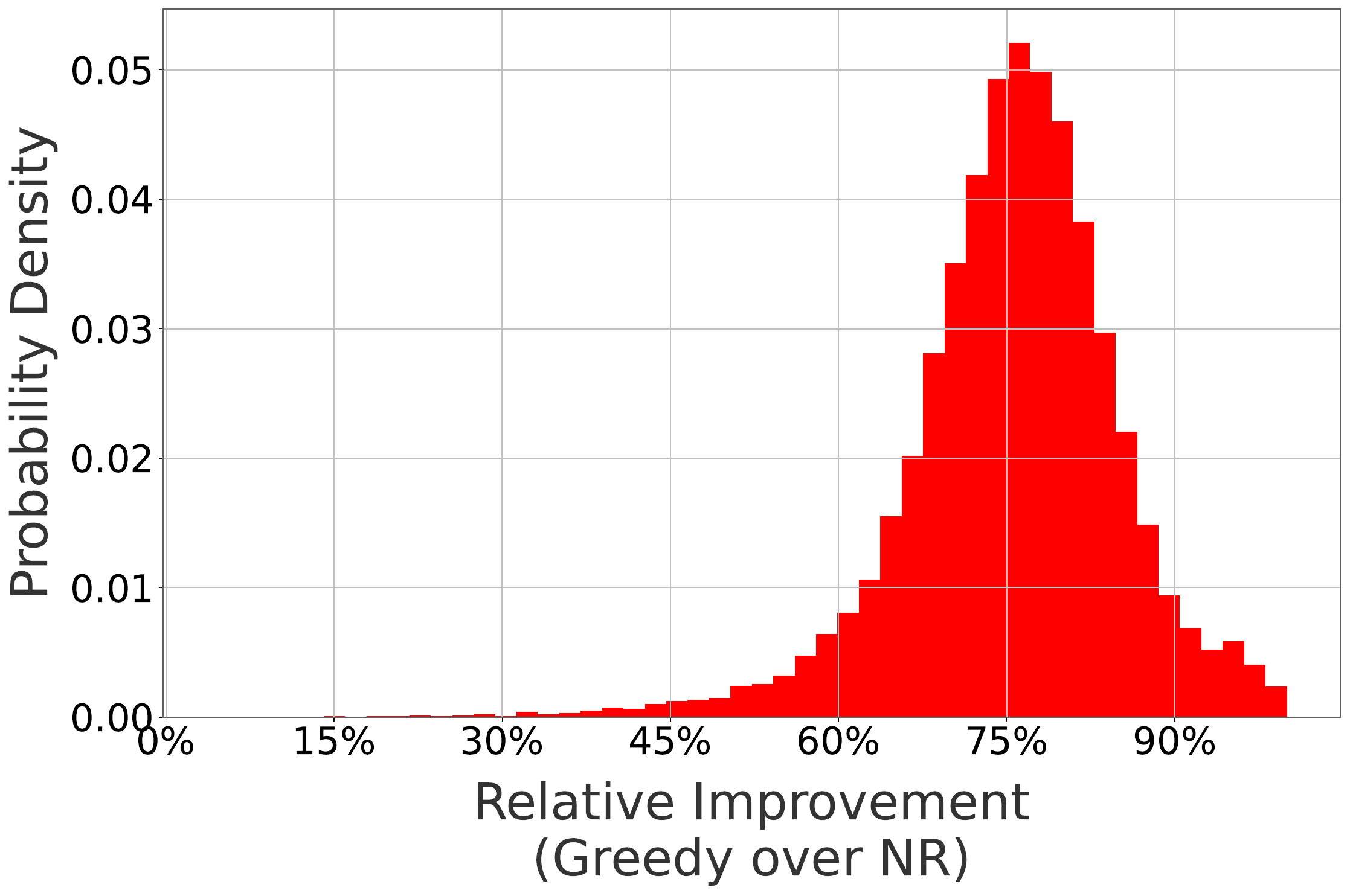}
  \caption{Greedy over NR}
  \label{fig:link_buffer_2}
\end{subfigure}
\caption{Distribution of Relative Improvement on Scheduling Buffer Size (US-Topo).}
\label{fig:link_buffer}
\end{figure*}

\highlight{\useColor}{Similar results are reported for US-Topo in~\figs{fig:link_buffer}{fig:reprofiling_buffer}.
}

\begin{figure*}[!h]
\centering
\begin{subfigure}{0.35\linewidth}
  \centering
  \includegraphics[width=\linewidth]{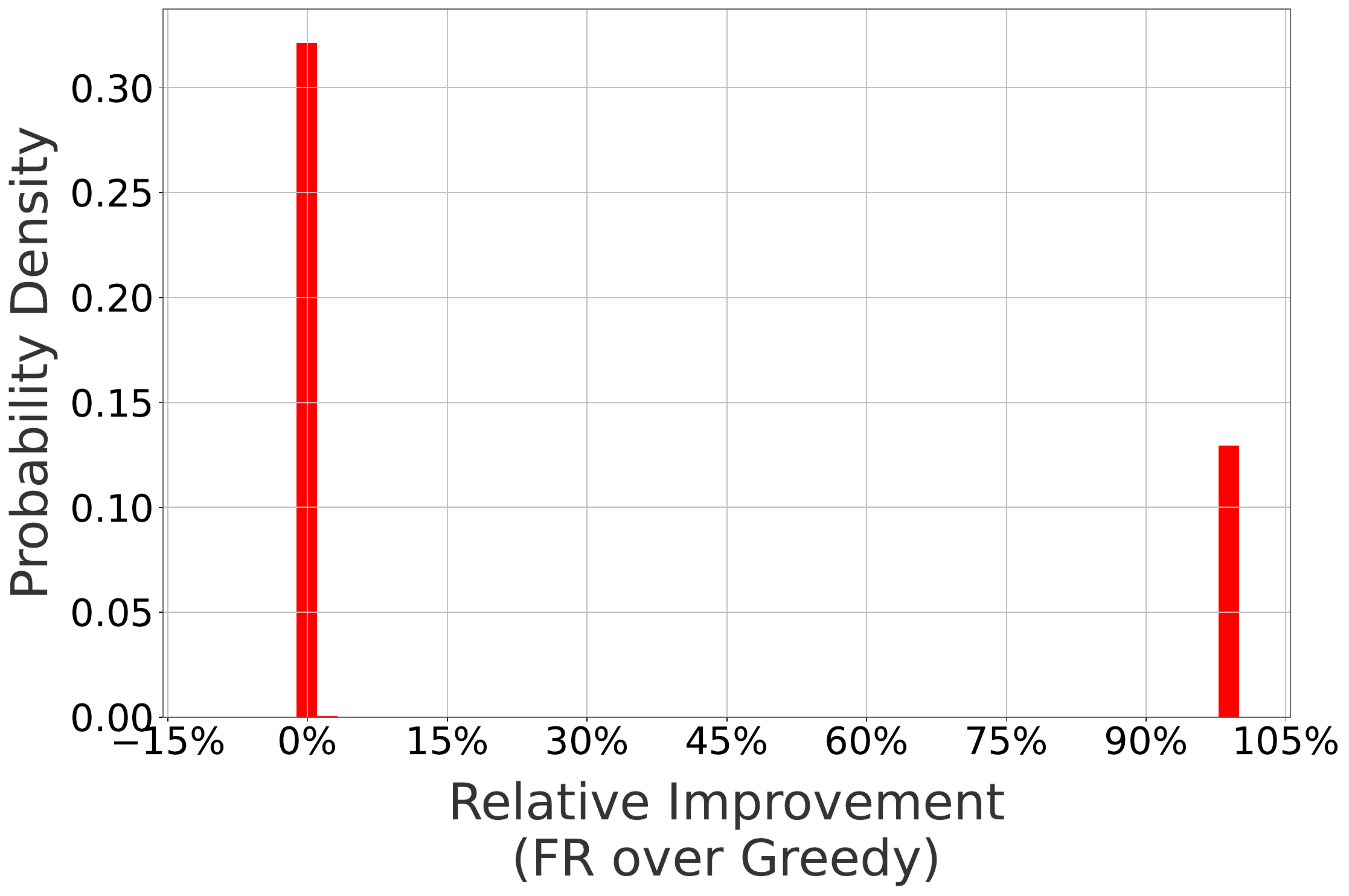}
  \caption{FR over Greedy}
  \label{fig:reprofiling_buffer_1}
\end{subfigure}
\begin{subfigure}{0.35\linewidth}
  \centering
  \includegraphics[width=\linewidth]{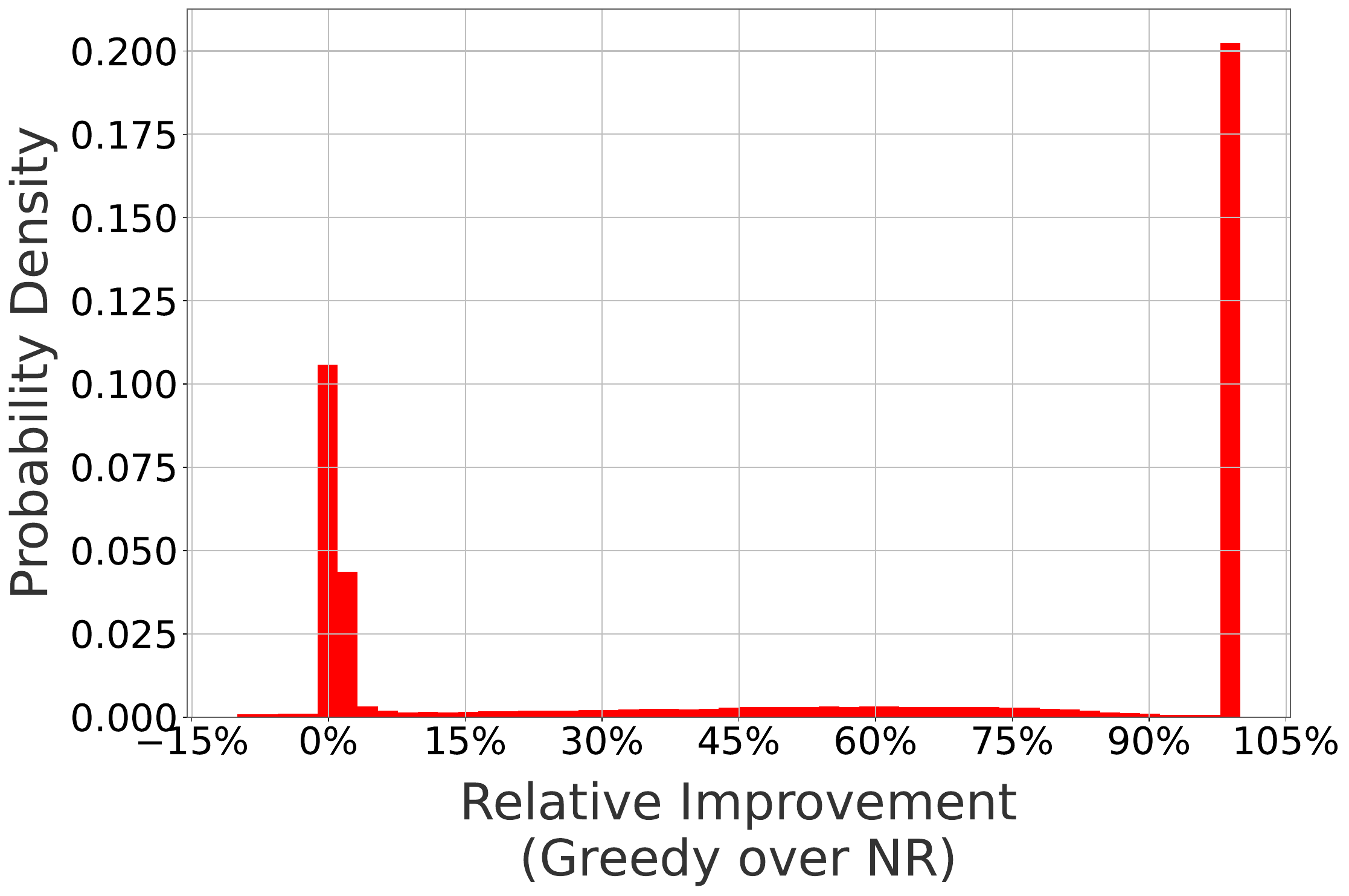}
  \caption{Greedy over NR}
  \label{fig:reprofiling_buffer_2}
\end{subfigure}
\caption{Distribution of Relative Improvement on Reprofiling Buffer Size (US-Topo).}
\label{fig:reprofiling_buffer}
\end{figure*}

\subsection{Linear Approximation of Network Bandwidth}
\label{sec:linear_assumption}
In this Section we show that the relative improvement $\overline{m}$ in the \emph{number of flows} a given network can accommodate by using Greedy can be approximated by $\overline{m} = x/(1-x)$, where $x$ the relative improvement in \emph{bandwidth} that Greedy affords.

When the number of flows a network carries is large enough and scales in equal proportion across classes, the total network bandwidth the network needs to carry a number $K$ of flows can be reasonably approximated by $K\times \overline{bw}$, where $\overline{bw}$ is the average bandwidth a random individual flow requires.  In other words, the network bandwidth is a linear function of the number of flows it carries.

Consider next that the network bandwidth requirements to carry $K$ flows under a baseline solution, \ie either FR or NR, and when using Greedy are $BW$ and $BW_G$, respectively, where $BW_G<BW$. The relative bandwidth improvement that Greedy affords is then $x=\frac{BW-BW_G}{BW}$.  This can be rewritten as $BW_G=(1-x)BW$, which under the assumption that the network bandwidth is a linear function of the number of flows it carries also implies that $\overline{bw}_G=\frac{BW_G}{K}=\frac{(1-x)BW}{K}=(1-x)\overline{bw}$. 

From this result, the number of flows $K_G$ that a network with bandwidth $BW$ but operated using Greedy would carry is then given by $K_G=\frac{BW}{\overline{bw}_G}=\frac{BW}{(1-x)\overline{bw}}=\frac{K}{1-x}$.  This in turn implies that the relative improvement in the number of flows the network (with bandwidth $BW$) carries when introducing Greedy is given by $m=\frac{K_G-K}{K}=\frac{x}{1-x}$.

\end{document}